\documentclass{article}

\onecolumn 

\usepackage[margin=2cm]{geometry}
\usepackage{multirow}
\usepackage{algpseudocode}

\usepackage{babel}
\usepackage{amsfonts}
\usepackage{amssymb,amsmath}
\usepackage[T1]{fontenc}  
\usepackage{hhline}
\usepackage{lmodern}
\usepackage{mathabx}
\usepackage{mathrsfs}
\usepackage{calrsfs}
\usepackage{hyperref}
\usepackage{tcolorbox}
\usepackage{calc}
\usepackage{booktabs}
\usepackage{dsfont}
\usepackage{color}
\usepackage{amsthm}
\usepackage{blkarray}
\usepackage{mathtools}
\usepackage{hyperref}
\usepackage{xcolor}
\usepackage[round]{natbib}
\numberwithin{equation}{section}
\usepackage{graphicx}

\usepackage{comment}
\usepackage{algorithm}
\usepackage{graphicx}
\usepackage{bm}
\usepackage[normalem]{ulem}

\newcommand{\mbf}{\mathbf}

\newcommand{\mb}{\mathbb}

\newcommand{\s}{\geq}
\newcommand{\m}{\leq}

\newcommand{\1}{\mathds{1}}

\newcommand{\ie}{\emph{i.e.}, }

\newcommand{\as}[1]{ {\color{violet} {\footnotesize A:}~#1 } }

\definecolor{blue}{rgb}{0,0,0}

\begin{document}

\title{Multi-site modelling and reconstruction of past extreme skew surges along the French Atlantic coast}

\author{
  Nathan Huet\textsuperscript{1},
  Philippe Naveau\textsuperscript{2},
  Anne Sabourin\textsuperscript{3} \vspace{0.5cm} \\
  \textsuperscript{1}{\small
  DAIS, LTCI, Università Ca’ Foscari Venezia, Télécom Paris, Venezia, 30170, Italia} \\
  \textsuperscript{2}{\small LSCE, ESTIMR, CEA-CNRS-UVSQ, IPSL \& Université Paris-Saclay, Saint-Aubin, 91190, France}\\
  \textsuperscript{3}{\small Université Paris Cité, Université Paris Saclay, ENS Paris Saclay, CNRS, SSA,} \\{\small INSERM \& Centre Borelli, Paris, F-75006, France}}

\maketitle

\abstract{Appropriate modelling of extreme skew surges is crucial, particularly for coastal risk management. Our study focuses on modelling extreme skew surges along the French Atlantic coast, with a particular emphasis on investigating the extremal dependence structure between stations. We employ the peak-over-threshold framework, where a multivariate extreme event is defined whenever at least one location records a large value, though not necessarily all stations simultaneously. A novel method for determining an appropriate level (threshold) above which observations can be classified as extreme is proposed. Two complementary approaches are explored. First, the multivariate generalized Pareto distribution is employed to model extremes, leveraging its properties to derive a generative model that predicts extreme skew surges at one station based on observed extremes at nearby stations. Second, a novel extreme regression framework is assessed for point predictions. This specific regression framework enables accurate point predictions using only the ‘angle’ of input variables, i.e., input variables divided by their norms. The ultimate objective is to reconstruct historical skew surge time series at stations with limited data. This is achieved by integrating extreme skew surge data from stations with longer records, such as Brest and Saint-Nazaire, which provide over 150 years of observations.}


\maketitle

\section{Introduction}

The rise in sea levels driven by global warming  \citep{jevrejeva2016coastal} increases the need to model and understand extreme sea level events, which can lead to devastating coastal flooding, such as the 1953 North Sea flood \citep{mcrobie2005big}. 
Assessing the intensity and frequency of these events is crucial for developing effective coastal risk management strategies aimed at reducing significant human and material losses
\citep{genovese2013storm, chadenas2014impact,karamouz2019building}. Classical studies on extreme sea levels often focus on past storm-induced floods and their impacts \citep{pineau2012sea,wadey2015comparison}, the influence of global warming and anthropogenic factors \citep{boumis2023coevolution,pan2020impact}, forecasting and projections of extreme sea levels \citep{vousdoukas2018global,bellinghausen2024using}, and long-term historical data analysis to identify   trends and estimate return periods of extreme events \citep{weisse2014changing,tebaldi2021extreme}. We refer the reader to  \citet{bernier2024storm} for a recent comprehensive review on the topic of extreme sea levels; see also \citet{D'Arcy23}.  

Aligned with the hydrological literature on extreme sea levels,  this paper focuses on extreme surges\footnote{A similar study on extreme sea levels is deferred to the Appendix.}. 
Sea levels can be decomposed into a deterministic tidal component and a stochastic non-tidal component called surge \citep{dixon1994extreme, idier2012tide}. 
Alternatively, sea levels can be deduced from skew surges by convolution with predicted tides, namely by using the joint probability method \citep{pugh1978extreme, pugh1980applications} or its refinement \citep{tawn1989extreme,tawn1992estimating}. 

\begin{figure}
    \centering
    \includegraphics[scale=0.5]{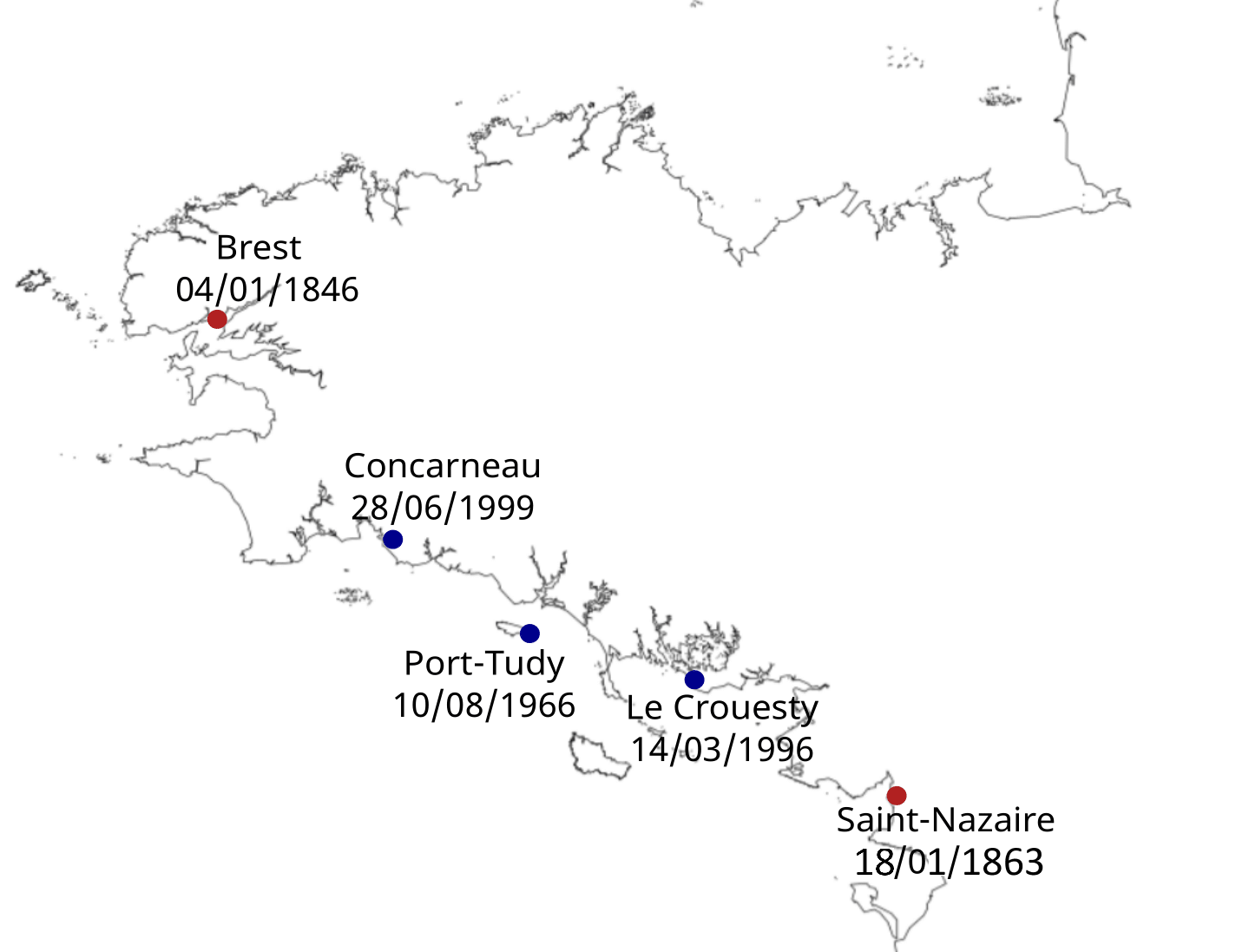}
    \caption{
    Five French   tide gauge locations along the French  Atlantic coast. 
    Brest and Saint-Nazaire (red dots) have long sea level measurements, while Concarneau, Port Tudy and Le Crouesty (blue dots) have much shorter  recordings. 
    The objective of this work is to reconstruct  missing surges in the past  given extreme values from stations with long historical records (red dots).
    \label{fig:carte_bretagne}}
\end{figure}

In this work, our main environmental objective is to propose statistical methods to complete  (reconstruct) coastal gauges recordings from other nearby stations that possess a much longer historical depth.  To illustrate our approach, we will focus on tide gauges located along the French Atlantic coast.   
For example, the Brest station in Figure \ref{fig:carte_bretagne} spans a period going back to 1846, while the nearby  station of Concarneau starts its recording in 1999. 
Although many of the cited references \cite[e.g.,][]{harter2024underestimation} consider the influence of covariates on extremes at a given location, the literature is scarce regarding the task of reconstructing    
extreme events and their intensity  
within a multivariate extreme value modelling; see, e.g., \citet{Beck20}.
One key element of the present work is  to precisely model    the statistical dependence among recorded extreme surges.
To leverage multivariate extreme value theory in our task of completing extreme surges   in short  time series from 
long historical records at     nearby stations, 
different regression models will be  trained and compared on the overlapping record periods during which both older and newer tide gauges have been operational. 
The input stations are located at Brest and Saint-Nazaire, while the target stations for prediction are located at Port Tudy, Concarneau, and Le Crouesty. 
%
The reconstruction of extreme sea level time series (or equivalently extreme surge time series) is an active area of research aimed at producing increasingly precise inferences at specific sites and plays an important role in understanding ocean changes; see, e.g., \citet{harter2024underestimation,Wang24}. Most studies in this domain, including those focused on forecasting (i.e., predicting future values),  use meteorological data, such as atmospheric pressure and wind, as predictors within various models, including multiple linear regression \citep{cid2018storm,ji2020historical}, random forests \citep{tadesse2021database} and deep learning \citep{bruneau2020estimation,tadesse2020data}. These approaches often disregard potential spatial statistical dependence among locations. Univariate models are fitted at each site.
 Regional Frequency Analysis approaches \citep{hosking1997regional,bardet2011regional, bernardara2011application, collings2024global}  do leverage available information at nearby locations, however this information concerns the parameters of the aforementioned univariate distributions.  
 To the best of our knowledge, the work which stands closest to ours (in the sense that spatial statistical dependence is used for reconstruction of extreme skew surges) is 
 \citet{calafat2020probabilistic} who employ Bayesian hierarchical models to capture the dependence structure between extreme surges measured across Europe, for the purpose of climate reanalysis. Unlike their block-maxima approach, our work belongs within a peaks-over-threshold framework. In addition to proposing a different fully parametric model, we also introduce a nonparametric machine-learning-based method to reduce reliance on strict model assumptions.



    The main methodological purpose of this work is to promote the use of two new approaches derived from the theory of multivariate extreme values for the analysis of extreme sea levels in geoscience and climate research; see, e.g., \citet{NaveauSegers25} for a recent review on multivariate extremes. 
    On the one hand, we propose a machine learning approach designed to minimize a pointwise prediction performance criterion. 
    On the other hand, we implement a parametric modelling approach that  enables comprehensive (conditional) distributional modelling of the missing component, providing prediction confidence intervals.  We carefully highlight the theoretical connections between the two approaches, emphasizing that, despite their methodological differences, the statistical assumptions regarding the distributional tail of the data are essentially equivalent.
To the best of our knowledge, such   comparisons between machine learning and parametric approaches  are   rare in the applied statistical literature dedicated to multivariate extreme analysis.  
In particular, sections~\ref{sec:background} and~\ref{sec:methods} explain  the fundamental relationships between these two approaches that may seem incompatible at first sight.  
Another contribution lies in proposing  a simple  and original  method for determining an appropriate level (threshold) above which observations can be classified as extreme.  

The  nonparametric machine learning method envisioned here is closely related to an active area of research focused on developing a sound  statistical learning and machine learning framework for multivariate extreme value settings, relying on minimal model assumptions.
Examples include clustering methods \citep{goix2016sparse,goix2017sparse,chiapino2019identifying,simpson2020determining,meyer2021sparse,janssen+w:2020,vignotto2021clustering,meyer2023multivariate}, principal component analysis \citep{cooley2019decompositions,drees2021principal,clemenccon2024regular}, graphical models \citep{engelke2020graphical}, classification \citep{jalalzai2018binary}. 
These machine learning algorithms dedicated to extreme value analysis are not black boxes; the understanding of statistical learning mechanisms and error margins is rapidly advancing, as seen in the recent review of \cite{clemenccon2025weak}. 
In this work, we employ the least-squares regression procedure proposed in \citet{huet2023regression}, designed for extreme-value predictive problems where extremality is assessed relative to covariates. In our case, the covariates are  surges recorded at neighbouring stations. The regression algorithm generates a predictive function that enables the estimation of  surges at the target station, given the covariate, particularly when the norm of the covariate is large. 

The parametric method implemented here leverages recent advances in parametric multivariate extreme value modelling. Specifically, we use the flexible multivariate generalized Pareto model developed in \citet{kiriliouk2019peaks,rootzen2018multivariate,ROOTZEN2018117}, building on earlier work from \citet{rootzen2006multivariate}. In applications, this framework has proved relevant for different purposes, such as the joint simulation of extreme waves \citep{legrand2023joint}, anomaly detection of extreme precipitation \citep{nezaki2023new}, and risk modelling associated with wildfires \citep{cisneros2024spatial}.
In this framework, an observation is considered extreme if at least one of its input components exceeds a critical threshold (see Section~\ref{sec:background}). This definition is well-suited to the present use case, as a single large skew surge can lead to flooding at the target station, regardless of conditions at neighbouring stations. With the addition of parametric assumptions, model fitting enables comprehensive modelling of the full conditional distribution of the skew surges at the target station, given large values at one of the ‘covariate' stations.  

Note that, in this work,    the terms \emph{parametric} and \emph{nonparametric}  
only characterize the multivariate  {dependence structure}, but not the marginal behaviours. 
At the marginal level, extreme value theory 
provides a strong justification for parametric models for exceedances. 
In contrast, asymptotic mathematical arguments point towards a non-parametric 
dependence structure for multivariate extremes; see, e.g., \citet{NaveauSegers25}. Still,  
parametric dependence family choices are  often  made for convenience and interpretation purposes. In addition,  
they facilitate the construction of confidence intervals.

This paper is structured as follows. In Section~\ref{sec:data}, we present the skew surge data. Section~\ref{sec:background} introduces the theoretical framework necessary to understand the proposed methods. In Section~\ref{sec:methods}, we outline the marginal preprocessing and threshold selection, followed by two prediction schemes, which are described using two algorithms. The results of applying these procedures to the skew surge datasets are presented in Section~\ref{sec:results}. Additional details are deferred to the Appendix, in particular, additional analyses conducted on sea levels and two other output stations are provided in Section~\ref{sec:add_stud}.

\section{Skew surge time series}\label{sec:data}





The dataset considered in our work is produced by the REFMAR (Réseaux de REFérence des observations MARégraphiques), a sea level observation network managed by the SHOM (Service Hydrographique et Océano\-graphique de la Marine).
Some tide gauges in the observation network have recorded over a century of hourly measurements, while others have only been operational for about twenty years. 
Figure~\ref{fig:carte_bretagne} shows the locations of the five tide gauges involved in our study, along with the dates of their first recordings. 
To avoid modelling tide-surge interactions in this dataset, which is beyond the scope of this study, we only focus on skew surges which are defined as the difference between the maximum observed sea level and the astronomically predicted tide. Skew surges have been shown to be independent of high tides \citep{williams2016tide}. This independence has been empirically validated for most tide gauges along the French coast, including the five stations considered in this study \citep{kergadallan2014improving, refmar2022}.

As mentioned in the introduction, the analysis focuses on the maximum skew surges observed during tides along the French Atlantic coast. The dataset comprises hourly validated records, each associated with a timestamp, from 11/08/1966 to 31/12/2023. The input measurements are provided by the long-range tide gauges located at Brest and Saint-Nazaire, while the output stations with limited historical records are located at Port Tudy, Concarneau, and Le Crouesty (see Figure~\ref{fig:carte_bretagne}). Three separate studies are conducted, each aimed at predicting values at one output station. The analysis with output skew surges measured at Port Tudy is presented in the main text, while the analyses for sea levels and the other two stations are deferred to Section~\ref{sec:add_stud} of the Appendix, as similar conclusions apply across all three stations.

In the sequel, the notation $Y$ stands for the response (target) variable,  while  $\bm{X}$  is the covariate vector. Here, $\bm{X} = (X_B,X_N)^\top$ represents the bivariate vector comprising skew surges at the Brest and Saint-Nazaire stations, while $Y$ represents skew surges at an output station.

To evaluate the performance of our methods, each dataset is divided into two subsets. The marginal and joint dependence structures are learnt on the training set, comprising the most recent observations, while the model performances are evaluated on the test set, consisting of the earliest observations. The training and test sets are divided on December 31, 1999, for Port Tudy, in 2010 for Concarneau, and in 2014 for Le Crouesty, ensuring similar sizes across the sets.

{\color{blue} 
To reduce computational costs and because our focus is on extremes, we discard measurements $(X_B, X_N, Y)$ where both $X_B$ and $X_N$ fall below their empirical medians. This is consistent with our definition of an extreme in this paper: an observation is considered extreme if at least one of the values from the two input stations exceeds a large threshold (see Section~\ref{sec:marg_model}), since the marginal observations at the output station are unknown in practice and since our goal is to reconstruct its extreme time series prior to systematic measurements. This pre-processing step also aims to improve the marginal EGP fit, which benefits from a more homogeneous dataset. Additionally, for convenience, we shift the origin so that the component-wise minimum value of the selected measurements is zero (see Section~\ref{sec:preprocess} in the Appendix for details). The data obtained after this initial thresholding and shifting are considered our “working dataset" (denoted as $\mathcal{D}_n$ in Section~\ref{sec:methods}). These operations on the training set yield the observations displayed in Figure~\ref{fig:scatterplot}.
}

{\color{blue} Section~\ref{sec:stationary} of the Appendix provides evidence that the time series become stationary once a linear temporal trend is removed. Therefore, this trend is systematically removed when fitting the model but added back for the visualization of the results.}
\begin{figure}[ht!]
\centering
    \includegraphics[width=.29\textwidth]{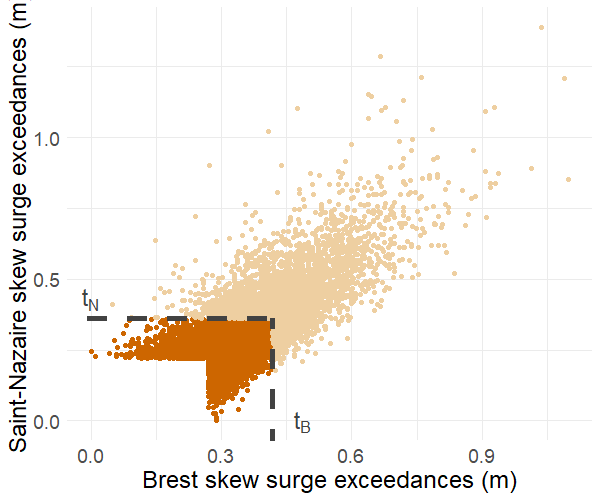}
  \hspace{0.7cm}
  \includegraphics[width=.29\textwidth]{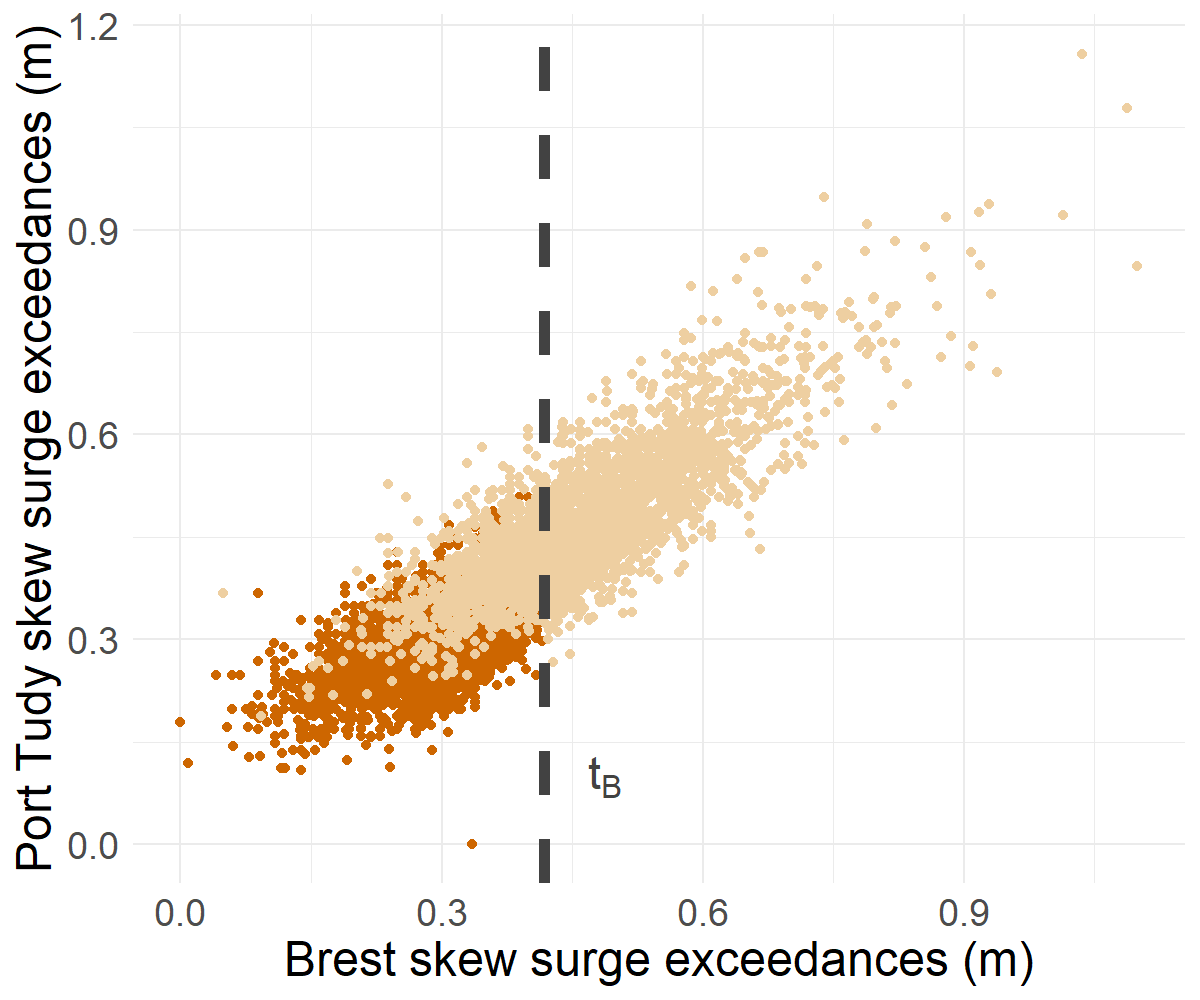}
  \hspace{0.7cm}
  \includegraphics[width=.29\textwidth]{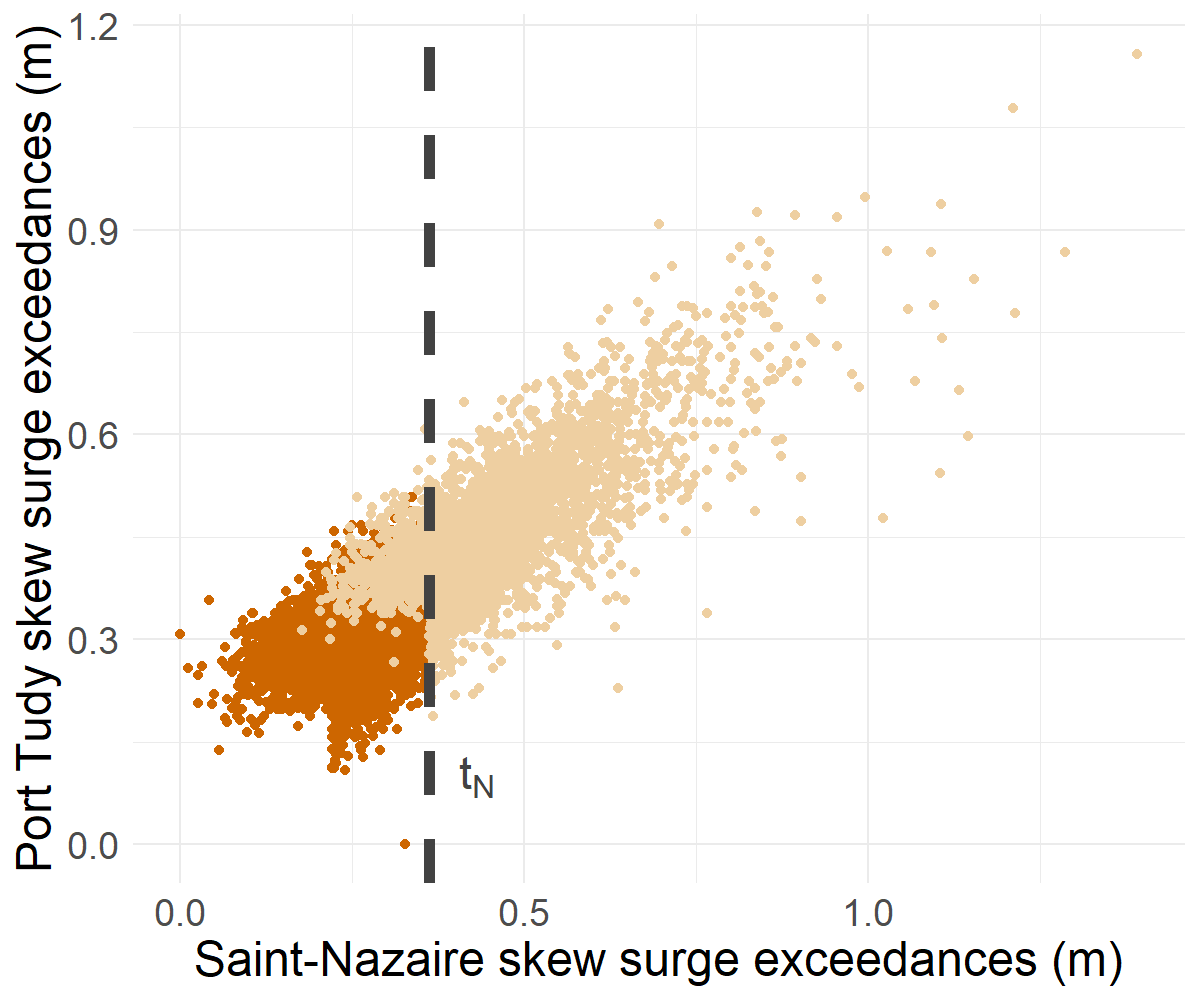}
  \caption{Pairwise skew surge exceedances for all pairs of stations from 01/12/2000 to 31/12/2023. Dark orange (resp. light orange) points represent the observations below (resp. above) the threshold specified in Table~\ref{tab:egpd_tudy}. The grey dotted lines represent the chosen marginal thresholds via Algorithm~\ref{algo:th_select} (see Section~\ref{sec:res_marg}).
  \label{fig:scatterplot}} 
\end{figure}

\section{Marginal and multivariate modelling of extremes: Background}\label{sec:background}

Multivariate extreme value models are asymptotically justified under mild regularity assumptions regarding their  tail behaviour. 
These assumptions imply that  after some  marginal standardization, the joint distribution enjoys a homogeneity property in the tails, making extrapolation possible. 
This section provides some minimal background in univariate and  multivariate  Extreme Value Theory (EVT) underlying the  methods developed in the following sections.
Here and throughout,  multivariate objects are denoted in bold font to avoid confusion, e.g., $\bm{z}=(z_1,...,z_{d+1})^\top \in \mb{R}^{d+1}$. To simplify notations, for $\bm{x} \in \mb{R}^d$ and $y\in \mb{R}$, we will write expressions like $(\bm{x}^\top,y)^\top \in \mb{R}^{d+1}$ more concisely as $(\bm{x},y)\in \mb{R}^{d+1}$.  Let $\bm{Z} = (\bm{X},Y)\in\mb R^{d+1}$ represent the considered random pair (covariate, target), with $\bm X \in \mb{R}^d$ and $Y \in \mb{R}$ and let $F_j$, for $1 \m j\le d+1$, denote the $j^{th}$ marginal distribution of the random vector $\bm{Z}$, $F_j(z)=  \mb{P}(Z_j\le z)$.  
 Univariate operations applied to multivariate objects should be understood componentwise, i.e., for $f_\theta:\mb{R} \rightarrow \mb{R}, \bm{\theta} \in \mb{R}^d$ and $\bm{x} \in \mb{R}^d$, $f_{\bm{\theta}}(\bm{x}) = (f_{\theta_1}(x_1),...,f_{\theta_d}(x_d))$.  The order relations $\le,<$ in $\mb R^d$ should be understood as the partial order resulting from componentwise comparisons, i.e., for $\bm x, \bm t \in \mb R^{d}$, say, where $\bm t$ typically represents a multivariate high threshold for the covariate $\bm X$,  the notation $\bm x \le \bm t$ means that for \emph{all} $1 \m j\le d$, $x_j\le t_{j}$, while $\bm x \not\le \bm t$ means that for \emph{some} $1 \m j\le d$, $x_j>t_{j}$. 
We denote by $\mathcal{L}(T)$ the distribution of a random element $T$ and $\mathcal{L}(T_n)\to\mathcal{L}(T)$ means that the random variables or vectors $T_n$ converge in distribution to $T$. 

\subsection{Marginal models for extremes}\label{sec:marg_model}
A key result in dimension one \citep{balkema1974residual} is that under mild assumptions, the distribution of excesses above high thresholds may be legitimately modelled by a Generalized Pareto (GP) distribution  $H_{\sigma,\xi}(z)=1 -  (1+\xi z/\sigma)_+^{-1/\xi}$, i.e., an asymptotically valid approximation of $(1-F_j(z+t_j))/(1 - F_j(t_j))$ is $1-H_{\sigma_j,\xi_j}(z) $ where $t_j$ is a large threshold and $ \xi_j,\sigma_j$ are respectively a shape and a scale parameter for the $j^{th}$ margin.
However, modelling excesses above high thresholds by a GP distribution  has some limitations. First, a pervasive challenge in extreme value analysis is selecting an appropriate threshold above which observations are considered extreme, so that  the GP distribution is in effect an appropriate model for excesses. Second, certain extreme-value frameworks necessitate modelling the entire distribution, not just the tail above a threshold. 
Indeed, as mentioned in Section~\ref{sec:data}, an observation is considered extreme if a large value is recorded at least at one of the input stations. Therefore, marginal observations of multivariate records considered as extreme  may include non-extreme components (e.g., a storm affecting only Saint-Nazaire but not Brest). These residual non-extreme observations fall into the left tail of the marginal distribution at each station (see Figure~\ref{fig:density_tudy}). 

In this paper,  in order to avoid imposing (or having to make a choice for) a threshold $t_j$, we consider a flexible parametric family, the Extended Generalized Pareto (EGP) distribution  \citep{papastathopoulos2013extended,naveau2016modeling},  which permits to model the entire positive part of the marginal distribution, while the conditional distribution of excesses above high threshold is asymptotically equivalent to a GP distribution. This modelling strategy has been  applied for modelling extreme precipitation \citep{tencaliec2020flexible,rivoire2021comparison,nanditha2024strong}, large wave heights \citep{legrand2023joint} or strong wind speed \citep{turkman2021calibration}. 
Specifically, we use the simplest EGP model (EGPD3 in \citet{papastathopoulos2013extended}), with distribution function 
\begin{equation}\label{eq:egpd_cdf}
    F_{\sigma,\xi,\kappa}(z) = \bigg(1-\Big(1+\frac{\xi z}{\sigma}\Big)^{-1/\xi}\bigg)^\kappa,
\end{equation}
where $\sigma>0$ is a  scale parameter, $\xi \in \mb{R}$ is a  shape parameter and $\kappa >0$ is a  lower tail shape parameter. Indeed  $F_{\sigma,\xi,\kappa}(z) \approx Cst \times z^\kappa$, as $z \rightarrow 0$. 
The associated density function $f_{\sigma,\xi,\kappa}$ is 
\begin{equation}\label{eq:egpd_dens}
    f_{\sigma,\xi,\kappa}(z) = \frac{\kappa}{\sigma}\Big(1+\frac{\xi z}{\sigma}\Big)^{-1/\xi-1}\bigg(1-\Big(1+\frac{\xi z}{\sigma}\Big)^{-1/\xi}\bigg)^{\kappa-1}.
\end{equation}
The EGP distribution has support $[0,+\infty[$ if $\xi \s 0$ and $[0, -\sigma/\xi]$ if $\xi < 0$.


\subsection{Multivariate modelling of extremes}\label{sec:background_reg}

As a  multivariate extension of the univariate  framework introduced above, an essential result is that   for a large class of  distributions, including, e.g., multivariate Gaussian, Cauchy or Student multivariate families, the joint distribution of properly rescaled excesses above  high thresholds  converges to a Multivariate Generalized Pareto (MGP) distribution \citep{rootzen2006multivariate,ROOTZEN2018117,rootzen2018multivariate,kiriliouk2019peaks}. In practice, this implies that for a sufficiently high multivariate threshold
$\bm{t} = (t_1,\ldots, t_{d+1})$,  
\begin{equation}
  \label{eq:mgpdLimit}
  \mathcal{L}( (\bm{Z} -\bm{t} )\vee \bm{\eta} \; | \; \bm{Z}\nleq \bm{t}) \approx 
  \mathcal{L}(\bm{Z}_\infty),
\end{equation}
where  $\bm{Z}_\infty$ is  a MGP random vector and $\bm{\eta}$ is the vector of   lower end-points of marginal distributions of $\bm{Z}_\infty$. 
The latter marginal distributions  are, by construction, univariate GP distributions above zero, namely 
\begin{equation*}\label{eq:pos_part}
    \mb{P}(Z_{\infty,j} \s z \mid Z_{\infty,j}> 0)= 1-H_{\sigma_j,\xi_j}(z) := \Big(1+\frac{\xi_j}{\sigma_j}z\Big)_+^{-1/\xi_j}.
  \end{equation*}
For modelling purposes a customary approach is to consider marginally standardized versions of $\bm{Z}$. This allows for convenient representations of the limit distribution amenable both to parametric modelling and statistical learning algorithms. Standard statistical practice in multivariate EVT involves modelling separately the tails of the marginal distributions $F_j$ and the extremal dependence structure encapsulated by the limiting distribution of marginally standardized excesses conditioned on an excess  above high (multivariate) thresholds. 
Of course, different marginalisation choices lead to different limit models and statistical methods, which are asymptotically equivalent after back-transformation to the original scale. The two methods implemented and compared in this work (resp. a nonparametric regression algorithm and a parametric model) are respectively formulated with unit Pareto margins, and unit Exponential margins, introduced next.

\subsubsection{Pareto margins modelling}
Consider the componentwise transformation
\begin{equation}\label{eq:paretoTransform}
  p(\bm{z}) = \left( \frac{1}{1-F_1(z_1)}, \ldots,  
    \frac{1}{1-F_{d+1}(z_{d+1})}\right)^\top
\end{equation}
and let $p(\bm{Z})$ denote the marginally transformed random observation. Because the marginal distributions are unknown, a preliminary step is to fit marginal models to each margin $Z_j$, as outlined in the previous paragraph. 
{\color{blue} This transformation is motivated by the fact that for any univariate random variable $Z$ with continuous cdf $F_Z$, the transformed random variable $1/(1-F_Z(Z))$ is a unit Pareto random variable. A key property of a unit Pareto $P$ random variable is that
\begin{equation*}
    \mathcal{L}(P/t  \,| \, P>t) \rightarrow \mathcal{L}(P),
\end{equation*}
as $t \rightarrow+\infty$ (in fact, the equality $\mathcal{L}(P/t  \,| \, P>t) = \mathcal{L}(P)$ holds exactly, but what matters here is that the rescaled law of a Pareto random variable conditioned on exceeding a threshold converges to a Pareto distribution as the threshold increases).
Hence, in the multivariate case, after the transformation \eqref{eq:paretoTransform}, all margins of $p(\bm{Z})$ are unit Pareto variables, and, analogously to the univariate case, the rescaled law of $p(\bm{Z})$, conditional on its norm exceeding a threshold, converges to a non-degenerate distribution as the threshold tends to infinity.}
Then, choosing any norm $\|\cdot\|$ on $\mb{R}^{d+1}$, and switching to pseudo-polar coordinates by letting $\bm \Theta  = \|p(\bm{Z})\|^{-1}p(\bm{Z})$, \eqref{eq:mgpdLimit}
implies in particular that
\begin{equation}
  \label{eq:cv-paretoScale}
  \mathcal{L} \big( (\|p(\bm{Z})\|/t, \bm{\Theta} ) | \,  \|p(\bm{Z})\|>t\big) \to \mathcal{L}
  (R_\infty, \bm{\Theta}_\infty) , 
\end{equation}
as $t\rightarrow +\infty$, where
$R_\infty $  
is a radial component, $\bm{\Theta}_{\infty} $ 
is a pseudo-angle, and $R_\infty$ and $\bm{\Theta_{\infty}}$ are independent. An equivalent writing of~\eqref{eq:cv-paretoScale} is that
\begin{equation}\label{eq:cv_cartesian_paretoscale}\mathcal{L} ( t^{-1} p(\bm{Z}) \, | \, \| p(\bm{Z}) \|>t) \to \mathcal{L}(\tilde{\bm{Z}}_\infty),\end{equation}
as $t\rightarrow +\infty$, where $\tilde{\bm{Z}}_\infty$ has a multiplicative structure $\tilde{\bm{Z}}_\infty = R_\infty.\bm{\Theta}_\infty$ with independent factors $R_\infty,\bm{\Theta}_\infty$ as in~\eqref{eq:cv-paretoScale}.
  In practice, a wealth of statistical analyses of extreme environmental events have been successfully conducted based on that assumption \citep{davison2015statistics}. 
  The reader may refer to \citet{resnick2008extreme} for a comprehensive presentation of  multivariate EVT and its connections with Condition~\eqref{eq:cv-paretoScale}. 

\subsubsection{Exponential margins in the multivariate generalized Pareto model}
As an alternative to the Pareto scaling depicted in the previous paragraph, recent advances in statistical modelling of multivariate extremes  \citep{rootzen2006multivariate,ROOTZEN2018117,rootzen2018multivariate,kiriliouk2019peaks} have revealed convenient limiting structures on the exponential scale, in the MGP distribution introduced at the beginning of this section. Observe first that if $p(Z_j)$ follows a unit Pareto distribution then $\log(p(Z_j))$ follows a unit exponential distribution. {\color{blue} Then, to work with exponential scale margins, we need to apply the marginal transformation $\log \circ\, p$.} More precisely, for $\bm{z}\in\mb{R}^{d+1}$, let
 \begin{equation}\label{eq:expo-transform-function}
 e(\bm{z}) =  \big(-\log(1-F_1(z_1)), \ldots, -\log(1-F_{d+1}(z_{d+1})\big)^\top = \log\circ \,p\,(\bm{z})
 \end{equation}
  denote a componentwise transformation mapping a random vector with margins $F_j$ to a vector with exponential margins, and 
 define accordingly a marginally transformed vector 
 \begin{equation*}
 e(\bm{Z})  
 =  \big(\log(p(Z_1)), \ldots, \log(p( Z_{d+1}))\big)^\top.   
 \end{equation*}
{\color{blue} By construction, the logarithm transformation turns multiplicative structures into additive ones. Hence, on the exponential scale, the convergence of the law towards an extreme random variable occurs not by dividing by the threshold, as on the Pareto scale, but by subtracting the threshold from the random variable.}
Then under the same assumptions as the ones underlying the multivariate tail model for $p(\bm{Z})$ with Pareto margins, namely~\eqref{eq:mgpdLimit},   considering a multivariate threshold $\bm{t}'$ tending to infinity,
\begin{equation}\label{eq:CV_excess_expo_margins}
     \mathcal{L}\big(e(\bm{Z}) -\bm{t}'  \mid e(\bm{Z})\not\le\bm{t}'\big) \rightarrow \mathcal{L}(\bm{Z}_\infty'),  \end{equation}
where  the  margins of $\Tilde{\bm{Z}}_\infty$, when conditioned on being positive, follow a unit exponential distribution,
\begin{equation*}
    \mb{P}(Z'_{\infty,j} \s z \mid Z'_{\infty,j} > 0) = \exp(-z), \qquad 1 \m j \m d+1. 
\end{equation*} 
The random vector $\bm{Z}'_\infty$ is called a  \emph{standard} MGP distribution \citep{rootzen2018multivariate,wan2024characterizing}.  
As hinted above, the multiplicative structure of $\Tilde{\bm{Z}}_\infty$ in~\eqref{eq:cv_cartesian_paretoscale} translates into an additional structure for $\bm{Z}'_\infty$ in~\eqref{eq:CV_excess_expo_margins},
\begin{equation}\label{eq:mgpd_decomp}
    \bm{Z}'_\infty = E + \bm{T} - \max \bm{T},
\end{equation}
where $E \in [0,+\infty[$ is a unit exponential random variable and $\bm{T}$ is  a random vector valued in $\mb{R}^{d+1}$, independent of $E$. 
Conversely, any random vector $\bm T$ in ${\mb R}^{d+1}$ gives rise to a  standard MGP vector $\bm{Z}'_\infty$ through~\eqref{eq:mgpd_decomp}. The additive structure~\eqref{eq:mgpd_decomp} 
permits to construct flexible models for $\bm{Z}_\infty'$ (after transformation to exponential margins) amenable to likelihood inference, based on any parametric model for $\bm T$ (see Appendix~\ref{app:density_model} for more details).

\section{Prediction of a missing component:  Methodology}\label{sec:methods}
This section presents a detailed description of the methodologies developed and implemented in this study to predict the target $Y$ based on observations of a covariate $\bm X$ exceeding a sufficiently large multivariate threshold $\bm t_X$, using training data where both the covariate and the target are recorded simultaneously. {\color{blue} With this definition of extremality, note that the extremality of an observation $(\bm X,Y)$ is measured solely with respect to the input variable $\bm{X}$. Consequently, the target variable $Y$ that we aim to predict is not necessarily extreme.}
We denote the training dataset by $\mathcal{D}_n=\{\bm{Z}_i\}_{1 \m i \m n}=\{(\bm{X}_i, Y_i)\}_{1 \m i \m n}$. Recall from the Introduction that $\mathcal{D}_n$ results from a pre-processing step fully described in Appendix~\ref{sec:preprocess}. 
Following the framework outlined in Section~\ref{sec:background}, Sections~\ref{sec:th_selec}, \ref{sec:reg_proc}, and~\ref{sec:mgp_proc} respectively detail our methodologies for univariate analysis, point prediction of a missing component using a Machine Learning algorithm, and distribution regression through parametric modelling and likelihood-based inference of the conditional distribution of a missing component.


\subsection{Univariate study and threshold selection}\label{sec:th_selec}
As outlined in Section~\ref{sec:marg_model},  we   model the marginal distributions by fitting an   EGP model~\eqref{eq:egpd_cdf} to each coordinate of the full training dataset $\mathcal{D}_n$. 
As illustrated in Section~\ref{sec:results} this specific marginal model  fits the considered data set reasonably well (see Figure~\ref{fig:density_tudy}). Note that alternative variants of the chosen EGP submodel, 
such as the EGP family introduced in \citet{naveau2016modeling}, could have been considered. Another possible approach  would be to use the traditional method of modelling with a GP distribution above a preselected threshold and the empirical cdf below  threshold  as discussed  in, e.g., \citet{heffernan2004conditional}. Since marginal modelling is not the central focus  of our work (our primary emphasis being on multivariate modelling), we do not provide a  detailed comparison  of all potential  univariate modelling approaches. 

The necessary step of univariate modelling, classically aimed at performing marginal transformations in~\eqref{eq:paretoTransform} and  \eqref{eq:expo-transform-function},  also enables us to select a multivariate threshold
 $\bm t$ in \eqref{eq:mgpdLimit}  which, in the multivariate analysis, allows for the identification of extreme training data. To the best of our knowledge, the method we propose here and below for threshold selection is novel. 
As noted in \citet{naveau2016modeling} (see also Section~\ref{sec:marg_model}), the EGP distribution behaves similarly to a GP distribution in the right tail. Since the GP density is strictly convex for $\xi > -1/2$ (a condition always met  here, see Table~\ref{tab:egpd_tudy}), the EGP density $f_{\sigma,\xi,\kappa}$ \eqref{eq:egpd_dens} is also strictly convex for sufficiently large values and closely resembles a GP distribution on this domain.  This suggests to choose the threshold as the lowest value above which the fitted EGP density is convex. Numerically, this is the largest zero of the second derivative $\frac{d^2 f_{\sigma, \xi,\kappa}(z)}{d z^2}$. 
Straightforward computations reported in Appendix~\ref{sec:appexdix_th} yield an explicit expression for the latter value which is given in Algorithm~\ref{algo:th_select}. The advantage of this automatic selection procedure is that it eliminates reliance on visual inspection of stability plots as discussed in \citet{kiriliouk2019peaks,legrand2023joint,huet2023regression}, which can sometimes be challenging to interpret, thereby reducing a layer of arbitrariness. The relevance of this approach is illustrated in Section~\ref{sec:results}, namely Figure \ref{fig:tudy_egpdvsgpd} shows that the EGP and GP distributions are indeed nearly indistinguishable above the chosen threshold. {\color{blue} Note that, our choice of multivariate threshold only relies on marginal properties, to provide univariate thresholds such that the marginal distribution above it are deemed into the GP model. Hence, multivariate threshold selection procedures, such as in \citet{kiriliouk2019peaks}, taking into account the multivariate extreme dependence structure could be more appropriate to assess that, not only the marginal distributions above this threshold are deemed into the GP model, but also that the multivariate distribution above it is deemed into the MGP model.}



\begin{algorithm}[t!]
   \caption{Marginal modelling and threshold selection.\label{algo:th_select}}
   
\begin{algorithmic}[1]
  \State {\bfseries INPUT:} Training dataset $\{X_1,\; \ldots,\;X_n\}$ with $X_i \in \mb{R}$ for all $1\m i \m n$.
    \State {\bfseries Marginal Fitting:} Fit an EGP distribution to the dataset to obtain a triplet of estimated parameters $(\sigma,\xi,\kappa) \in [0,+\infty[ \times \mb{R} \times [0,+\infty[$.
    \State {\bfseries Threshold computation:} Compute the threshold $t$ according to the EGP estimated terms
    \begin{equation*}
        \! t=\frac{\sigma}{\xi}\bigg( \Big(\frac{4\xi^2+3(\kappa\xi+\kappa+\xi) -1-\sqrt{(4\xi^2+3(\kappa\xi+\kappa+\xi) -1)^2-4(\kappa^2+2\xi^2+3\kappa \xi)(2\xi^2+3\xi+1)}}{2(\kappa^2+2\xi^2+3\kappa\xi)}\Big)^{-\xi}-1\bigg).    \end{equation*}
\State {\bfseries OUTPUT:} estimated marginal EGP parameters $(\sigma,\xi,\kappa)$ and a threshold $t$.
\end{algorithmic}
\end{algorithm}

In the sequel, we denote by  $\bm{\sigma},\bm{\xi},\bm{\kappa} = (\bm{\sigma}_X,\sigma_Y),(\bm{\xi}_X,\xi_Y),(\bm{\kappa}_X,\kappa_Y) \in \mb{R}^{d+1}$, with $\bm{\sigma}_X,\bm{\xi}_X,\bm{\kappa}_X = (\sigma_1,...,\sigma_d)^\top, \\(\xi_1,...,\xi_d)^\top,(\kappa_1,...,\kappa_d)^\top\in \mb{R}^d$  the EGP parameters, and $\bm{t} = (\bm{t}_X,t_Y) \in \mb{R}^{d+1}$, with $\bm{t}_X = (t_1,...,t_d)^\top\in \mb{R}^d$, the multivariate extreme threshold,  resulting from Algorithm~\ref{algo:th_select} applied independently to each marginal dataset $\{Z_{i,j}\}_{1 \m i \m n}$, for $1 \le j\le d+1$.
{\color{blue} It is important to distinguish this threshold from the one introduced in Section~\ref{sec:data} (the median), which serves exclusively to filter out observations that are clearly non-extreme; the remaining observations, however, are not guaranteed to be multivariate extremes.} The latter threshold $\bm t$ allows to define a subset of extreme observations $\mathcal{D}_{ext}$ for subsequent multivariate analysis, in terms of an excess of $\bm t$ in at least one component of the covariable. {\color{blue} We recall that, according to this definition of extremality, the univariate target variables $Y_i$ in $\mathcal{D}_{ext}$ are not necessarily extreme.} We let $k$ denote the number of such extreme training observations and without loss of generality (up to reordering the indices), to alleviate notations, we assume that these extreme observations form the first $k$ entries of the training set, namely 
\begin{equation*}
    \mathcal{D}_{ext} = \{\bm{Z}_i = (\bm X_i, Y_i), \text{ such that } \bm{X}_i \nleqslant \bm{t}_X \}_{1 \m i \m n} = \{\bm{Z}_i\}_{1\le i\le k}.
\end{equation*}

\subsection{Point prediction based on Machine Learning}\label{sec:reg_proc}
The marginal modelling step from Section~\ref{sec:marg_model} is crucially exploited both for estimating the marginal distributions $F_j$ involved in~\eqref{eq:paretoTransform} and for threshold selection in~\eqref{eq:mgpdLimit} via Algorithm~\ref{algo:th_select}.
In the sequel for clarity we denote by $p_{\sigma,\xi,\kappa}$ the marginal transformation based on  EGP-modelled marginals $F_{\sigma,\xi,\kappa}$,

\begin{equation}\label{eq:egpd_pareto_transf}    p_{\sigma,\xi,\kappa}(x) = \frac{1}{1-F_{\sigma,\xi,\kappa}(x)}, \end{equation}
and we use boldface notation for  the multivariate mapping
$p_{\bm \sigma,\bm \xi,\bm \kappa}(\bm{z}) = (p_{\bm{\sigma}_X,\bm{\xi}_X,\bm{\kappa}_X}((z_1,...,z_d)), p_{\sigma_{Y},\xi_{Y},\kappa_{Y}}(z_{d+1}))$ so that  $ p_{\bm \sigma,\bm \xi,\bm \kappa}(\bm Z)$ is the Pareto-transformed variable, where $\bm{\sigma},\bm{\xi},\bm{\kappa} = (\bm{\sigma}_X,\sigma_Y),(\bm{\xi}_X,\xi_Y),(\bm{\kappa}_X,\kappa_Y)$ are the  parameters estimated  with Algorithm~\ref{algo:th_select}. For conciseness, and when clear from the context, we denote $p_{\bm{\sigma}_X,\bm{\xi}_X,\bm{\kappa}_X}$, $p_{\sigma_Y,\xi_Y,\kappa_Y}$ and $p_{\bm{\sigma},\bm{\xi},\bm{\kappa}}$ simply as $p$. Note that the former always applies to a quantity homogeneous to $\bm X$, the second to a quantity homogeneous to $Y$, and the latter to a quantity homogeneous to $\bm Z$.

Getting back to the prediction task considered in this paper, recall that the goal is to predict $Y = Z_{d+1}$ based on $\bm{X} = (Z_1,\ldots, Z_d)^\top$. This may be done by
predicting $p(Y)$ based on $p(\bm X)$
  and then transforming back the predicted value for $p(Y)$ through an 
  inversion of \eqref{eq:paretoTransform}.

  For
  mathematical reasons rooted in statistical learning theory,
  prediction tasks associated with bounded target variables are more
  convenient to handle.  The intuition behind is that boundedness
  prevents spurious behaviours of the least square estimator arising
  with large residuals.  With this in mind, following in the footsteps of~\cite{huet2023regression} (Example 2 in the cited reference), 
    our strategy here is to learn a 
  regression function $\hat f$ to predict an intermediate, bounded, ‘angular' target 
\begin{equation}\label{eq:angulartarget}
      \Theta_Y = p(Y)/\|p(\bm Z)\|,
  \end{equation}  
  using any off-the-shelf machine learning algorithm trained on the extreme dataset $\mathcal{D}_{ext}$. 
  The learnt regression function $\hat f$ for $\Theta_Y$ then allows to define 
  a prediction function $\hat g$ for the original target $Y$ by plugging the predicted value
  $\hat \Theta_Y = \hat f(\bm{ X})$ into the straightforward inversion of the above display (with $\|\,\cdot\,\|$
  chosen as the Euclidean norm)
  \begin{equation*}\label{eq:equi_roxane}
p(Y) = \Theta_{Y}\| p(\bm X)\|  / \sqrt{(1- \Theta_{Y}^2)}.
\end{equation*}
In the end, the target quantity $Y$ is obtained by transforming back to the natural marginal scale using an inversion $p_{\sigma_Y,\xi_Y,\kappa_Y}^{-1}$ of~\eqref{eq:egpd_pareto_transf}.
Summarizing, the final prediction function $\hat g$ makes predictions $\hat Y$ defined by 
$$\hat Y = \hat g(\bm{ X} )  = p_{\sigma_Y, \xi_Y, \kappa_Y}^{-1}\Big(\hat f(\bm X)\| p(\bm X)\|  / \sqrt{(1- \hat f(\bm X)^2)}  \Big). 
$$

Regarding the training step aiming at constructing $\hat f$ to  predict $\Theta_Y$ as described above, a key finding in~\cite{huet2023regression} is
that it is advisable to leverage
an independence  property, above high thresholds, regarding angular and radial components  of the  rescaled pair (covariate, target); see~\eqref{eq:cv-paretoScale}.
More precisely,  under 
mild additional regularity assumptions regarding the density functions (See  Example 2 of \cite{huet2023regression} for details), letting 
$\bm{\Theta}_{X} =\| p(\bm X) \|^{-1}p(\bm X)$, and defining $\Theta_Y$ as in~\eqref{eq:angulartarget}, it holds that
\begin{equation}\label{eq:reg_variation}
  \mathcal{L}\Big( 
  ( \| p(\bm X) \|/t , 
  \bm{\Theta}_{X}, 
  \Theta_Y) \,|\, \|p(\bm X)\|>t \Big)\xrightarrow[t\to+\infty]{} \mathcal{L}(
  \|\tilde {\bm{X}}_\infty\|, \bm{\Theta}_{X,\infty}, \Theta_{Y,\infty}),  
\end{equation}
where $\tilde{\bm{X}}_\infty = \| \tilde{\bm{X}}_\infty \| \bm{\Theta}_{X,\infty}$ is the limit in distribution of $\mathcal{L}(t^{-1}p(\bm X)\,|\, \| p(\bm X)\|>t)$ as $t\to+\infty$. Importantly, in the  right-hand side of the above display, the  radial component $\| \tilde{\bm{X}}_\infty \|$ is independent of the angular components  $(\bm{\Theta}_{X,\infty},\Theta_{Y,\infty})$. 
This suggests  learning a prediction function for
$\Theta_{Y}$ (conditional on $\|p(\bm X)\|$ being large) taking as input only the ‘angular' component
$\bm{\Theta}_X$ 
of the largest observations, not their
radial component $\|p(\bm X)\|$. {\color{blue} Equation~\eqref{eq:reg_variation} also clearly illustrates that the output component $Y$ is not necessarily extreme in our framework, since the conditional extreme event involves only $\|p(\bm X)\|$, and the rescaling by the threshold $t$ in the law applies exclusively to $\|p(\bm X)\|$.}

This strategy is  embodied in the ROXANE
Algorithm~\ref{algo:rox_proc}, originally proposed by \cite{huet2023regression}.   ROXANE permits to learn a prediction function $\hat f(\bm{X}) := \hat f_\Theta(\bm{\Theta}_X)$  that depends only on the angular part $\bm{\Theta}_X$ of $\bm X$ and is asymptotically optimal in terms of mean squared error. 
In practice, ROXANE should be viewed as a meta-algorithm 
which may be implemented with any  machine learning prediction algorithm $\mathcal{A}_{pred}$,  suitable for prediction of a real-valued, continuous target. The algorithm $\mathcal{A}_{pred}$ is trained on the (covariate,target)  pairs $(\bm{\Theta}_{X,i},\Theta_{Y,i} )$, which are obtained by applying  the  previously described transformations to the extreme observations $\bm Z_i\in\mathcal{D}_{ext}$.  
The uniqueness of ROXANE lies in the pre-processing and rescaling steps described above. 
In our experiments (Section~\ref{sec:results}), we present results obtained using two specific algorithms $\mathcal{A}_{pred}$, namely Ordinary Least Squares (OLS) and Random Forest (RF), although other choices for the algorithm $\mathcal{A}_{pred}$ would have been equally valid.
It is rigorously demonstrated in~\cite{huet2023regression} that ROXANE enjoys strong statistical guarantees when implemented with a stylized example of a prediction machine learning algorithm, namely a least squares algorithm searching for a minimizer of the squared error over a class of predictors with controlled complexity, with arguments pertaining to Vapnik-Chervonenkis theory and concentration inequalities.

Numerical experiments on a toy financial dataset are presented in \citet{huet2023regression}, where (multivariate) extreme value theory is generally recognized as providing suitable tail models.  To our best knowledge the practical performance of ROXANE on environmental data such as skew surge time series has never been investigated. In particular, no comparative study has yet been conducted between this model-agnostic method and parametric modelling approaches, such as  the one proposed in the next section.



\begin{algorithm}[t!]
   \caption{ROXANE regression algorithm. }
   \label{algo:rox_proc}
\begin{algorithmic}[1]
\State{\bf TRAINING STEP}
\State{\hspace{0.5cm}\bfseries INPUT:} 
  \begin{itemize}
      \item The output of Algorithm~\ref{algo:th_select} applied independently to each marginal dataset $\{Z_{i, j}\}_{1 \m i \m n}$ for $1 \m j\m d+1$: EGP parameters $\bm{\sigma},\bm{\xi},\bm{\kappa} = (\bm{\sigma}_X,\sigma_Y),(\bm{\xi}_X,\xi_Y),(\bm{\kappa}_X,\kappa_Y) \in \mb{R}^{d+1}$; A multivariate threshold $\bm{t}_X\in \mb{R}^d$ for the covariate;  
      Extreme training dataset made of observations such that $\bm{X}\not\le \bm{t}_X$, $\mathcal{D}_{ext}=\{\bm{Z}_1,...,\bm{Z}_k\}$ where $\bm{Z}_i = (\bm{X}_i,Y_i) \in \mb{R}^{d+1}$ represents a covariate-target pair.
      \item A machine learning algorithm $\mathcal{A}_{pred}$ suitable for prediction of a real target (e.g., least squares, random forest, support vector regression, k-nn, gradient boosting, neural networks,\dots)
  \end{itemize}  
    \State{\hspace{0.5cm}\bfseries Extreme data transformation to Pareto margins and angular rescaling:} Apply the componentwise Pareto transformation~\eqref{eq:egpd_pareto_transf} to extreme observations 
    \begin{equation*}
p_{\bm{\sigma},\bm{\xi},\bm{\kappa}}(\bm{Z}_i)=\big(p_{\bm{\sigma}_X,\bm{\xi}_X,\bm{\kappa}_X}(\bm{X}_i),p_{\sigma_Y,\xi_Y,\kappa_Y}(Y_i)\big), \quad \text{for } 1 \m i\le k  \; (\text{equivalently for } \bm{Z}_i\in\mathcal{D}_{ext}).
    \end{equation*}
    Form the angular components of the covariates and the target, 
    \begin{equation*}
        \bm{\Theta}_{X,i} = \|p_{\bm{\sigma}_X,\bm{\xi}_X,\bm{\kappa}_X}(\bm X_i)\|^{-1}p_{\bm{\sigma}_X,\bm{\xi}_X,\bm{\kappa}_X}(\bm X_i) \; ,
   \qquad
        \Theta_{Y,i} = p_{\sigma_Y,\xi_Y,\kappa_Y}(Y_i) /\| p_{\bm{\sigma},\bm{\xi},\bm{\kappa}}(\bm Z_i)\|.
    \end{equation*}
\State{\hspace{0.5cm}\bfseries Training the angular prediction model: } Train Algorithm $\mathcal{A}_{pred}$ on the angular training dataset $\mathcal{D}_{ang}$ made of  angular pairs (covariates, target) 
$$
\mathcal{D}_{ang} = 
\{ \bm{\Theta}_{X,i}, \Theta_{Y,i}\}_{ 1 \m i \le k}.
$$
\State{\hspace{0.5cm}\bf Intermediate output: } An angular   prediction function $\hat f_\Theta$, and an associated prediction function $\hat f$:  
For a new (test) extreme covariate $\bm{x}$ such that $\bm{x}\not\le \bm{t}_X$, the prediction of $\mathcal{A}_{pred}$ regarding  
$\Theta_y = y/\|(\bm x,y)\|$ is 
 $$\hat f(\bm x) = \hat f_\Theta(\bm{\Theta_x}), $$
where  $\bm{\Theta}_x =  p_{\bm{\sigma}_X,\bm{\xi}_X,\bm{\kappa}_X}(\bm{x})/ \| p_{\bm{\sigma}_X,\bm{\xi}_X,\bm{\kappa}_X}(\bm{x})\|$. 
     
\State{\hspace{0.5cm}\bfseries OUTPUT of the training step:} A prediction function 
 $\hat g$ defined for $\bm x\not\le \bm{t}_X$, 
\begin{equation*}
     \hat g(\bm{x} )=     p^{-1}_{\sigma_Y,\xi_Y,\kappa_Y}
     \Big(
     \hat f(\bm{x})\|p_{\bm{\sigma}_X,\bm{\xi}_X,\bm{\kappa}_X}(\bm{x})\| / 
     \sqrt{1-\hat f(\bm{x})^2}
     \Big),     
\end{equation*}
where 
$p^{-1}_{\sigma_Y,\xi_Y,\kappa_Y}$ is the inverse of the Pareto transformation~\eqref{eq:egpd_pareto_transf}. 
~\\~\\
\State {\bfseries PREDICTION STEP } 
\State{\hspace{0.5cm}\bf INPUT: }
New observation $\bm{X}_{new}$ (with unobserved associated target $Y_{new}$) such that $\bm{X}_{new}\not\le \bm{t}_X$.  

\State{\hspace{0.5cm}\bfseries OUTPUT of the prediction step: } 
Predicted target $\hat Y = \hat g(\bm{X}_{new})$. 
\end{algorithmic}
\end{algorithm}

\subsection{Distribution regression based on parametric multivariate generalized Pareto modelling}\label{sec:mgp_proc}
As an alternative to the point prediction strategy described in the previous paragraph,  we consider a distribution regression strategy relying on  parametric modelling of  extreme observations transformed to exponential scale.   
The procedure is outlined in Algorithm~\ref{algo:mgp}, which we refer to as MGPRED (Multivariate Generalized Pareto modelling for Prediction).  Given the parameters $\bm{\sigma},\bm{\xi},\bm{\kappa} = (\bm{\sigma}_X,\sigma_Y),(\bm{\xi}_X,\xi_Y),(\bm{\kappa}_X,\kappa_Y)$ estimated by Algorithm~\ref{algo:th_select}, we denote by $e_{\bm{\sigma}, \bm{\xi}, \bm{\kappa}}$ the specific instance of the componentwise standardization function $e$  defined in~\eqref{eq:expo-transform-function}, with marginal distributions chosen as the fitted EGP distributions, namely 
\begin{equation}\label{eq:expo_transform_egpd}
    e_{\bm \sigma, \bm\xi, \bm \kappa}(\bm z) = 
    \big(-\log(1- F_{\bm \sigma_X,\bm \xi_X,\bm \kappa_X}((z_1,...,z_d))), 
    -\log(1- F_{\sigma_Y,\xi_Y,\kappa_Y}(z_{d+1}))\big). 
\end{equation}
Similarly to the Pareto marginal transformation $p$, we denote $e_{\bm{\sigma}_X,\bm{\xi}_X,\bm{\kappa}_X}$ and $e_{\sigma_Y,\xi_Y,\kappa_Y}$ simply as $e$ for conciseness when clear from the context. Note that the former always applies to a quantity homogeneous to $\bm X$, while the latter applies to a quantity homogeneous to $Y$.
In view of~\eqref{eq:CV_excess_expo_margins}, it is asymptotically justified to fit a standard MGP model (i.e., a model for $\bm{Z}'_\infty$ in the latter equation) to  exceedances of extreme data in $\mathcal{D}_{ext}$ transformed to exponential margins,  above the transformed threshold  $e_{\bm \sigma, \bm\xi, \bm \kappa}(\bm{t})$, where we recall that $\bm t$ is the multivariate threshold selected \emph{via} Algorithm~\ref{algo:th_select}. 
Thus, inference in MGPRED use as a training transformed dataset   $$\mathcal{D}_{expo} = \{e_{\bm \sigma, \bm\xi, \bm \kappa}(\bm{Z}_i) - e_{\bm \sigma, \bm\xi, \bm \kappa}(\bm{t})\}_{1 \m i\le k} $$ 
used for fitting standard MGP models. 
Here following the approach developed in~\citet{kiriliouk2019peaks}, we consider a panel $\mathcal{H} = \{ \mathcal{H}_1, \ldots,  \mathcal{H}_M\}$ of parametric models accounting for density models for $\bm T$ in \eqref{eq:mgpd_decomp}. Using the same families, an alternative method to construct standard MGP densities is explored (related to vector $\bm U$ in~\citet{kiriliouk2019peaks}, see Appendix~\ref{app:density_model} for details).

 During the training step of MGPRED, following \citet{kiriliouk2019peaks}, a censored likelihood approach is employed to fit each parametric model $\mathcal{H}_m$. This means that only observations $\bm{Z}_i$ such that all $d+1$ components  $Z_{i,j}$ are larger than $t_j$ are used to fit the models. 
  This censored likelihood inference step  results in a family of $M$    fitted joint densities $\{\hat h_m\}_{1 \m m\le M}$, modelling standardized and shifted  excesses above threshold, namely, modelling  the random vector $e_{\bm \sigma,\bm \xi, \bm\kappa}(\bm Z) - e_{\bm \sigma,\bm \xi, \bm\kappa}(\bm t)$, as a standard MGP random vector $\bm{Z}'_\infty\in{\mb R}^{d+1}$. 
 Model selection is then performed \emph{via} a simple information criterion, e.g., the AIC, and the final output of the training step is a density function $\hat h$. 
 
During the prediction step, MGPRED provides  an estimate of the conditional distribution of $Y_{new}$ given $\bm{X}_{new}$, under the form of a Monte Carlo sample $\mathcal{D}_Y^{MC} = \{\hat Y_\ell\}_{1\le\ell \le L}$. As an immediate by-product, a point prediction  $\hat Y$ can be derived as the conditional expectation of $Y_{new}$ given $\bm{X}_{new}$, which  is estimated by simply averaging the Monte Carlo sample. The sample $\mathcal{D}_Y^{MC}$ 
is obtained from the  following steps: First the new observation is transformed to exponential margins and shifted by the same operations as the ones performed during the training step. The resulting variable is given by $e(\bm{X}_{new}) - e(\bm{t_X})$. Then the conditional density  of $ e(Y)-e(t_Y)$ conditional on  $e(\bm{X}_{new}) - e(\bm{t_X})$ is given by  $\hat h(\cdot| e(\bm{X}_{new}) - e(\bm{t_X})) = \hat h( e(\bm{X}_{new}) - e(\bm{t_X}), \cdot) / C$, where $C = \int_{\mb{R}} \hat h( e(\bm{X}_{new}) - e(\bm{t_X}), s )\, \mathrm{ d} s $ is a normalizing constant. A practical Monte Carlo method for sampling from a density known only up to a normalizing constant is the Accept-Reject algorithm (see, e.g., \citet[Chapter 2]{robert2010introducing}). Notably, the constant $C$ does not need to be explicitly computed to apply this method. 
Finally, the latter Monte Carlo sample is transformed back to the original scale by inverting the transformation steps using $e_{\sigma_Y, \xi_Y,\kappa_Y}^{-1}$.

In this work, MGPRED  serves as a natural example of parametric modelling for conditional inference of extremes, offering a point of comparison with the Machine Learning-based  ROXANE algorithm described in the previous section. During the training step, it may be viewed  as a simplified version of the approach developed in \citet{kiriliouk2019peaks} which simultaneously fits the parameters of the marginal distribution, together with the parameters of the MGP model. The advantage of the latter joint fitting strategy is that uncertainty stemming from marginal estimation may  be taken into account. However preliminary experiments (not shown)  on the dataset considered here demonstrated poorer performance compared to the two-step strategy implemented in this study. This was observed in two aspects.  First, as expected,  the computational cost of the joint fitting procedure was significantly higher, requiring  optimization steps over $7$ to $11$ parameters simultaneously. 
Second, the model fit was less satisfactory. We attribute this lack of fit to potential failures of the  optimization routine which may depend on its starting point, given that the joint likelihood function is not convex. For this reason, we restrict our presentation to the two-stage model fitting process: marginal fitting using Algorithm~\ref{algo:th_select}, followed by multivariate model fitting on exponential marginal scale during the training phase of MGPRED.
During the prediction step, MGPRED  bears similarities to the methodology developed by~\citet{legrand2023joint} in a different context, namely for joint and conditional simulation of extreme waves. The key distinctions in our implementation for the specific use case considered here are: (i) The threshold selection is driven by marginal EGP distribution fitting;  
(ii) Our procedure is capable of handling three or more stations, whereas their approach heavily depends on the assumption of working with only two measurement stations;
(iii) The downstream task is to reconstruct missing data based on their conditional distribution given observed neighbouring gauges, not to generate synthetic data, although Monte Carlo sampling is used as a technical tool in our approach. Also our final averaging step for missing data imputation is not  required in~\citet{legrand2023joint}.

\begin{algorithm}[t!]
   \caption{MGPRED algorithm.\label{algo:mgp}}
\begin{algorithmic}[1]
    \State {\bfseries TRAINING STEP }    

  \State{\hspace{0.5cm}\bfseries INPUT:}
  \begin{itemize}
      \item  The output of Algorithm~\ref{algo:th_select} applied independently to each marginal dataset $\{Z_{i, j}\}_{1 \m i\le n}$ for $1 \m j\m d+1$: EGP parameters $\bm{\sigma},\bm{\xi},\bm{\kappa} = (\bm{\sigma}_X,\sigma_Y),(\bm{\xi}_X,\xi_Y),(\bm{\kappa}_X,\kappa_Y)\in \mb{R}^{d+1}$; A multivariate threshold $\bm{t} = (\bm{t}_X,t_Y)\in \mb{R}^{d+1}$; 
      Extreme training dataset made of observations such that $\bm{X}\not\le \bm{t}_X$, $\mathcal{D}_{ext}=\{\bm{Z}_1,...,\bm{Z}_k\}$ where $\bm{Z}_i = (\bm{X}_i,Y_i) \in \mb{R}^{d+1}$ represents a covariate-target pair.
      \item A panel of $M$ candidate parametric models for MGP  data  with exponential margins as in~\eqref{eq:mgpd_decomp}, namely a panel $\mathcal{H}=\{\mathcal{H}_1,...,\mathcal{H}_M\}$ of $M$ parametric families of density functions. 
  \end{itemize}
    \State{\hspace{0.5cm}\bfseries Extreme data transformation to exponential margins and shifting:} Apply the exponential componentwise transformation $e_{\bm \sigma,\bm \xi, \bm \kappa}$ defined in~\eqref{eq:expo_transform_egpd}  
    to 
    extreme observations in $\mathcal{D}_{ext}$
    \begin{equation*}
    e_{\bm \sigma,\bm \xi, \bm \kappa} (\bm Z_i), \qquad \text{for } 1 \m i\le k  \; (\text{equivalently for } \bm{Z}_i\in\mathcal{D}_{ext}). 
    \end{equation*}
    Transform the threshold $\bm t$ as
    $
    e_{\bm \sigma,\bm \xi, \bm \kappa} (\bm t). $ 
    
    
    \State{\hspace{0.5cm}\bfseries Model fitting and model selection: } 
          Fit (through a maximum-censored likelihood method) each parametric model $\mathcal{H}_m$, for all $1 \m m \m M$, to the shifted  exponential margin, extreme data 
          $$\mathcal{D}_{expo} =\{ e_{\bm \sigma,\bm \xi, \bm \kappa} (\bm Z_i) - e_{\bm \sigma,\bm \xi, \bm \kappa} (\bm t)\}_{1\le i\le k}.$$ 
        Among the          fitted density functions $\hat h_1, \ldots, \hat h_M$,           
        select the density $\hat{h}\in\{\hat h_1,\ldots, \hat h_M\}$ with the smallest AIC.     

 \State{\hspace{0.5cm}\bfseries OUTPUT of the training step: }Fitted joint density function $\hat{h}$ in $\mathcal{H}$ defined on $\mb R^{d+1}$.~\\~\\
     
   
\State {\bfseries PREDICTION STEP}
\State{\hspace{0.5cm}\bf INPUT: } A new observation   
$\bm{X}_{new} \in \mb R^d$ such that $\bm{X}_{new} \not\le \bm{t}_X$, and a Monte Carlo sample size $L$.    
   \State{\hspace{0.5cm}\bf Conditional  Monte Carlo sampling: }
    Generate a sample of size $L$,  
     $\{Y^{e}_1, \ldots, Y^{e}_L \}  $
    from the conditional density of  $\hat h(\cdot\, | e_{\bm \sigma_X,\bm \xi_X, \bm \kappa_X} (\bm{ X}_{new}) - e_{\bm \sigma_X,\bm \xi_X, \bm \kappa_X} (\bm{t}_X) ) $ defined above. 
  
    \State{\hspace{0.5cm}\bfseries Back-transformation: }  Define
  $ 
 \hat Y_{\ell} = e_{\sigma_Y,\xi_Y,\kappa_Y}^{-1}
 (Y^e_{\ell} + e_{\sigma_Y,\xi_Y,\kappa_Y}(t_Y)), \; 1\le \ell \le L ; 
$ 
and form the Monte Carlo sample on the original scale,  
$
\mathcal{D}_{Y}^{MC} = \{ \hat Y_\ell\}_{1 \m \ell \m L}.$


 \State{\hspace{0.5cm}\bfseries OUTPUT of the prediction step: } 
    \begin{itemize}
        \item Conditional distribution of $Y_{new}$ given the observed covariate $\bm{X}_{new}$ represented by the Monte Carlo sample $\mathcal{D}_{Y}^{MC}$ (amenable to prediction intervals such as the  interquantile range)
        \item Predicted value $\hat Y$ for $Y_{new}$  by Monte Carlo average 
         $\hat{Y}  = L^{-1} \sum_{\ell=1}^L \hat{Y}_\ell.$  
    \end{itemize}
\end{algorithmic}
\end{algorithm}

\section{Results}\label{sec:results}
The two algorithmic procedures described in Section~\ref{sec:methods} are applied to the skew surge exceedance datasets. The training dataset consists of 9,213 observations spanning the period from 10 August 1966 to 31 December 1999, while the test dataset includes 6,024 observations covering 1 January 2000 to 31 December 2023. Sections~\ref{sec:res_marg} and~\ref{sec:mgpd_results} present the results for the fitted marginal models and the trained prediction models, respectively. Additional analyses involving similar studies on sea levels, as well as data from the Concarneau and Le Crouesty stations as output locations, are deferred to Section~\ref{sec:add_stud} of the Appendix. Section~\ref{sec:discussion} provides a discussion and comparison of the various models. Finally, in Section~\ref{sec:recons}, the models are applied to reconstruct the extreme skew surge time series at Port Tudy prior to the deployment of its tide gauge.

\subsection{Marginal fitting and threshold selection}\label{sec:res_marg}
\begin{figure}[t!]
\centering
  \includegraphics[width=.315\textwidth]{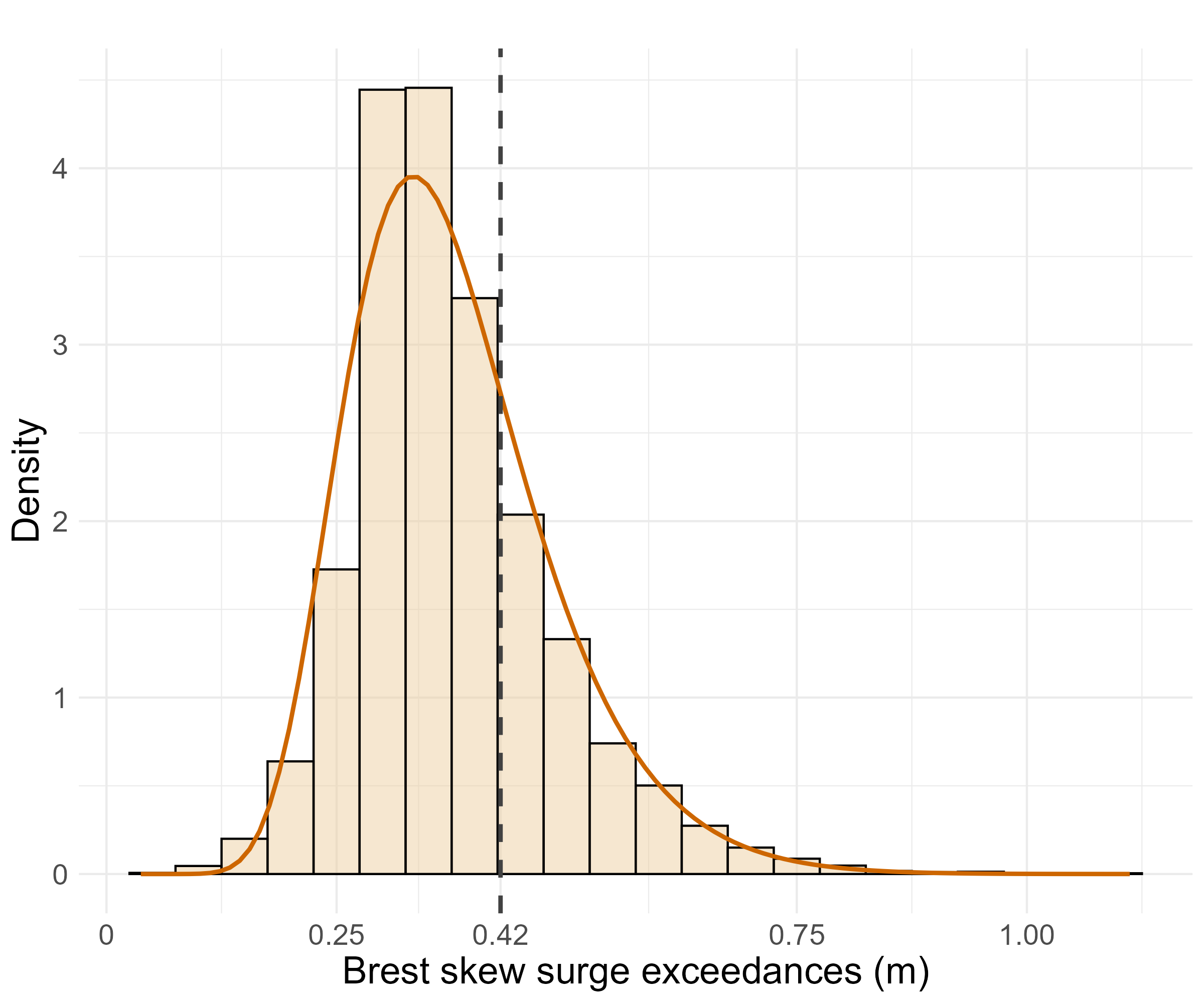}
  \hspace{0.2cm}
  \includegraphics[width=.315\textwidth]{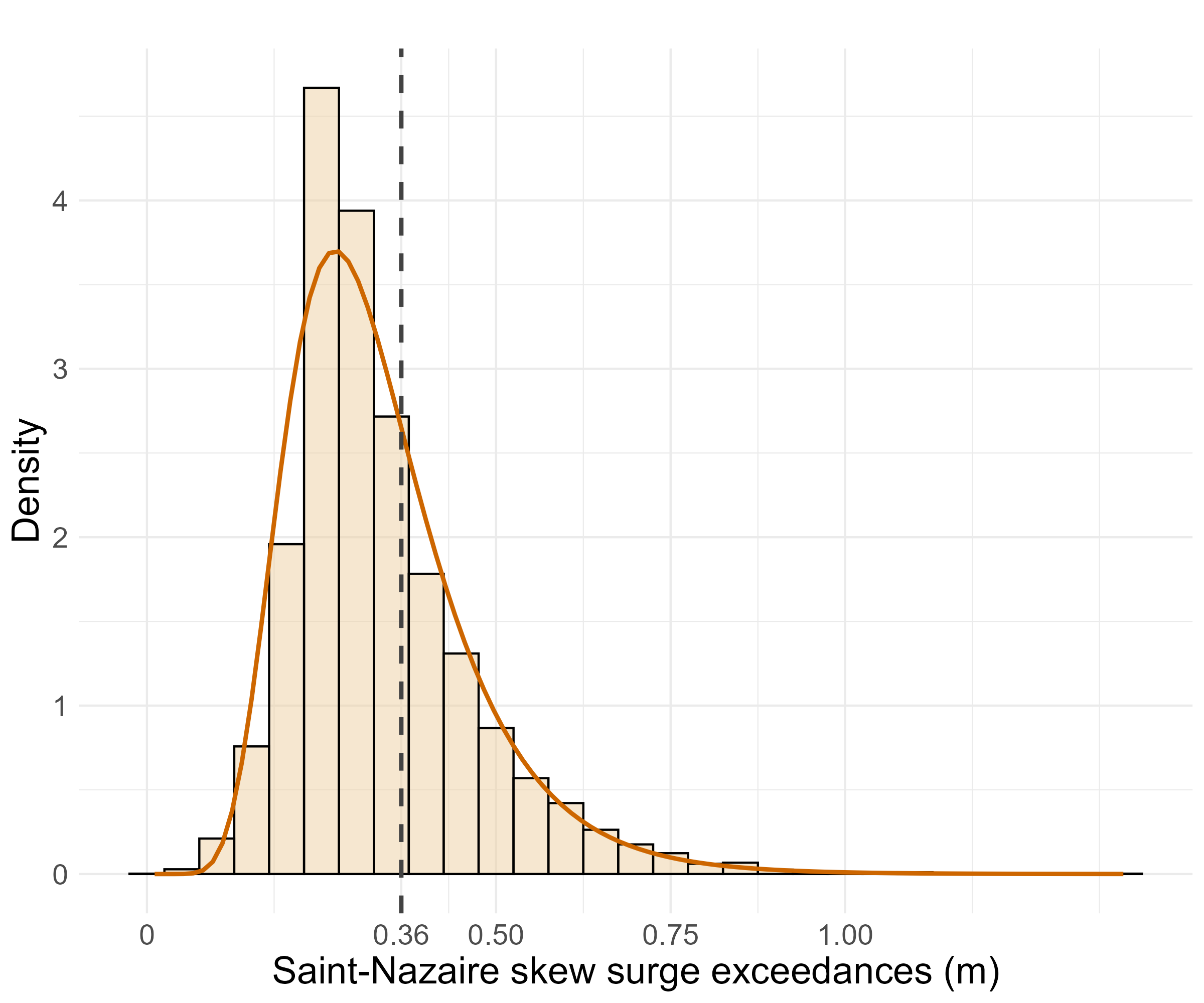}
  \hspace{0.2cm}
  \includegraphics[width=.315\textwidth]{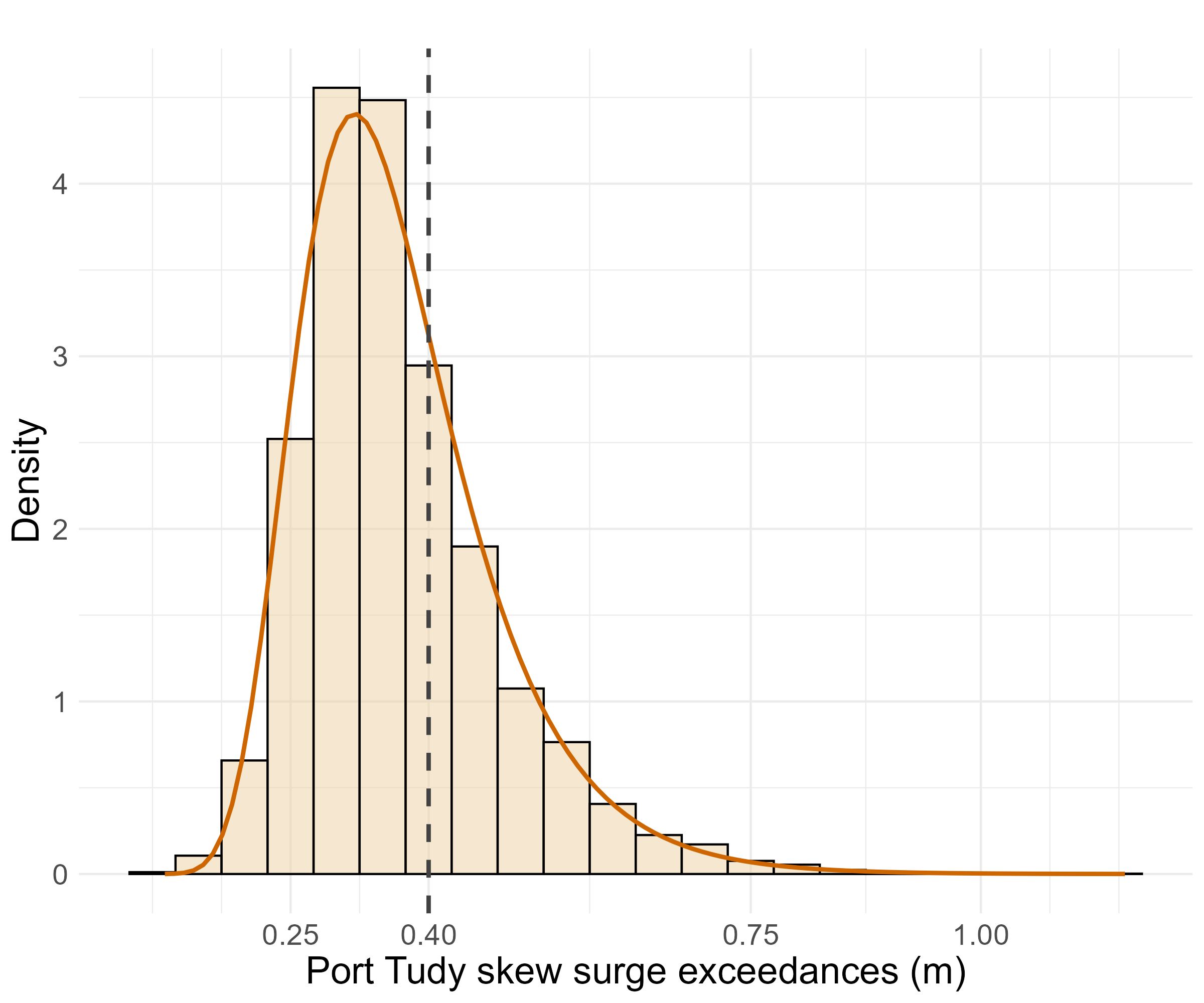}

  \caption{Histograms of skew surge exceedances at the three stations Brest (left), Saint-Nazaire (middle) and Port Tudy (right), from 01/12/2000 to 31/12/2023. The orange curves represent the fitted EGP densities, with parameters specified in Table~\ref{tab:egpd_tudy}. The dotted vertical grey lines represent the smallest point above which each fitted density is convex, which  represent the chosen marginal thresholds via Algorithm~\ref{algo:th_select} (see Section~\ref{sec:res_marg}). \label{fig:density_tudy}}
\end{figure}

The initial step common to both procedures involves modelling the marginal distributions using the EGP distribution and computing the thresholds as described in Algorithm~\ref{algo:th_select}. {\color{blue}
The marginal fit of the training data is performed with the function \textsf{fit.egp} from the \textsf{mev} package \citep{belzile2025package}.
}

We denote by $\bm{t} := (t_B,t_N,t_Y)^\top$ the multivariate threshold and by $\bm{\sigma} := (\sigma_B,\sigma_N,\sigma_Y)^\top$, $\bm{\xi} := (\xi_B,\xi_N,\xi_Y)^\top$ and  $\bm{\kappa} := (\kappa_B,\kappa_N,\kappa_Y)^\top$ the EGP parameters resulting from the application of Algorithm~\ref{algo:th_select}. The estimated parameters are presented in Table~\ref{tab:egpd_tudy}. Histogram plots with the corresponding fitted densities are shown in Figure~\ref{fig:density_tudy}. After thresholding, the final training sets consist of 3,310 skew surge exceedances, while final test sets consists of 2,507 skew surge exceedances.

We recall that Figure~\ref{fig:scatterplot} presents pairwise plots of skew surge exceedances of stations. These plots illustrate the strong correlation between stations, although this linear dependence is notably weaker for the most extreme observations.

\begin{table}[ht!]
\caption{Point estimates of the parameters of the fitted EGP distribution for skew surge exceedances at the three stations. The chosen thresholds, determined using via Algorithm~\ref{algo:th_select}, are shown in the $t$ rows. The data used for inference are from the training set ranging from 01/12/2000 to 31/12/2023. \label{tab:egpd_tudy}}
\vspace{0.2cm}
\begin{center}
\begin{small}
\begin{sc}
\begin{tabular}{cccccr}
\toprule
Parameters/Stations & Brest & Saint-Nazaire & Port Tudy  \\
\midrule
$\sigma$ & 0.12 & 0.10 & 0.09 \\
$\xi$ & -0.076 & 0.012 & -0.005 \\
$\kappa$ & 20.70 & 14.40 & 41.93 \\
$t$ & 0.42 & 0.36  & 0.40\\
\bottomrule
\end{tabular}
\end{sc}
\end{small}
\end{center}
\end{table}

\subsection{Modelling multivariate dependencies}\label{sec:mgpd_results}

The multivariate procedures presented in the sections~\ref{sec:reg_proc} and \ref{sec:mgp_proc} are now applied to the extreme skew surge exceedance observations. Recall that the selected thresholds and EGP parameters for each margin, computed on the training set are summarized in Table~\ref{tab:egpd_tudy}.

\subsubsection{Regression method: ROXANE}
Following the steps of Algorithm~\ref{algo:rox_proc}, we transformed the data to the Pareto scale so that
\begin{equation*}
p_{\bm{\sigma},\bm{\xi},\bm{\kappa}}(\bm{Z}_i)=\big(p_{\sigma_{X_B},\xi_{X_B},\kappa_{X_B}}({X_{B,i}}),p_{\sigma_{X_N},\xi_{X_N},\kappa_{X_N}}({X_{N,i}}),p_{\sigma_Y,\xi_Y,\kappa_Y}(Y_i)\big)^\top,
\end{equation*}
which is shortly denoted by $p_{\bm{\sigma},\bm{\xi},\bm{\kappa}}(\bm{Z}_i) = \big(p({X_{B,i}}),p({X_{N,i}}),p(Y_i)\big)^\top$ when clear from the context.
We then consider the angular parts of the input and the output variables w.r.t. the $L^2$-norm which are given by
    \begin{equation*}
        \bm{\Theta}_{X,i}:=(\Theta_{B,i},\Theta_{N,i})^\top := \frac{\big(p({X_{B,i}}),p({X_{N,i}})\big)^\top}{\sqrt{p({X_{B,i}})^2 + p({X_{N,i}})^2}},
    \end{equation*}
    \begin{equation*}
        \Theta_{Y,i} := \frac{p({Y_{i}})}{\sqrt{p({X_{B,i}})^2 + p({X_{N,i}})^2 + p({Y_{i}})^2}}. 
    \end{equation*}

We employ two regression algorithms to learn a prediction function mapping the angular input observations $\bm{\Theta}_{X,i}$ to the angular output observations $\Theta_{Y,i}$ : Ordinary Least Squares (OLS), implemented in the \textsc{stats} package, and Random Forest (RF), implemented in the \textsc{randomForest} package. Based on the scatterplot shown in Figure~\ref{fig:scatterplot}, a linear model, as assumed in the OLS approach, appears to be the most suitable choice. However, for completeness and to highlight the flexibility of the ROXANE procedure in higher dimensions, we also present results obtained using the RF algorithm, which is designed for high-dimensional data. The depth of the trees in the RF algorithm is selected using a cross-validation procedure on the training set.

Following the equivalence~\eqref{eq:equi_roxane}, the predicted angular values $\hat{\Theta}_{Y,i}$ on the test set are then backtransformed to the Pareto scale via
\begin{equation*}
    \widehat{p({Y_{i}})} = \frac{\hat{\Theta}_{Y,i}\sqrt{p({X_{B,i}})^2+p({X_{N,i}})^2}}{\sqrt{1-\hat{\Theta}_{Y,i}^2}},
\end{equation*}
and then backtransformed to the original scale via the inverse function $p^{-1}_{\sigma_Y,\xi_Y,\kappa_Y}$.
As visual goodness-of-fit assessments, the left and middle columns of Figure~\ref{fig:qqplot_tudy_ss} depict QQ-plots of the observed skew surge exceedances against the estimated values with the OLS algorithm and RF algorithm on the test set. In addition, Figure~\ref{fig:pred_tudy} shows the predicted curves 
on the years $1989$, $1978$ and $1977$, years characterized by significant skew surges at the Port Tudy tide gauge.

\subsubsection{Plug-in method: MGPRED}
The routine of MGPRED is described in two stages: the modelling step, followed by the prediction step. The modelling step used an adaptation of the procedure in \citet{kiriliouk2019peaks} to obtain a joint density model for our data, while the prediction step employs a Monte Carlo procedure to obtain an estimate using the density learnt in the modelling step.

\textbf{Modelling of the data.} As suggested in Algorithm~\ref{algo:mgp}, the extreme observations are transformed to an exponential scale and then relocated by subtracting the corresponding threshold on exponential scale $e_{\bm \sigma, \bm\xi, \bm \kappa}(\bm{t})$ such that
\begin{equation*}
    e_{\bm \sigma, \bm\xi, \bm \kappa}(\bm{Z}_i) - e_{\bm \sigma, \bm\xi, \bm \kappa}(\bm{t})= \big(e({X_{B,i}}) - e(t_B),e({X_{N,i}})-e(t_N),e(Y_i) - e(t_Y)\big)^\top
\end{equation*}
where indices are again omitted for clarity.

The density models used to construct standard MGP densities are the same as those in \citet{kiriliouk2019peaks}, stemming from multivariate distributions with independent components, where the marginal distributions are either all reverse exponential or all Gumbel distributions. To facilitate sampling in the prediction step from these distributions and avoid complex approximations, we limit our analysis to densities that do not involve unknown integral quantities. Section~\ref{app:density_model} of the Appendix provides details on the construction of MGP densities, while Section~\ref{sec:dens_model} presents the closed-form expressions for the considered densities. Again, as in \citet{kiriliouk2019peaks}, we select a density using a censored likelihood criterion, maximizing the classic product likelihood function using only uncensored observations, where an observation is considered censored if any of its components is negative. This choice focuses the fit on the most extreme observations, which are of prime interest. After application of this procedure, the selected density model for $e_{\bm \sigma, \bm\xi, \bm \kappa}(\bm{Z}) - e_{\bm \sigma, \bm\xi, \bm \kappa}(\bm{t})$ is associated with independent Gumbel components with a unique dependence parameter $\alpha=1.91$ and varying locations parameters $\bm{\beta} = (\beta_B,\beta_N,\beta_Y)^\top = (-0.22,0.05,0.00)^\top \in \mb{R}^3$ given by
\begin{equation*}
    h(\bm{x}) = \1 \{\max(\bm{x}) > 0 \} \alpha^2 \Gamma(3)\exp\Big(-\max(\bm{x}) \frac{\prod_{j \in \{B,N,Y\}}\exp \big(\alpha(x_j - \beta_j)\big)}{\sum_{j \in \{B,N,Y\}}\exp \big(\alpha(x_j - \beta_j)\big)}\Big),
\end{equation*}
for $\bm{x} = (x_B,x_N,x_Y)^\top$ and where $\Gamma$ is the classic $\Gamma$-function. Note that $\beta_Y=0$ for identifiability purpose.

\textbf{Predictions on the test set.}  To obtain predictions for the extreme Port Tudy values in the test set, we sample $100$ values through rejection sampling for each conditional densities
\begin{equation*}
    h\big(\cdot \mid e({X_{B,i}}) - e(t_B),e({X_{N,i}}) - e(t_N)\big) = \frac{h\big((e({X_{B,i}}) - e(t_B),e({X_{N,i}}) - e(t_N),\cdot)^\top\big)}{\int_\mb{R} h\big((e({X_{B,i}}) - e(t_B),e({X_{N,i}}) - e(t_N),s)^\top\big)ds}.
\end{equation*}

As outlined in Section~\ref{sec:mgp_proc}, computing the normalizing constant in the denominator is not required for applying the rejection sampling method. However, approximating this constant can significantly enhance computational efficiency by ensuring the resulting function closely approximates a valid density. In our implementation, we use the numerical routine \textsc{integrate} to compute this approximation. For the dominant densities required in the rejection sampling method, we employ Student-t distributions, which have heavier tails than the densities of $h_T$. These distributions are centred at the mean of $e({X_{B,i}}) - e(t_B)$ and $e({X_{N,i}}) - e(t_N)$, that is, at $\big(e({X_{B,i}}) - e(t_B) + e({X_{N,i}}) - e(t_N)\big)/2$.

The generated values on exponential scale are then backtransformed to original scale via the inverse function $e^{-1}_{\sigma_Y,\xi_Y,\kappa_Y}(\cdot+e(t_Y))$.
Point estimates are obtained as Monte Carlo averages over the 100 generated values. The right column of Figure~\ref{fig:qqplot_tudy_ss} presents a QQ-plot comparing the observed values with the predicted skew surge exceedances, while Figure~\ref{fig:pred_tudy} illustrates the predicted curves for the years 1989, 1978, and 1977{\color{blue}, along with 0.95 simulation-based predictive intervals. These intervals correspond to the 0.025 and 0.975 empirical quantiles computed from the set of predictions produced by the MGPRED procedure at each point.} The 0.95-coverage probability for these intervals is 0.89 for skew surges, where the 0.95-coverage probability is defined as the proportion of test observations that fall within the computed 0.95-prediction confidence intervals.

\subsection{Discussion and comparison of the methods} \label{sec:discussion}

\begin{figure}[t!]
  \centering
  \includegraphics[width=.315\textwidth]{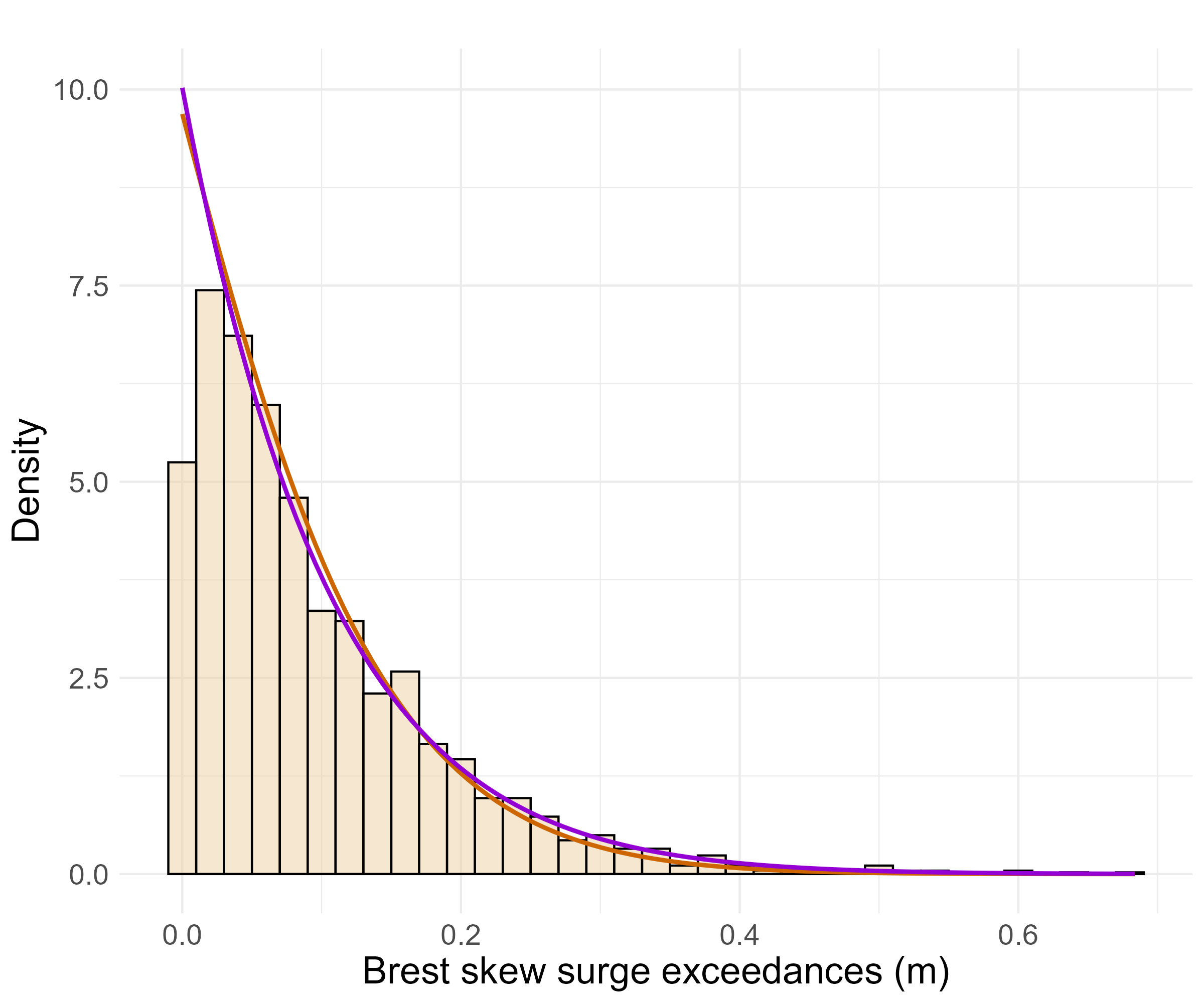}
  \hspace{0.2cm}
  \includegraphics[width=.315\textwidth]{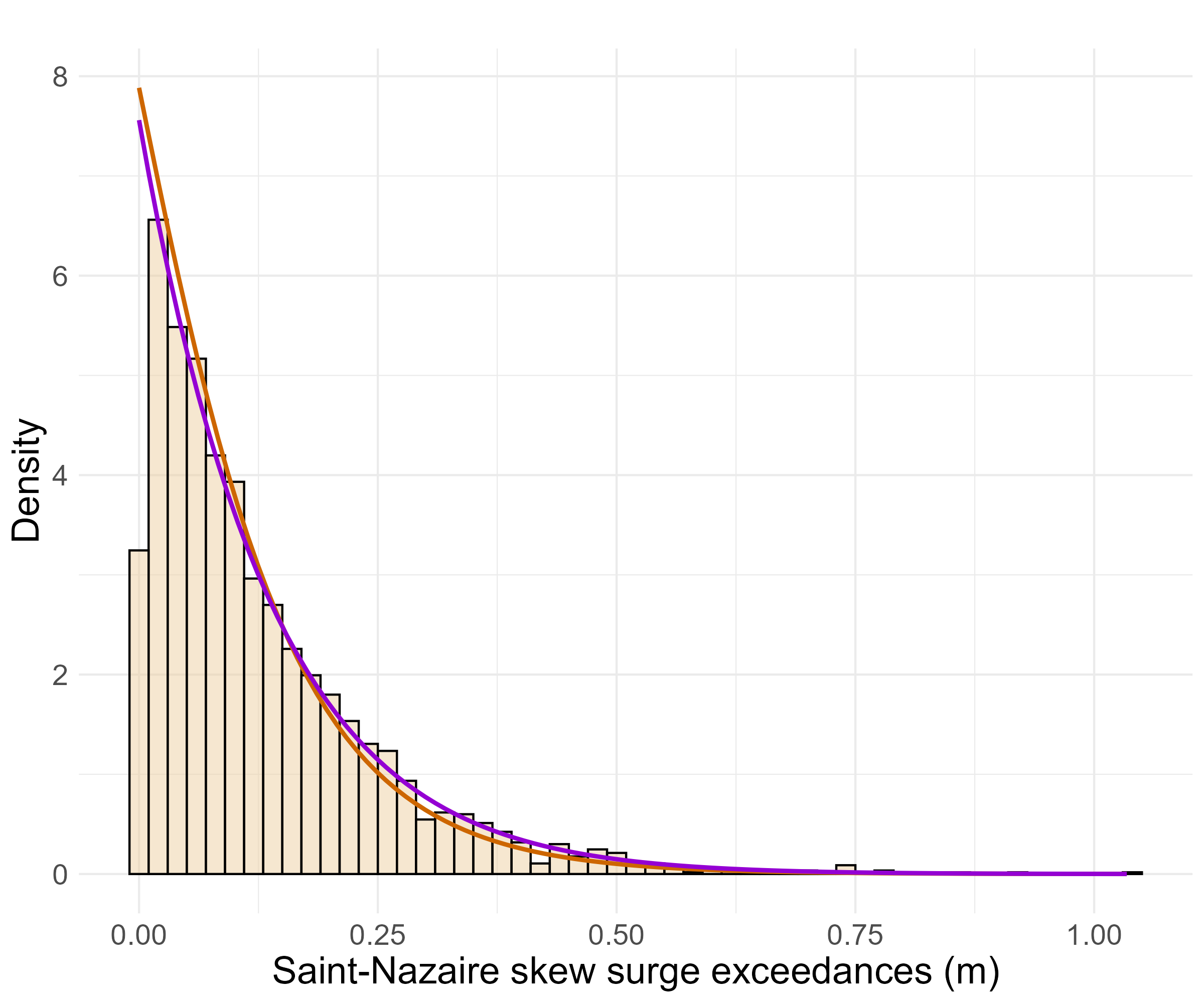}
  \hspace{0.2cm}
  \includegraphics[width=.315\textwidth]{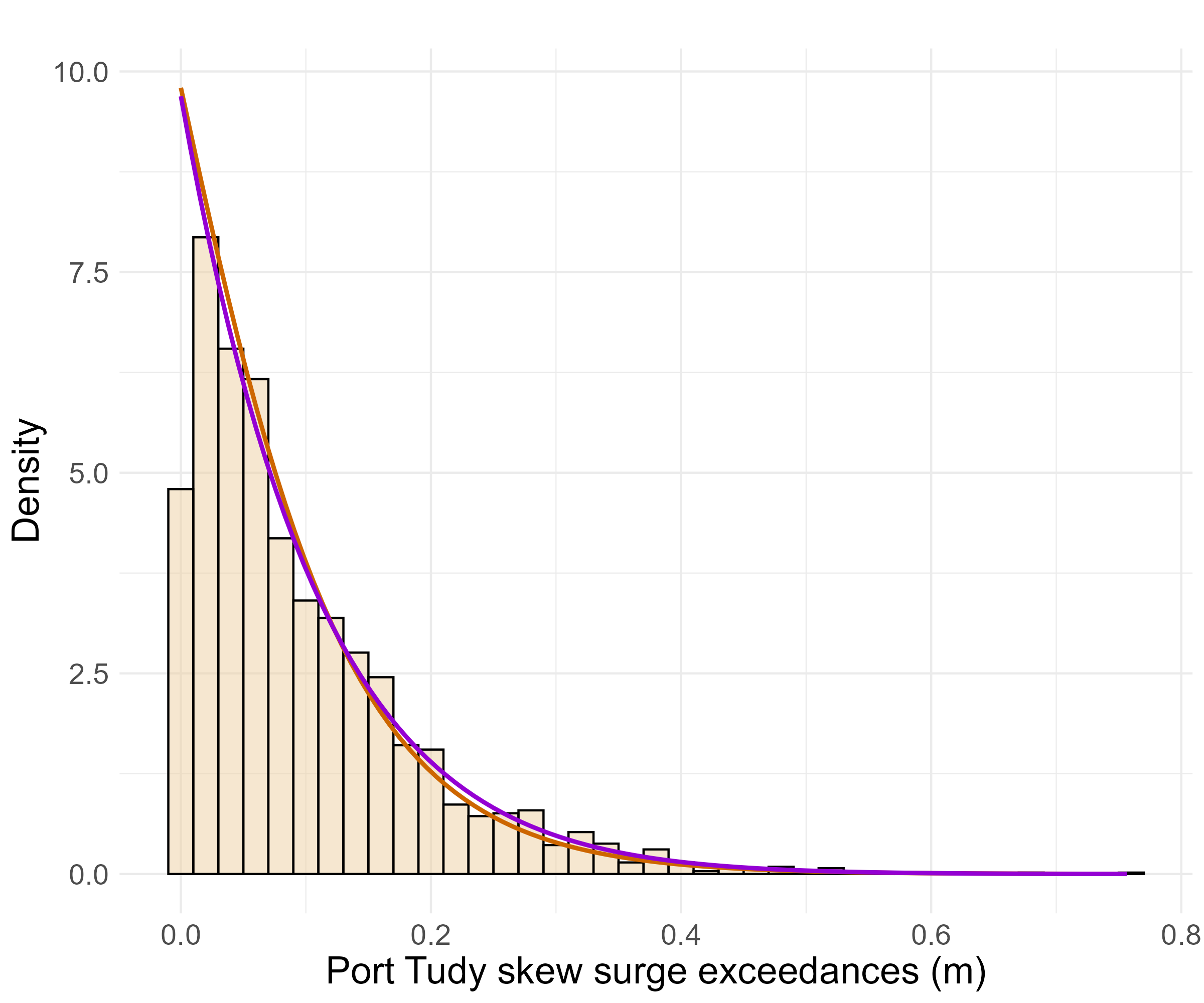}

  \caption{Histograms of skew surge exceedances above the threshold specified in Table~\ref{tab:egpd_tudy} at the three stations Brest (left), Saint-Nazaire (middle) and Port Tudy (right), from 01/12/2000 to 31/12/2023. The orange curves represent the fitted EGP densities above the thresholds, with parameters specified in Table~\ref{tab:egpd_tudy}. The purple curves represent the fitted GP densities.\label{fig:tudy_egpdvsgpd}}
\end{figure}

In this section, we discuss the results of the marginal modelling and of both multivariate procedures. First, the EGP distribution successfully models the margins, particularly in the right tail of the distribution for data exceeding the selected thresholds, as illustrated in Figure~\ref{fig:density_tudy}. 
The threshold is selected as the lowest value above which the EGP density becomes strictly convex, based on the property that the GP distribution is strictly convex for $\xi > -1/2$ and that the EGP distribution closely behaves as the GP distribution in the right tail. It is therefore reasonable to consider the GP distribution as a suitable model for the data above this threshold. To validate this approach, Figure~\ref{fig:tudy_egpdvsgpd} presents a comparison between the EGP-fitted and GP-fitted densities above the thresholds. The two densities are remarkably similar; in the Port Tudy case, they are nearly indistinguishable, assessing the reliability of our method for selecting thresholds above which data could be considered in the extreme regime, i.e., above which the GP distribution is an appropriate model.

\begin{figure}[ht!] 
  \centering  \includegraphics[width=.315\textwidth]{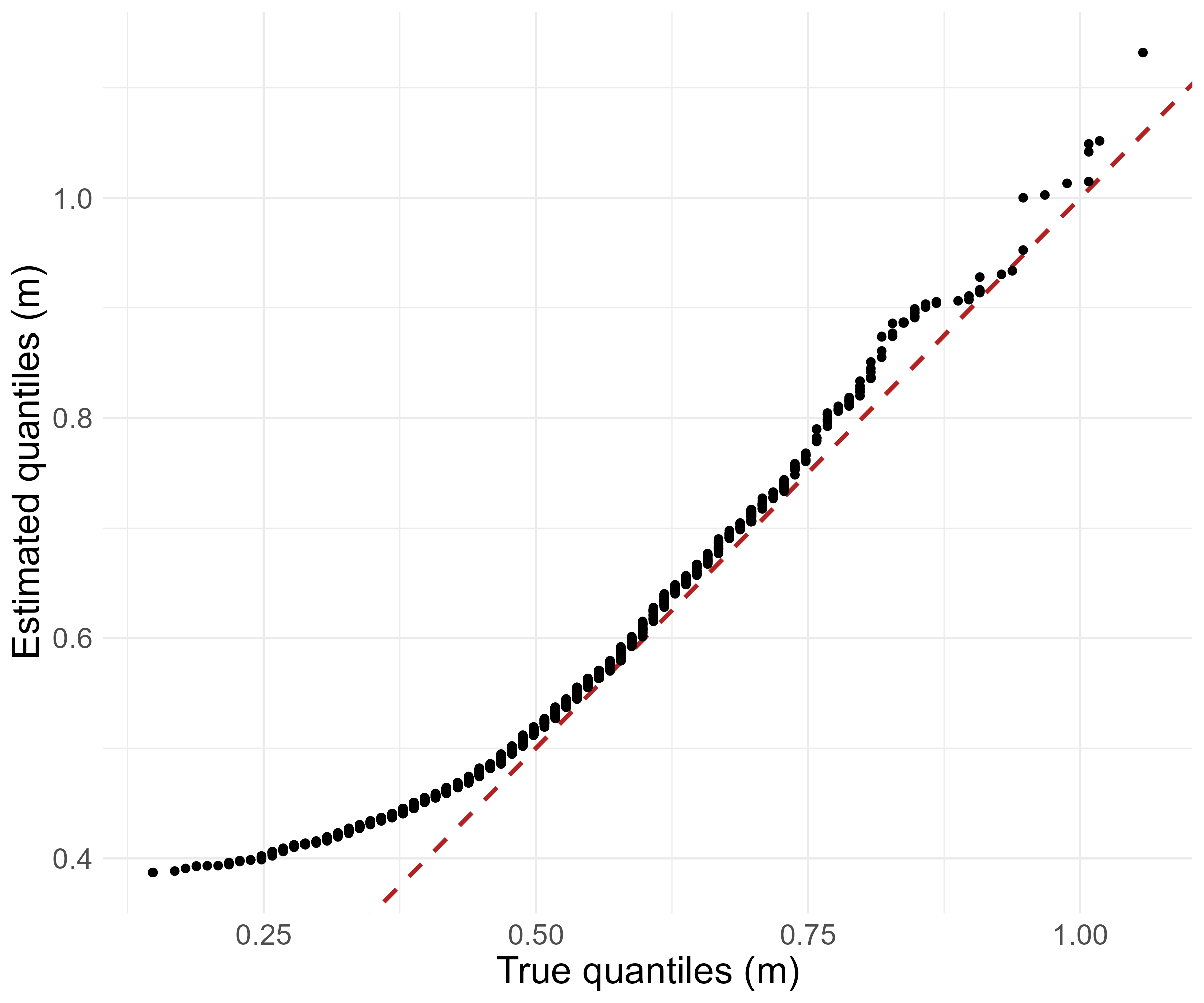}
  \hspace{0.2cm}
  \includegraphics[width=.315\textwidth]{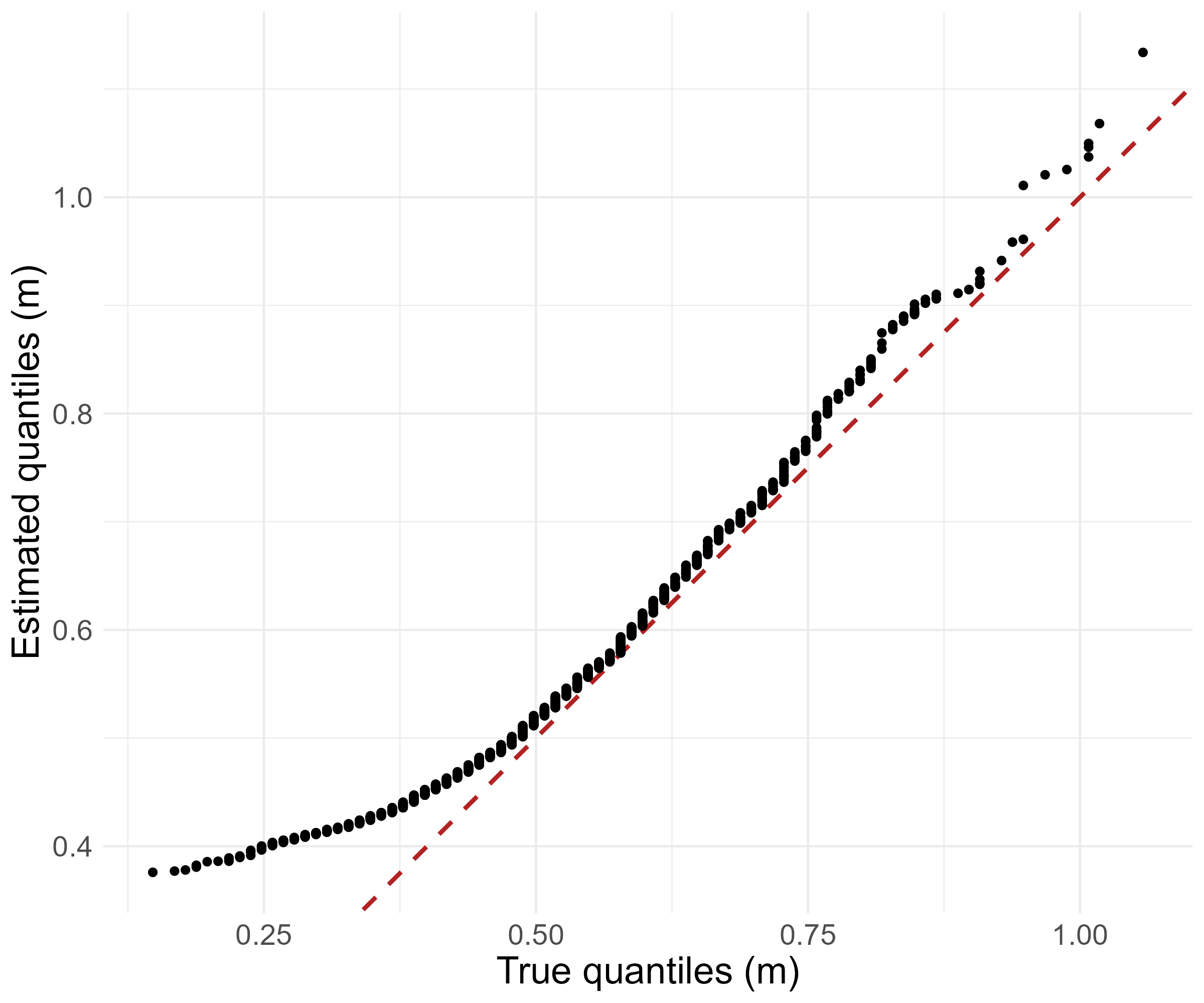}
  \hspace{0.2cm}
  \includegraphics[width=.315\textwidth]{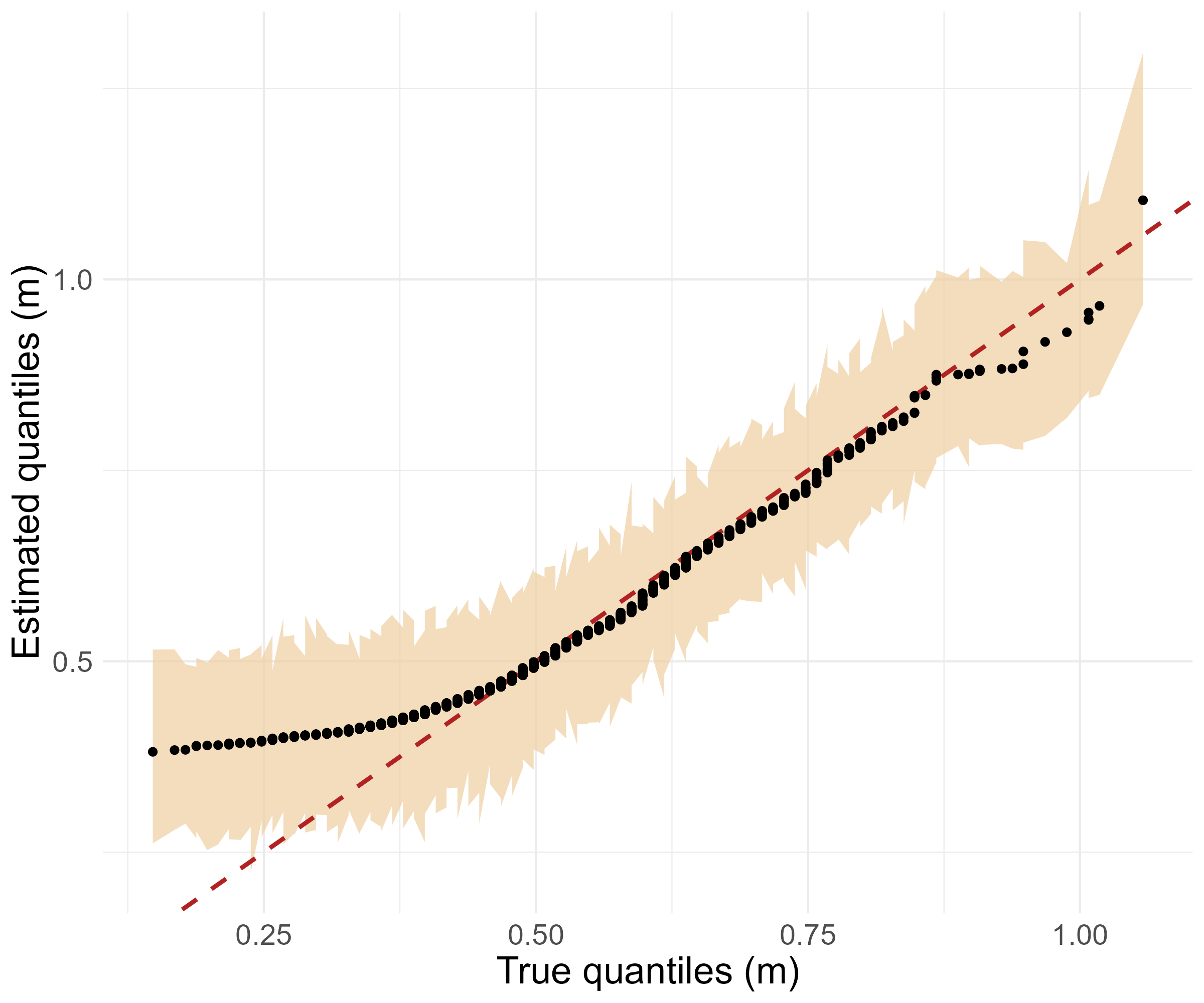}
\caption{QQ-plots comparing observed skew surge exceedances of the Port Tudy test set (x-axis), ranging from 10/08/1966 to 31/12/1999, to predicted data (y-axis) from the algorithms of Sections~\ref{sec:reg_proc} and \ref{sec:mgp_proc}. The plots show results from the ROXANE procedure with RF regression (left), ROXANE procedure with OLS regression (middle), and MGPRED (right) with 0.95-confidence bands (lightorange). The dotted red line represents the identity line $x=y$.\label{fig:qqplot_tudy_ss}}
\end{figure}

\begin{figure}[ht]
  \centering
  \includegraphics[width=.315\textwidth]{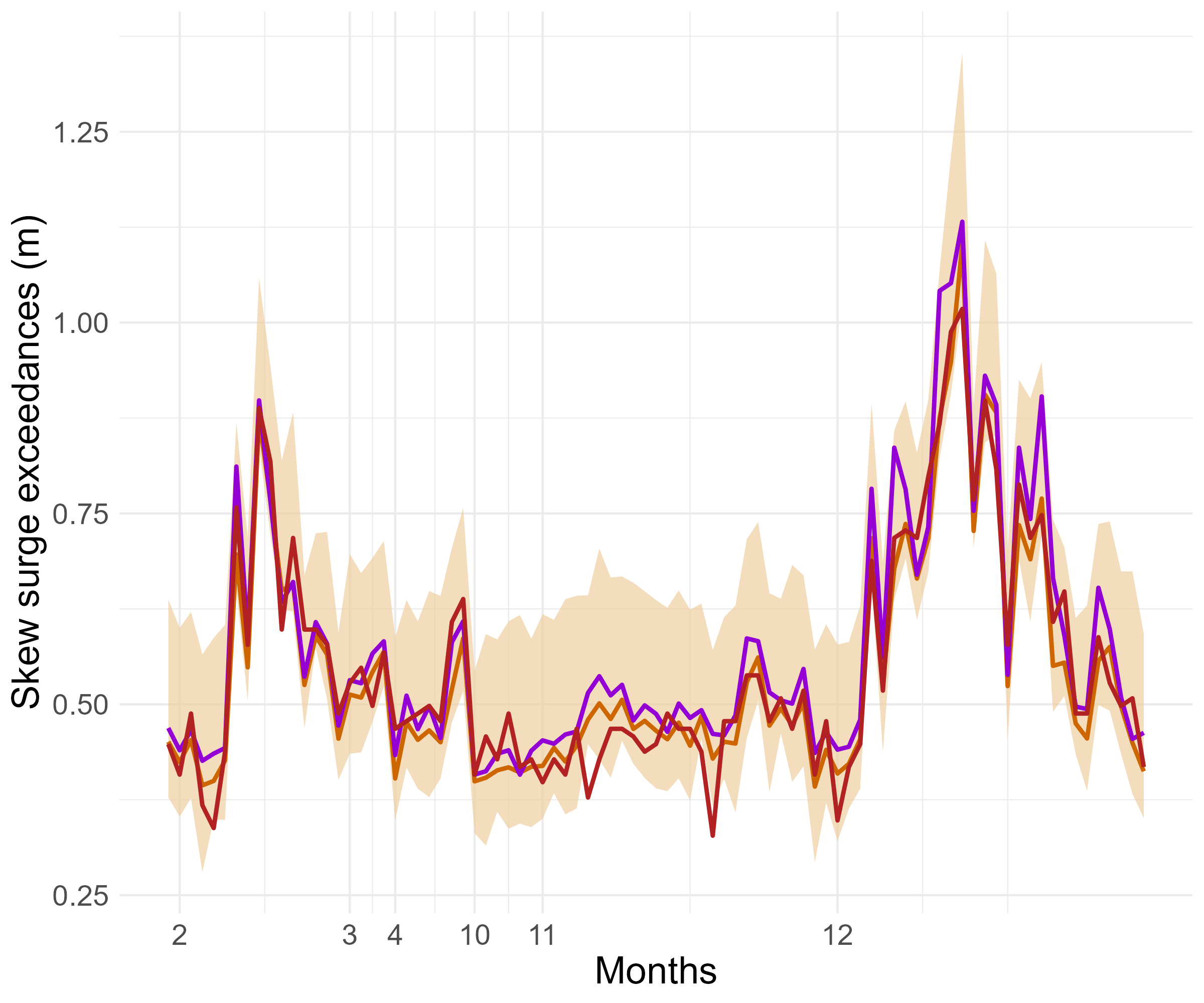}
  \hspace{0.2cm}
  \includegraphics[width=.315\textwidth]{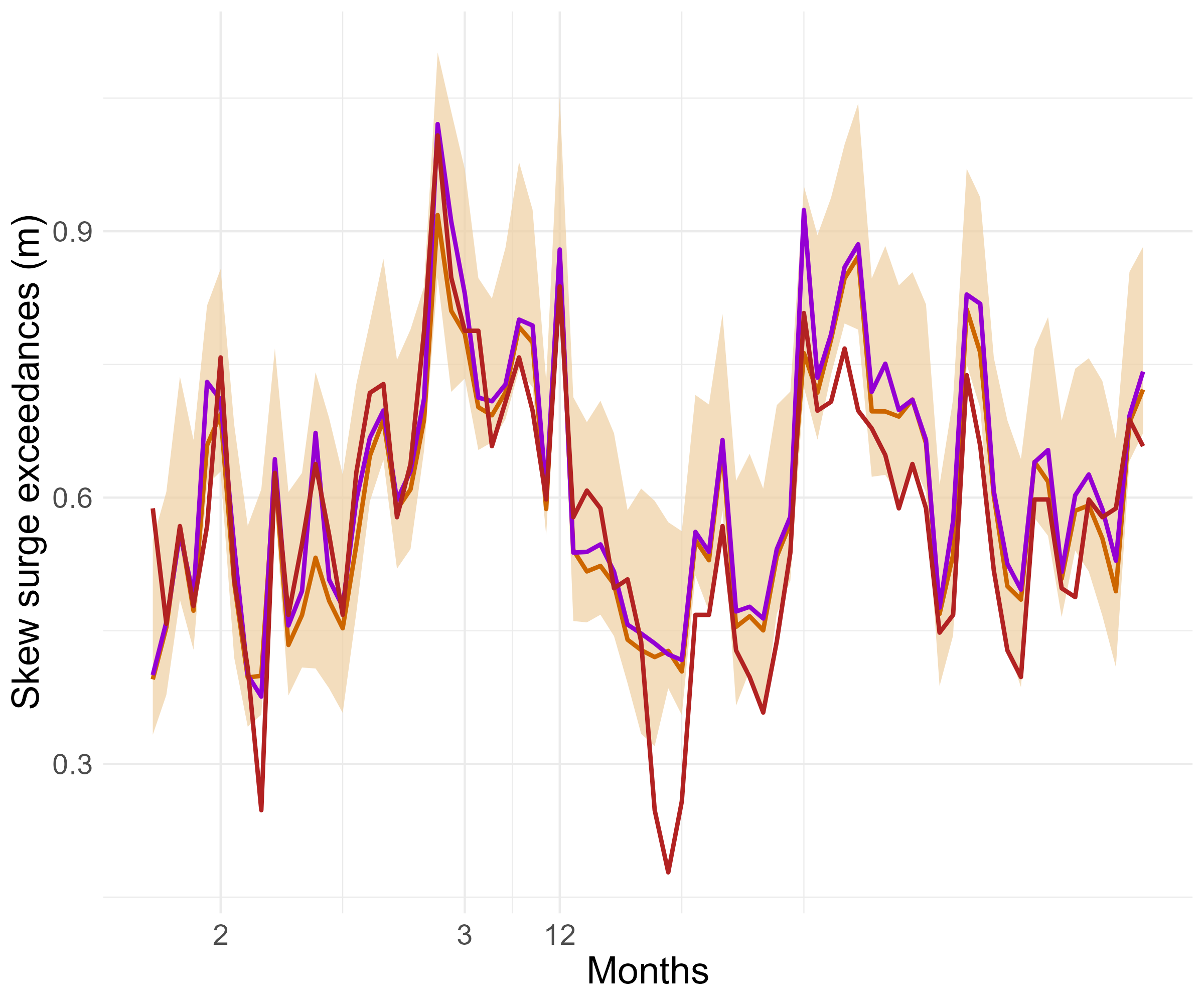}
  \hspace{0.2cm}
  \includegraphics[width=.315\textwidth]{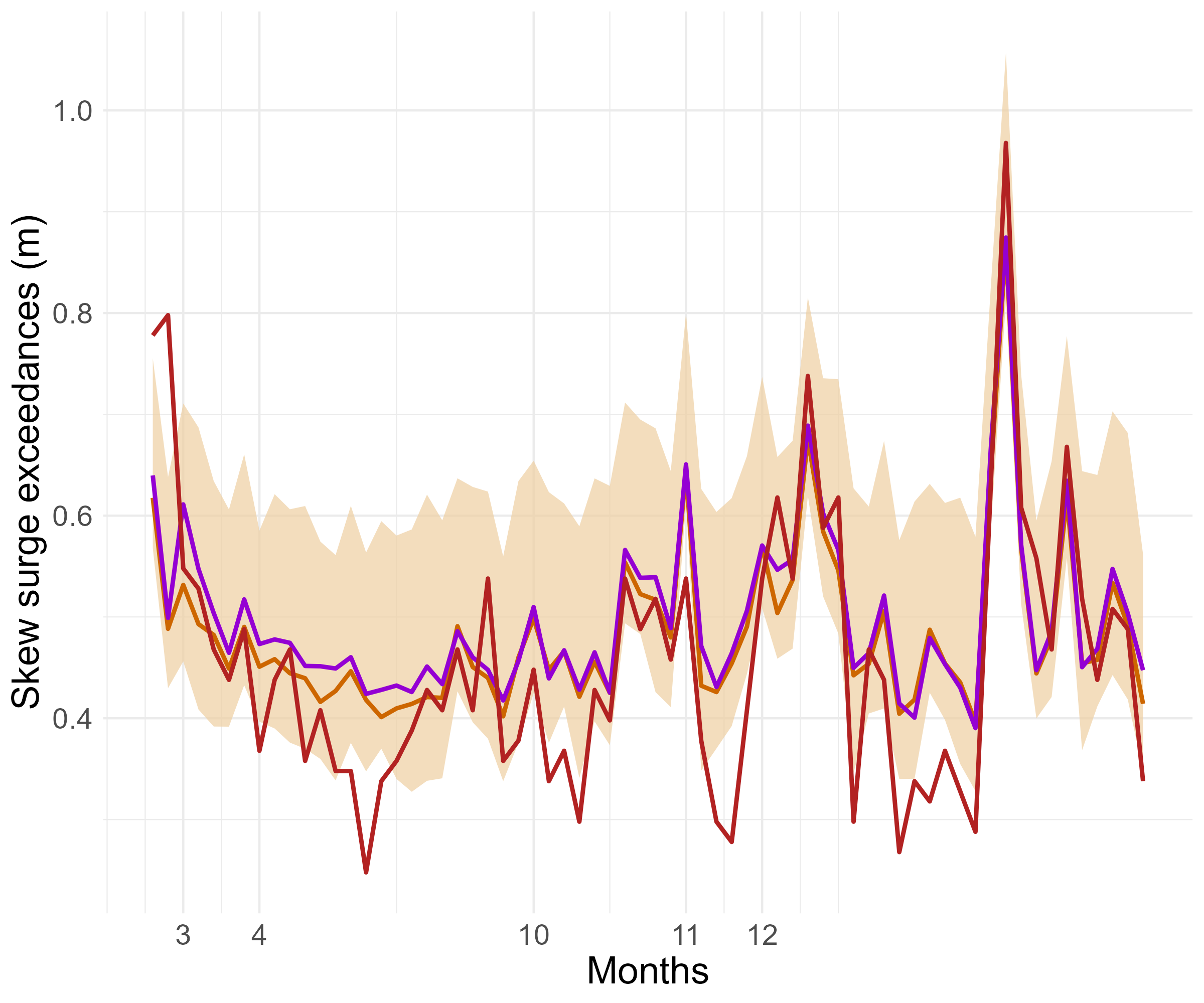}

  \caption{Predicted skew surge exceedances at Port Tudy station for the years 1989 (left), 1978 (middle), 1977 (right). Red curves represent the true values; purple curves represent the predicted values by the ROXANE procedure with OLS algorithm; orange curves represent the predicted values by MGPRED with 0.95-confidence bands (lightorange). 
  \label{fig:pred_tudy}}
\end{figure}

Regarding the joint models, both procedures exhibit reasonably good performance, as evidenced by Figures~\ref{fig:qqplot_tudy_ss} and~\ref{fig:pred_tudy}. According to the QQ-plots, the generated distributions closely mimic the behaviour of the observed distributions in the most extreme regions. 
For the left tail of the extreme observations, the QQ-plots in Figure~\ref{fig:qqplot_tudy_ss} indicate poor model performance. This issue is further illustrated in Figure~\ref{fig:pred_tudy}, where the largest misestimations occur for the smallest values. Several factors contribute to these shortcomings. First, in MGPRED, a censored likelihood is used to focus on accurately predicting the most extreme observations, at the expense of smaller ones. Second, common to both procedures, the estimation of the $\kappa$ parameter lacks robustness and is prone to significant overestimation. Although this parameter does not affect the most extreme observations, it can have a considerable impact on the smallest ones. Since the back-transformation $p^{-1}_{\sigma_Y,\xi_Y,\kappa_Y}$ behaves as $x^{1/\kappa}$ near zero, a small estimation error on Pareto scale could lead to an enormous estimation error on original scale. For instance, with $\kappa = 41.93$ (corresponding to $ \kappa_Y$ in Table~\ref{tab:egpd_tudy}), an error of just 0.01 on the Pareto scale results in a discrepancy of 0.90 meters on the original scale. Finally, another limitation common to both procedures is that an extreme observation may be extreme at Brest or Saint-Nazaire but not at the other two stations. Consequently, the models struggle to distinguish these cases from “typical” extreme events, where values at Port Tudy and another station are both large. This limitation could be addressed by incorporating wind-related data into the models, as wind strongly influences skew surges \citep{pugh2014sea}, and wind direction plays a crucial role in storm propagation and the formation of large skew surges. {\color{blue} Another way to mitigate these pathological cases would be to increase the threshold $\bm{t}$ so as to retain only the ‘truly extreme' observations. However, as in any extreme-value analysis, increasing the threshold generally improves model performance at the cost of higher uncertainty and reduced robustness, due to the resulting smaller sample size.}
Given these issues with the lower tail of the extreme distributions, the performance of both models should be evaluated primarily for the largest observations, which aligns with the goal of EVT. 


Figure~\ref{fig:pred_tudy} also shows that our models are less precise for older years. One possible reason is the failure of our models to account for the time trend due to, e.g., global warming \citep{seneviratne2021weather}. Although we assume in our study that this trend is present for all three stations and can be ignored, if the trend behaviour differs across stations, it must be considered.

To emphasise the respective strengths of both procedures, a comparison is essential. We assess the performance of our models using the Root Mean Squared Error (RMSE) (and a Squared Standard Error (SSE)\footnote{SSE $=\frac{1}{\sqrt{n}}\Big(\frac{1}{n}\sum_{i=1}^n \Big((Y_i-\hat{Y}_i)^2-\big(\frac{1}{n}\sum_{j=1}^n(Y_j-\hat{Y}_j)^2\big)^2\Big)\Big)^{1/4}$.}) and Mean Absolute Error (MAE) (and a Absolute Standard Error (ASE)\footnote{ASE$=\frac{1}{\sqrt{n}}\Big(\frac{1}{n}\sum_{i=1}^n \Big(|Y_i-\hat{Y}_i|-\big(\frac{1}{n}\sum_{j=1}^n|Y_j-\hat{Y}_j|\big)^2\Big)\Big)^{1/2}$.}) as measures of predictive performance. Table~\ref{tab:tudy} contains the errors computed on the entire test set and on its most extreme half w.r.t. the Port Tudy observations, defined as the observations for which $Y_i$ exceeds its empirical median computed on the extreme test set. Overall, the MGPRED procedure performs better: despite using a censored likelihood criterion, it models the smallest values more accurately than the ROXANE procedure. However, the ROXANE procedure, with its OLS predictive algorithm, performs better in the most extreme regions. This tendency is generally observed in sea level studies and when Concarneau and Le Crouesty are used as output stations (see Tables~\ref{tab:tudy_sl}, \ref{tab:concarneau}, and \ref{tab:lecrouesty} in the Appendix).
Figure~\ref{fig:qqplot_tudy_comparison} presents QQ-plots comparing predicted values from different models. All three routines produce values of the same order, with larger values estimated by the ROXANE algorithm in the most extreme regions, which appear to better match the true ones, as shown in Figure~\ref{fig:qqplot_tudy_ss}. All predicted values from the ROXANE algorithm fall within the 0.95-prediction confidence interval derived from MGPRED.

{\color{blue} In light of the above discussion, for predicting extremes in contexts similar to ours, we recommend using the MGPRED procedure to achieve better overall performance, along with prediction intervals, while the ROXANE procedure should be preferred when more accurate predictions of the most extreme observations are required. Regarding the regression algorithm used within the ROXANE framework, general guidelines apply. For instance, the OLS algorithm is suitable when the data exhibit linear relationships, whereas RF or Support Vector Regression algorithms may be more appropriate in high-dimensional or nonlinear settings.

The construction of prediction intervals remains a hot topic in the machine learning community \citep[see, e.g.,][]{sluijterman2024evaluate,tyralis2024review}. Even if several approaches exist, such as bootstrap-based, Bayesian, or conformal inference methods, they usually rely on strong distributional assumptions or asymptotic arguments that are difficult to justify in our case study. For these reasons, we leave the question of how to construct fully justified prediction intervals to further research, while we illustrate the parameter sensitivity of our model through Figures~\ref{fig:stability_margins_ss},~\ref{fig:stability_margins_sl},~\ref{fig:stability_rl_ss}, and~\ref{fig:stability_rl_sl} in the Appendix, thus providing heuristic insights on the model uncertainty and robustness.
}
\begin{table}[ht!]
\caption{RMSE$\times10^2$(SSE$\times10^2$) and MAE$\times10^2$(ASE$\times10^2$) of predicted skew surge exceedances at Port Tudy station from the ROXANE procedure with RF regression (ROX RF), ROXANE procedure with OLS regression (ROX OLS) and MGPRED. Errors are measured on the test set covering the period from 10/08/1966 to 31/12/1999. Errors are computed on the entire test set (columns: RMSE and MSE) and on its most extreme half, i.e., for observations with $Y_i$ exceeding their empirical median computed on the extreme test set (columns: $\mbox{RMSE}_{\mbox{EXT}}$ and $\mbox{MAE}_{\mbox{EXT}}$). \label{tab:tudy}}
\vspace{0.2cm}
\begin{center}
\begin{small}
\begin{sc}
\begin{tabular}{ccccccr}
\toprule
Training models/Errors & RMSE & MAE & $\mbox{RMSE}_{\mbox{ext}}$ & $\mbox{MAE}_{\mbox{ext}}$ \\
\midrule
ROX RF & 8.4(0.3) & 6.1(0.1) & \textbf{7.1(0.3)} & \textbf{5.1(0.1)}\\
ROX OLS & 8.4(0.3) & 6.1(0.1) & 7.2(0.3) & 5.2(0.1) \\
MGPRED & \textbf{7.5(0.2)} & \textbf{5.5(0.1)} & 7.3(0.3) & 5.4(0.1)\\
\bottomrule
\end{tabular}
\end{sc}
\end{small}
\end{center}
\end{table}

\begin{figure}[ht!] 
  \centering
  \includegraphics[width=.315\textwidth]{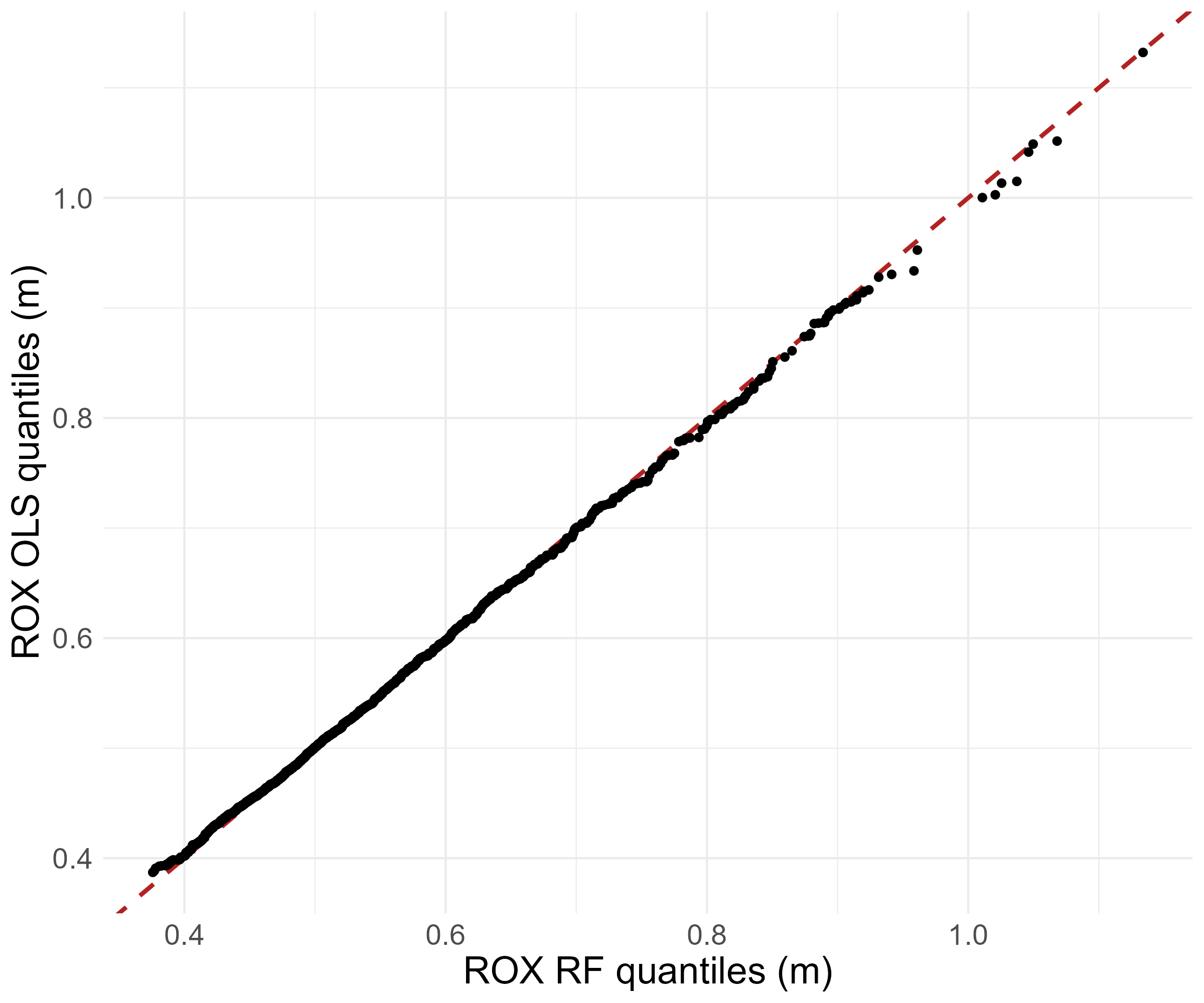}
  \hspace{0.2cm}
  \includegraphics[width=.315\textwidth]{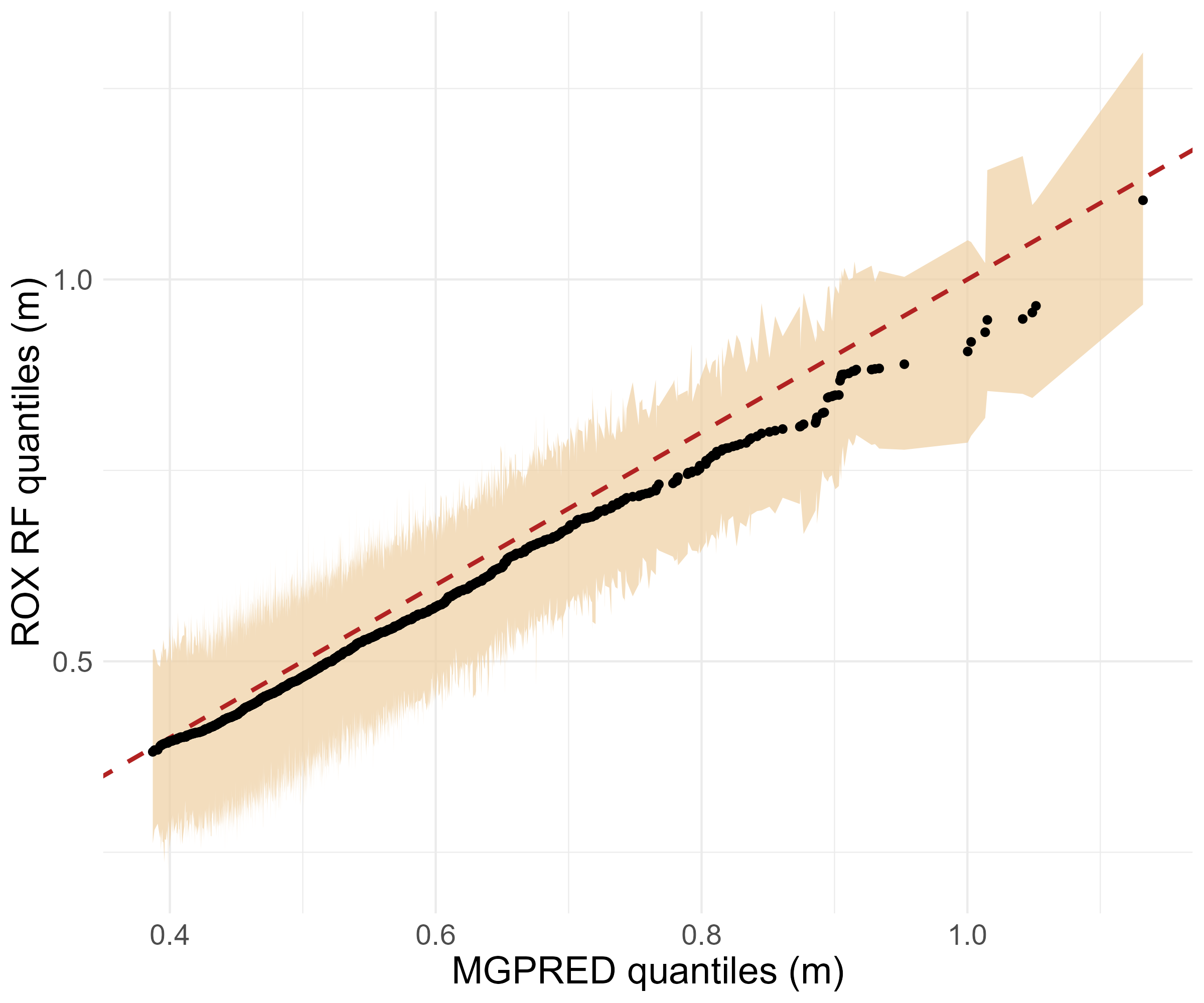}
  \hspace{0.2cm}
  \includegraphics[width=.315\textwidth]{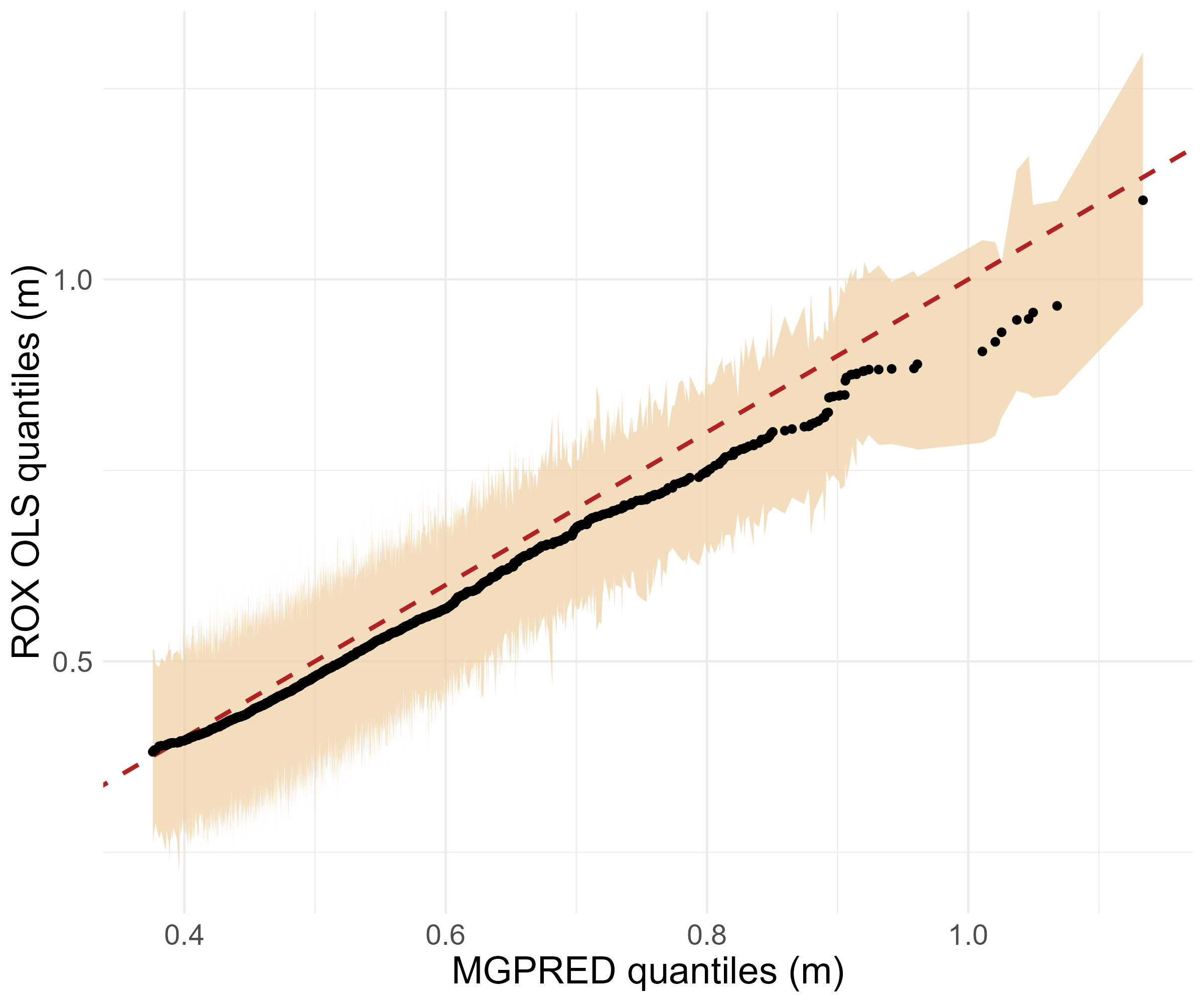}
\caption{QQ-plots comparing predicted skew surge exceedances of the Port Tudy test set, ranging from 10/08/1966 to 31/12/1999, from the algorithms of Sections~\ref{sec:reg_proc} and \ref{sec:mgp_proc}: ROX RF \emph{vs} ROX RF (left), MGPRED \emph{vs} ROX RF (middle), and MGPRED \emph{vs} ROX OLS (middle), with 0.95-confidence bands (lightorange) from MGPRED. The dotted red line represents the identity line $x=y$.\label{fig:qqplot_tudy_comparison}}
\end{figure}













\subsection{Extreme skew surges time series reconstruction} \label{sec:recons}
This section presents the results of reconstructing extreme skew surge time series at the Port Tudy tide gauge prior to its deployment on 10 August 1966. The reconstruction uses historical data from the Brest and Saint-Nazaire tide gauges, the latter of which has recorded measurements since 18 January 1863. New predictive models were trained using observations collected during the overlapping operational period of the three tide gauges, spanning from 10 August 1966, to 31 December 2023. This period corresponds to what was previously designated as the training and test sets into a unified training dataset. The training procedures described in the sections~\ref{sec:res_marg} and~\ref{sec:mgpd_results}, as well as preceding sections, are strictly applied.

\begin{figure}[hb!] 
  \centering  \includegraphics[width=.45\textwidth]{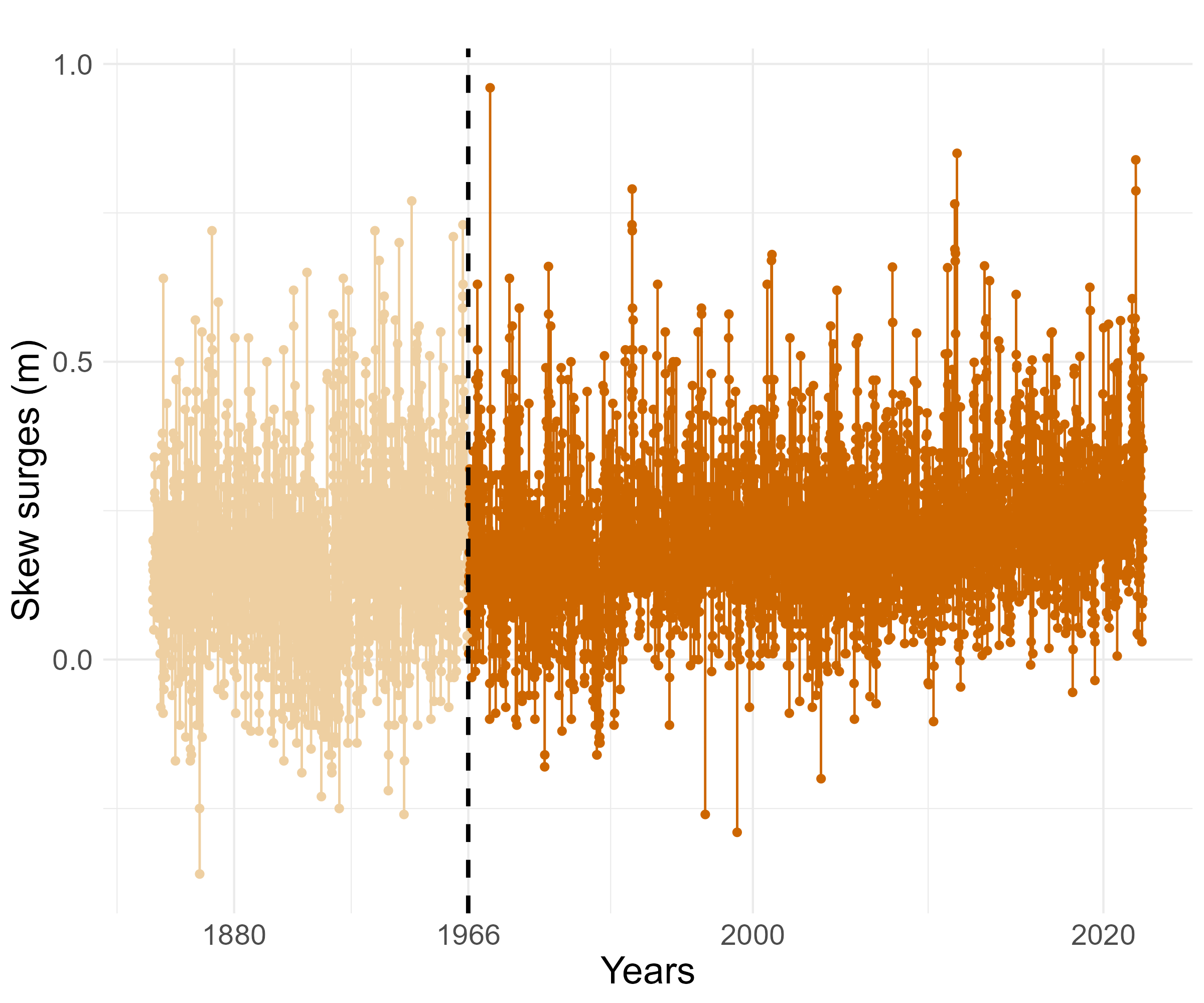}
  \hspace{0.7cm}
  \includegraphics[width=.45\textwidth]{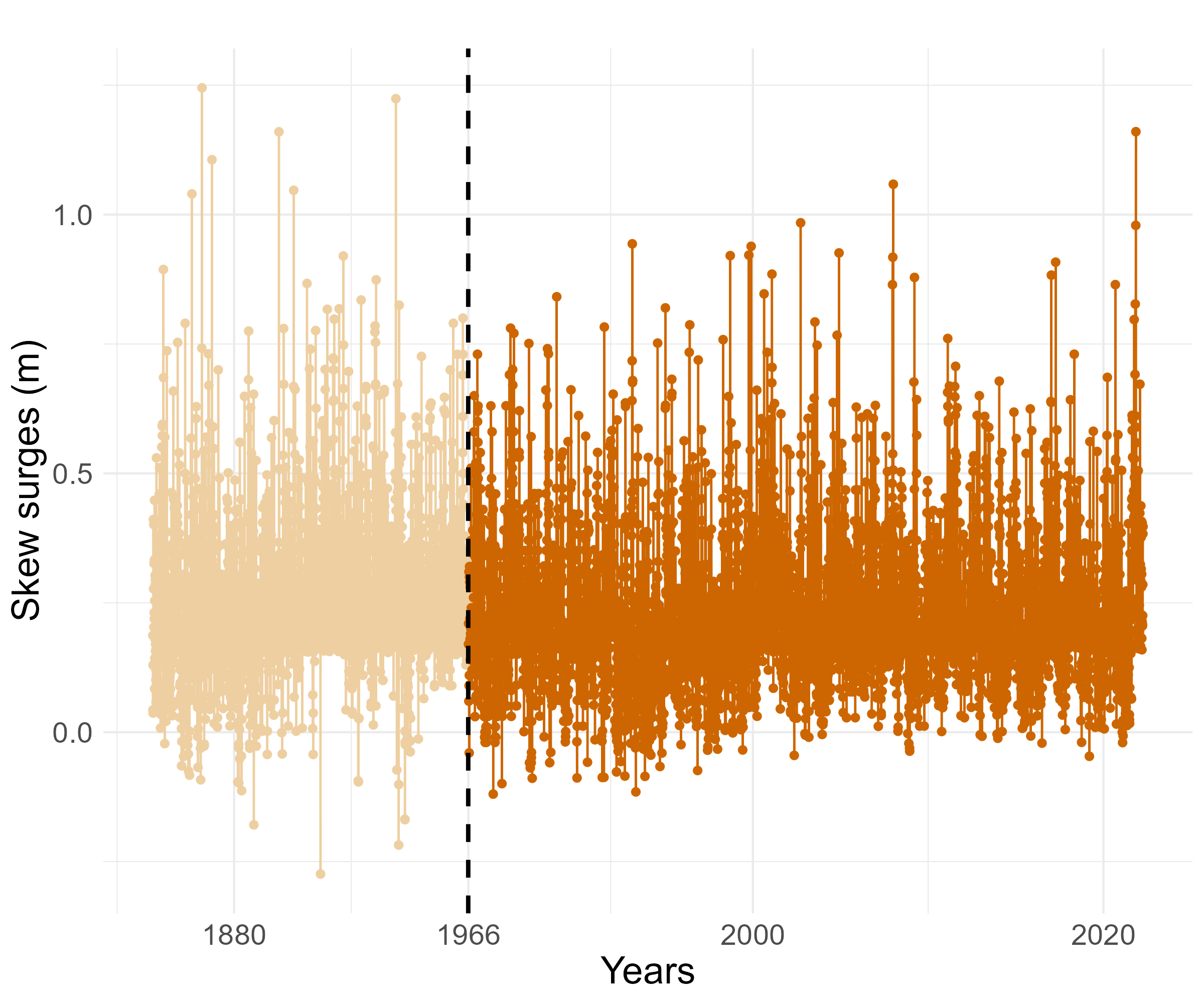}
\caption{Extreme skew surge time series at Brest (left) and Saint-Nazaire (right). The dotted vertical black line represents the deployment date of the tide gauge at Port Tudy (10/08/1966). The dark orange points represent values after 10/08/1966, used for training the prediction models; the light orange points represent values before 10/08/1966, corresponding to the reconstruction period of the Port Tudy time series. \label{fig:recons_ts}}
\end{figure}

Figure~\ref{fig:recons_ts} illustrates the extreme skew surge time series at Brest and Saint-Nazaire, distinguishing between the training and reconstruction periods. Although the reconstruction period is longer than the training period, it contains fewer data points, likely due to gaps in historical records at Brest and Saint-Nazaire before 1966. Specifically, the average number of annual measurements prior to August 10, 1966, is 319, compared to 539 per year thereafter. 
As in the experiments presented in the previous sections, a positive temporal trend is removed, to end up with stationary data. However, no discernible trend is observed for the most extreme skew surges, as evidenced by the Saint-Nazaire time series in Figure~\ref{fig:recons_ts}, in which the largest observations were recorded before 1966.

\begin{figure}[ht!] 
  \centering
  \includegraphics[width=.315\textwidth]{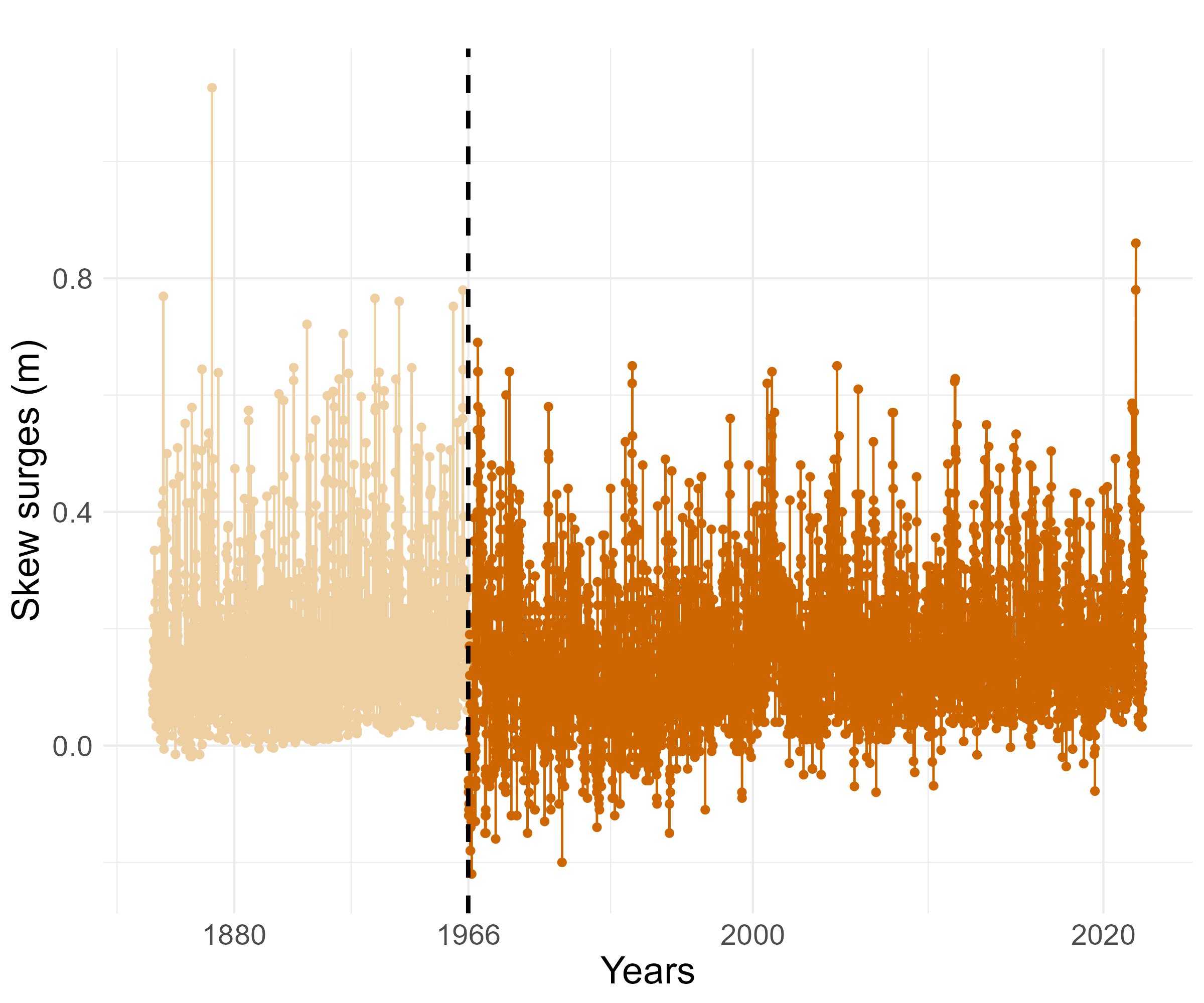}
  \hspace{0.2cm}
  \includegraphics[width=.315\textwidth]{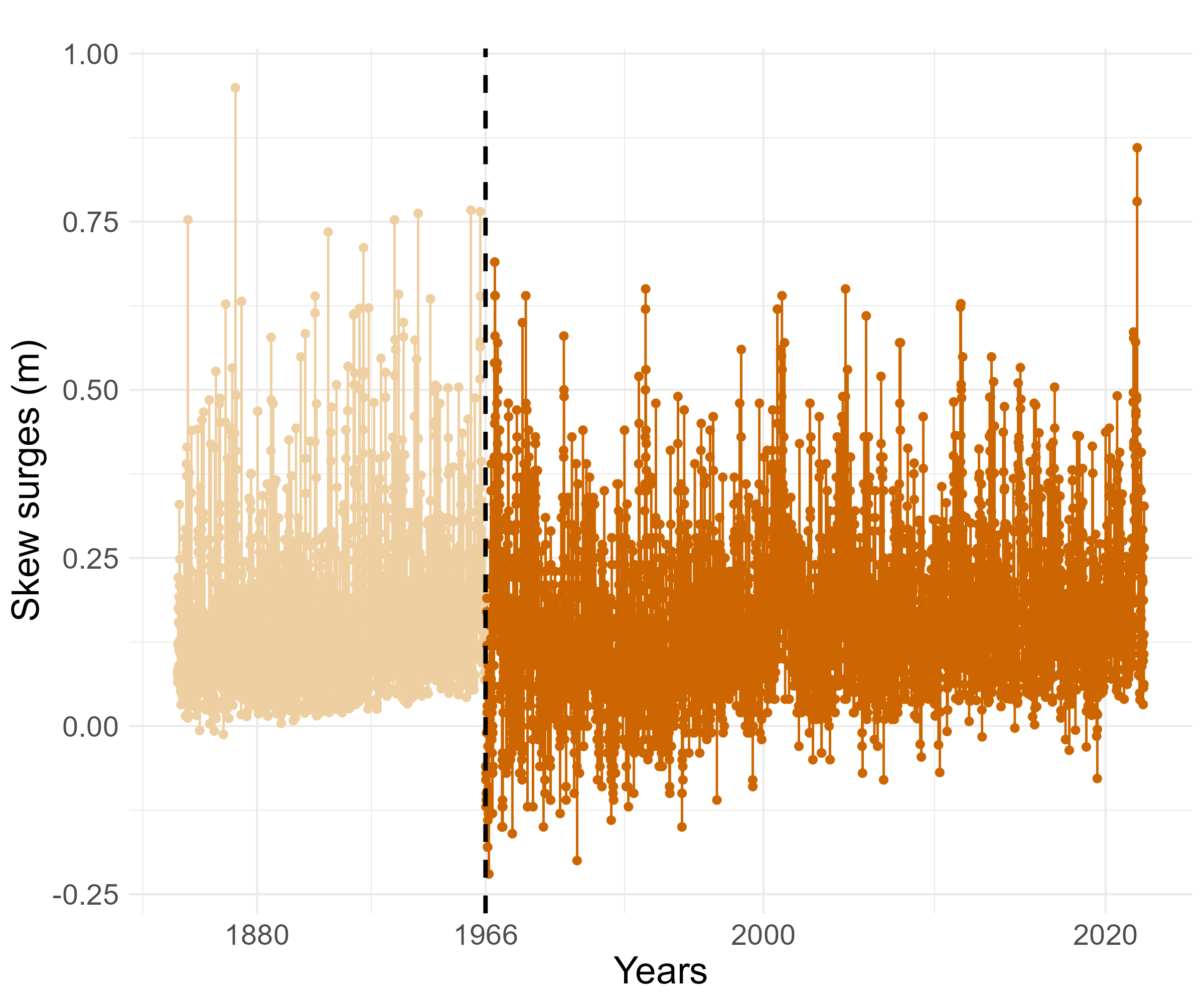}
  \hspace{0.2cm}
  \includegraphics[width=.315\textwidth]{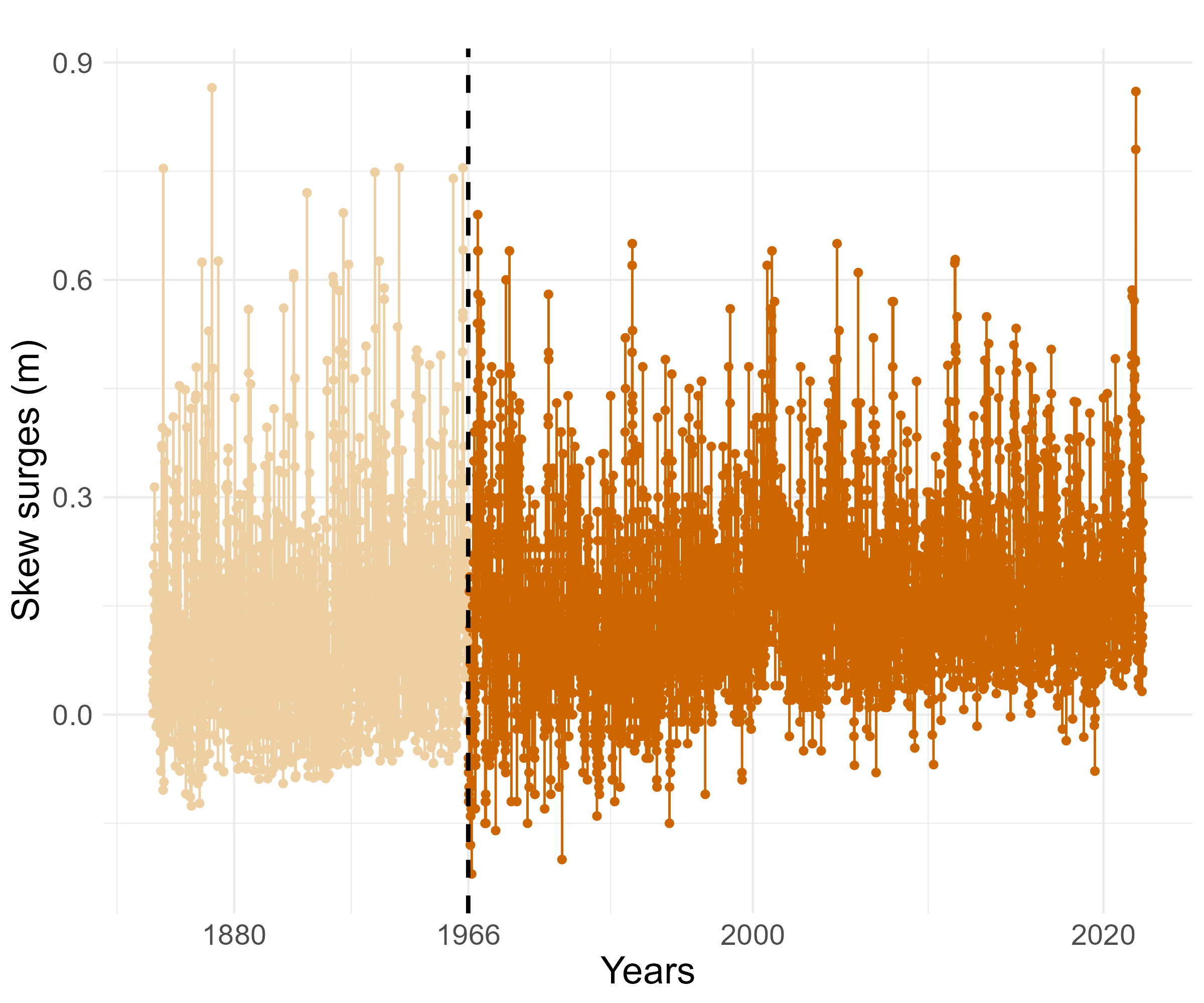}
\caption{Extreme skew surge time series at Port Tudy. The dotted vertical black line represents the deployment date of the tide gauge at Port Tudy (10/08/1966). The dark orange points represent values after 10/08/1966, used for training the prediction models; the light orange points represent the predicted values by the ROXANE procedure with OLS (left), RF (middle) algorithms and by MGPRED (right) before 10/08/1966.\label{fig:ts_reconstrution}}
\end{figure}

\begin{figure}[hb!] 
  \centering
\includegraphics[width=.45\textwidth]{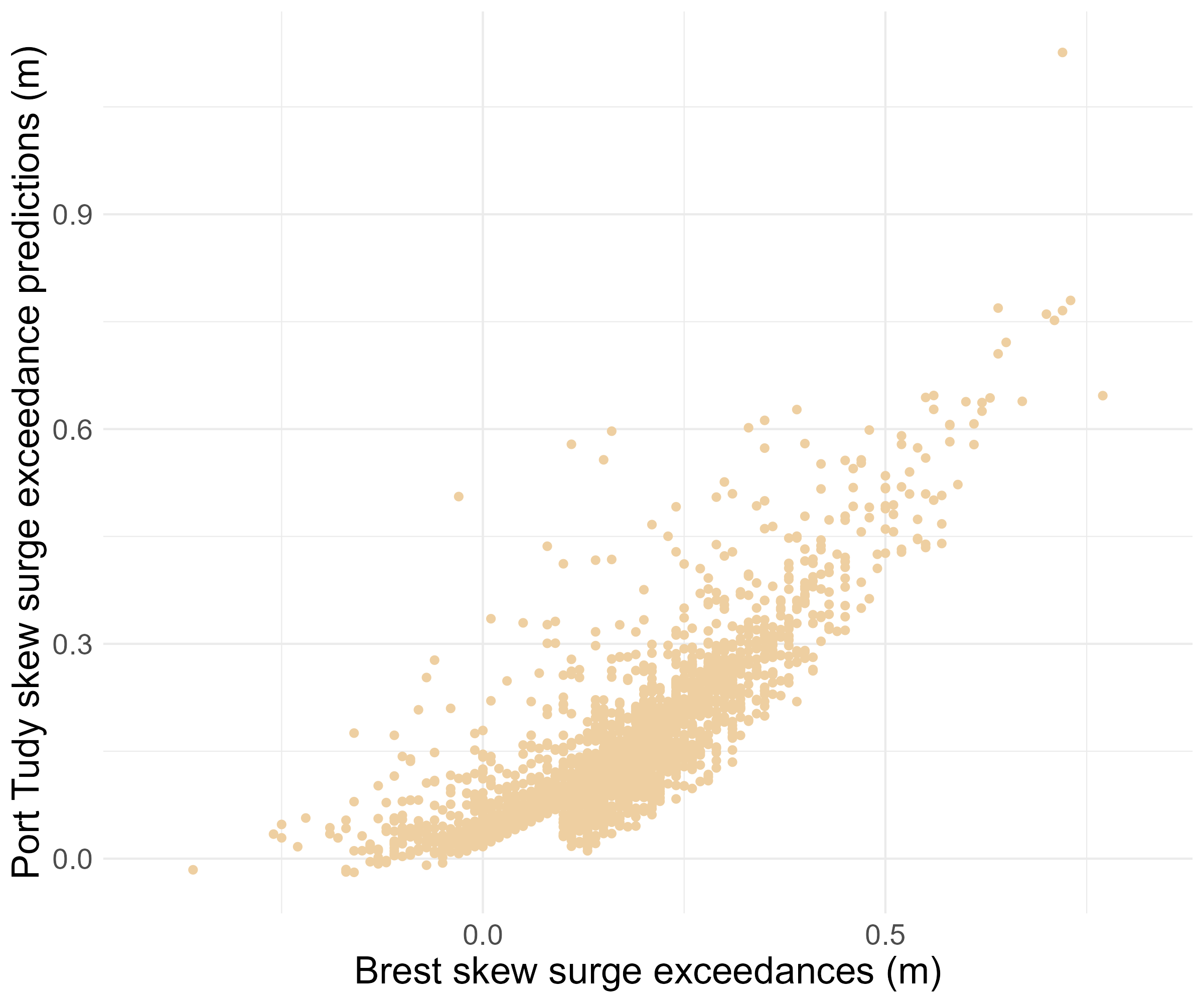}
  \hspace{0.7cm}
  \includegraphics[width=.45\textwidth]{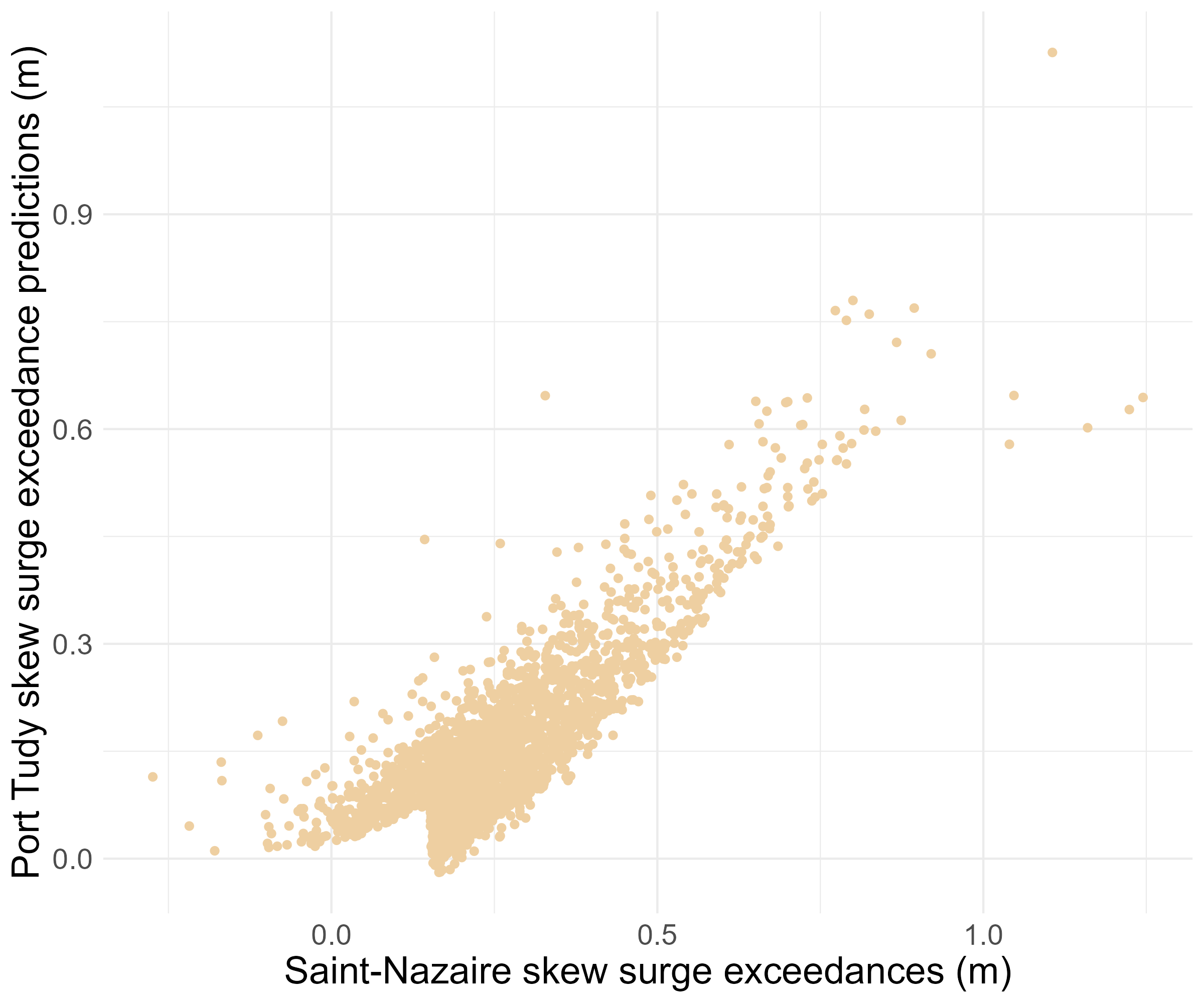}
  
    \includegraphics[width=.45\textwidth]{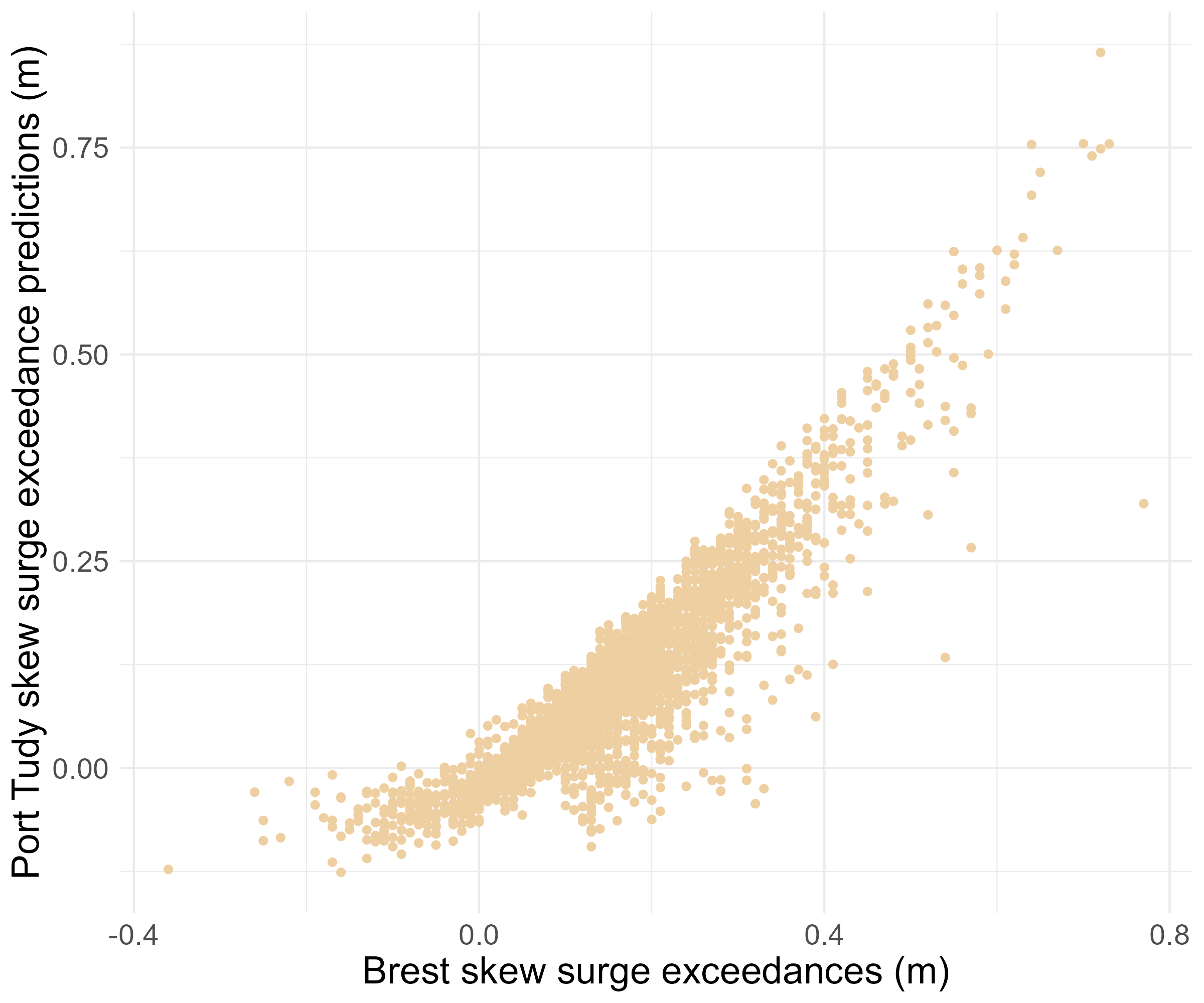}
  \hspace{0.7cm}
  \includegraphics[width=.45\textwidth]{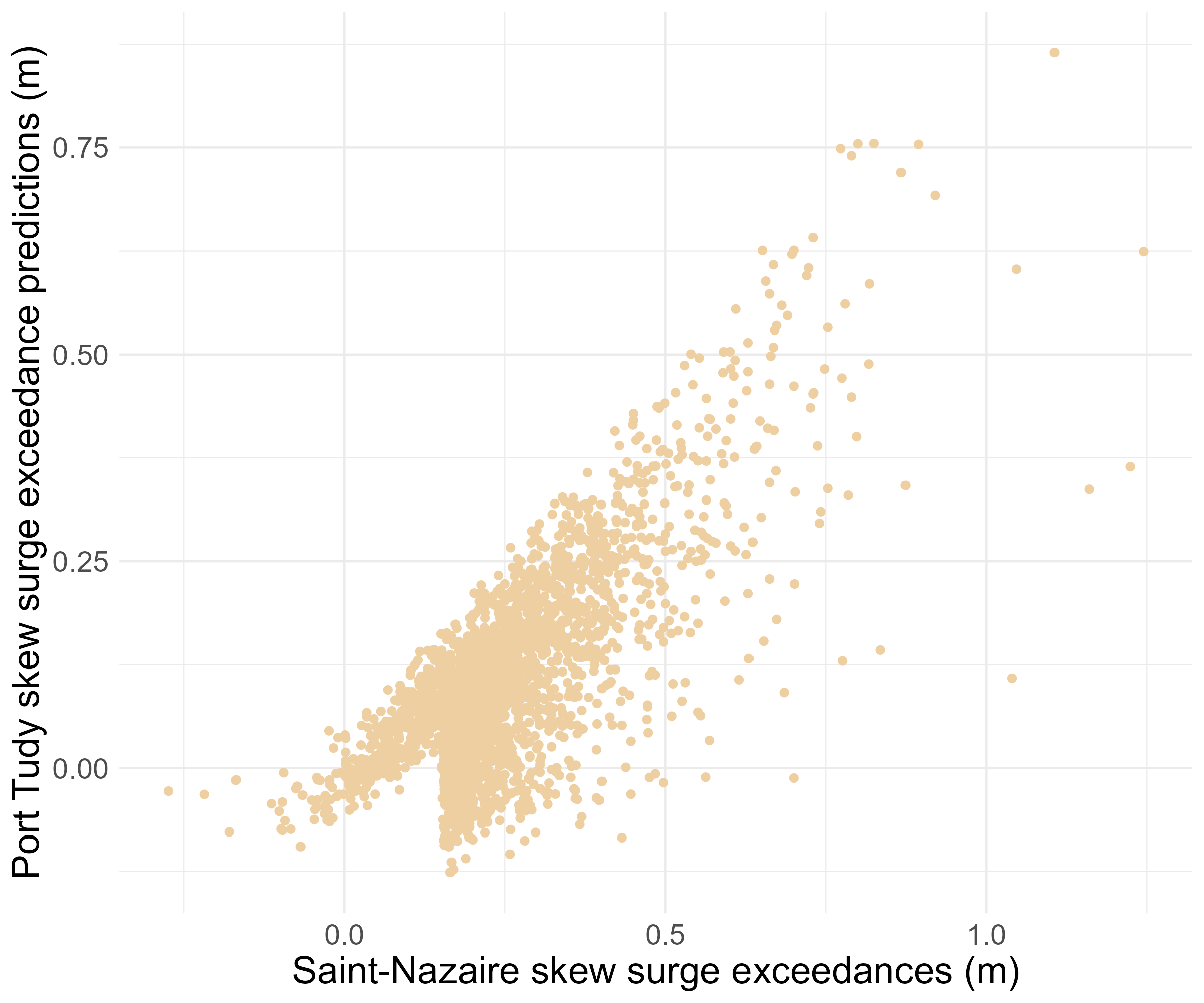}
  
\caption{Predicted skew surges (y-axis), by the ROXANE procedure with OLS algorithm (top row) and the MGP procedure (bottom row), at Port Tudy before 10/08/1966 against the true values (x-axis) at Brest (left) and at Saint-Nazaire (right). The black dotted lines represent the chosen marginal threshold via Algorithm~\ref{algo:th_select}.\label{fig:scatter_recons}}
\end{figure}

Figure~\ref{fig:ts_reconstrution} presents the predicted time series generated by the ROXANE routine using both the OLS and RF algorithms, and by MGPRED. All methods exhibit substantial limitations in predicting smaller extreme values, as discussed in Section~\ref{sec:discussion}, with only a few slightly negative skew surge predictions. However, the estimated largest skew surges align with observed values in the training set and with the input data from the reconstruction period, as illustrated in Figures~\ref{fig:recons_ts} and~\ref{fig:ts_reconstrution}. This consistency in predicting extreme values is further corroborated by Figure~\ref{fig:scatter_recons}, which depicts pairwise scatter plots of predictions against observed input values, revealing patterns similar to those observed in the training set scatter plot (Figure~\ref{fig:scatterplot}).

Consistent with previous findings, the ROXANE procedure tends to predict larger extreme surges than MGPRED. The largest skew surges predicted by both models coincide with significant skew surges recorded at Brest and/or Saint-Nazaire at the same time. In particular, the largest skew surge predicted by both models occurs during the night of 31 December 1876, to 1 January 1877, corresponding to two of the highest skew surges ever recorded at Brest and Saint-Nazaire. For the ROXANE procedure using the OLS algorithm, this value appears to be considerably overestimated. Nevertheless, such a large value was to be expected, since a major storm struck that night, causing massive flooding along the southern coast of Brittany and significant material losses \citep[see the report of ][]{storm1887}.
A more comprehensive evaluation of the reconstruction accuracy could be achieved through a detailed examination of historical meteorological records and archives at Port Tudy to identify correspondences between notable climatic events and predicted extreme surges. However, such an in-depth analysis lies beyond the scope of the present study and is left for future research.

\clearpage

\section{Conclusion and Perspectives}\label{sec:ccl}
Accurately reconstructing extreme skew surge time series is essential for refining the computation of long-range return periods, particularly at tide gauges with limited observational data. Leveraging robust theoretical results from extreme value theory, we explore a nonparametric and a parametric multivariate methods to address the challenge of predicting sea levels and skew surges measured at tide gauges along the French Atlantic coast. These methods are comprehensively compared and analysed, demonstrating that both yield valid and practically significant results. Each approach offers distinct advantages: one excels in providing precise point estimates, while the other delivers confidence intervals for the estimates and a robust generative model. Additionally, through the use of the EGP distribution for marginal modelling, we introduce an innovative and simple method for selecting a univariate threshold above which observations belong to an extreme regime.

The proposed methods, including the marginal modelling and both multivariate prediction techniques, are broadly applicable to other contexts involving spatial extremes.
Furthermore, these approaches can be extended to incorporate additional covariates to enhance prediction accuracy. For instance, the ROXANE procedure is compatible with any regression algorithm, including high-dimensional models like Random Forest regression. Incorporating wind-related measurements at input stations could address the challenges associated with estimating smaller extreme observations. {\color{blue} With such extensions, it could also be possible to adapt our methods to the task of missing data imputation. Another possible application of the methodology proposed in this work concerns the reconstruction, over periods when systematic meteorological measurements were not yet available, of other covariates such as wind or pressure. These covariates are typically strongly correlated with sea levels and skew surges, and the problem of reconstructing pressure data given sea-level data has already been successfully addressed in \cite{platzer2024could}. A natural application and extension of our work would be to use the reconstructed sea-level and skew surge data in extreme regimes with the EVT models described in this paper, in order to improve the reconstruction of  historical records of wind and pressure associated with extreme sea events. Finally, a promising avenue for future research would be to develop statistically sound methods dedicated to the construction of prediction (confidence) intervals in a context where covariates are extreme.}
These possible extensions are left to further research.


\section*{Acknowledgments}
We thank Dr. Gael Andr\'e (SHOM)  and  Dr. Xavier Kerdadallan  (CEREMA) or their insightful discussions on sea levels and tides, and in particular for the guided visit to one of the world’s oldest tide gauges located in Brest, France.

\section*{Funding}
Part of Anne Sabourin's work was funded by ANR EXSTA grant ANR-23-CE40-0009-01.
Part of Naveau’s research work was supported by the   Agence Nationale de la Recherche via:  the SICIM and SHARE PEPR Maths-Vives project (France 2030 ANR-24-EXMA-0008), EXSTA grant (ANR-23-CE40-0009-01), PORC-EPIC, the PEPR TRACCS program  (PC4 EXTENDING, ANR-22-EXTR-0005), and  the PEPR   IRIMONT (France 2030 ANR-22-EXIR-0003). He has also benefited  from the Geolearning research chair.

\section*{Data and code  availability}
The dataset, alongside with the code developed for the proposed methodology, is available at \url{github.com/HuetNathan/extremesealevels}.
 
\clearpage

\bibliographystyle{abbrvnat}
\bibliography{bibliowave}


\clearpage

\appendix 
\begin{center}
{\large\bfseries Appendix}
\end{center}

The appendix is organized as follows: Section~\ref{sec:preprocess} provides a detailed explanation of the preprocessing steps outlined in Section~\ref{sec:data}. Section~\ref{sec:appexdix_th} presents a proof of the threshold formula used in Algorithm~\ref{algo:th_select}. Section~\ref{sec:appexdix_dens} compiles details and closed-form expressions for the density models employed in MGPRED. Finally, Section~\ref{sec:add_stud} presents two additional analyses, analogous to those in Section~\ref{sec:results}, using sea level and skew surge data at Concarneau and Le Crouesty as the output stations, along with an additional study using sea level data at Port Tudy as the output.

\section{Preprocessing}\label{sec:preprocess}
\begin{figure}[ht!]
  \centering
  \includegraphics[width=.4\textwidth]{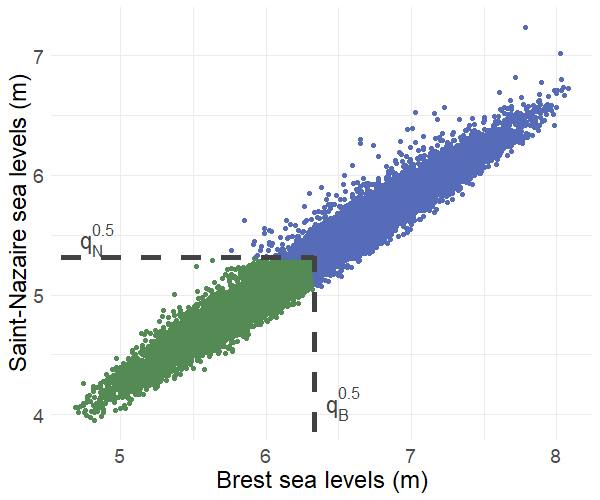}
  \hspace{1cm}
  \includegraphics[width=.4\textwidth]{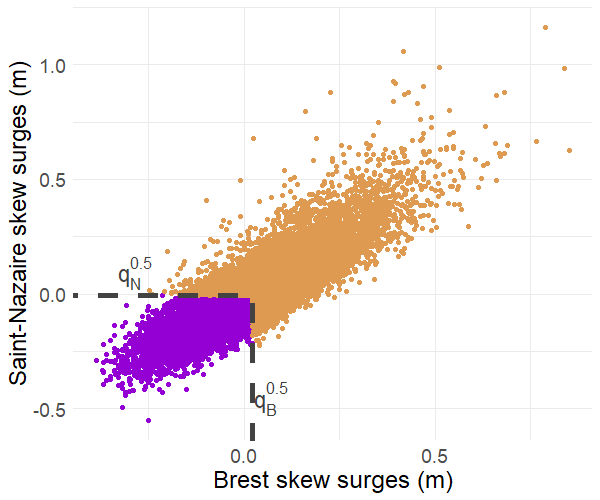}

  \vspace{1cm}

  \includegraphics[width=.4\textwidth]{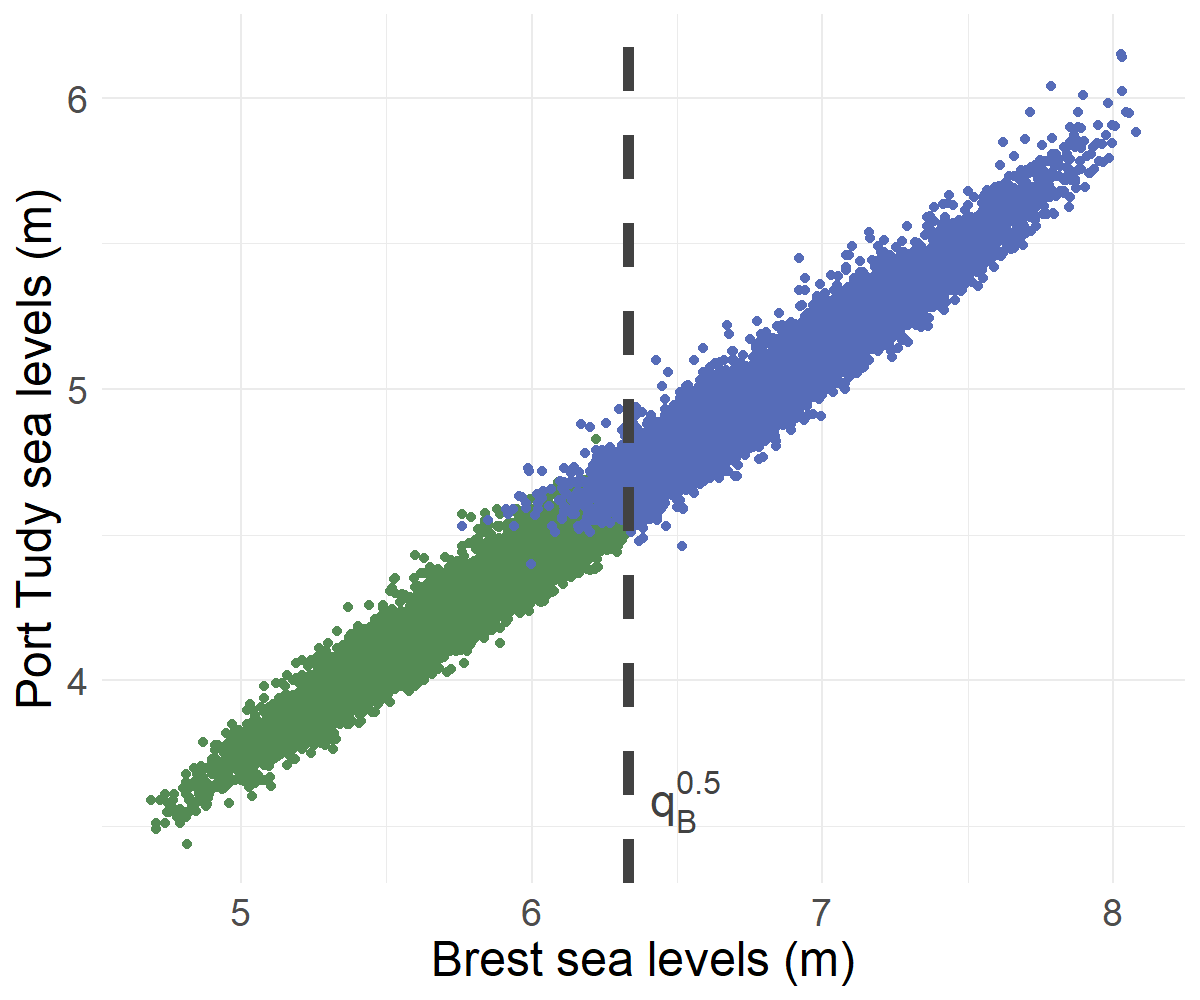}
  \hspace{1cm}
  \includegraphics[width=.4\textwidth]{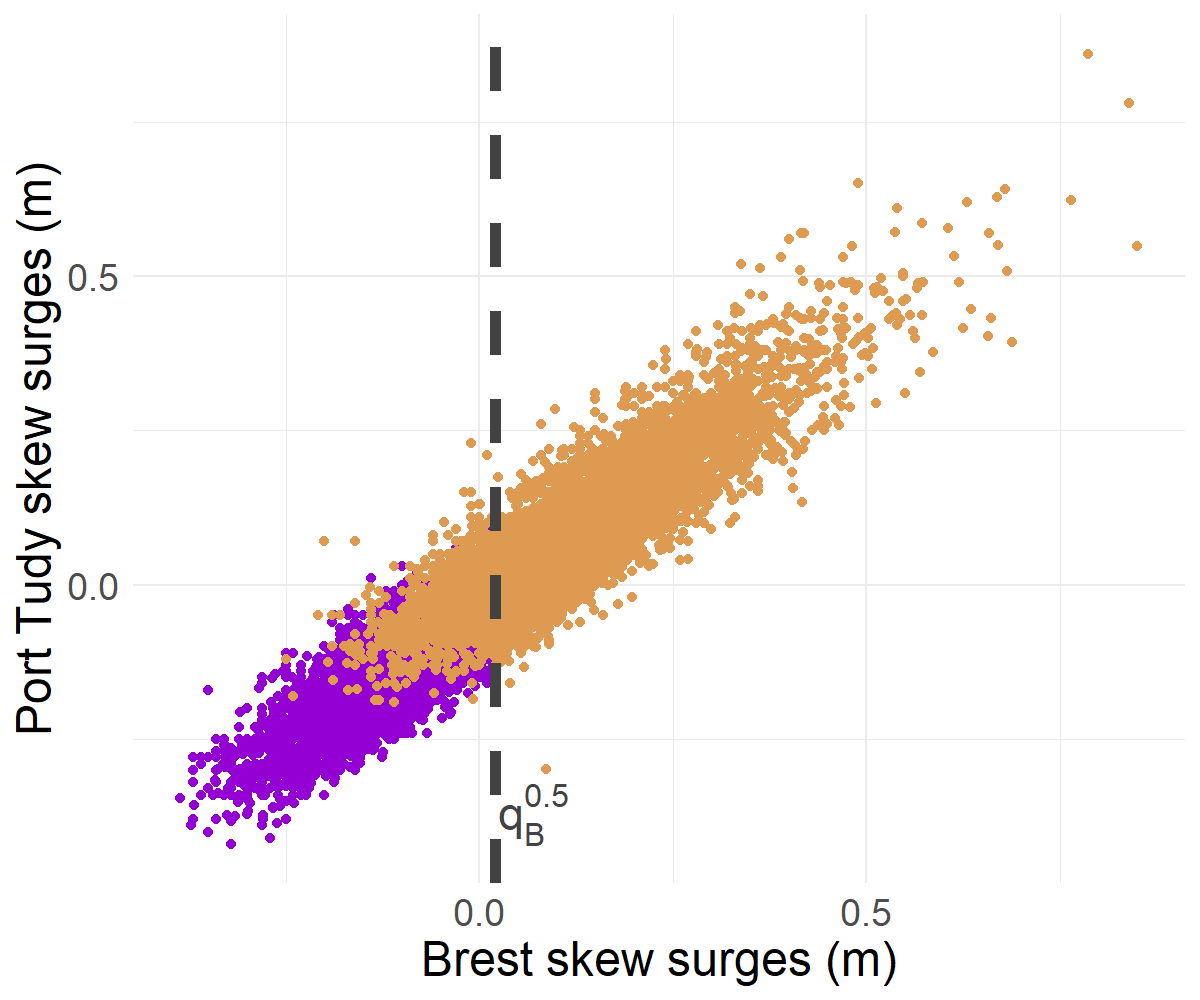}

  \vspace{1cm}

  \includegraphics[width=.4\textwidth]{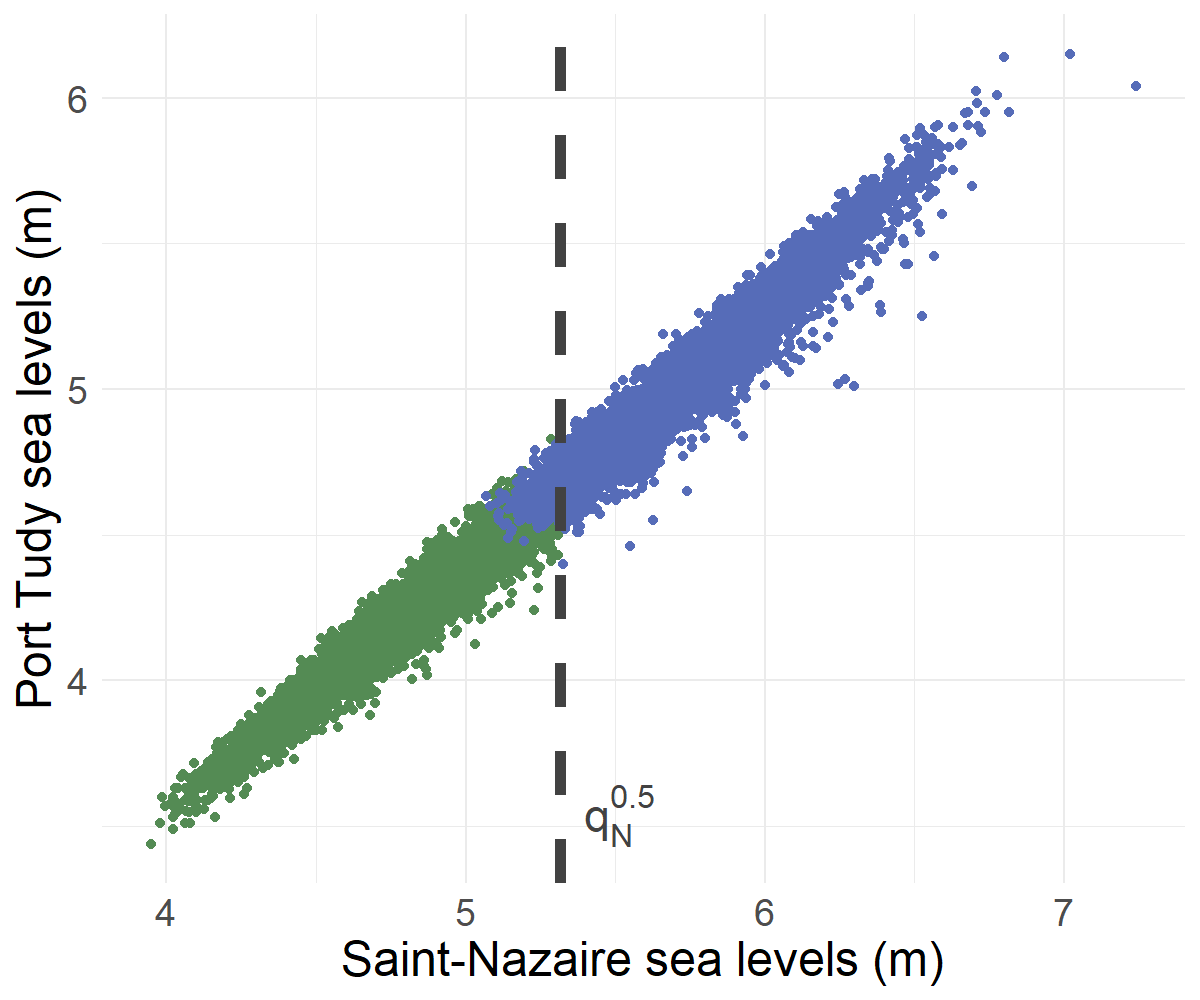}
  \hspace{1cm}
  \includegraphics[width=.4\textwidth]{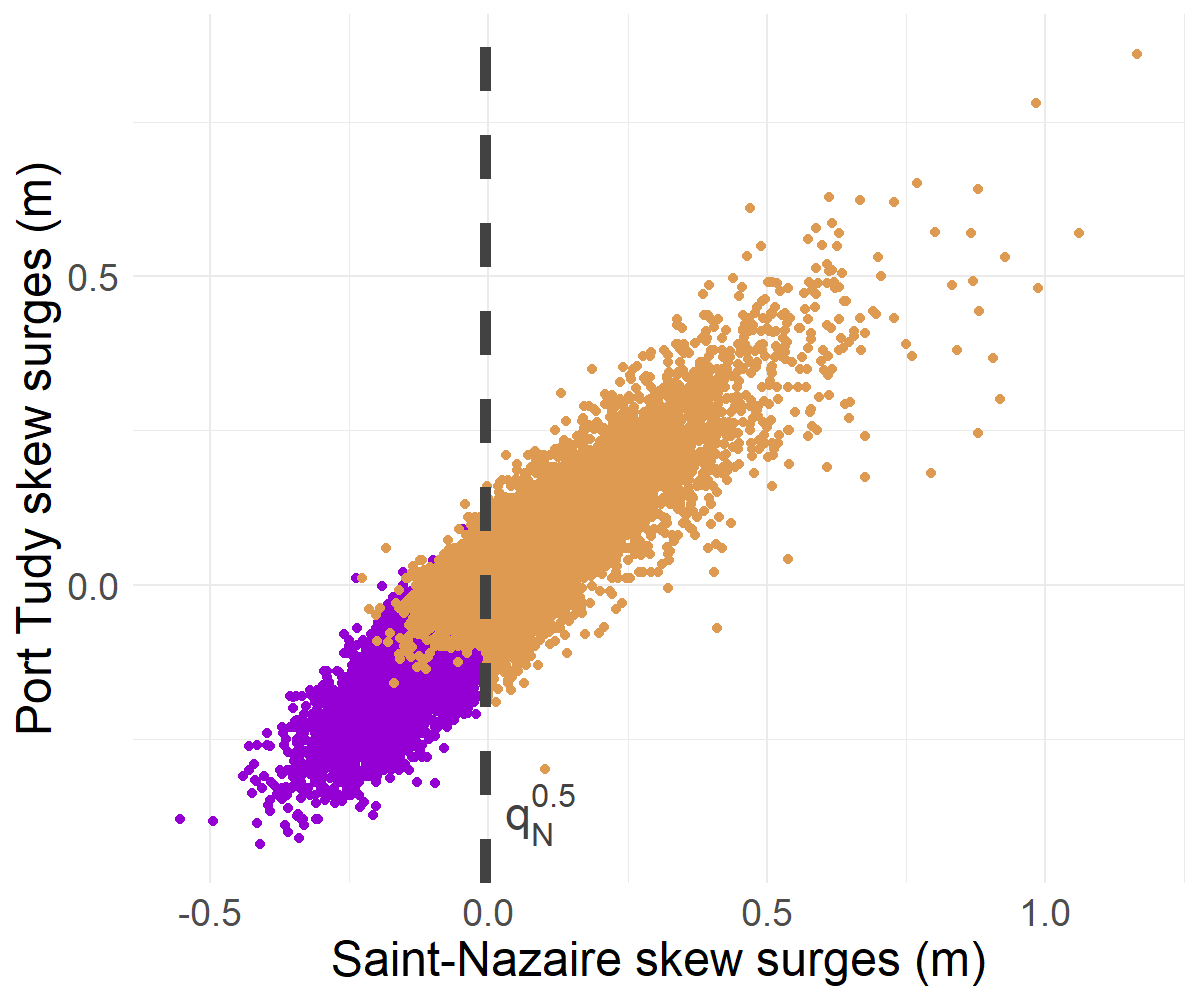}

  \caption{Scatter plots showing sea levels (left) and skew surges (right) at each station plotted against one another in the training set. Green and purple points represent the observations such that $X_B$ and $X_N$ are both below their empirical medians. Blue and orange points represent the observations such that $X_B$ or $X_N$ are above their empirical medians. The grey dotted lines represent the empirical medians.\label{fig:scatterplot_appendix}}
\end{figure}

The original dataset comprises triplets $(X^{ori}_{B,i}, X^{ori}_{N,i},Y^{ori}_i)_{1 \m i \m m}$ where $X^{ori}_{B,i}, X^{ori}_{N,i}, Y^{ori}_i$ are the raw sea levels or skew surges measured respectively at Brest, Saint-Nazaire and Port Tudy. A first pre-selection is performed such that we retain only triplets $(X^{ori}_{B,i}, X^{ori}_{N,i},Y^{ori}_i)$ such that $X^{ori}_{B,i}$ exceed its empirical median or $X^{ori}_{N,i}$ exceed its empirical median, i.e., we retain triplets for $i \in I$, where the set $I$ is defined as
\begin{equation*}\label{eq:set_I}
    I = \Big\{i \in 1,...,m , X^{ori}_{B,i} \s q_B^{ori,0.5} \mbox{ or } X^{ori}_{N,i} \s q_N^{ori,0.5} \Big\},
\end{equation*}
where $q_B^{ori,\rho}$ (resp. $q_N^{ori,\rho}$) is the original empirical $\rho$-quantile at Brest (resp. at Saint-Nazaire). This operation is depicted in Figure~\ref{fig:scatterplot_appendix}. For conciseness, we assume $I = \{1,...,n\}$ in the main text. Finally, the origin is shifted: 
we subtract, from each component of the observations,  the minimum value recorded over $I$, i.e., the observations considered in the main paper are 
\begin{equation*}\label{eq:final_set}
    (X_{B,i}, X_{N,i}, Y_i) := (X^{ori}_{B,i}-m_B, X^{ori}_{N,i}-m_N, Y^{ori}_i-m_Y),
\end{equation*}
for $i \in I$ and where $m_B = \min_{i \in I}X^{ori}_{B,i}$, $m_N = \min_{i \in I}X^{ori}_{N,i}$ and $m_Y = \min_{i \in I}Y^{ori}_{i}$. 

{\color{blue}
\section{Stationarity tests \label{sec:stationary}}
To study the stationarity of the surge time series, we perform two complementary tests: the Augmented Dickey–Fuller (ADF) test \citep{dickey1979distribution,said1984testing} and the Kwiatkowski–Phillips–Schmidt–Shin (KPSS) test \citep{kwiatkowski1992testing}. The ADF test evaluates the null hypothesis that the time series is non-stationary, whereas the KPSS test evaluates the null hypothesis that the time series is stationary. Both tests are implemented using functions from the \textsf{aTSA} package \citep{qiu2015package}. For additional details about these tests, see, e.g., \citep{fuller1995introduction} or the \textsf{aTSA} package documentation.

Table~\ref{tab:stationary_test_before_ss} (Table~\ref{tab:stationary_test_before_sl} for sea level, respectively) reports the p-values of the tests at the different stations. The ADF test results show no evidence against stationarity, while the KPSS test results suggest some reservations about assuming stationarity.

Based on these findings, temporal trends are removed, and the ADF and KPSS tests are repeated on the detrended data. Table~\ref{tab:stationary_test_after_ss} (resp. Table~\ref{tab:stationary_test_after_sl}) presents the results, which now provide no evidence against the stationarity of the detrended series according to both tests. Figure~\ref{fig:rolling_fig_ss} (resp. Figure~\ref{fig:rolling_fig_sl}) shows the rolling means and standard deviations at each station for the detrended data, visually confirming that the stationarity assumption is reasonable.

\begin{table}[ht!]
\caption{p-values of the ADF test (alternative hypothesis: stationarity) and the KPSS test (alternative hypothesis: non-stationarity) for the time series at the three stations. Bold values indicate strong evidence against the stationarity of the time series.\label{tab:stationary_test_before_ss}}
\vspace{0.2cm}
\begin{center}
\begin{small}
\begin{sc}
\begin{tabular}{c|c|ccc}
\toprule
Stations & Tests & no drift no trend & with drift no trend & with drift and trend   \\
\midrule
\multirow{2}{*}{Brest} & ADF & <0.01 & <0.01 & <0.01 \\
 & KPSS & \textbf{<0.01} & \textbf{<0.01} & >0.1  \\
 \midrule 
 \multirow{2}{*}{Saint-Nazaire} & ADF & <0.01 & <0.01 & <0.01 \\
 & KPSS & \textbf{<0.01} & >0.1 & >0.1  \\
 \midrule
 \multirow{2}{*}{Port Tudy} & ADF & <0.01 & <0.01 & <0.01 \\
 & KPSS & \textbf{<0.01} & >0.1 & 0.03  \\
\bottomrule
\end{tabular}
\end{sc}
\end{small}
\end{center}
\end{table}

\begin{figure}[ht!] 
  \centering
  \includegraphics[width=.315\textwidth]{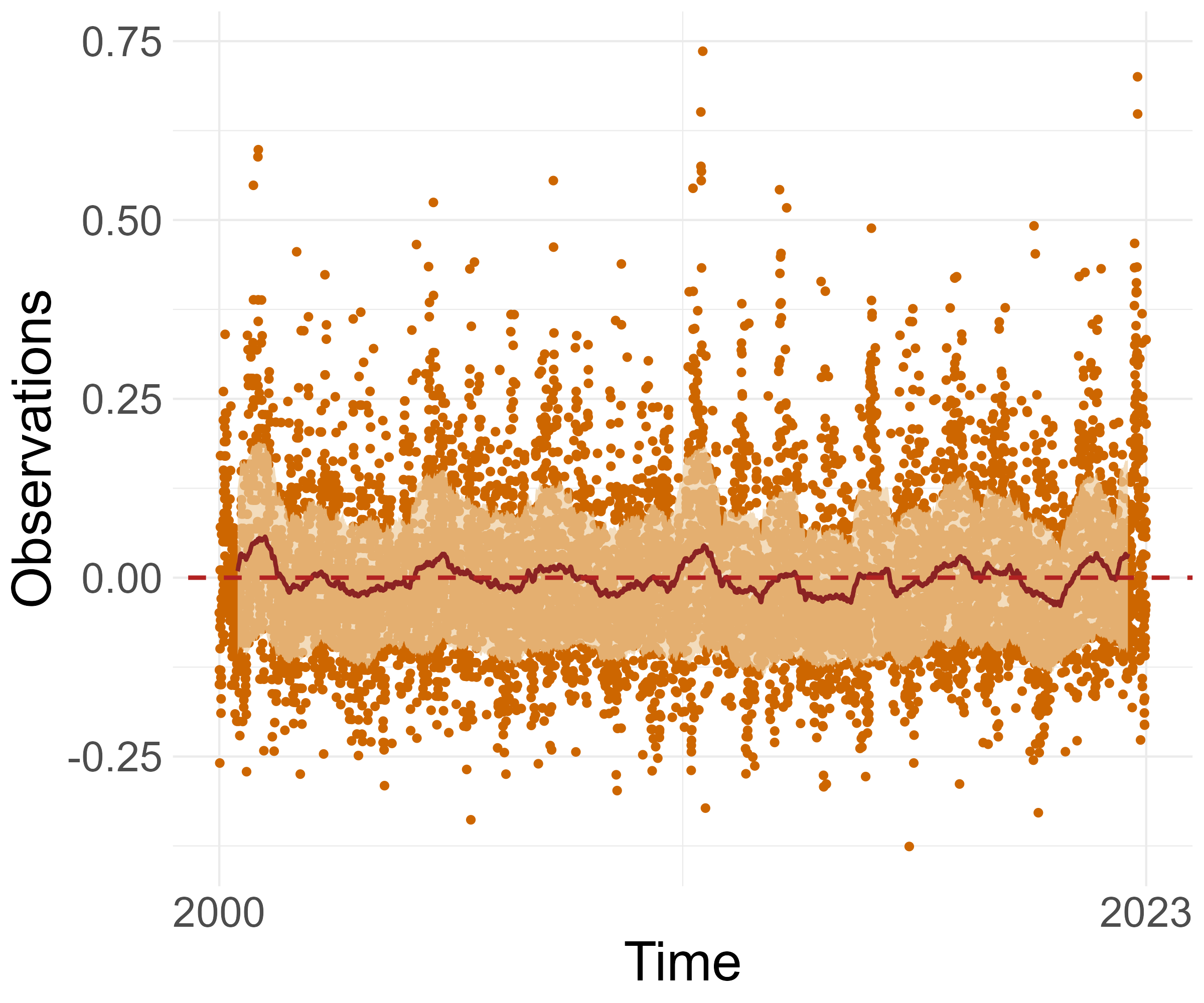}
  \hspace{0.2cm}
  \includegraphics[width=.315\textwidth]{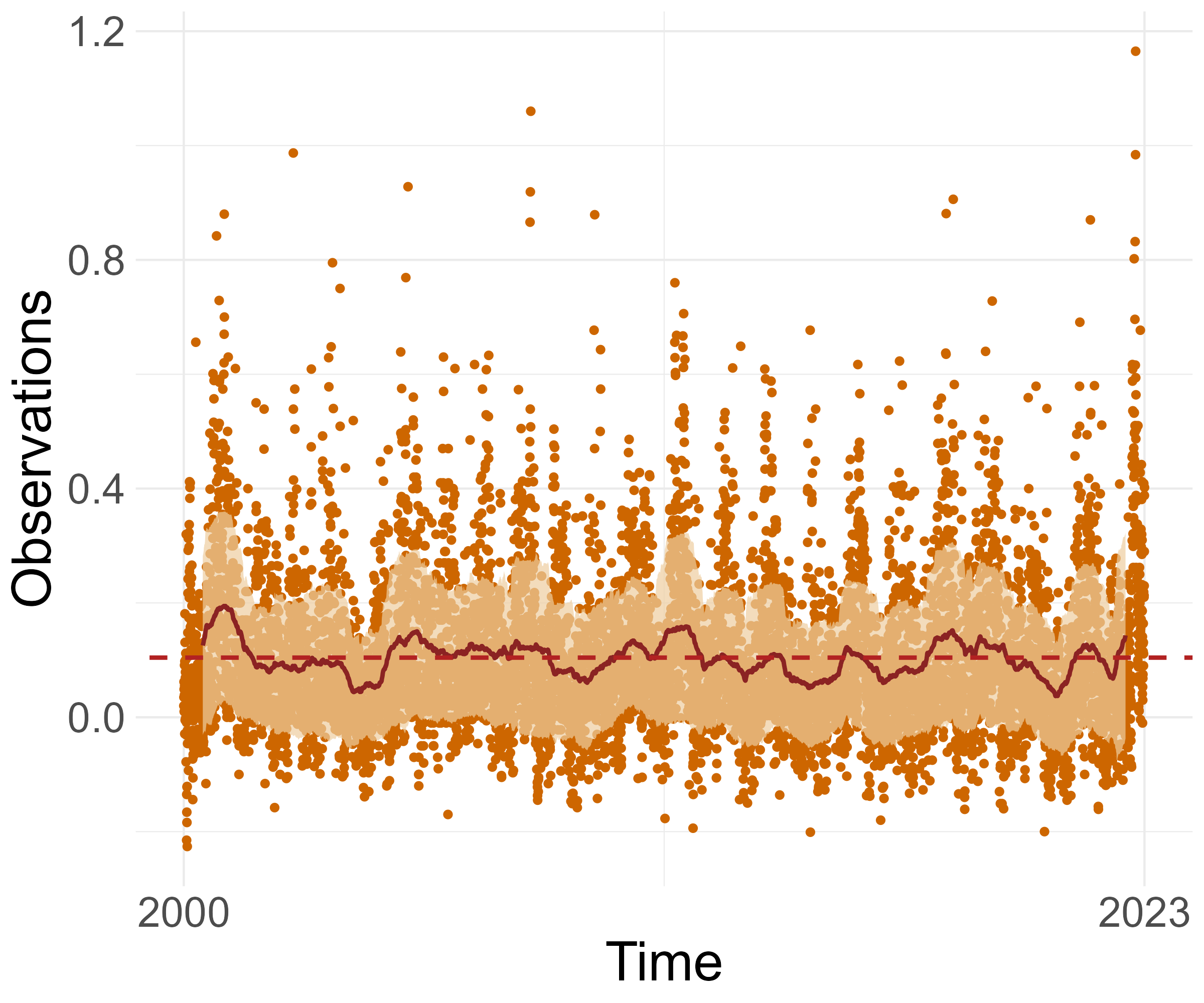}
  \hspace{0.2cm}
  \includegraphics[width=.315\textwidth]{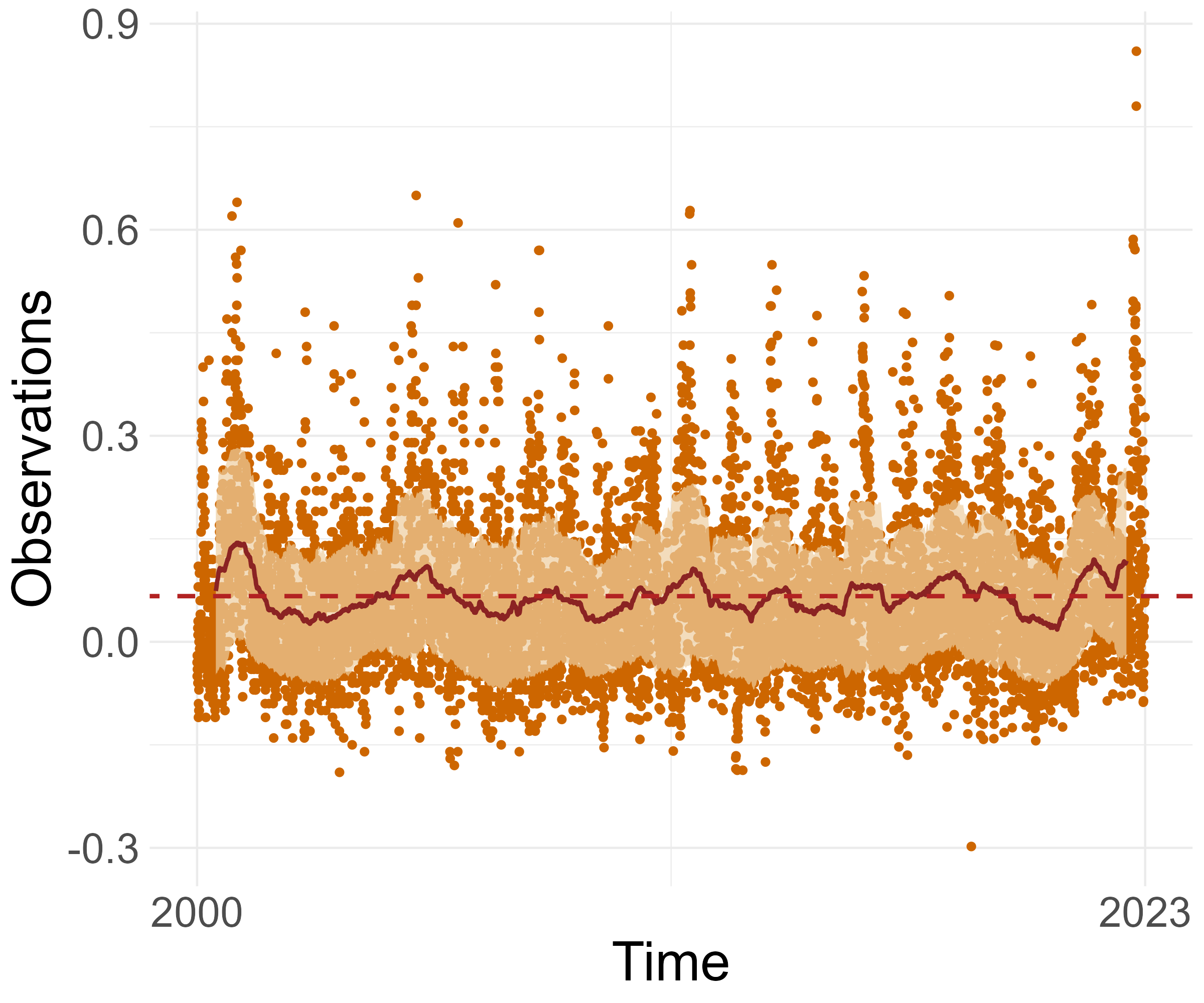}
\caption{Rolling means (brown) and rolling standard deviations (light orange) of the detrended skew surge time series at the three stations (Brest: left; Saint-Nazaire: middle; Port Tudy: right. The removed temporal slopes are all smaller than 1e-5 (Brest: 3.8e-6; Saint-Nazaire: -4.8e-07; Port Tudy 6.8e-07).\label{fig:rolling_fig_ss}}
\end{figure}

\begin{table}[ht!]
\caption{p-values of the ADF test (alternative hypothesis: stationarity) and the KPSS test (alternative hypothesis: non-stationarity) for the detrended time series at the three stations. No strong evidence against the stationarity of the time series is observed.\label{tab:stationary_test_after_ss}}
\vspace{0.2cm}
\begin{center}
\begin{small}
\begin{sc}
\begin{tabular}{c|c|ccc}
\toprule
Stations & Tests & no drift no trend & with drift no trend & with drift and trend   \\
\midrule
\multirow{2}{*}{Brest} & ADF & <0.01 & <0.01 & <0.01 \\
 & KPSS & >0.1 & >0.1 & 0.06  \\
 \midrule 
 \multirow{2}{*}{Saint-Nazaire} & ADF & <0.01 & <0.01 & <0.01 \\
 & KPSS & >0.1 & >0.1 & >0.1  \\
 \midrule
 \multirow{2}{*}{Port Tudy} & ADF & <0.01 & <0.01 & <0.01 \\
 & KPSS & >0.1 & >0.1 & 0.03  \\
\bottomrule
\end{tabular}
\end{sc}
\end{small}
\end{center}
\end{table}
}

\clearpage
\section{Threshold Formula}\label{sec:appexdix_th}
A proof of the formula for the threshold given in Algorithm~\ref{algo:th_select} is presented in this section.

We need to compute the second derivative of the density function $f_{\sigma,\xi,\kappa}$ \eqref{eq:egpd_dens} and then compute the zeros of this second derivative. Recall that the function is given by
\begin{equation*}
    f_{\sigma,\xi,\kappa}(x) = \frac{\kappa}{\sigma}\Big(1+\frac{\xi x}{\sigma}\Big)^{-1/\xi-1}\bigg(1-\Big(1+\frac{\xi x}{\sigma}\Big)^{-1/\xi}\bigg)^{\kappa-1}.
\end{equation*}
The first and second derivatives of $f_{\sigma,\xi,\kappa}$ are
\begin{equation*}
    \frac{df_{\sigma,\xi,\kappa}(x)}{dx} = \frac{\kappa}{\sigma^2}\bigg((\kappa-1)\Big(1+\frac{\xi x}{\sigma}\Big)^{-2/\xi-2}\Big(1-\Big(1+\frac{\xi x}{\sigma}\Big)^{-1/\xi}\Big)^{\kappa-2}-(\xi+1)\Big(1+\frac{\xi x}{\sigma}\Big)^{-1/\xi-2}\Big(1-\Big(1+\frac{\xi x}{\sigma}\Big)^{-1/\xi}\Big)^{\kappa-1}\bigg),
\end{equation*}
and
\begin{align*}
    \frac{d^2f_{\sigma,\xi,\kappa}(x)}{dx^2} = &\frac{\kappa}{\sigma^3}\Big(1+\frac{\xi x}{\sigma}\Big)^{-1/\xi-3}\Big(1-\Big(1+\frac{\xi x}{\sigma}\Big)^{-1/\xi}\Big)^{\kappa-3} \bigg[(\kappa-1)(\kappa-2)\Big(1+\frac{\xi x}{\sigma}\Big)^{-2/\xi}- \\
    &(\kappa-1)(3+3\xi)\Big(1+\frac{\xi x}{\sigma}\Big)^{-1/\xi}\Big(1-\Big(1+\frac{\xi x}{\sigma}\Big)^{-1/\xi}\Big)+(\xi+1)(1+2\xi)\Big(1-\Big(1+\frac{\xi x}{\sigma}\Big)^{-1/\xi}\Big)^2\bigg] \\
    = &\frac{\kappa}{\sigma^3}\Big(1+\frac{\xi x}{\sigma}\Big)^{-1/\xi-3}\Big(1-\Big(1+\frac{\xi x}{\sigma}\Big)^{-1/\xi}\Big)^{\kappa-3} A\Big(\Big(1+\frac{\xi x}{\sigma}\Big)^{-1/\xi}\Big),
\end{align*}
with $A(X) = (\kappa-1)(\kappa-2)X^2-(\kappa-1)(3+3\xi)X(1-X)+(\xi+1)(1+2\xi)(1-X)^2$. The two zeros of $\frac{d^2f_{\sigma,\xi,\kappa}(x)}{dx^2}$ associated with the two first factors of the above equation are $0$ and $-\sigma/\xi$ corresponding to the lower and upper bounds of the support of the density. Therefore, the desired point lies in the zeros of $A\Big(\Big(1+\frac{\xi x}{\sigma}\Big)^{-1/\xi}\Big)$. Note that if $X_0$ is a zero of $A$ then $(\sigma/\xi)\big(X_0^{-\xi}-1\big)$ is a zero of $A\Big(\Big(1+\frac{\xi x}{\sigma}\Big)^{-1/\xi}\Big)$.
The canonical form of $A$ is
\begin{equation*}
    A(X) = (\kappa^2 + 2\xi^2+3\kappa \xi)X^2 + (4\xi^2+3\kappa +3\xi + 3\kappa \xi -1)X + (2\xi^2+3\xi+1).
\end{equation*}
Solving the second-degree polynomial gives the largest zero of $A$
\begin{equation*}
    X_0 = \frac{4\xi^2+3\kappa\xi+3\kappa+3\xi -1-\sqrt{(4\xi^2+3\kappa\xi+3\kappa+3\xi -1)^2-4(\kappa^2+2\xi^2+3\kappa \xi)(2\xi^2+3\xi+1)}}{2(\kappa^2+2\xi^2+3\kappa\xi)},
\end{equation*}
and the threshold formula in Algorithm~\ref{algo:th_select} follows.

\section{Density models}\label{sec:appexdix_dens}
For the sake of completeness, this section provides detailed steps for constructing MGP densities along with the explicit formulas for the proposed densities used in Section~\ref{sec:mgpd_results}.
\subsection{MGP density construction}\label{app:density_model}
Theorem~5 in \cite{ROOTZEN2018117} provides the theoretical foundation for constructing MGP densities from a variety of underlying density functions.

First, recall the decomposition \eqref{eq:mgpd_decomp} of a standard MGP random vector $\bm Z_\infty'$ as defined in \eqref{eq:CV_excess_expo_margins}
\begin{equation*}
     \bm Z_\infty' = E + \bm{T} - \max \bm{T},
\end{equation*}
where $E \in [0,+\infty[$ is a unit exponential random variable and $\bm{T}$ is  a random vector valued in $\mb{R}^{d+1}$, independent of $E$.
Suppose that $\bm T$ has a density $f_T$. Based on the above decomposition, $\bm Z_\infty'$ has also a density $h_T$ given by 
\begin{equation*}\label{eq:dens_T}
    h_T(\bm{z}) = \1 \{\max(\bm{z}) > 0\} \exp(-\max(\bm{z}))\int_0^{+\infty} \frac{f_T(\bm{z}+\log t)}{t}dt.
\end{equation*}

Second, Theorem~5 in \cite{ROOTZEN2018117} also introduces an alternative method to construct a standard MGP density. Let $\bm{U} \in\mb{R}^{d+1}$ a random vector such that $0 < \mb{E}[\exp(U_j)] < \infty$, for $1\m j \m d+1$. Suppose $\bm{U}$ has a density $f_U$. Then, from this density, an another standard MGP density is given by the following function
\begin{equation*}
    h_U(\bm{z}) = \frac{\1 \{ \max(\bm{z})>0\}}{\mb{E}[\exp(\max(\bm{U}))]} \int_0^{+\infty} f_U(\bm{z} + \log t) dt.
\end{equation*}

Both methods demonstrate how a standard MGP density can be systematically derived from various general density functions.

\subsection{Density models}\label{sec:dens_model}
The explicit formulas for the density models used in the MGPRED procedure, namely with independent Gumbel components and with independent reverse exponential components, are gathered in this section.
\subsubsection{Densities with independent Gumbel components}
The density function of a Gumbel distribution with parameters $\alpha_j>0, \beta_j \in \mathbb{R}$ expressed as
\begin{equation*}
f_j(z_j)=\alpha \exp \big(-\alpha(z_j-\beta_j)\big) \exp \Big(-\exp \big(-\alpha_j(z_j-\beta_j)\big)\Big).
\end{equation*}
Given this, the joint density function for the random vectors $\bm T$ and $\bm U$ is the product of the marginal densities
\begin{equation*}
    f(\bm{z}) = \prod_{j=1}^nf_j(z_j).
\end{equation*}
To simplify computations and avoid numerical approximations of integral quantities, we restrict the study to the case $\alpha_1 = ... = \alpha_d = \alpha$. Under this simplification, the standard MGP density for the construction using $\bm T$ is given by
\begin{equation*}
    h_T(\bm{z}) = \exp(-\max (\bm{z})) \alpha^{d-1} \frac{\Gamma(d) \prod_{j=1}^d \exp\big(-\alpha(z_j-\beta_j)\big)}{\Big(\sum_{j=1}^d \exp\big(-\alpha(z_j-\beta_j)\big)\Big)^d},
\end{equation*}
Similarly, the standard MGP density for the construction using $\bm U$ is given by
\begin{equation*}
    h_U(\bm{z}) = \frac{\alpha^{d-1}\Gamma(d-1/\alpha)}{\Gamma(1-1/\alpha)\Big(\sum_{j=1}^d\exp(\beta_j\alpha)\Big)^{1/\alpha}}\frac{\prod_{j=1}^d\exp\big(-\alpha(z_j-\beta_j)\big)}{\Big(\sum_{j=1}^d\exp\big(-\alpha(z_j-\beta_j)\big)\Big)^{d-1/\alpha}}.
\end{equation*}

\subsubsection{Densities with independent reverse exponential components}
The density function of a reverse exponential distribution with parameters $\alpha_j>0, \beta_j \in \mathbb{R}$ expressed as
\begin{equation*}
f_j(z_j)=\alpha_j \exp\big(\alpha_j(z_j+\beta_j)\big),
\end{equation*}
for $z_j \in (-\infty,-\beta_j)$.
Given this, the joint density function for the random vectors $\bm T$ and $\bm U$ is the product of the marginal densities
\begin{equation*}
    f(\bm{z}) = \prod_{j=1}^nf_j(z_j)
\end{equation*}
The standard MGP density for the construction using $\bm T$ is then given by
\begin{equation*}
    h_T(\bm{z}) = \frac{\exp\big(-\max (\bm{z})-\max(\bm{z}+\bm{\beta})\sum_{j=1}^d\alpha_j\big)}{\sum_{j=1}^d\alpha_j}\prod_{j=1}^d\alpha_j\exp\big(\alpha_j(z_j+\beta_j)\big).
\end{equation*}
Similarly, the standard MGP density for the construction using $\bm U$ is given by
\begin{equation*}
    h_U(\bm{z}) = \frac{\exp\big(-\max(\bm{z}+\bm{\beta})(\sum_{j=1}^d\alpha_j+1)\big)}{\Big(1+\sum_{j=1}^d\alpha_j\Big)\mb{E}[\exp(\max(\bm{U}))]}\prod_{j=1}^d \alpha_j\exp\big(\alpha_j(z_j+\beta_j)\big),
\end{equation*}

\section{Additional studies}\label{sec:add_stud}
\begin{figure}[h!]
  \centering
  \includegraphics[width=.36\textwidth]{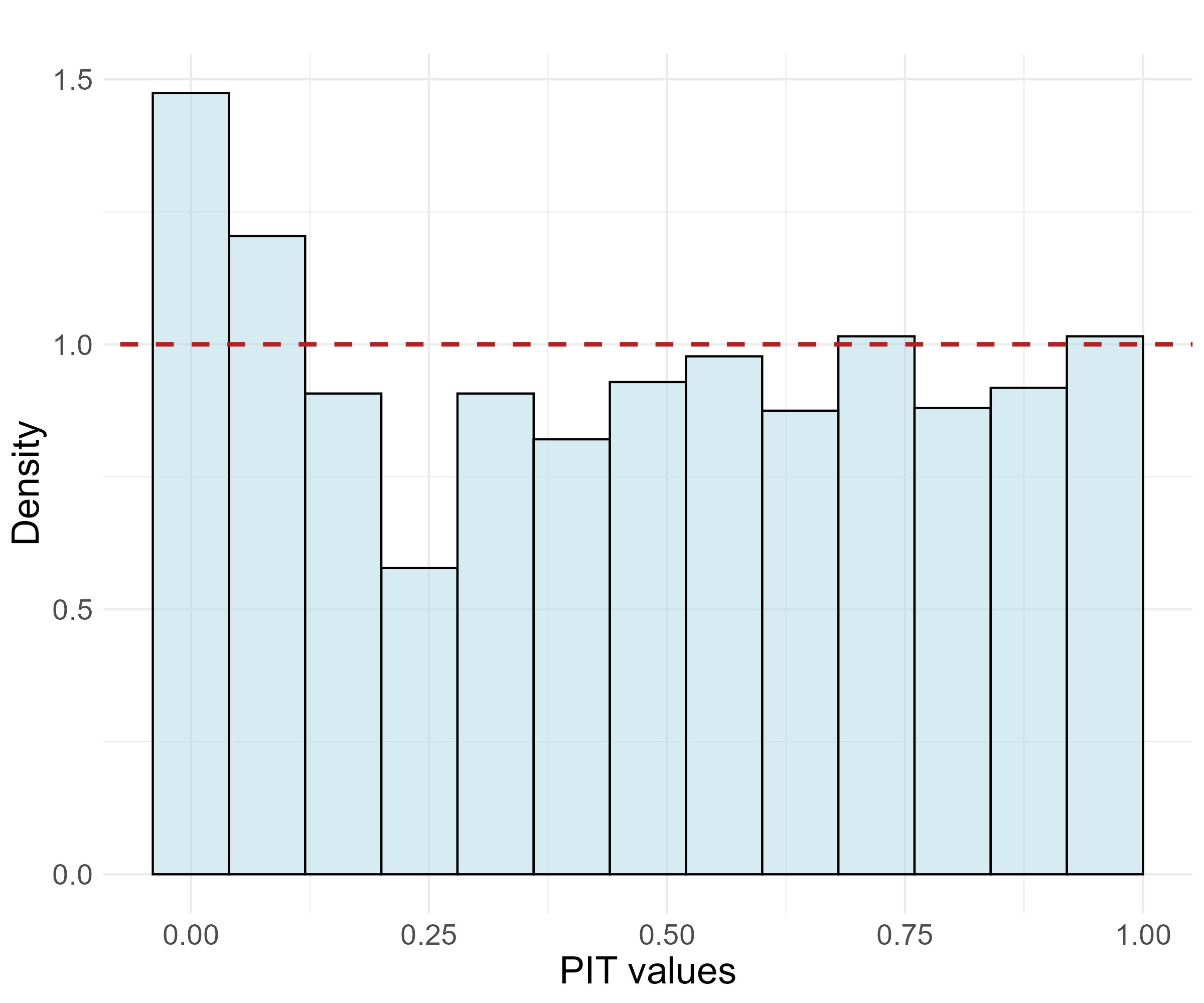}
	\quad
\includegraphics[width=.36\textwidth]{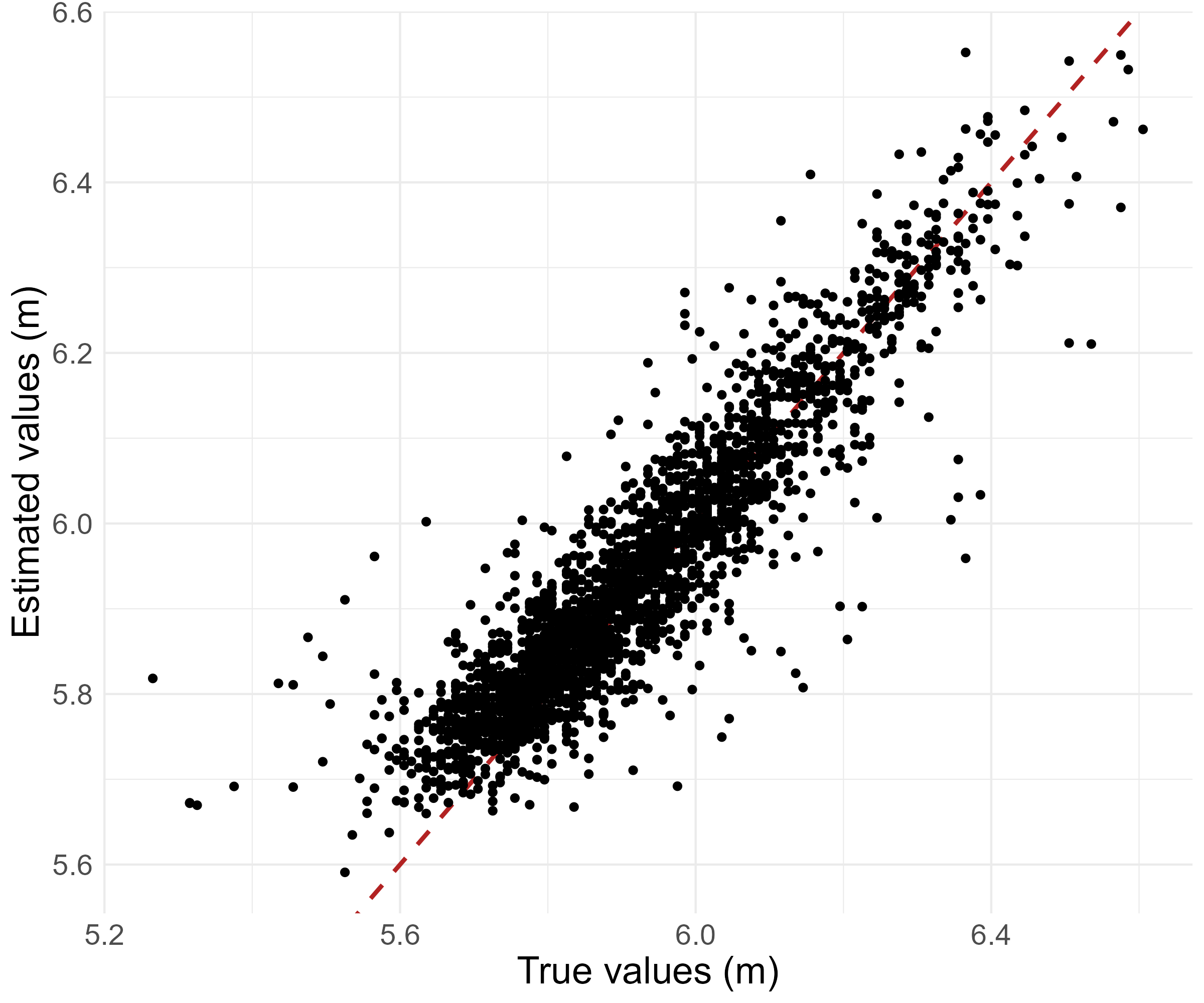}
  \quad
  \includegraphics[width=.18\textwidth]{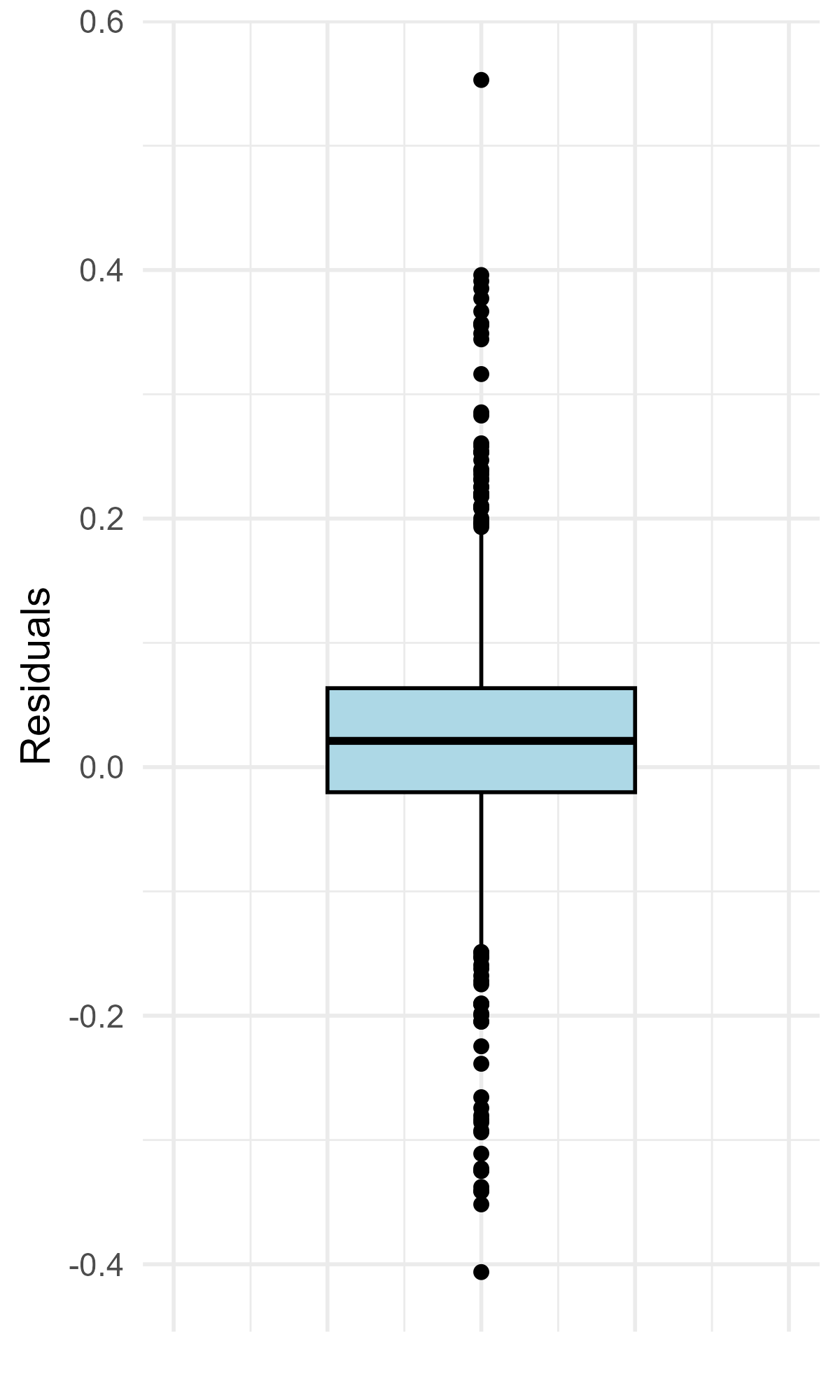}
  \caption{Goodness-of-fit diagnostics for the prediction of extreme sea levels using the ROXANE routine with the OLS regression algorithm: PIT plot (left), predictions vs. observations plot (middle), and boxplot of residuals (right). In this setting, the correlation coefficient is 0.91. Results are shown on the subset of the test set comprising the most extreme observations w.r.t. the Port Tudy observations, i.e., observations such that $Y_i \s q^{ext,0.5}_Y$.
  \label{fig:gof_predictions_ols_sl}}
\end{figure}

\begin{figure}[ht!]
  \centering
  \includegraphics[width=.36\textwidth]{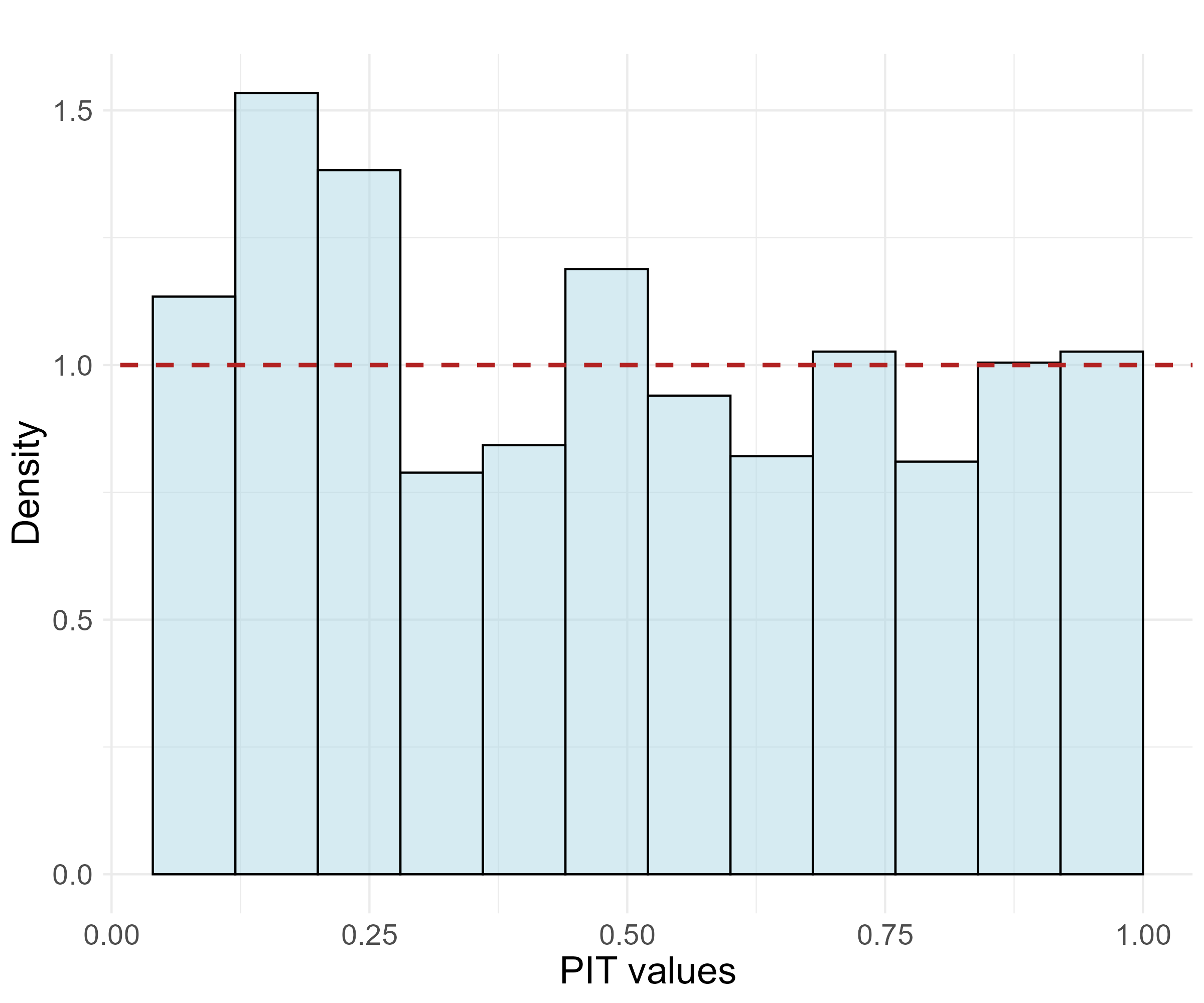}
	\quad
\includegraphics[width=.36\textwidth]{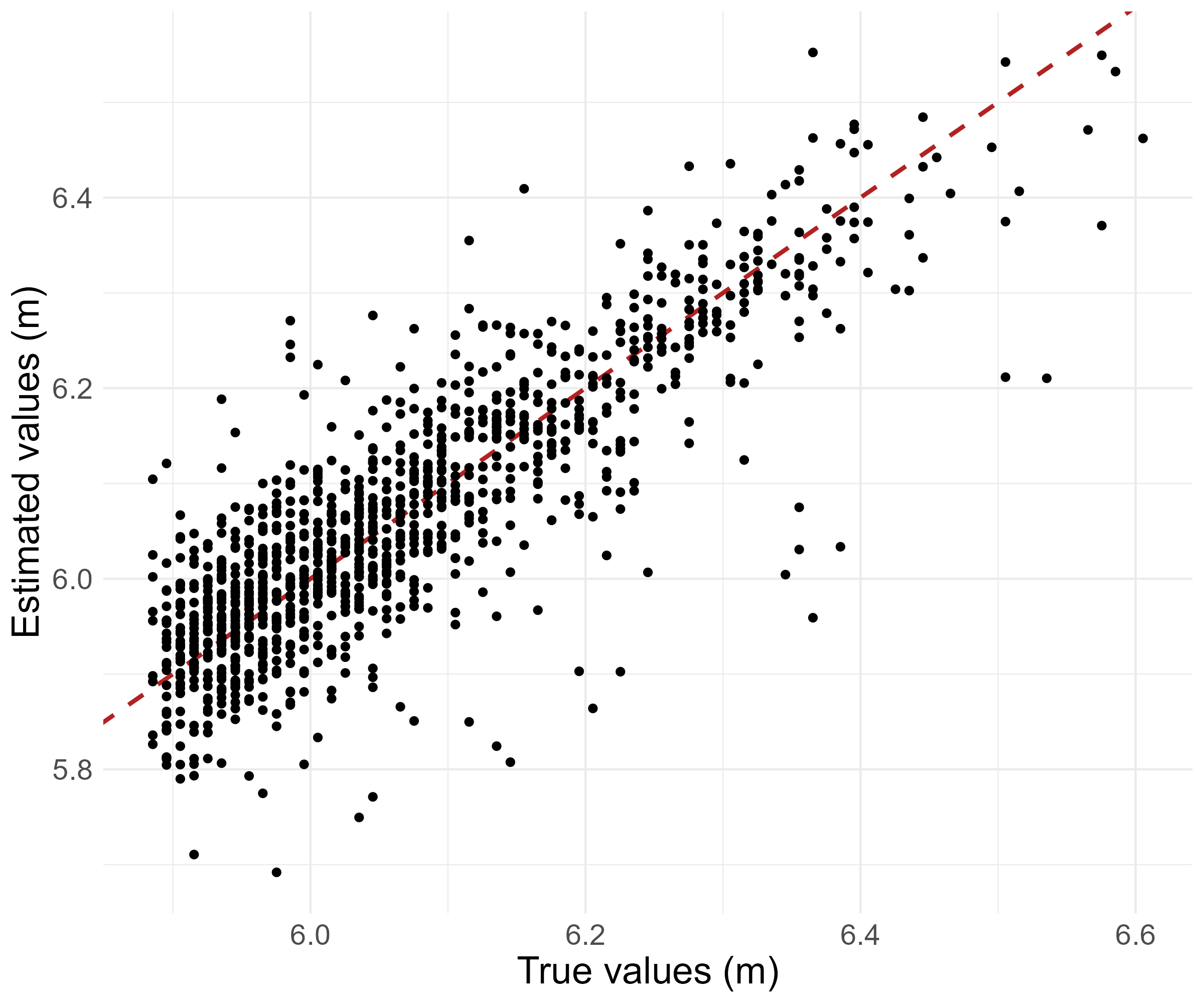}
  \quad
  \includegraphics[width=.18\textwidth]{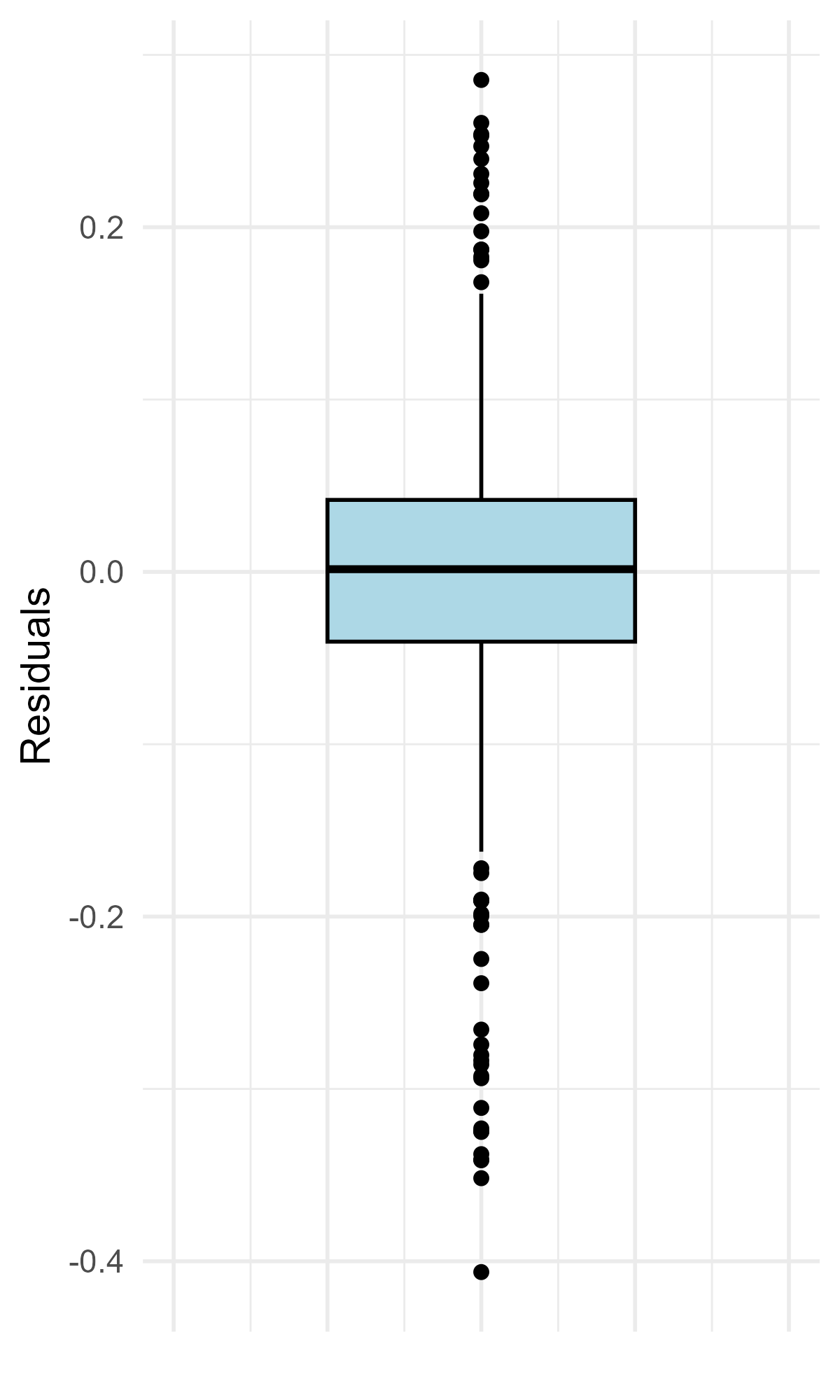}
  \caption{Goodness-of-fit diagnostics for the prediction of very extreme sea levels using the ROXANE routine with the OLS regression algorithm: PIT plot (left), predictions vs. observations plot (middle), and boxplot of residuals (right). In this setting, the correlation coefficient is 0.85.
  \label{fig:gof_predictions_ols_sl_th}}
\end{figure}

\begin{figure}[h!]
  \centering
  \includegraphics[width=.36\textwidth]{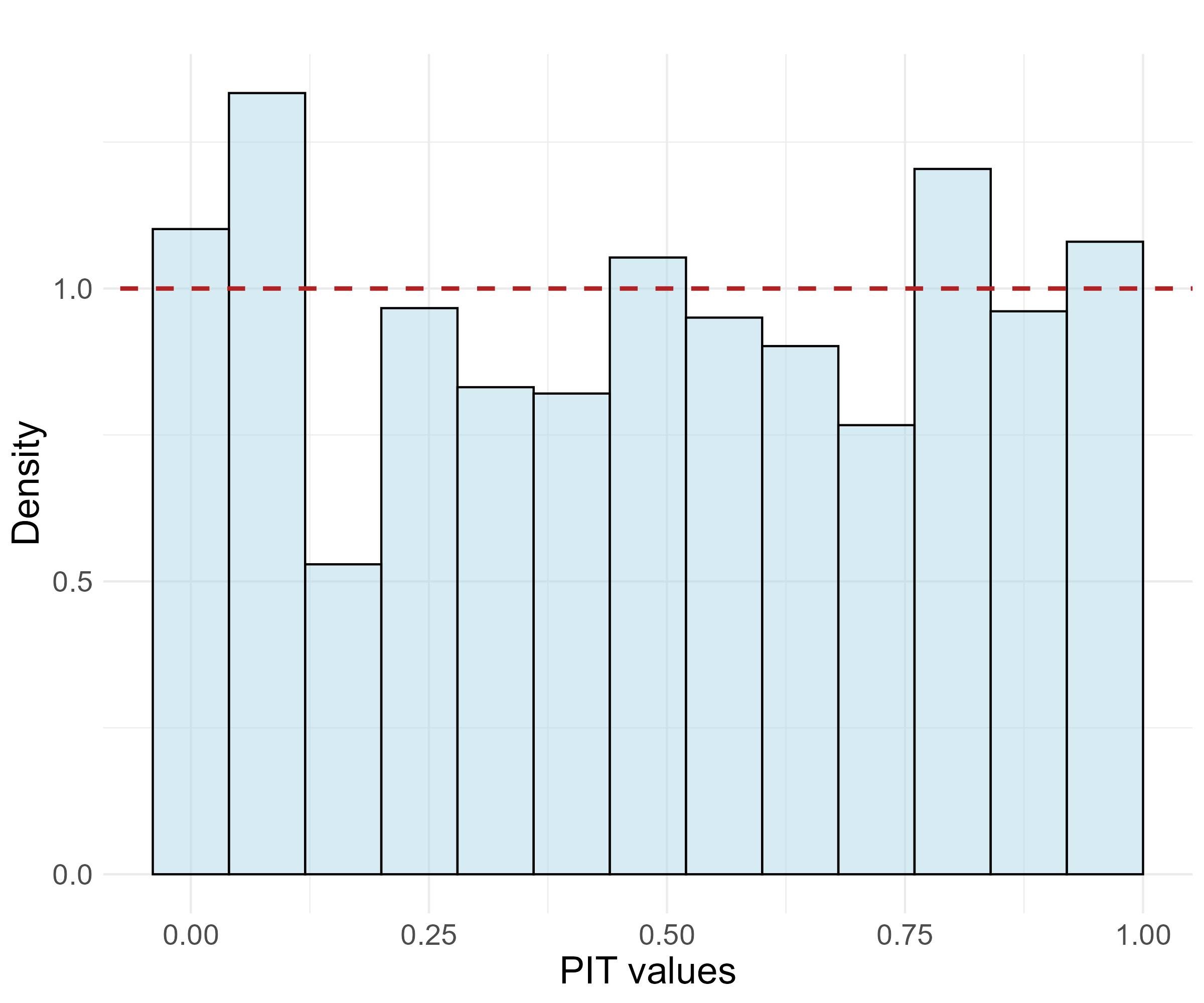}
	\quad
\includegraphics[width=.36\textwidth]{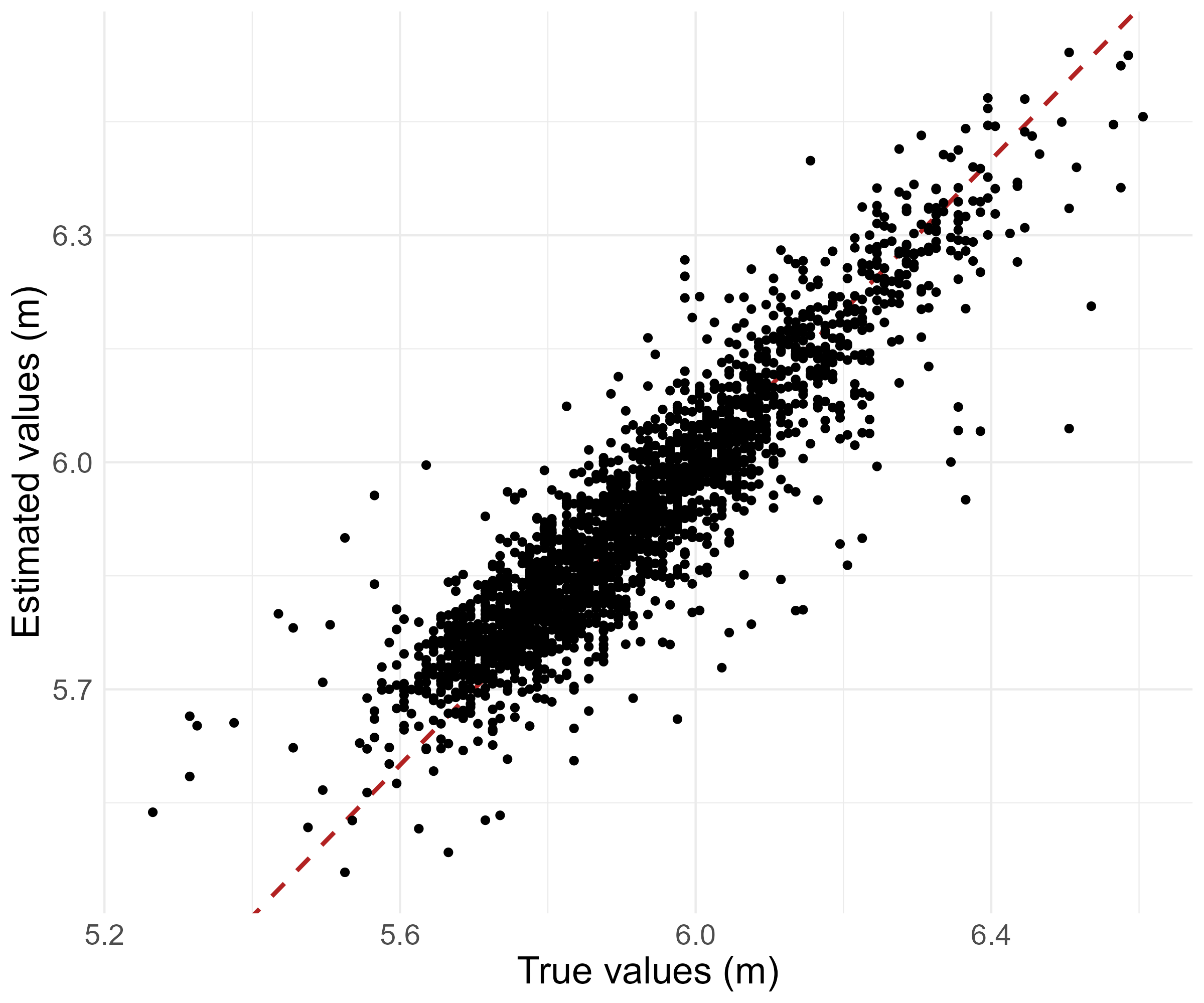}
  \quad
  \includegraphics[width=.18\textwidth]{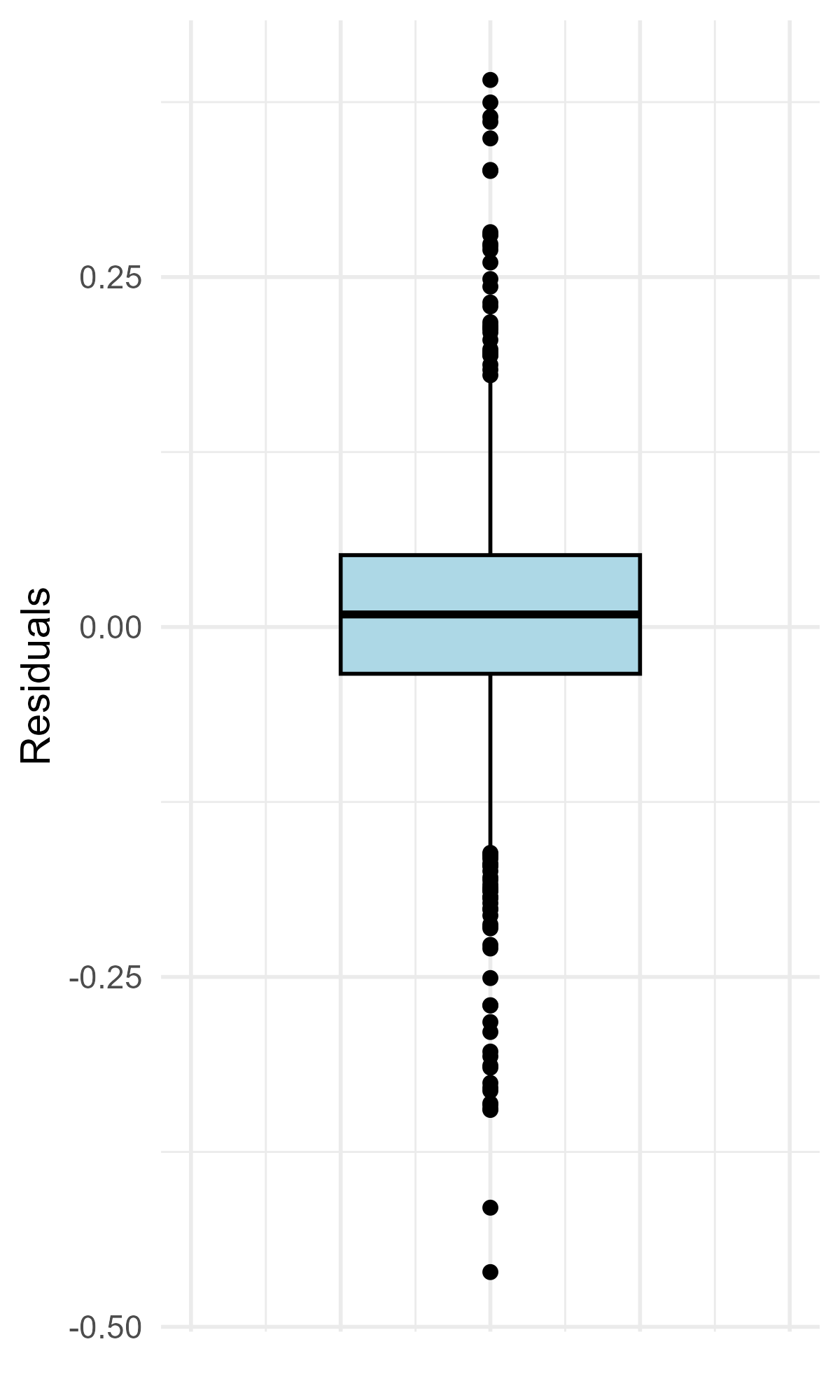}
  \caption{Goodness-of-fit diagnostics for the prediction of extreme sea levels using the MGPRED procedure: PIT plot (left), predictions vs. observations plot (middle), and boxplot of residuals (right). In this setting, the correlation coefficient is 0.78.
  \label{fig:gof_predictions_mgpd_sl}}
\end{figure}

\begin{figure}[h!]
  \centering
  \includegraphics[width=.36\textwidth]{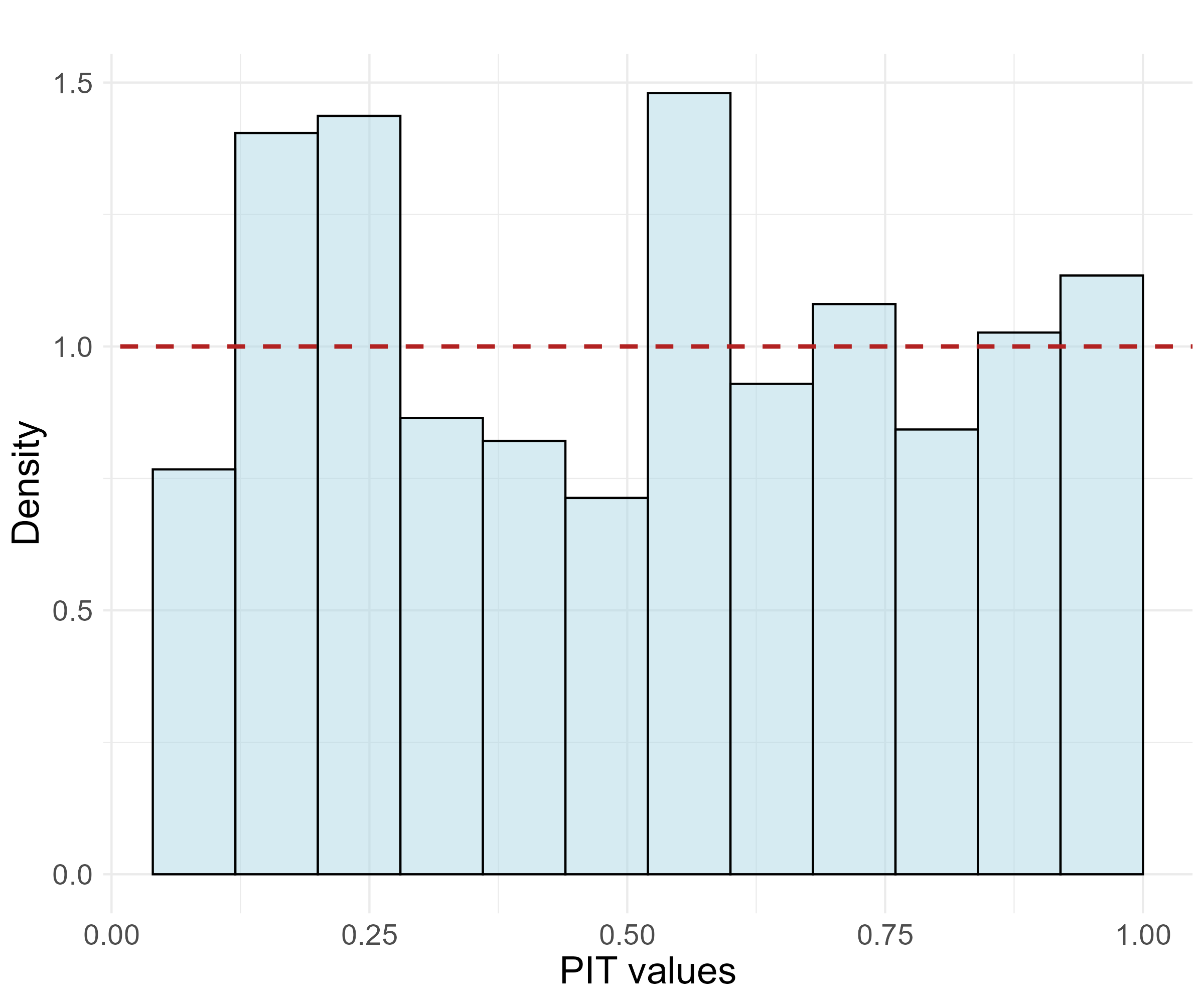}
	\quad
\includegraphics[width=.36\textwidth]{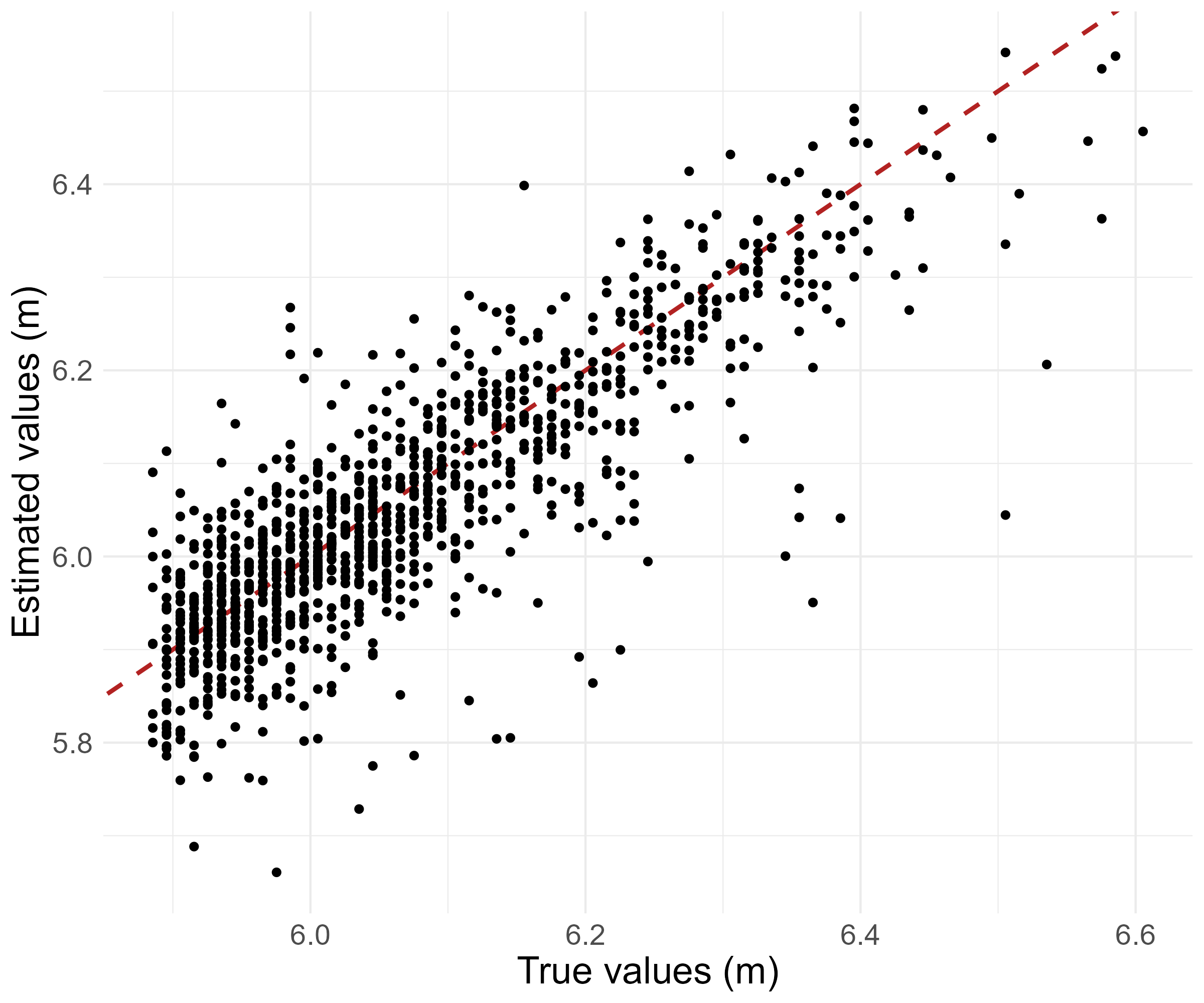}
  \quad
  \includegraphics[width=.18\textwidth]{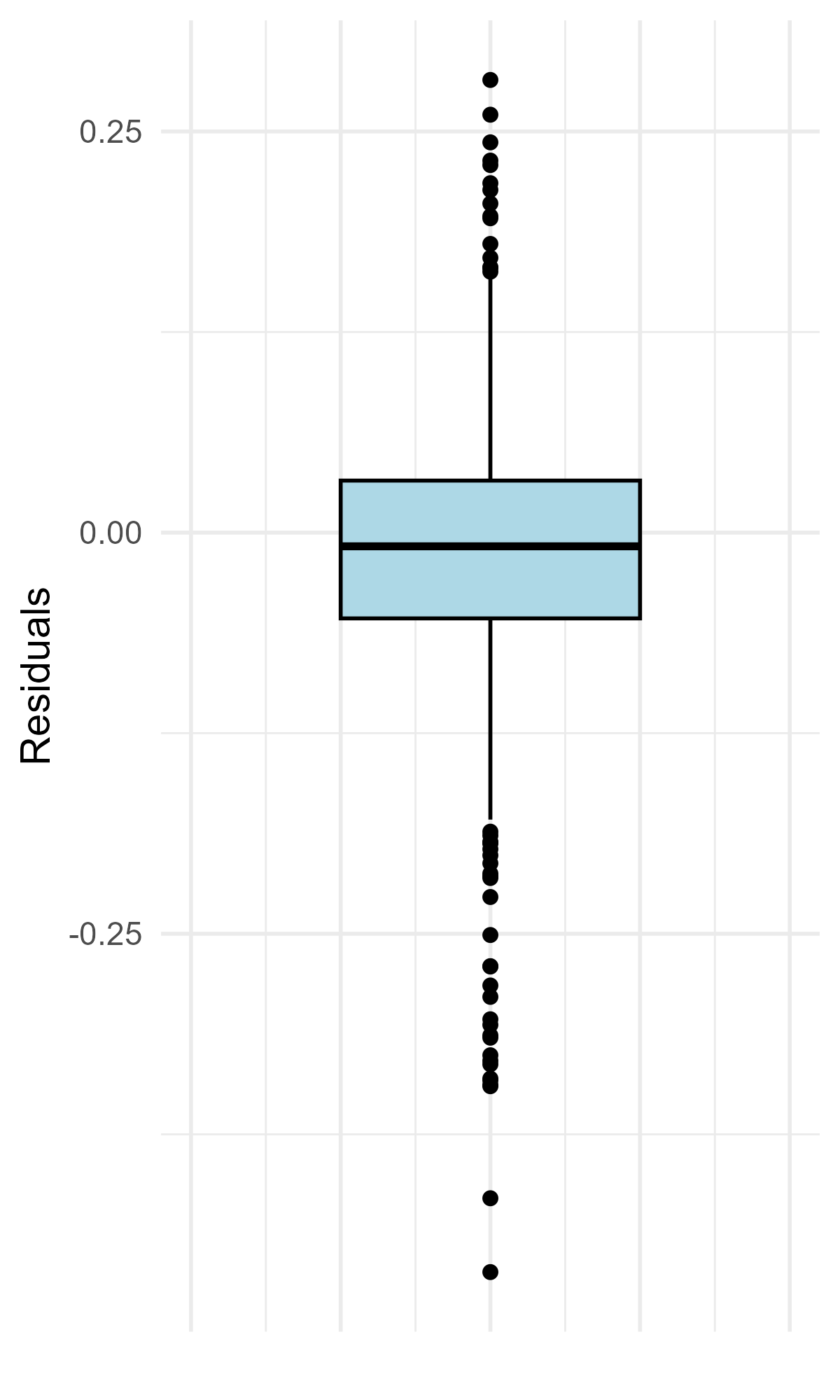}
  \caption{Goodness-of-fit diagnostics for the prediction of very extreme sea levels using the MGPRED procedure: PIT plot (left), predictions vs. observations plot (middle), and boxplot of residuals (right). In this setting, the correlation coefficient is 0.75. Results are shown on the subset of the test set comprising the most extreme observations w.r.t. the Port Tudy observations, i.e., observations such that $Y_i \s q^{ext,0.5}_Y$.
  \label{fig:gof_predictions_mgpd_sl_th}}
\end{figure}

In this section, we present first additional diagnostics using the skew surges at Port Tudy and the complete study using the sea levels at Port Tudy as output variable.
Then, we present more briefly the studies using Concarneau and Le Crouesty as output tide gauges. 
{\color{blue}
\subsection{Skew surge study at Port Tudy}\label{sec:tudy_appendix}

\subsubsection{Additional diagnostics}
We include three additional formal diagnostics to assess the performance of our models on skew surge data (and on sea level data at the end of Section~\ref{sec:sl_tudy}): PIT plots, prediction vs. observation plots, and residual boxplots computed on the test set for both skew surge and sea level data. Given the model’s trend to overestimate small values, we present these diagnostics both for the full test set (Figures \ref{fig:gof_predictions_ols_ss}, \ref{fig:gof_predictions_ols_sl}, \ref{fig:gof_predictions_mgpd_ss}, \ref{fig:gof_predictions_mgpd_sl}) and for the subset of the most extreme observations at Port Tudy (Figures \ref{fig:gof_predictions_ols_ss_th}, \ref{fig:gof_predictions_ols_sl_th}, \ref{fig:gof_predictions_mgpd_ss_th}, \ref{fig:gof_predictions_mgpd_sl_th}), as for the results in Table~\ref{fig:qqplot_tudy_ss}. These diagnostics are shown for the MGPRED procedure and the ROXANE routine with the OLS algorithm; results for the RF algorithm are omitted, as they are very similar to OLS.
    For the diagnostics on the full test set, the overestimation of small values is clearly visible, although the boxplot means remain close to zero. For the subset of the most extreme observations, errors are more evenly distributed, and residuals appear centred and less dispersed. Notably, model performance is best in the most extreme range, where larger errors might have been expected, which supports the strength of the model in this region. Regarding figures of Section~\ref{sec:sl_tudy}, performance is generally more satisfactory for sea levels than for skew surges. However, no substantial differences are observed between the ROXANE and MGPRED procedures.
\begin{figure}[h!]
  \centering
  \includegraphics[width=.36\textwidth]{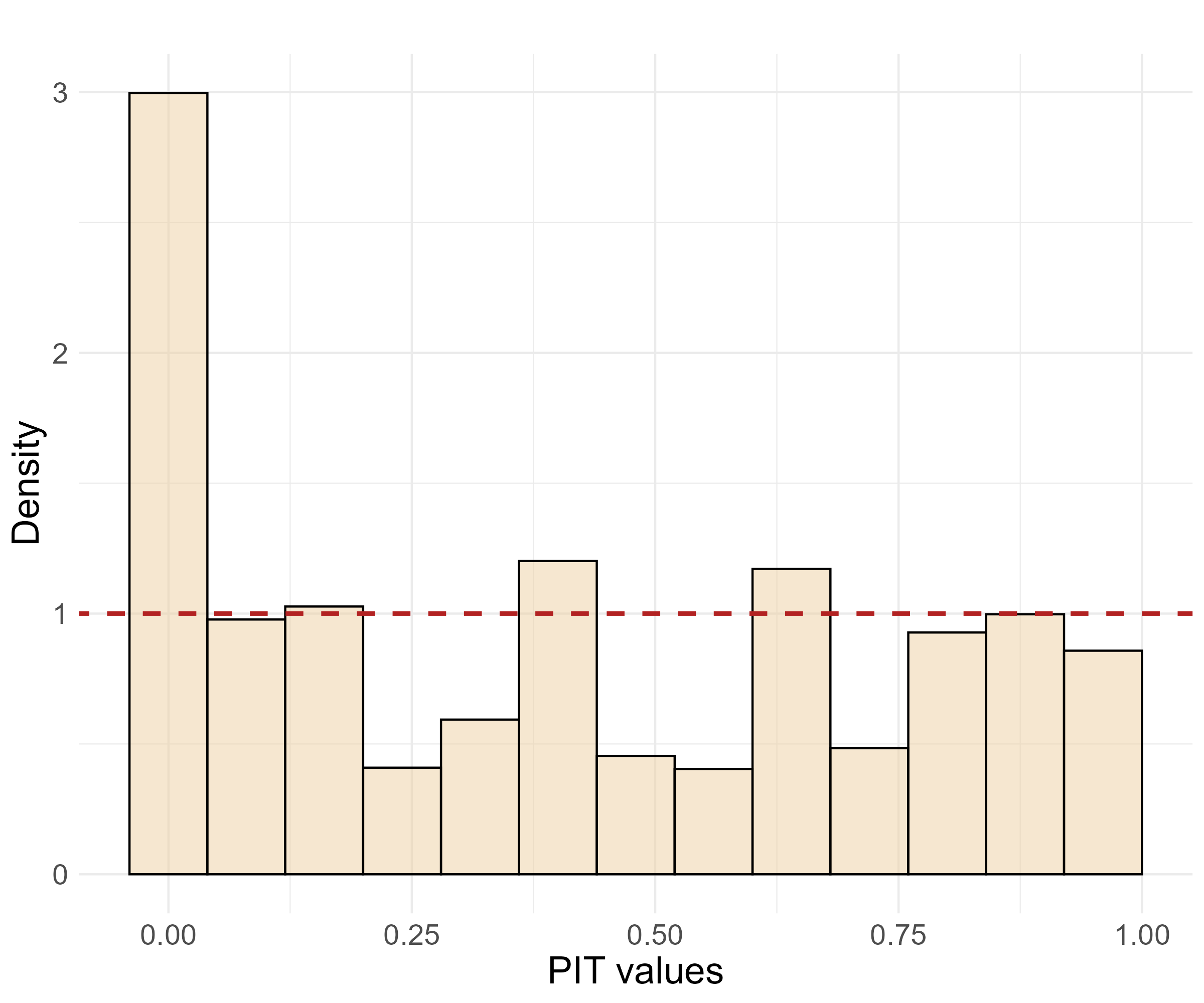}
	\quad
\includegraphics[width=.36\textwidth]{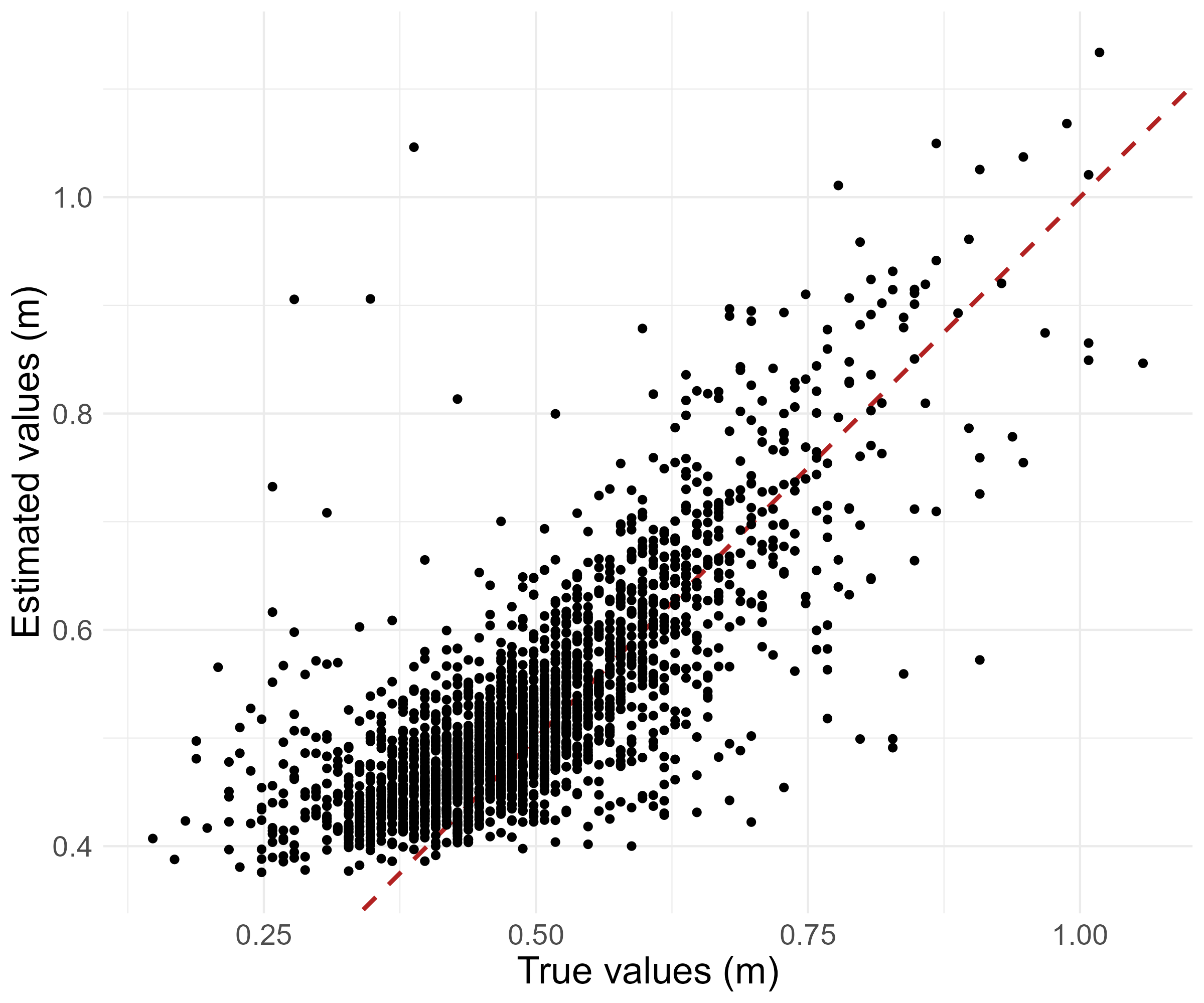}
  \quad
  \includegraphics[width=.18\textwidth]{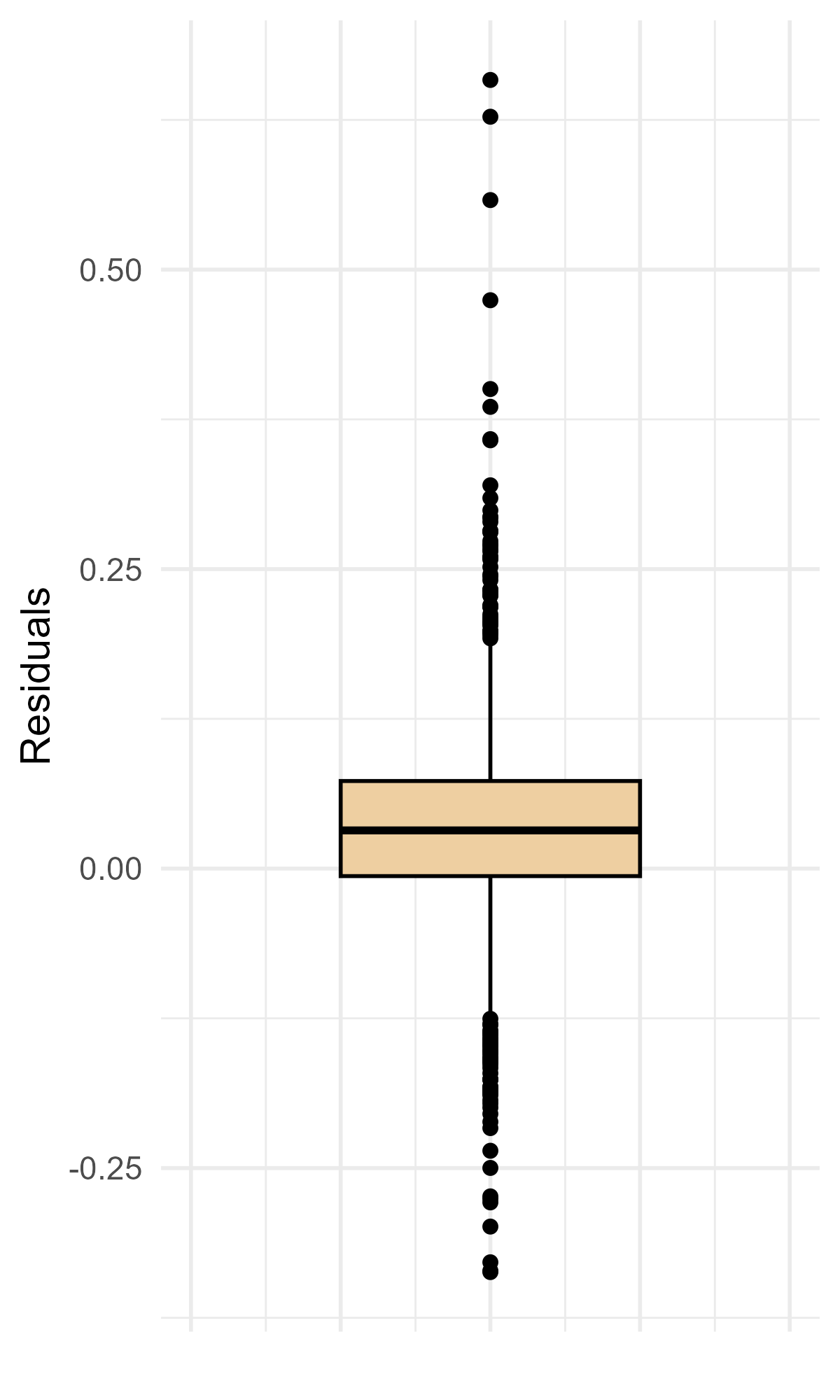}
  \caption{Goodness-of-fit diagnostics for the prediction of extreme skew surges using ROX OLS: PIT plot (left), predictions vs. observations plot (middle), and boxplot of residuals (right). In this setting, the correlation coefficient is 0.76.
  \label{fig:gof_predictions_ols_ss}}
\end{figure}  

\begin{figure}[h!]
  \centering
  \includegraphics[width=.36\textwidth]{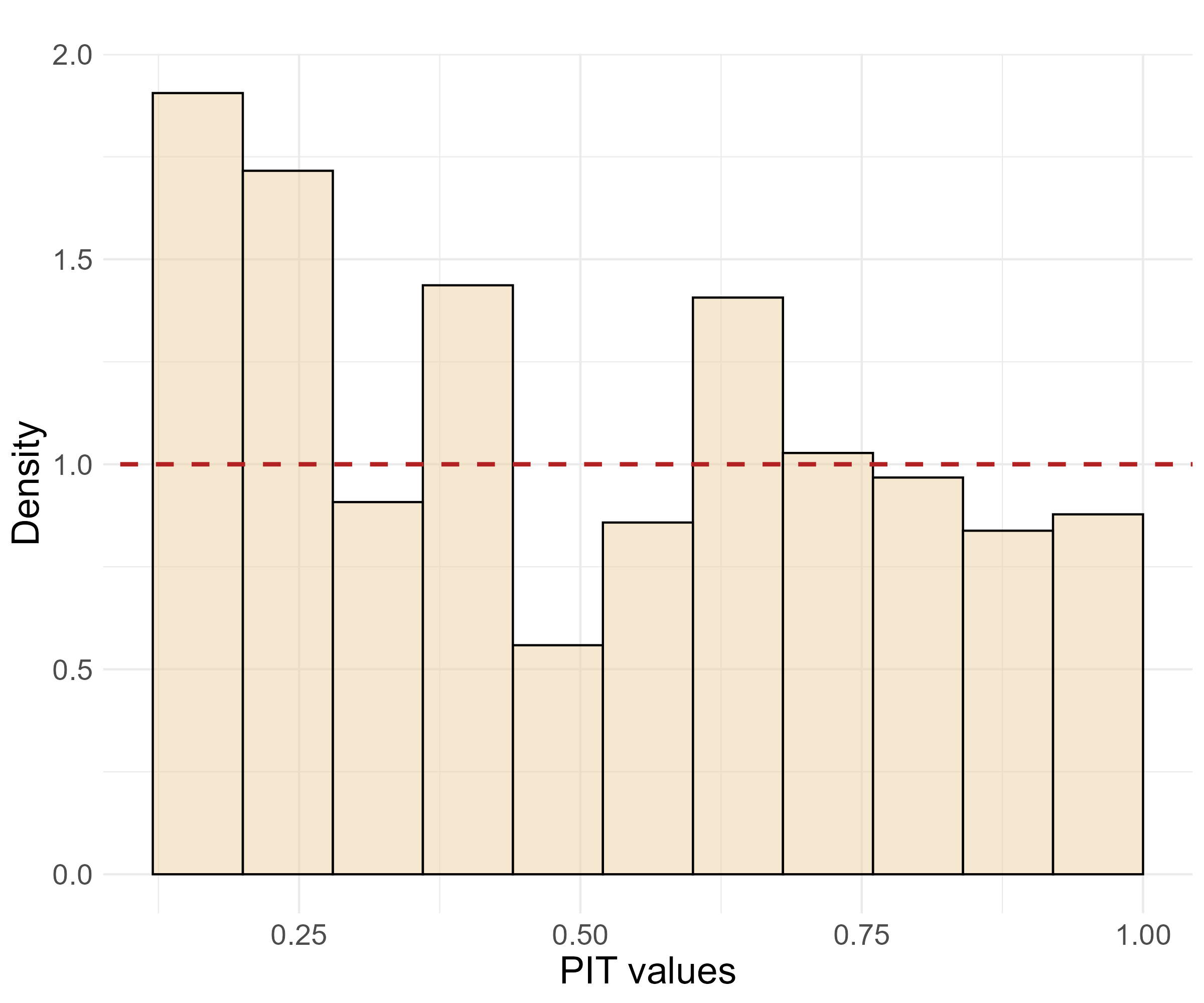}
	\quad
\includegraphics[width=.36\textwidth]{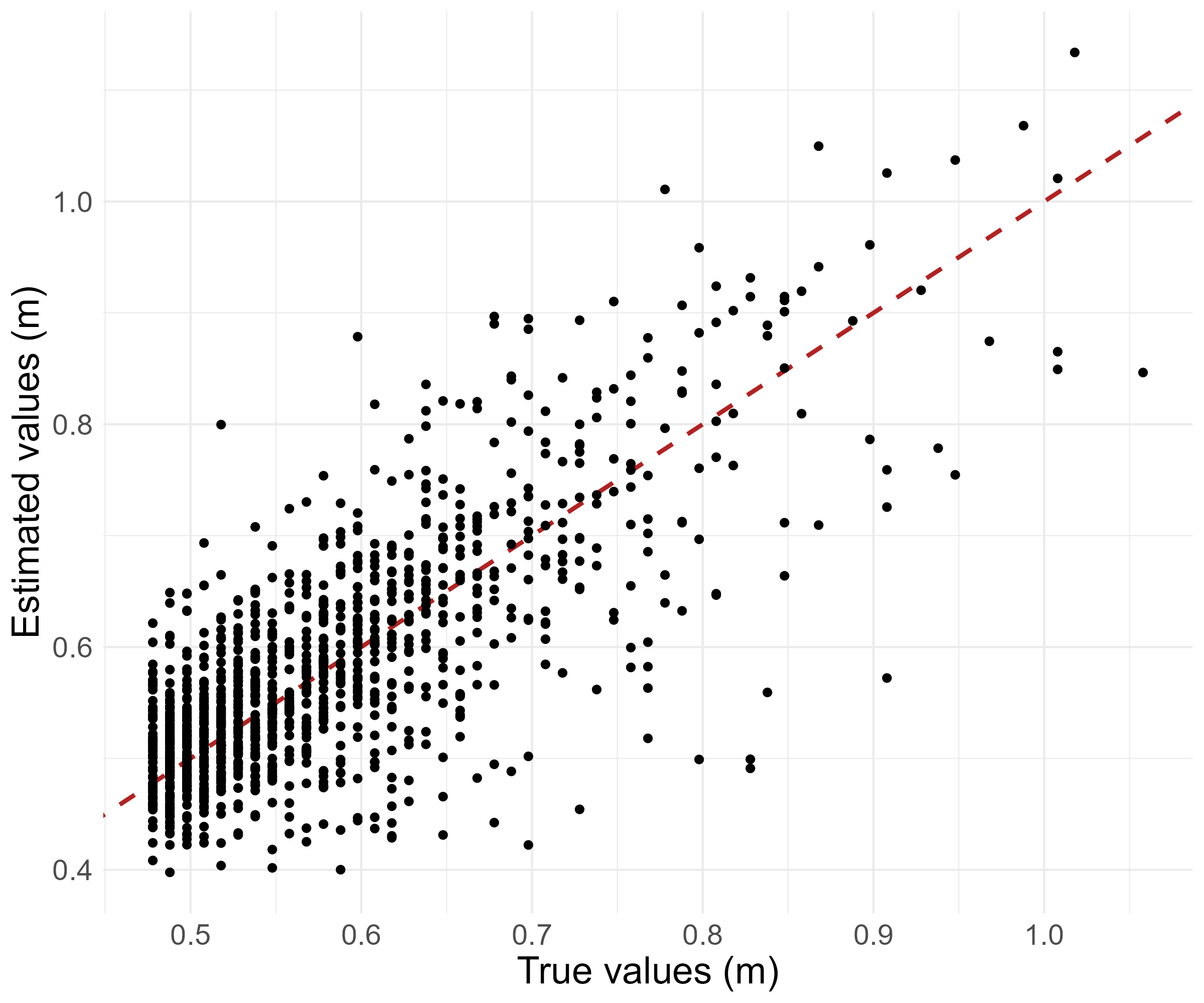}
  \quad
  \includegraphics[width=.18\textwidth]{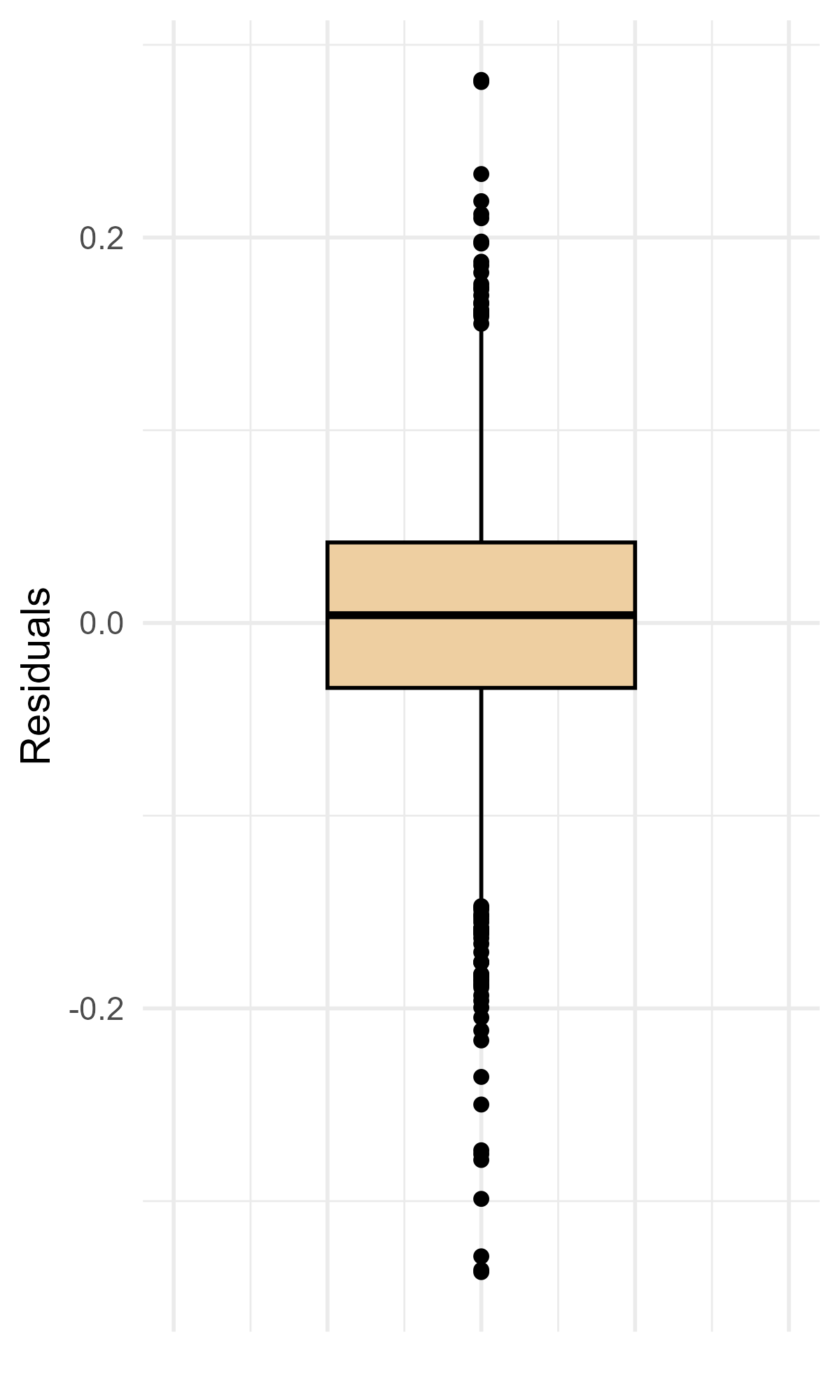}
  \caption{Goodness-of-fit diagnostics for the prediction of very extreme skew surges using ROX OLS: PIT plot (left), predictions vs. observations plot (middle), and boxplot of residuals (right). In this setting, the correlation coefficient is 0.76. Results are shown on the subset of the test set comprising the most extreme observations w.r.t. the Port Tudy observations, i.e., observations such that $Y_i \s q^{ext,0.5}_Y$.
  \label{fig:gof_predictions_ols_ss_th}}
\end{figure}

\begin{figure}[h!]
  \centering
  \includegraphics[width=.36\textwidth]{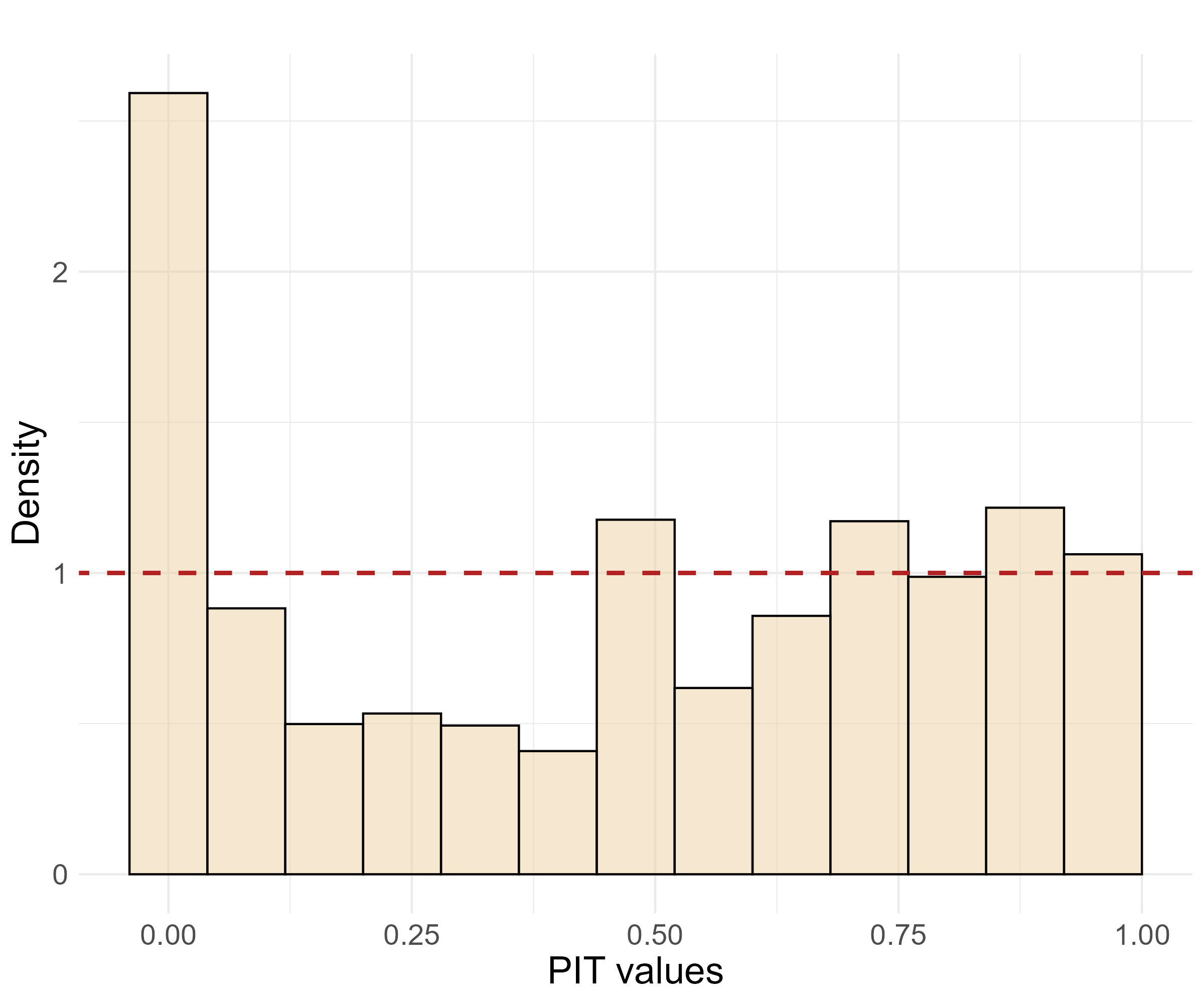}
	\quad
\includegraphics[width=.36\textwidth]{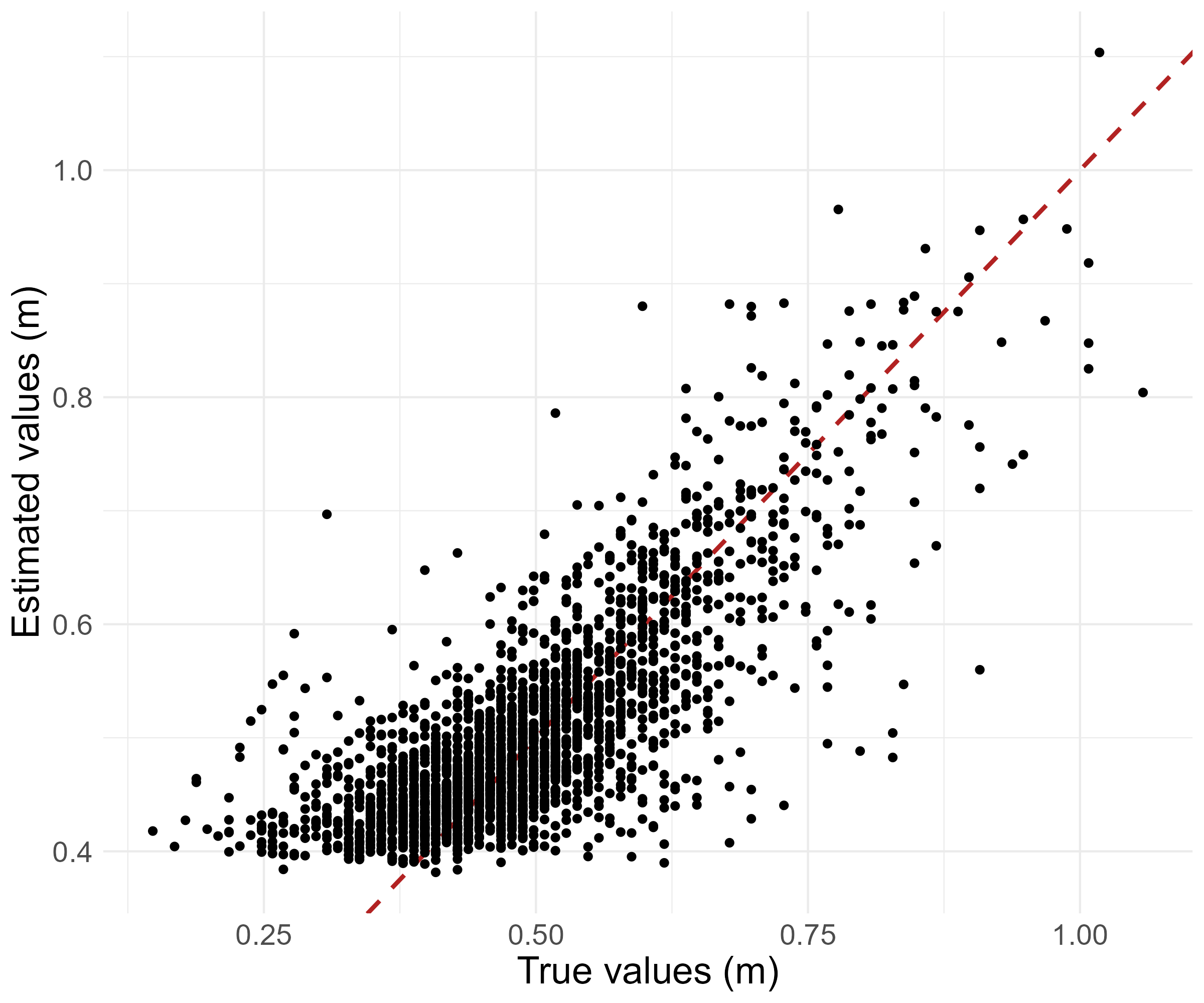}
  \quad
  \includegraphics[width=.18\textwidth]{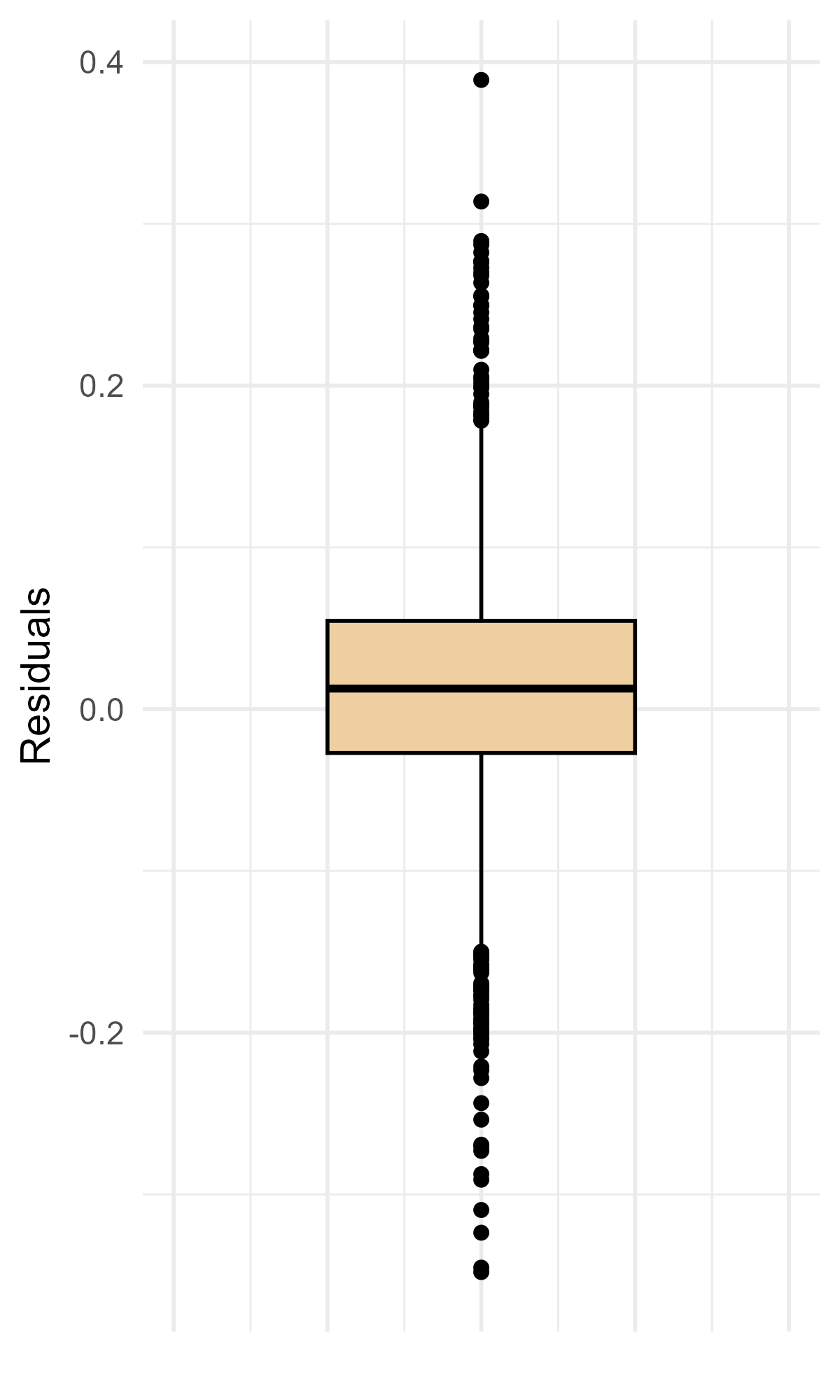}
  \caption{Goodness-of-fit diagnostics for the prediction of extreme skew surges using the MGPRED procedure: PIT plot (left), predictions vs. observations plot (middle), and boxplot of residuals (right). In this setting, the correlation coefficient is 0.78.
  \label{fig:gof_predictions_mgpd_ss}}
\end{figure}

\begin{figure}[h!]
  \centering
  \includegraphics[width=.36\textwidth]{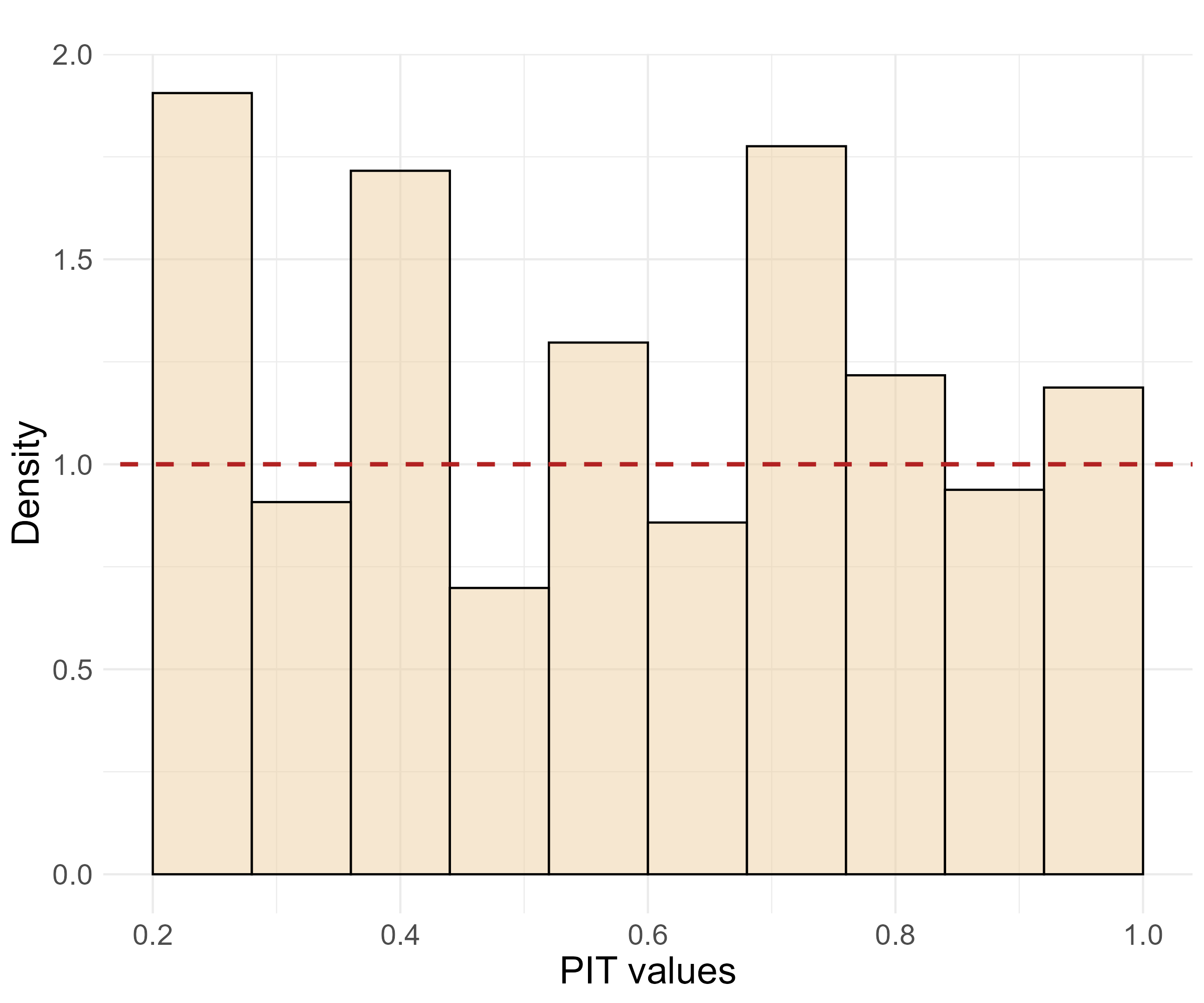}
	\quad
\includegraphics[width=.36\textwidth]{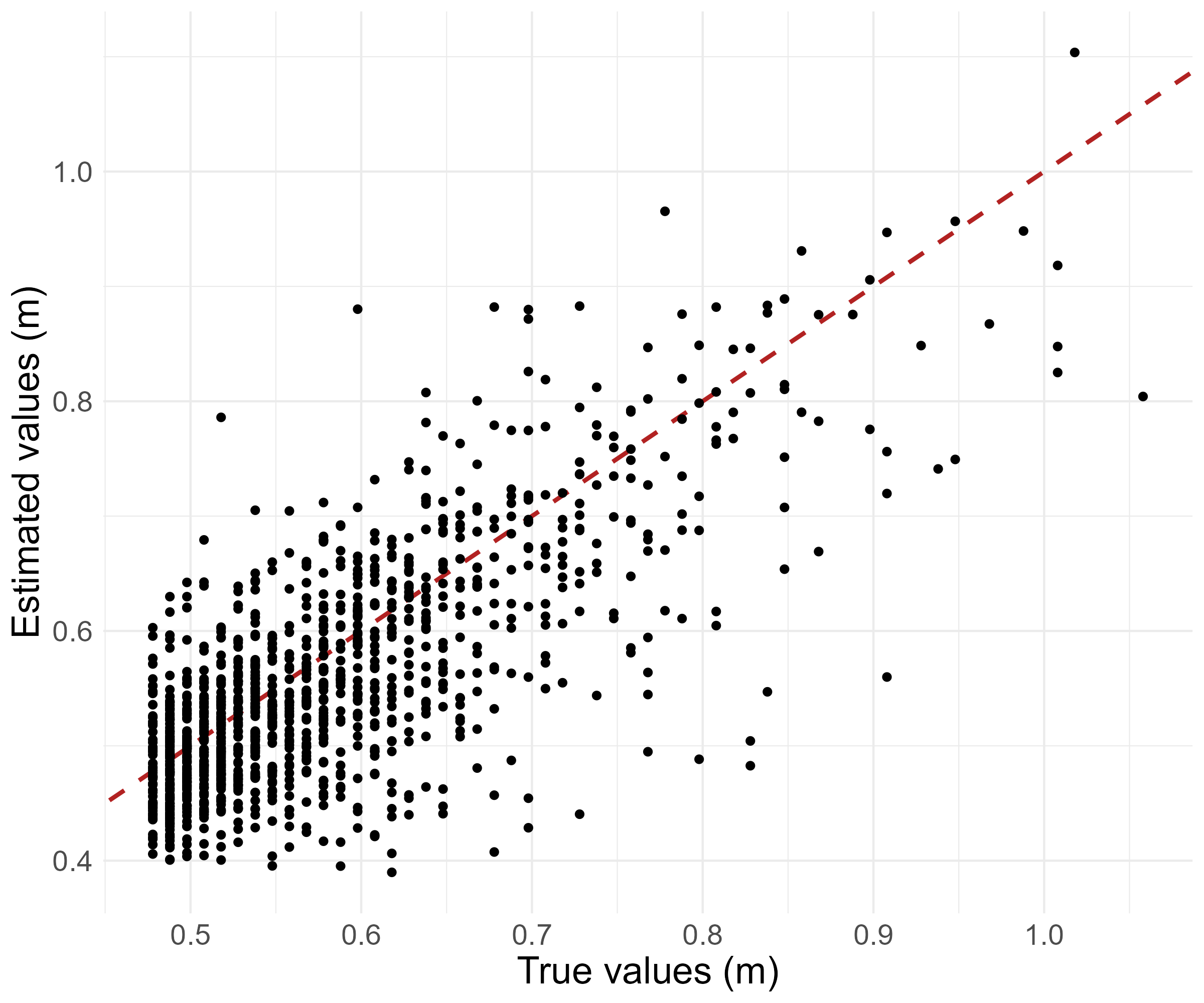}
  \quad
  \includegraphics[width=.18\textwidth]{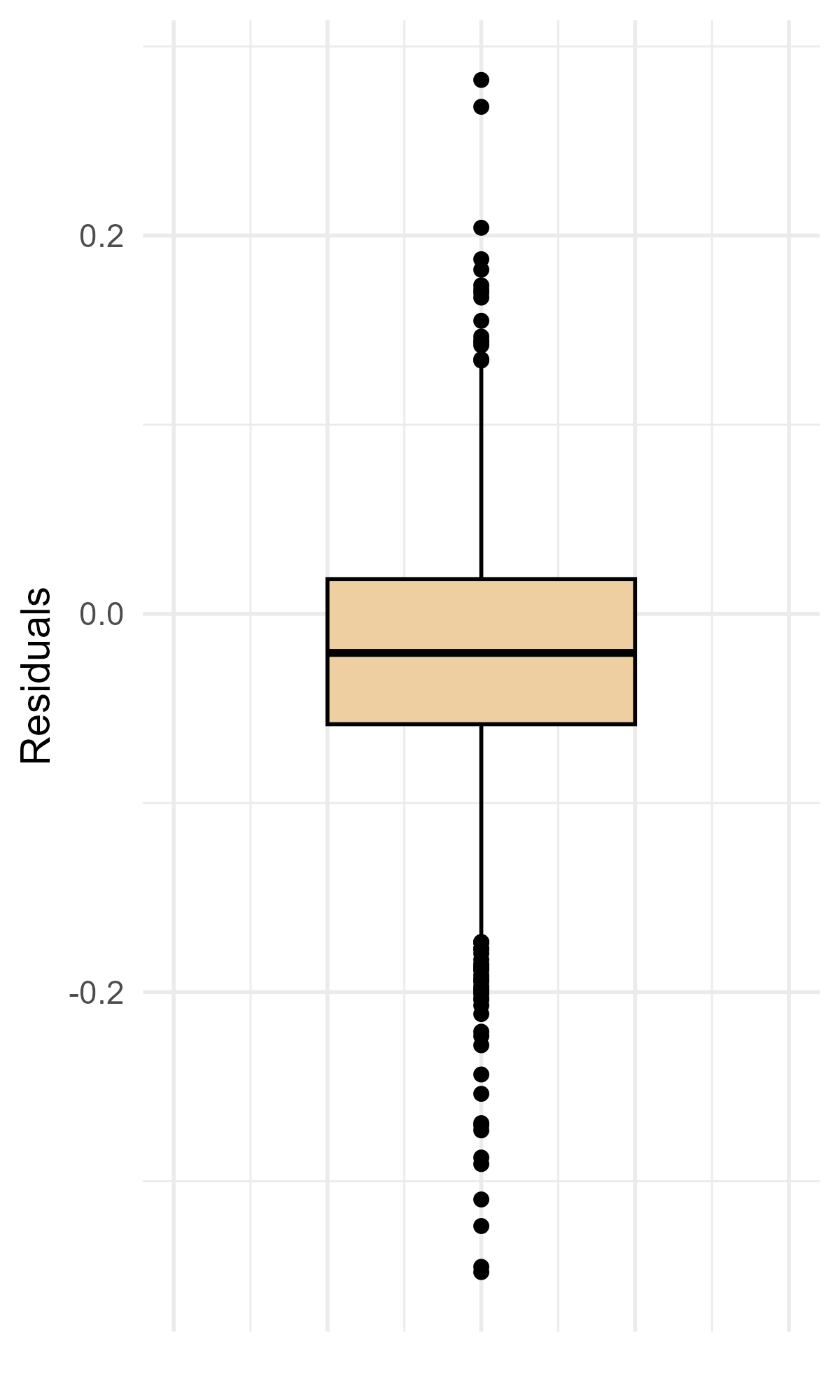}
  \caption{Goodness-of-fit diagnostics for the prediction of very extreme skew surges using the MGPRED procedure: PIT plot (left), predictions vs. observations plot (middle), and boxplot of residuals (right). In this setting, the correlation coefficient is 0.75. Results are shown on the subset of the test set comprising the most extreme observations w.r.t. the Port Tudy observations, i.e., observations such that $Y_i \s q^{ext,0.5}_Y$.
  \label{fig:gof_predictions_mgpd_ss_th}}
\end{figure} 

\subsubsection{Sensitivity analysis}

To study the sensitivity of parameter estimates, we examine their variability across bootstrap resamples. Because the data are time series, we use block bootstrap methods \cite[see, e.g.,][]{lahiri1999theoretical}, specifically the geometric block bootstrap of \cite{politis1994stationary}, with mean block size $l = 50$, determined from the marginal autocorrelation functions. For each station, we generate $R = 1000$ bootstrap samples and re-estimate the EGP parameters. Figure \ref{fig:stability_margins_ss} (Figure~\ref{fig:stability_margins_sl} for sea level data) displays histograms of the parameter estimates, while Figure \ref{fig:stability_rl_ss} (Figure~\ref{fig:stability_rl_sl} for sea level data) presents histograms of the corresponding 100-year return levels and of the RMSE values computed on the fixed test set but with varying training samples (ROXANE-OLS routine), as global measure of sensitivity.

The scale $\sigma$ and shape $\xi$ parameters are generally stable across bootstrap samples, with reasonable standard errors. The only exception occurs for the skew surge data at Port Tudy, where two regimes appear: a dominant one around $\xi \approx 0$ and $\sigma \approx 0.9$, and a secondary one with $\xi \approx 0.18$ and $\sigma \approx 0.03$, which likely arises from atypical bootstrap samples containing many extreme values. The $\kappa$ parameter shows greater variability, especially for skew surges.
Moreover, as with the GEV or GP distributions, the EGP parameters are known to interact and compensate each other, in the sense that different parameter combinations produce similar tail quantiles. Therefore, it is more meaningful to assess the sensitivity of the full distribution rather than individual parameters. A natural and commonly used global indicator of marginal robustness is the return level, and in particular, the 100-year return level. The standard errors across bootstrap samples are about 0.02 m for Brest and Saint-Nazaire, and 0.06 m for Port Tudy, for mean skew surges around 0.6 m. For sea levels, the standard errors are below 0.047 m for mean levels above 1 m, confirming the robustness of the model to resampling.
Finally, the multivariate measure of robustness, RMSE computed on the same test set but with varying bootstrap-based parameter fits, is also satisfactory, with standard errors of 0.015 m for skew surges and 0.001 m for sea levels. Although the RMSE for skew surges is less stable than for sea levels, the variability remains limited, and the errors are in general relatively small. This indicates that, while individual parameter estimates may vary moderately, their joint effect yields very stable model performance.

\begin{figure}[ht!] 
  \centering
  \includegraphics[width=.315\textwidth]{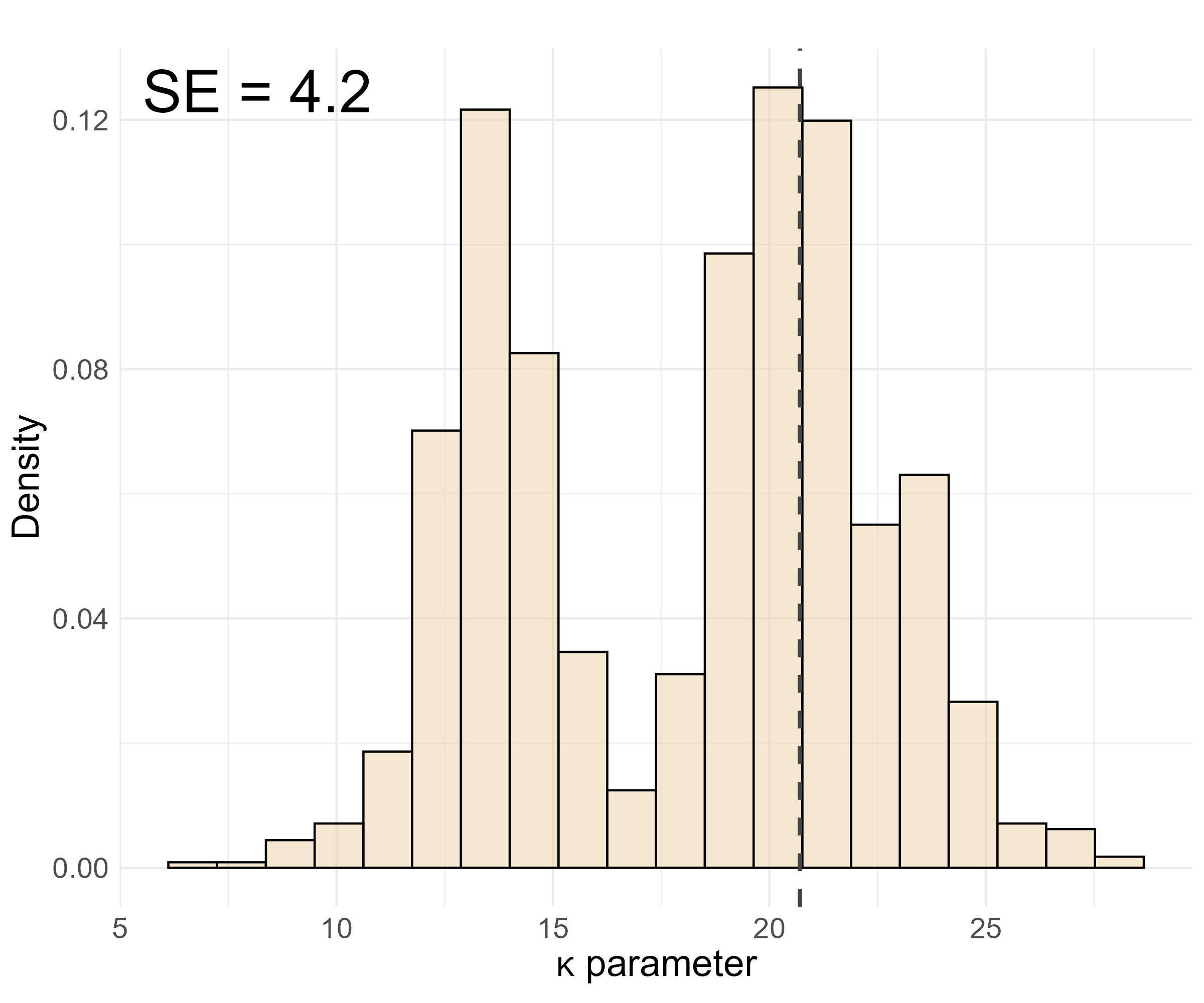}
  \hspace{0.2cm}
  \includegraphics[width=.315\textwidth]{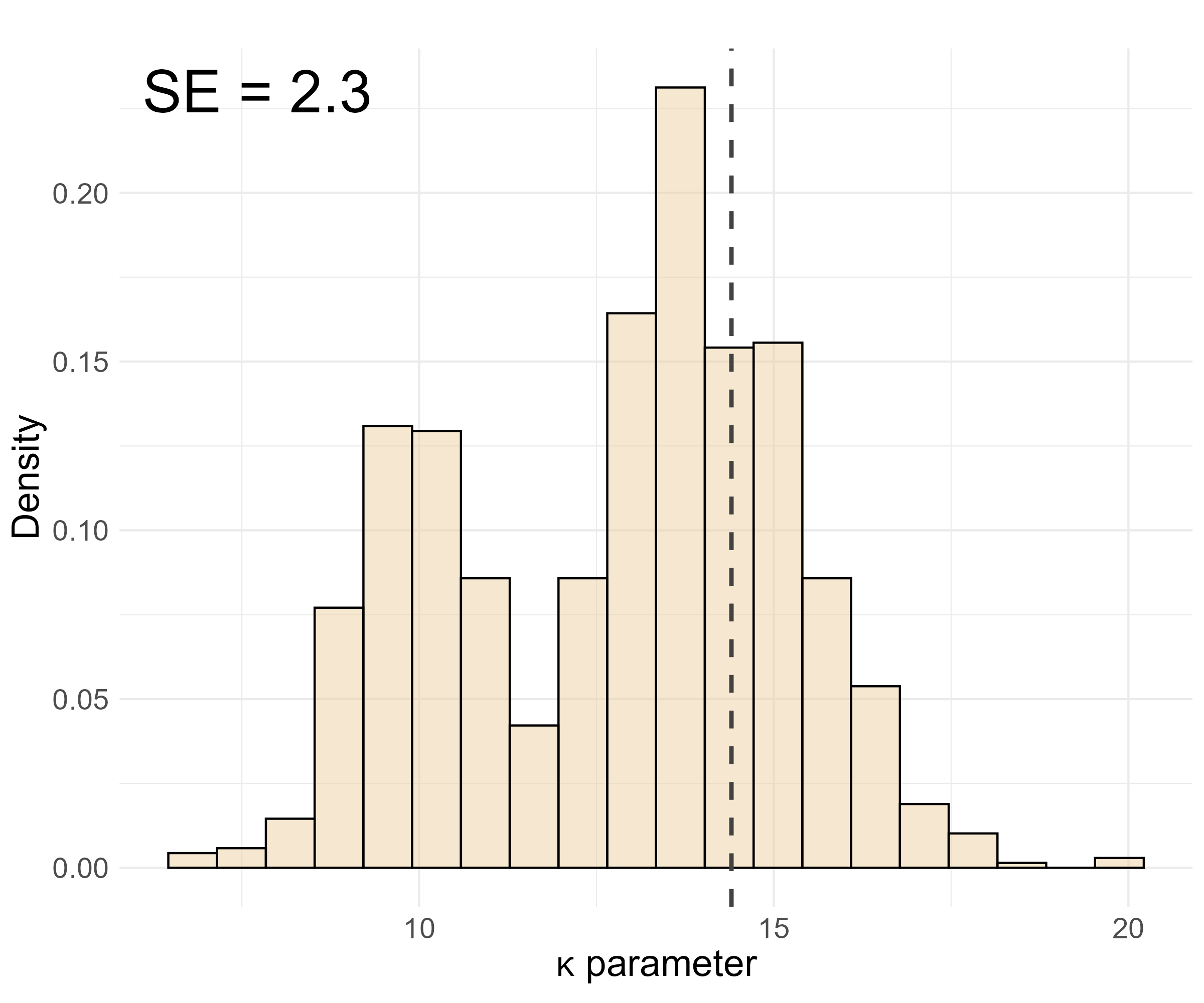}
  \hspace{0.2cm}
  \includegraphics[width=.315\textwidth]{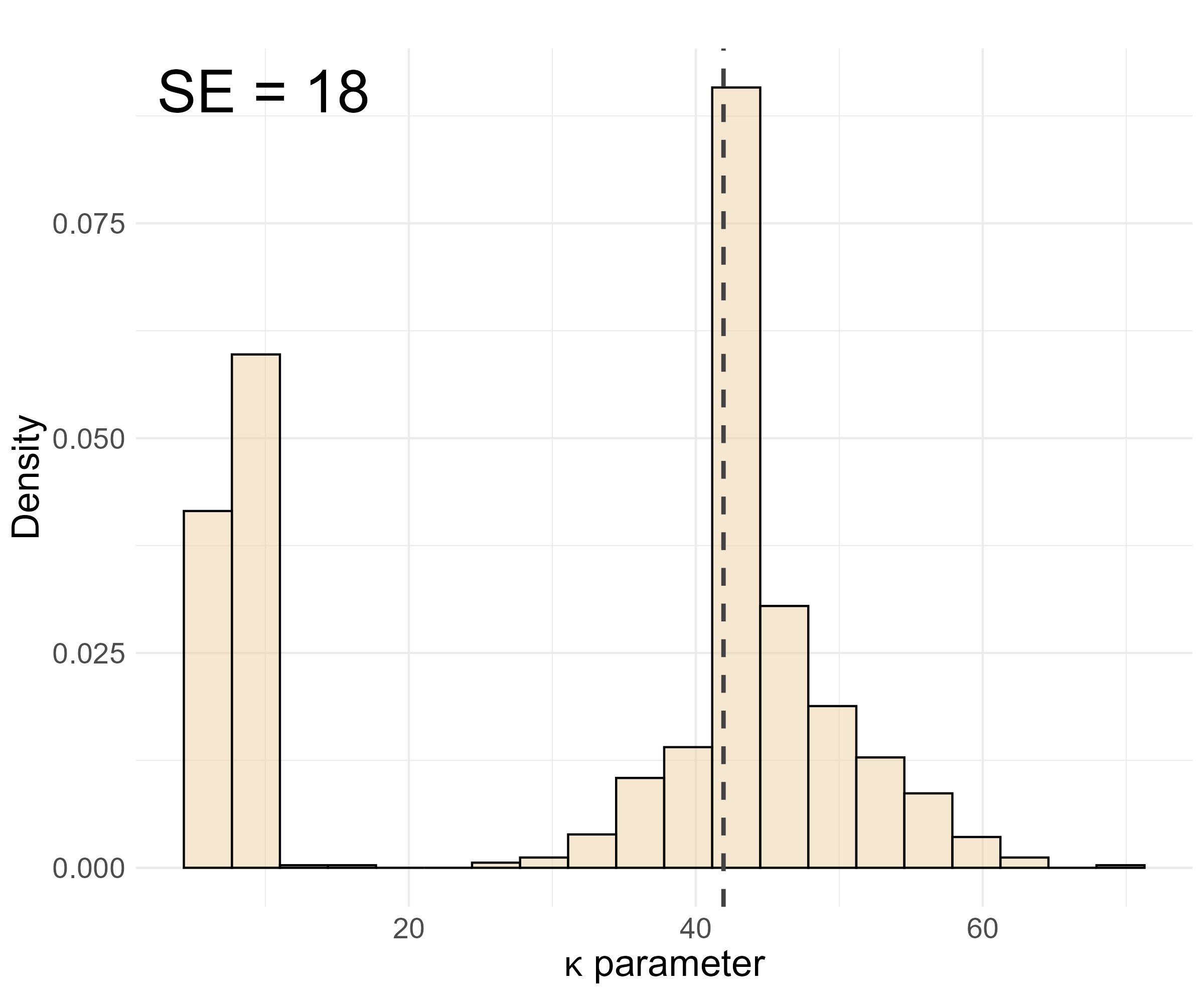}
  \vspace{0.2cm}
  \includegraphics[width=.315\textwidth]{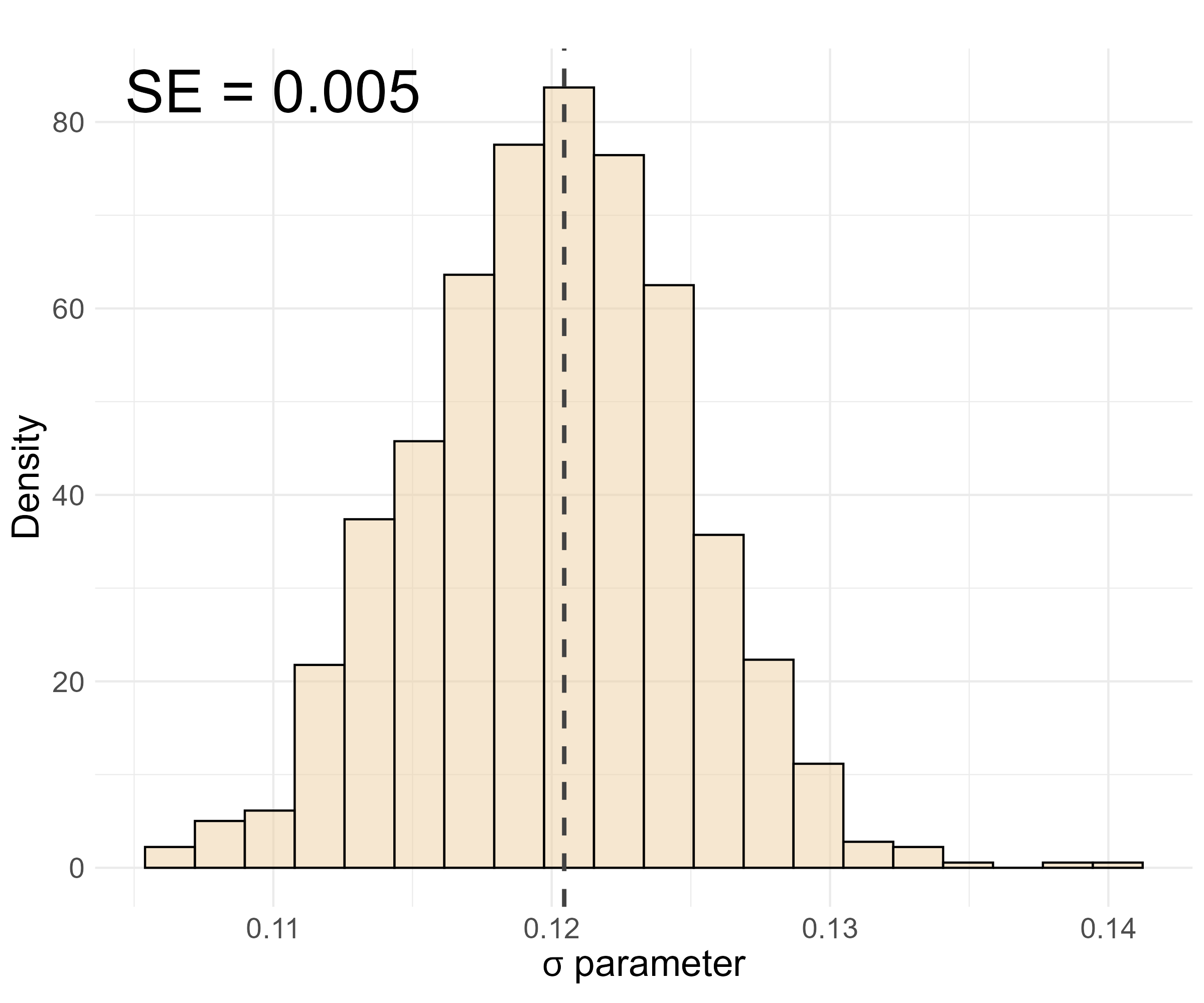}
  \hspace{0.2cm}
  \includegraphics[width=.315\textwidth]{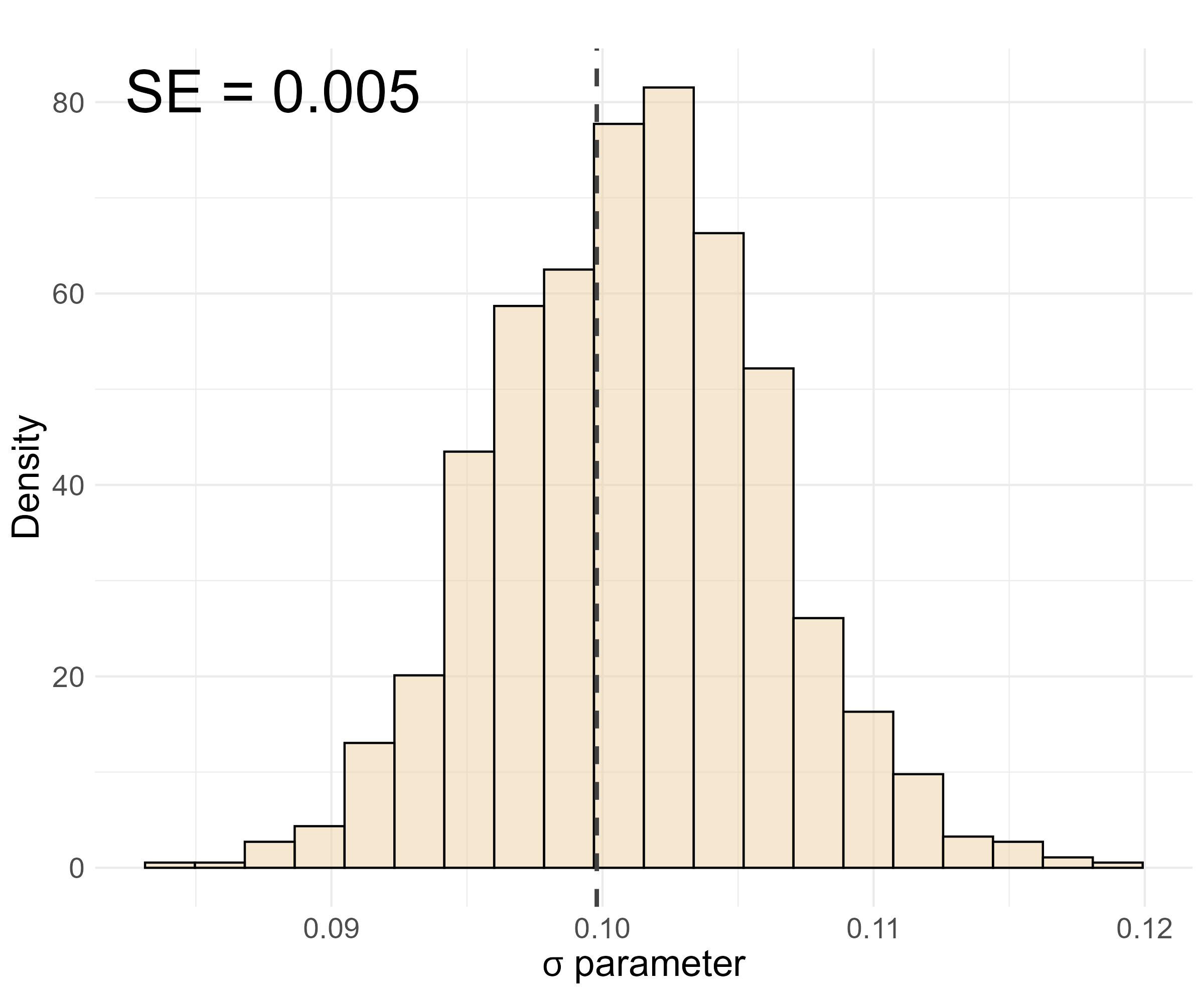}
  \hspace{0.2cm}
  \includegraphics[width=.315\textwidth]{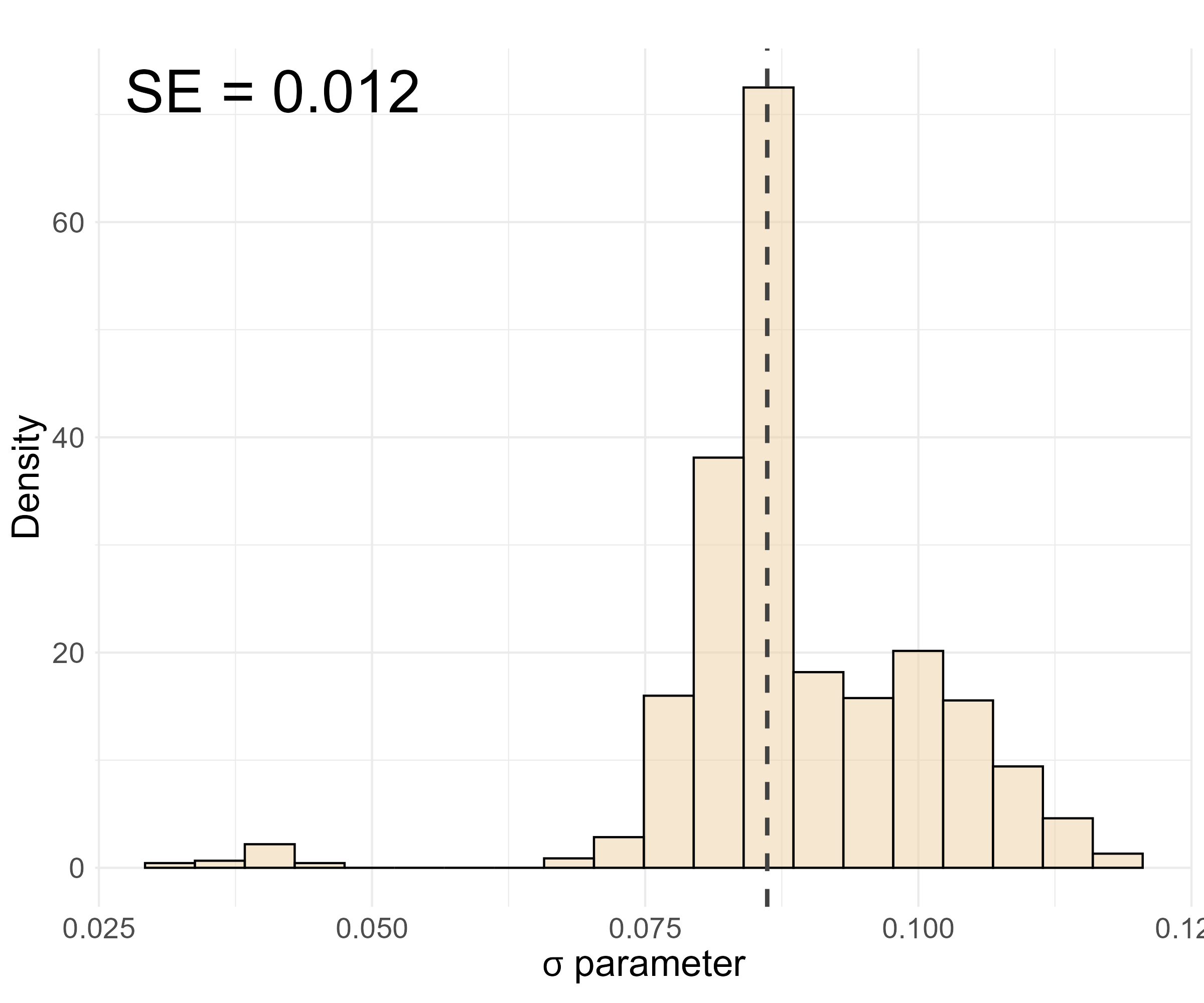}
  \vspace{0.2cm}
  \includegraphics[width=.315\textwidth]{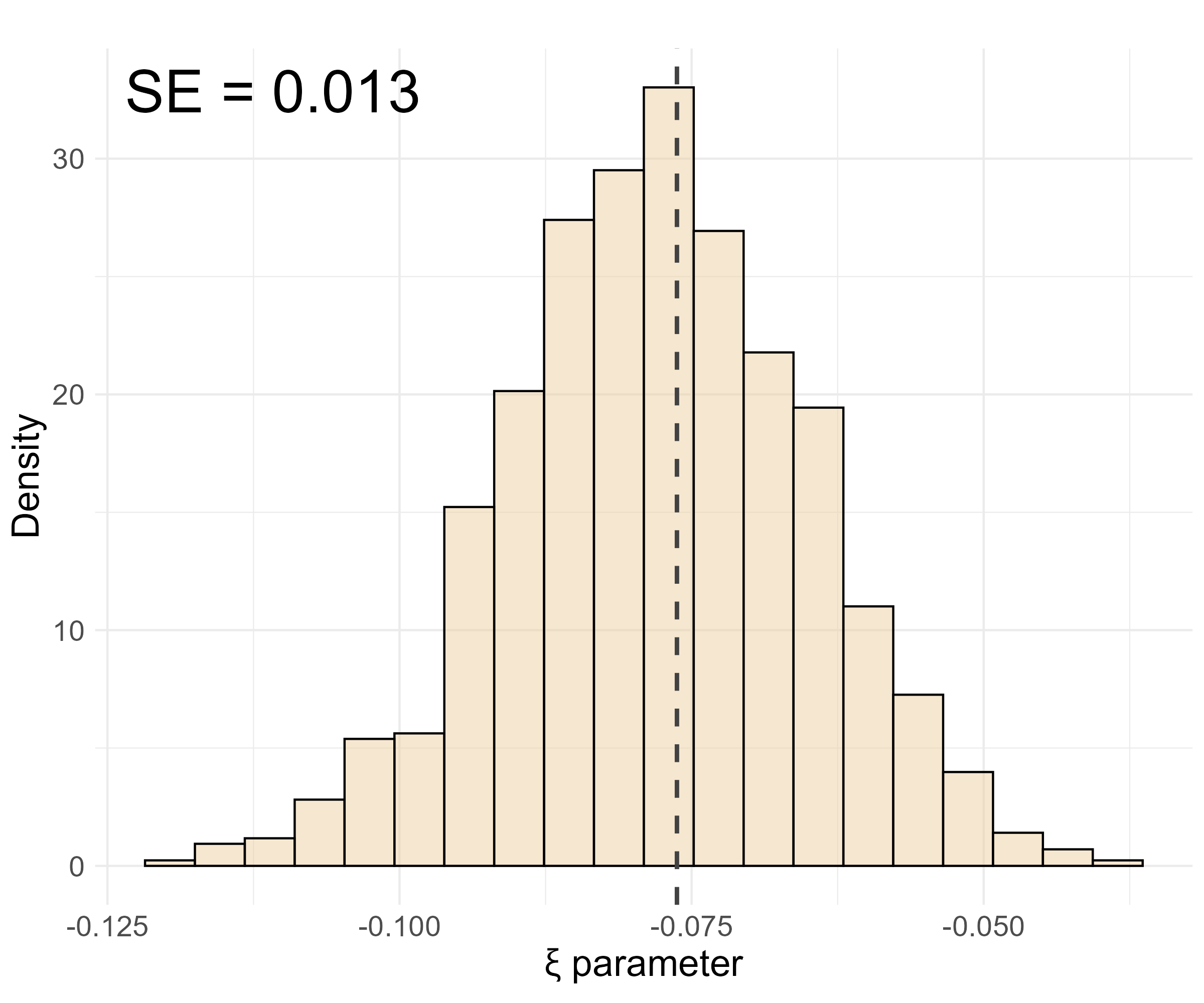}
  \hspace{0.2cm}
  \includegraphics[width=.315\textwidth]{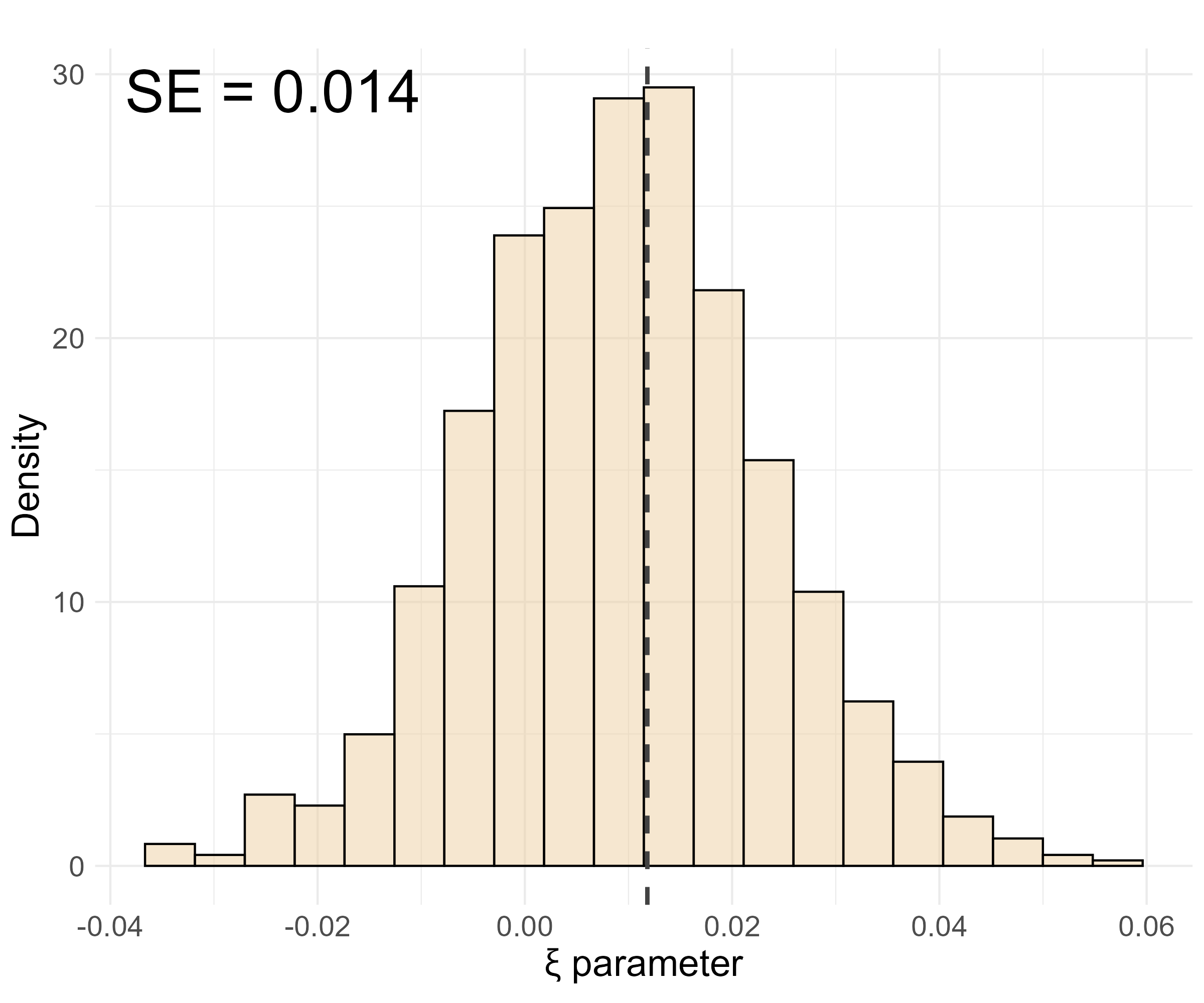}
  \hspace{0.2cm}
  \includegraphics[width=.315\textwidth]{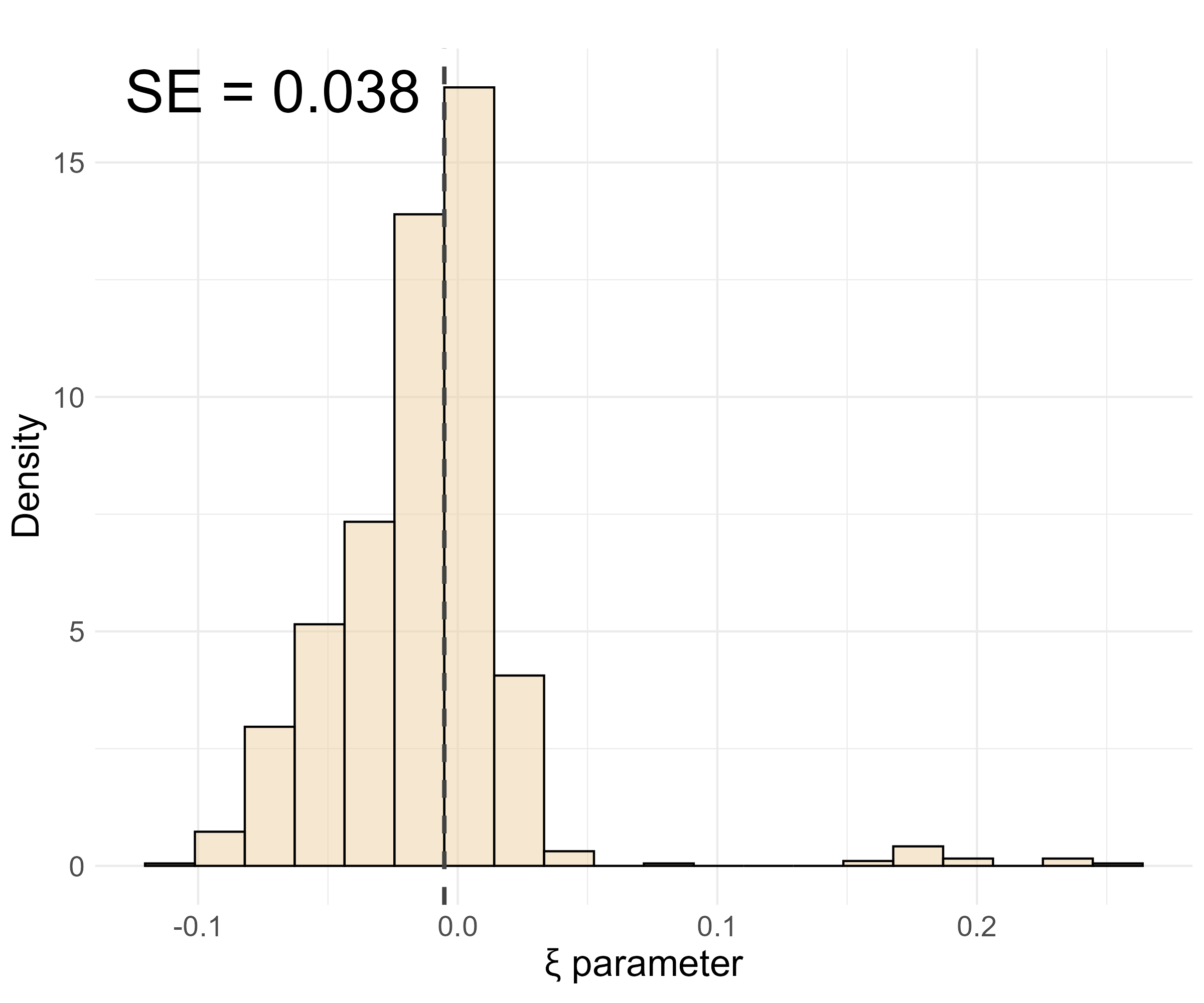}
\caption{Histograms of the EGP parameter estimators computed from $R = 1000$ bootstrap samples, using geometric block bootstrap with mean block length $l = 50$, for the skew surge data at each station. The grey vertical dashed line represents the estimate obtained from the full dataset (without resampling). The standard error is indicated in top-left corner of each figure.\label{fig:stability_margins_ss}}
\end{figure}

\begin{figure}[ht!] 
  \centering
  \includegraphics[width=.23\textwidth]{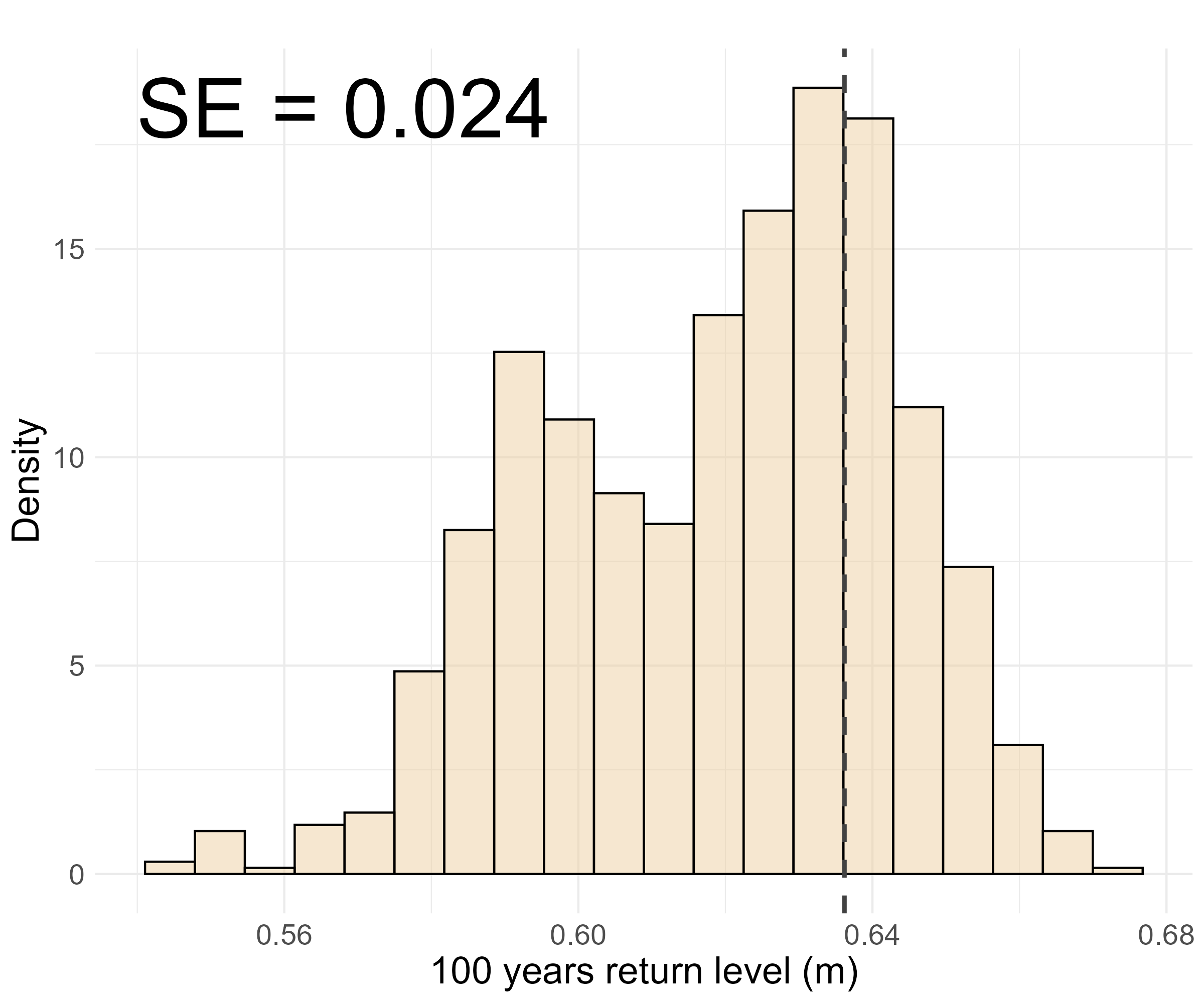}
  \hspace{0.2cm}
  \includegraphics[width=.23\textwidth]{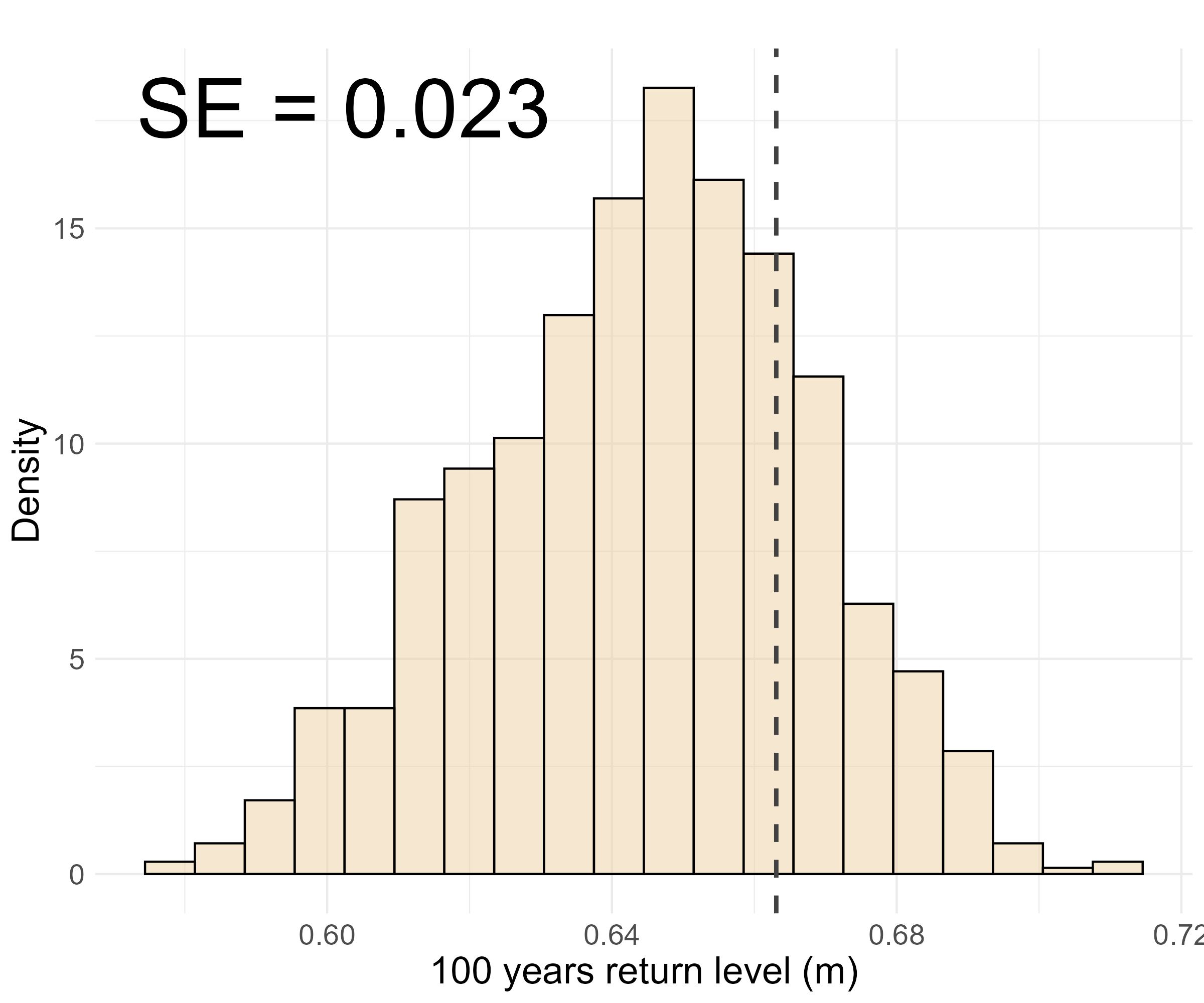}
  \hspace{0.2cm}
  \includegraphics[width=.23\textwidth]{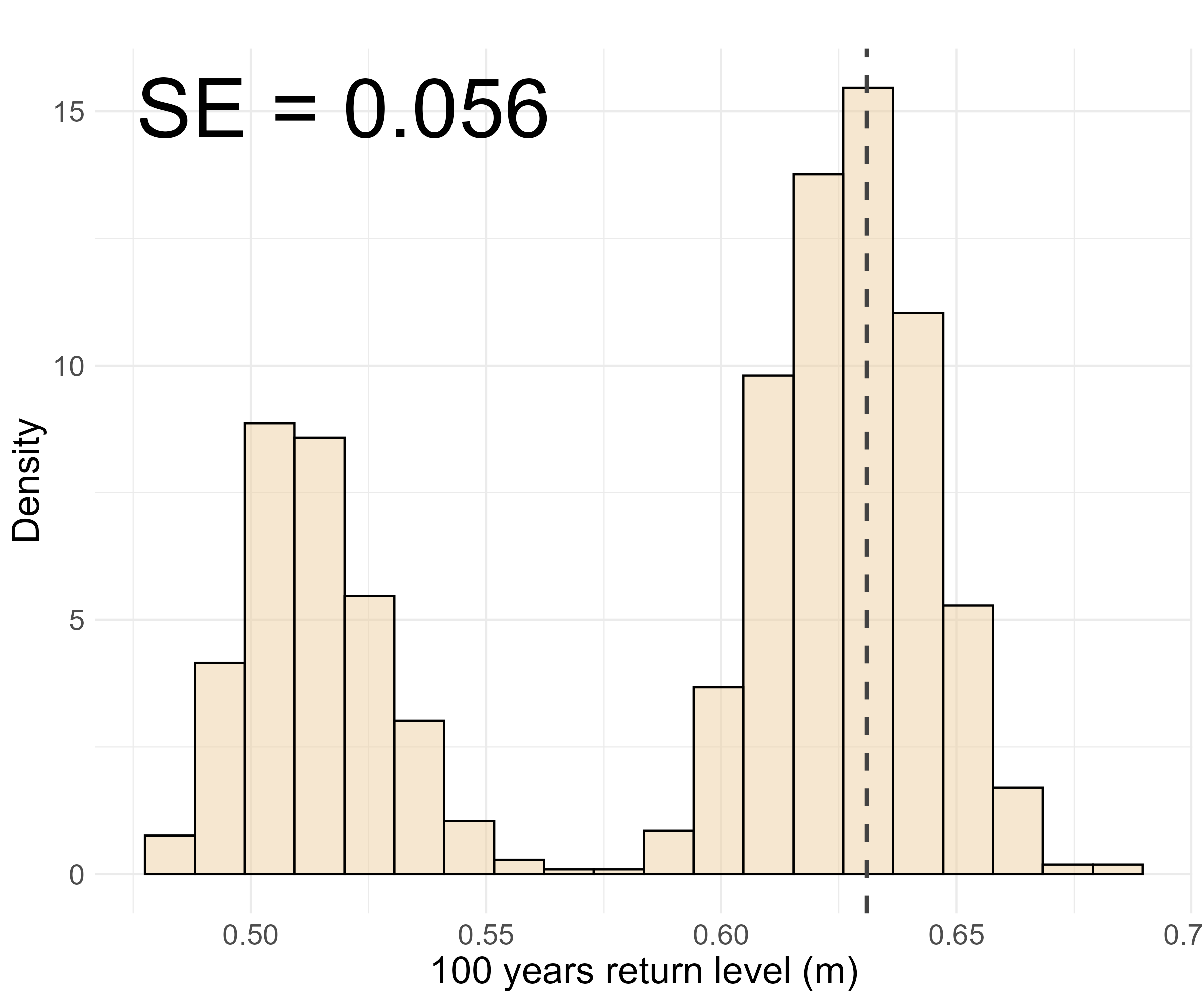}
\hspace{0.2cm}
  \includegraphics[width=.23\textwidth]{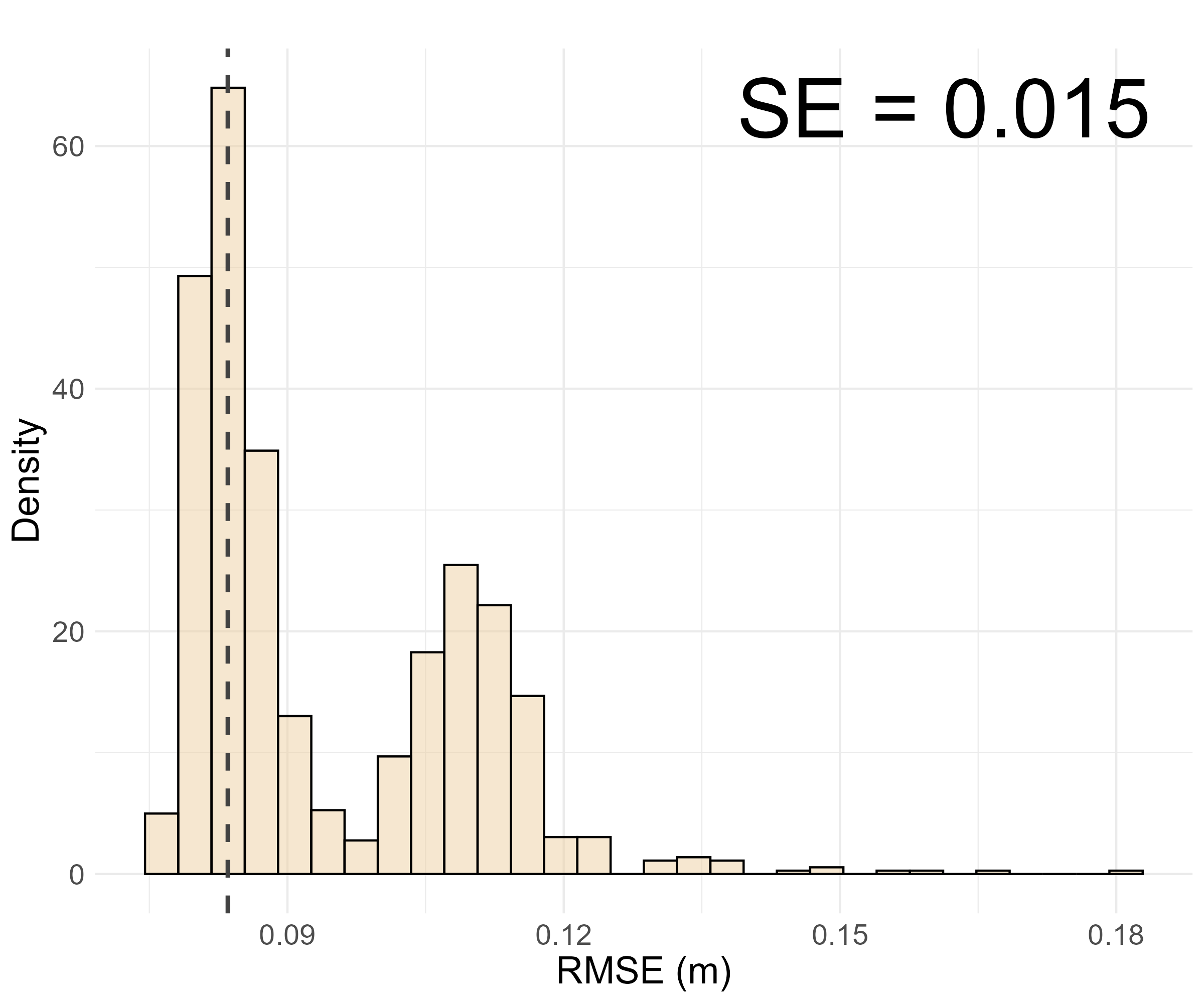}

\caption{Histograms of the 100-year return level (left and middle), computed using the EGP parameter estimates from $R = 1000$ bootstrap samples, using block
bootstrap with block length $l = 50$, for the skew surge data at each station (Brest: left; Saint-Nazaire: middle-left; Port Tudy: middle-right), and histogram of the RMSE (right) computed on the fixed test set using the ROXANE routine with the OLS algorithm and the different bootstrap training samples. The grey vertical dashed lines represent the estimates obtained from the full dataset (without resampling). The standard error is indicated in top-left corner of each figure.\label{fig:stability_rl_ss}}
\end{figure}
}

\clearpage
\subsection{Sea level study at Port Tudy}\label{sec:sl_tudy}
\begin{figure}[ht!]
\centering
  \includegraphics[width=.315\textwidth]{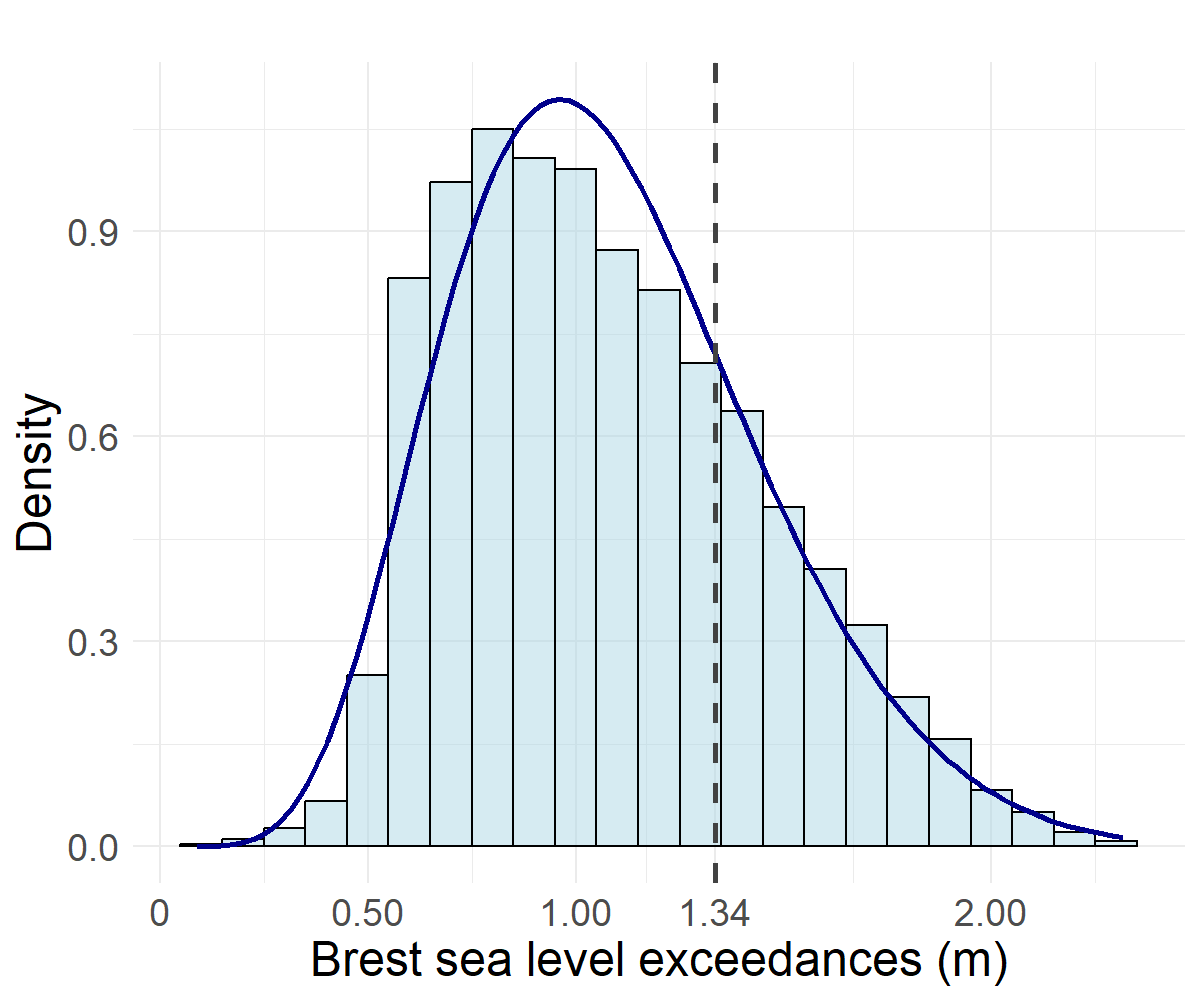}
  \hspace{0.2cm}
  \includegraphics[width=.315\textwidth]{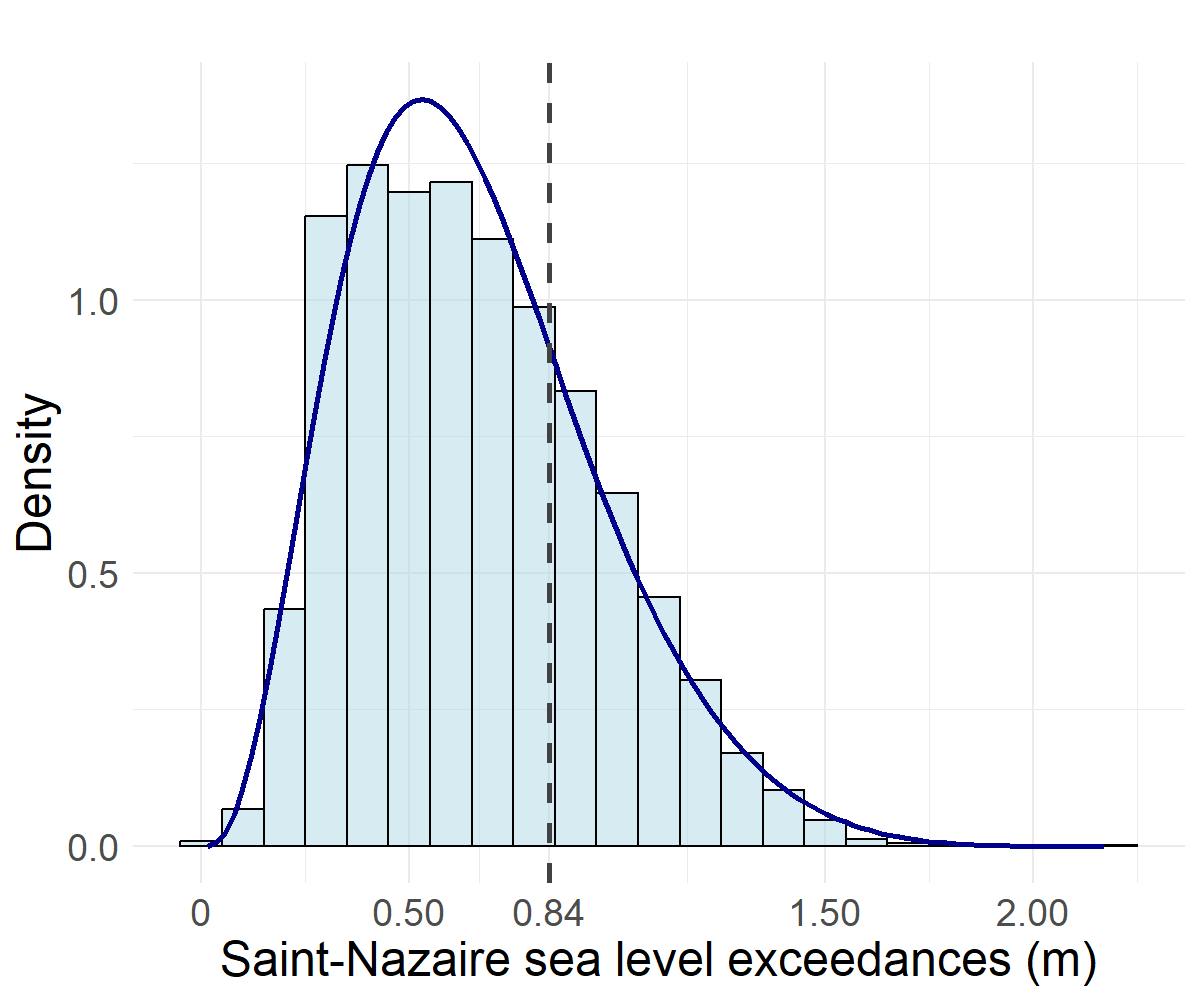}
  \hspace{0.2cm}
  \includegraphics[width=.315\textwidth]{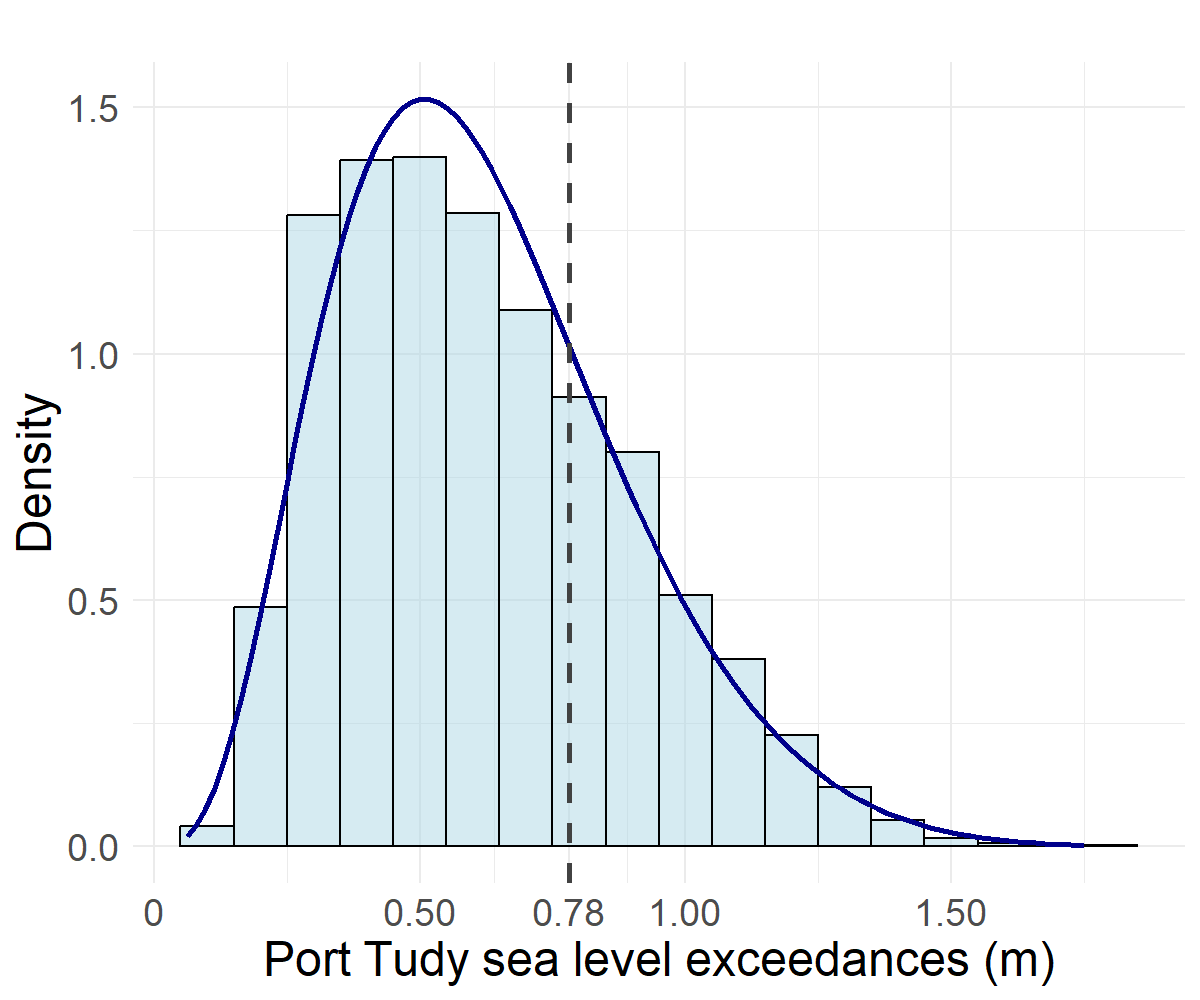}

  \caption{Histograms of sea level exceedances at the three stations Brest (left), Saint-Nazaire (middle) and Port Tudy (right), from 01/12/2000 to 31/12/2023. The darkblue curves represent the fitted EGP densities, with parameters specified in Table~\ref{tab:egpd_tudy}. The dotted vertical grey lines represent the smallest point above which each fitted density is convex, which  represent the chosen marginal thresholds via Algorithm~\ref{algo:th_select}. \label{fig:density_tudy_sl}}
\end{figure}

The training set comprises 8,271 observations ranging from 10/08/1966 to 31/12/1999 and the test set comprises 7,483 observations ranging from 01/01/2000 to 31/12/2023. Histogram plots with the corresponding fitted densities are shown in Figure~\ref{fig:density_tudy_sl}. After thresholding, the final training set consists of 2,465 sea level exceedances, while final test set consists of 2,315 sea level exceedances. {\color{blue} Refer to Sections~\ref{sec:results},~\ref{sec:stationary} and~\ref{sec:tudy_appendix} for details about the figures.}

\subsubsection{Stationarity tests}
{\color{blue}
\begin{table}[ht!]
\caption{p-values of the ADF test (alternative hypothesis: stationarity) and the KPSS test (alternative hypothesis: non-stationarity) for the time series at the three stations. Bold values indicate strong evidence against the stationarity of the time series.\label{tab:stationary_test_before_sl}}
\vspace{0.2cm}
\begin{center}
\begin{small}
\begin{sc}
\begin{tabular}{c|c|ccc}
\toprule
Stations & Tests & no drift no trend & with drift no trend & with drift and trend   \\
\midrule
\multirow{2}{*}{Brest} & ADF & \textbf{0.41} & <0.01 & <0.01 \\
 & KPSS & \textbf{<0.01} & \textbf{<0.01} & >0.1  \\
 \midrule 
 \multirow{2}{*}{Saint-Nazaire} & ADF & \textbf{0.44} & <0.01 & <0.01 \\
 & KPSS & \textbf{<0.01} & >0.1 & >0.1  \\
 \midrule
 \multirow{2}{*}{Port Tudy} & ADF & \textbf{0.42} & <0.01 & <0.01 \\
 & KPSS & \textbf{<0.01} & 0.04 & >0.1  \\
\bottomrule
\end{tabular}
\end{sc}
\end{small}
\end{center}
\end{table}

\begin{figure}[ht!] 
  \centering
  \includegraphics[width=.315\textwidth]{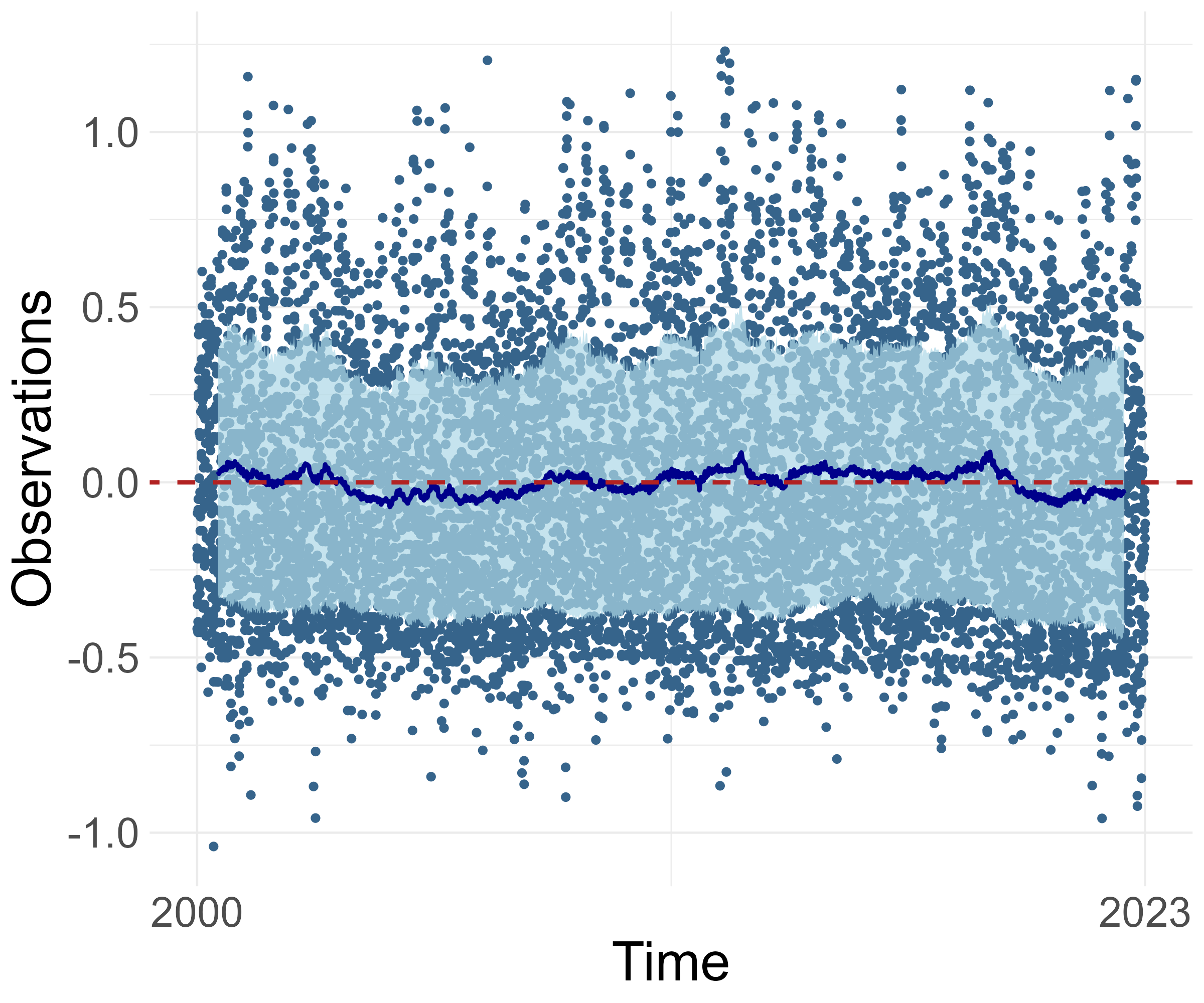}
  \hspace{0.2cm}
  \includegraphics[width=0.315\textwidth]{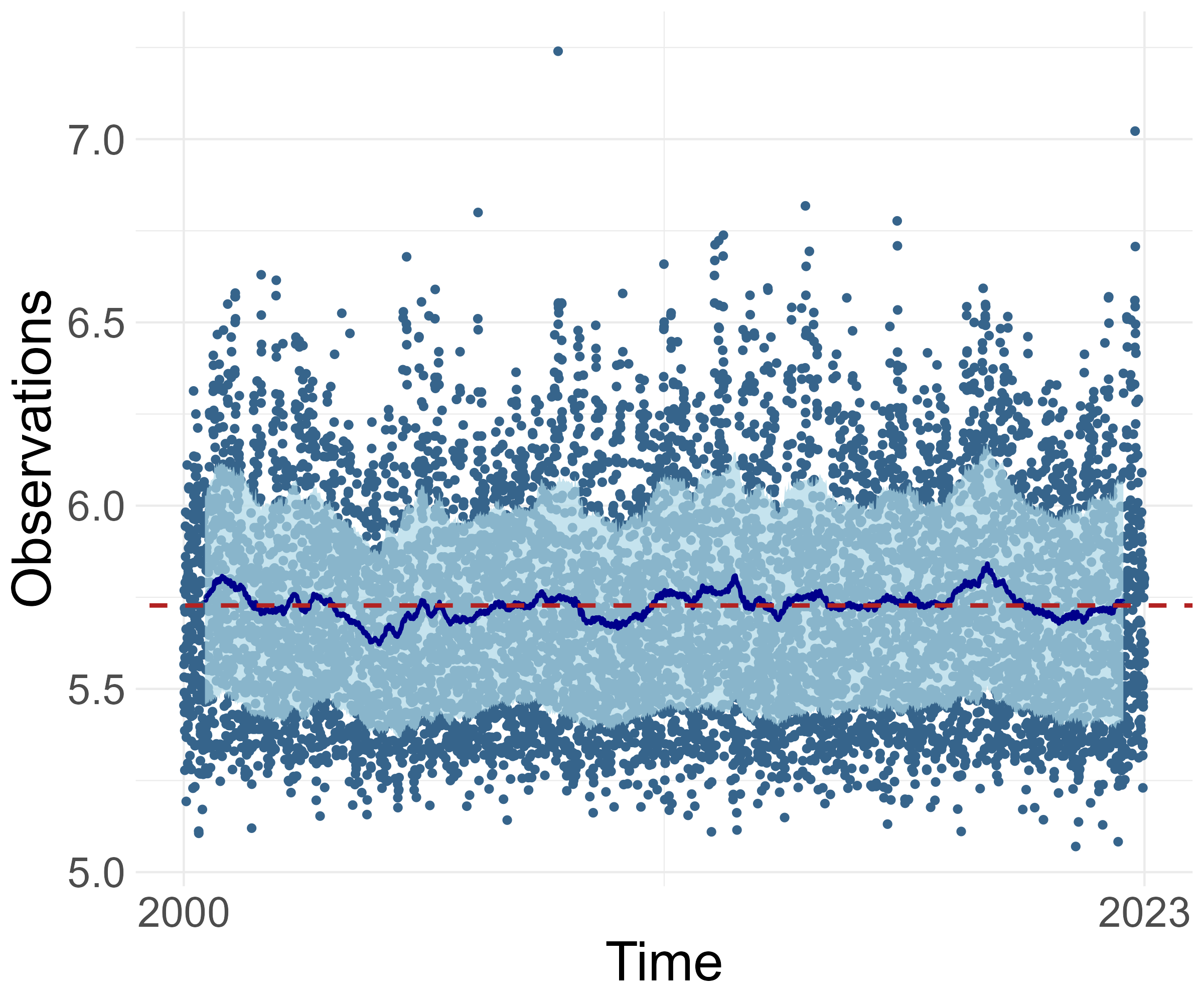}
  \hspace{0.2cm}
  \includegraphics[width=.315\textwidth]{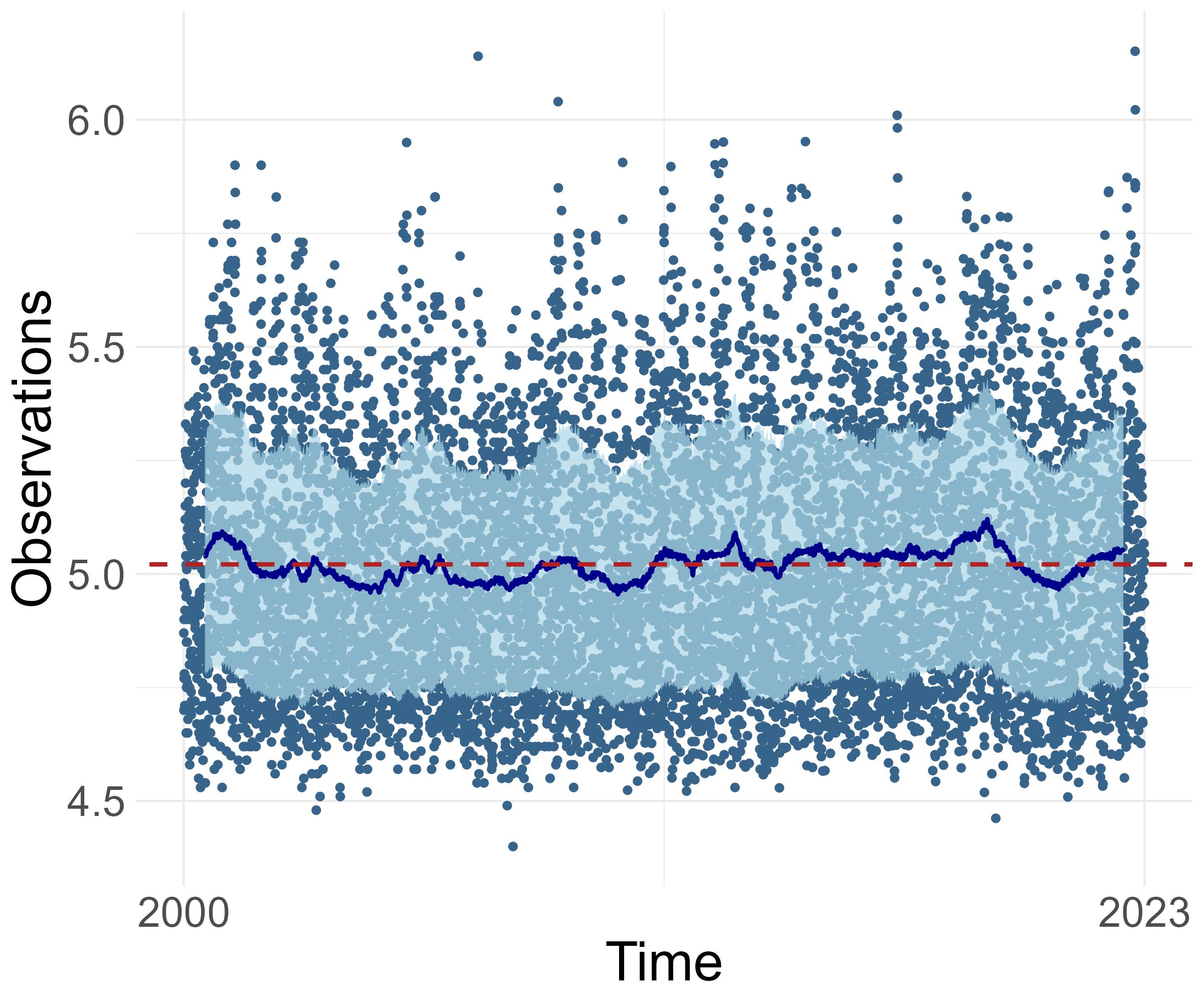}
\caption{Rolling means (dark blue) and rolling standard deviations (light blue) of the detrended skew surge time series at the three stations (Brest: left; Saint-Nazaire: middle; Port Tudy: right). The removed temporal slopes are all smaller than 10e-5 (Brest: 5.3e-06; Saint-Nazaire: 1.6e-06 ; Port Tudy 2.5e-06).\label{fig:rolling_fig_sl}}
\end{figure}

\begin{table}[ht!]
\caption{p-values of the ADF test (alternative hypothesis: stationarity) and the KPSS test (alternative hypothesis: non-stationarity) for the detrended time series at the three stations. No strong evidence against the stationarity of the time series is observed.\label{tab:stationary_test_after_sl}}
\vspace{0.2cm}
\begin{center}
\begin{small}
\begin{sc}
\begin{tabular}{c|c|ccc}
\toprule
Stations & Tests & no drift no trend & with drift no trend & with drift and trend   \\
\midrule
\multirow{2}{*}{Brest} & ADF & <0.01 & <0.01 & <0.01 \\
 & KPSS & >0.1 & >0.1 & >0.1  \\
 \midrule 
 \multirow{2}{*}{Saint-Nazaire} & ADF & <0.01 & <0.01 & <0.01 \\
 & KPSS & >0.1 & >0.1 & >0.1  \\
 \midrule
 \multirow{2}{*}{Port Tudy} & ADF & <0.01 & <0.01 & <0.01 \\
 & KPSS & >0.1 & >0.1 & >0.1  \\
\bottomrule
\end{tabular}
\end{sc}
\end{small}
\end{center}
\end{table}

}

\subsubsection{Results}

\begin{figure}[ht!]
  \centering
  \includegraphics[width=.315\textwidth]{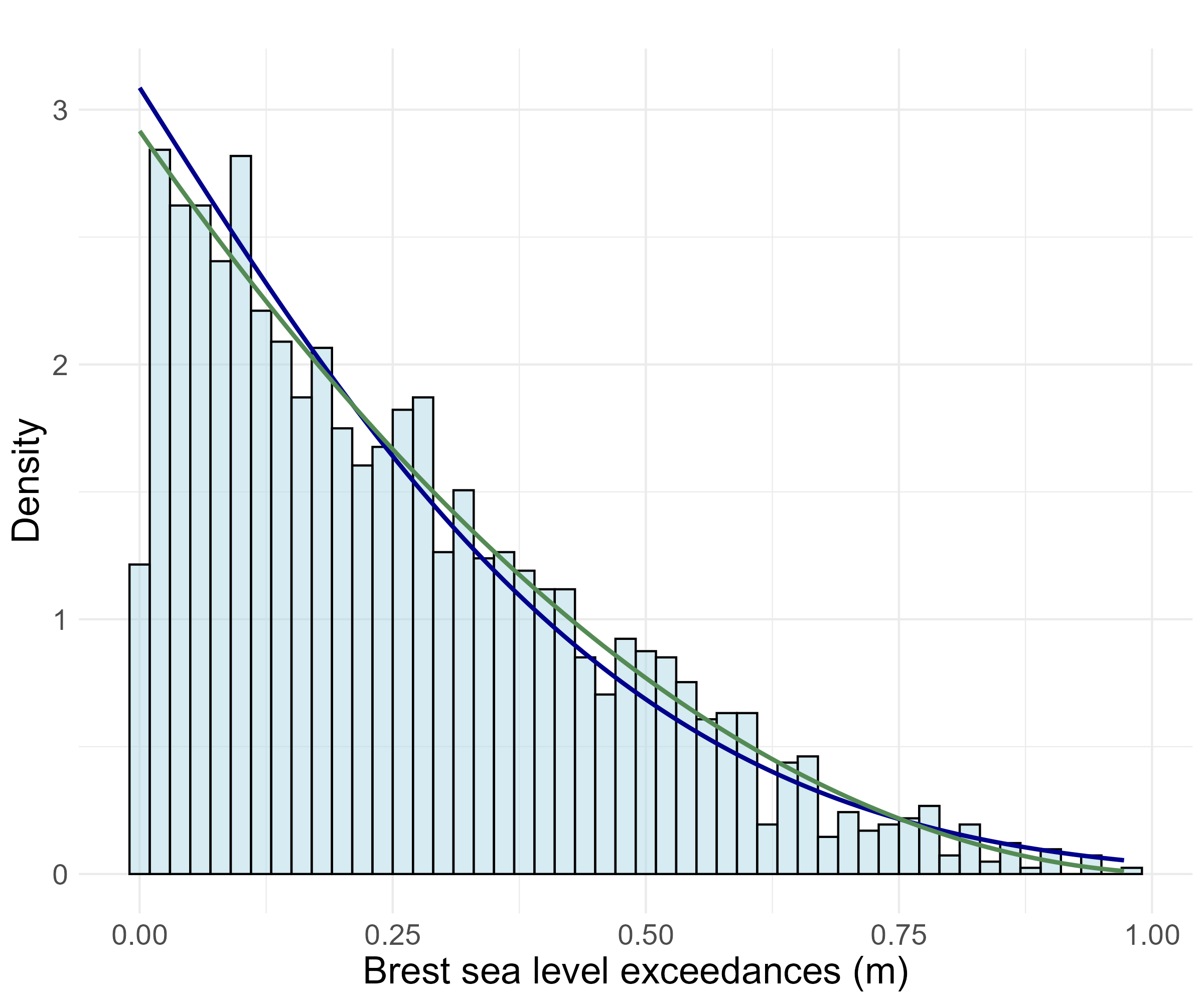}
  \hspace{0.2cm}
  \includegraphics[width=.315\textwidth]{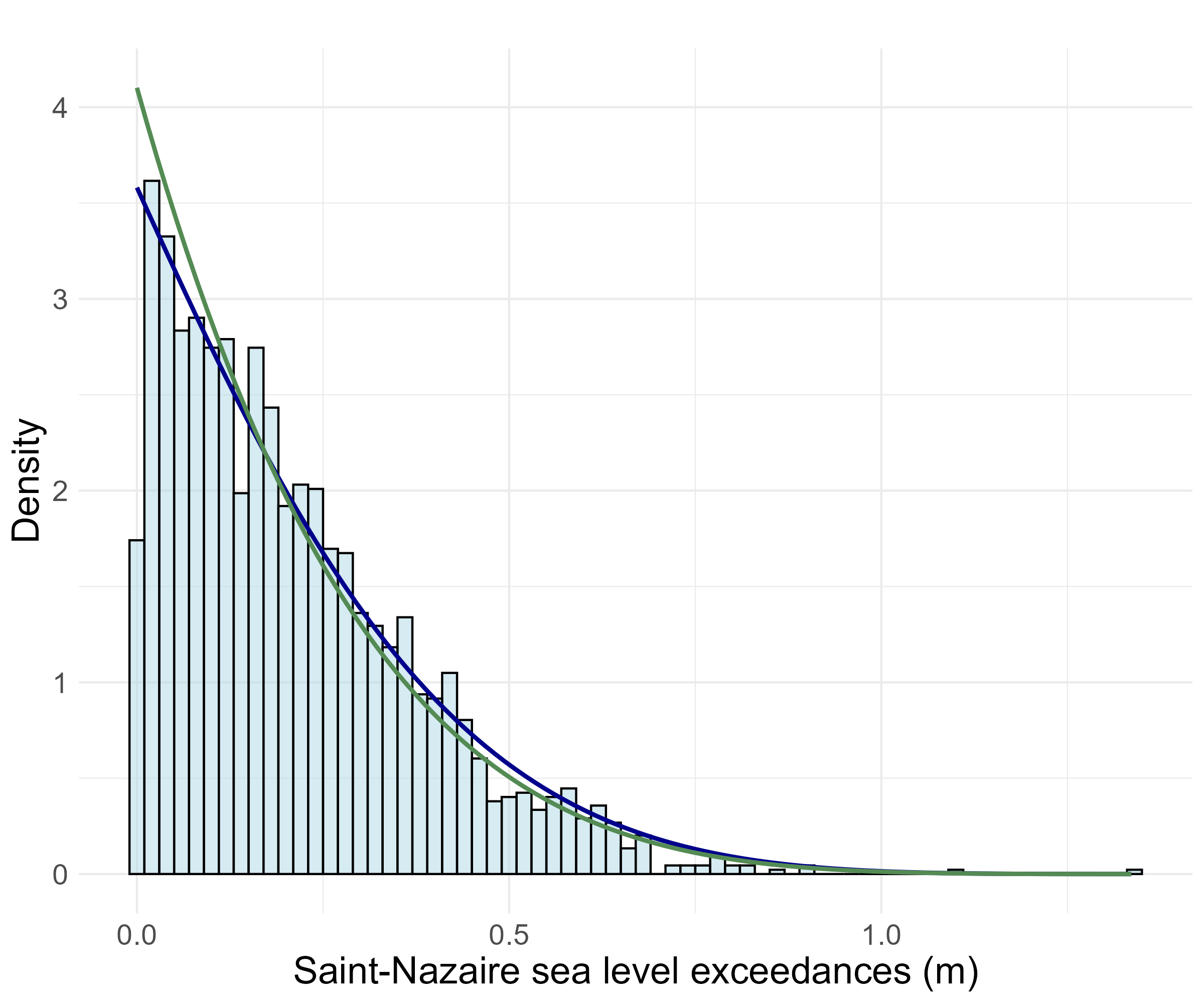}
  \hspace{0.2cm}
  \includegraphics[width=.315\textwidth]{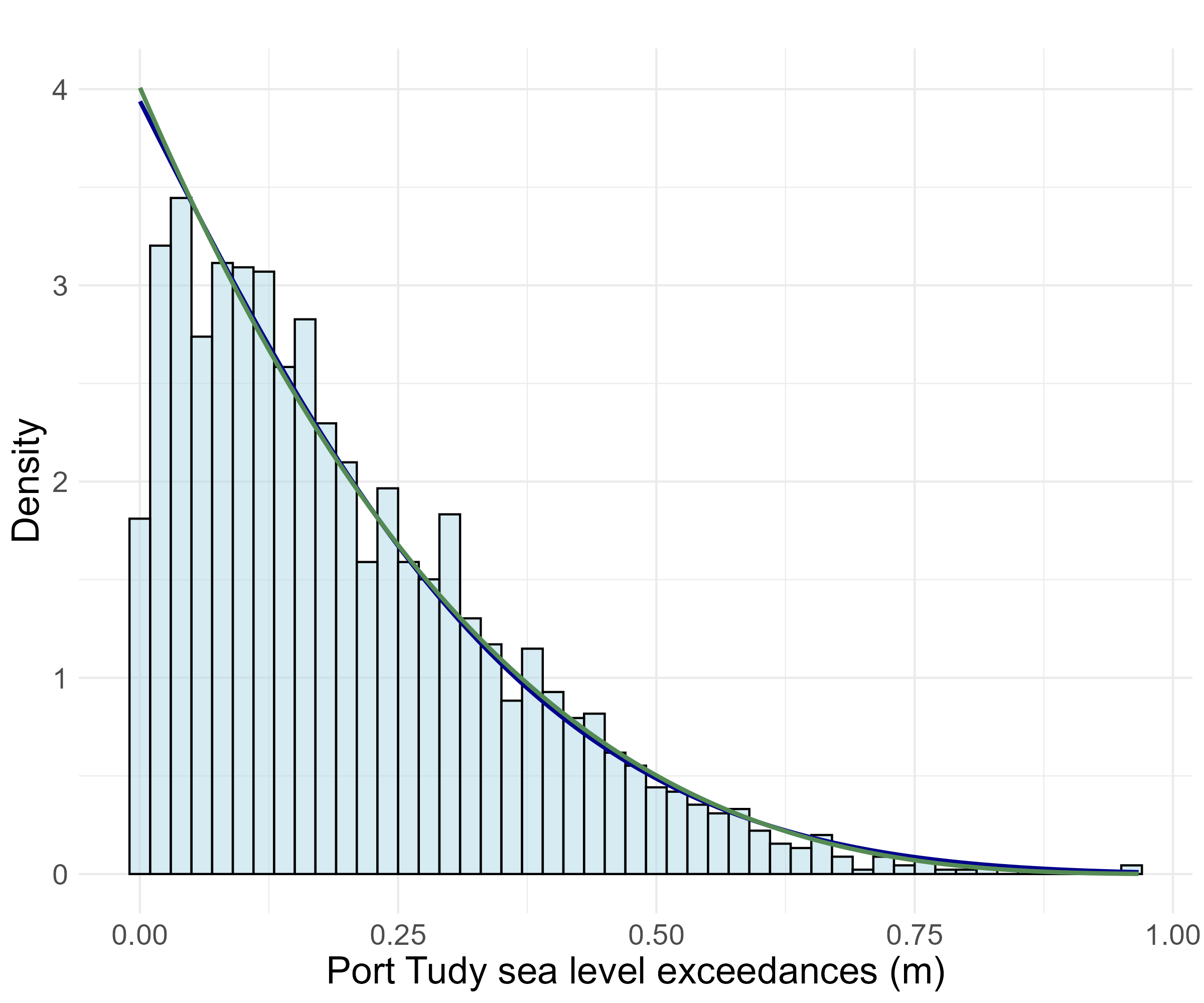}

  \caption{Histograms of sea level exceedances above the threshold specified in Table~\ref{tab:egpd_tudy_sl} at the three stations Brest (left), Saint-Nazaire (middle) and Port Tudy (right), from 01/12/2000 to 31/12/2023. The darkblue curves represent the fitted EGP densities above the thresholds, with parameters specified in Table~\ref{tab:egpd_tudy_sl}. The green curves represent the fitted GP densities.\label{fig:tudy_egpdvsgpd_sl}}
\end{figure}
\begin{table}[ht!]
\caption{Point estimates of the parameters of the fitted EGP distribution for sea level exceedances at the three stations. The chosen thresholds, determined using via Algorithm~\ref{algo:th_select}, are shown in the $t$ rows. The data used for inference are from the training set ranging from 01/12/2000 to 31/12/2023.\label{tab:egpd_tudy_sl}}
\vspace{0.2cm}
\begin{center}
\begin{small}
\begin{sc}
\begin{tabular}{cccccr}
\toprule
Parameters/Stations & Brest & Saint-Nazaire & Port Tudy  \\
\midrule
$\sigma$ & 0.53  & 0.40 & 0.36 \\
$\xi$ & -0.18  & -0.18  & -0.17\\
$\kappa$ & 6.89  & 4.09  & 4.33\\
$t$ & 1.30  & 0.85  & 0.77\\
\bottomrule
\end{tabular}
\end{sc}
\end{small}
\end{center}
\end{table}

\begin{table}[ht!]
\caption{RMSE$\times10^2$(SSE$\times10^2$) and MAE$\times10^2$(ASE$\times10^2$) of predicted sea level exceedances at Port Tudy station from the ROXANE procedure with RF regression (ROX RF), ROXANE procedure with OLS regression (ROX OLS) and MGPRED. Errors are measured on the test set covering the period from 10/08/1966 to 31/12/1999. Errors are computed on the entire test set (columns: RMSE and MSE) and on its most extreme half, i.e., for observations with $Y_i$ exceeding their empirical median computed on the extreme test set (columns: $\mbox{RMSE}_{\mbox{EXT}}$ and $\mbox{MAE}_{\mbox{EXT}}$). \label{tab:tudy_sl}}
\vspace{0.2cm}
\begin{center}
\begin{small}
\begin{sc}
\begin{tabular}{ccccccr}
\toprule
Training models/Errors & RMSE & MAE & $\mbox{RMSE}_{\mbox{ext}}$ & $\mbox{MAE}_{\mbox{ext}}$ \\
\midrule
ROX RF & 8.1(0.3) & 5.9(0.1) & 7.8(0.4) & 5.6(0.2) \\
ROX OLS & 8.0(0.3) & 5.9(0.1) & \textbf{7.7(0.4)} & \textbf{5.5(0.2)} \\
MGPRED & \textbf{7.8(0.2)} & \textbf{5.6(0.1)} & 8.0(0.4) & 5.7(0.2) \\
\bottomrule
\end{tabular}
\end{sc}
\end{small}
\end{center}
\end{table}

\begin{figure}[ht!] 
  \centering
  \includegraphics[width=.315\textwidth]{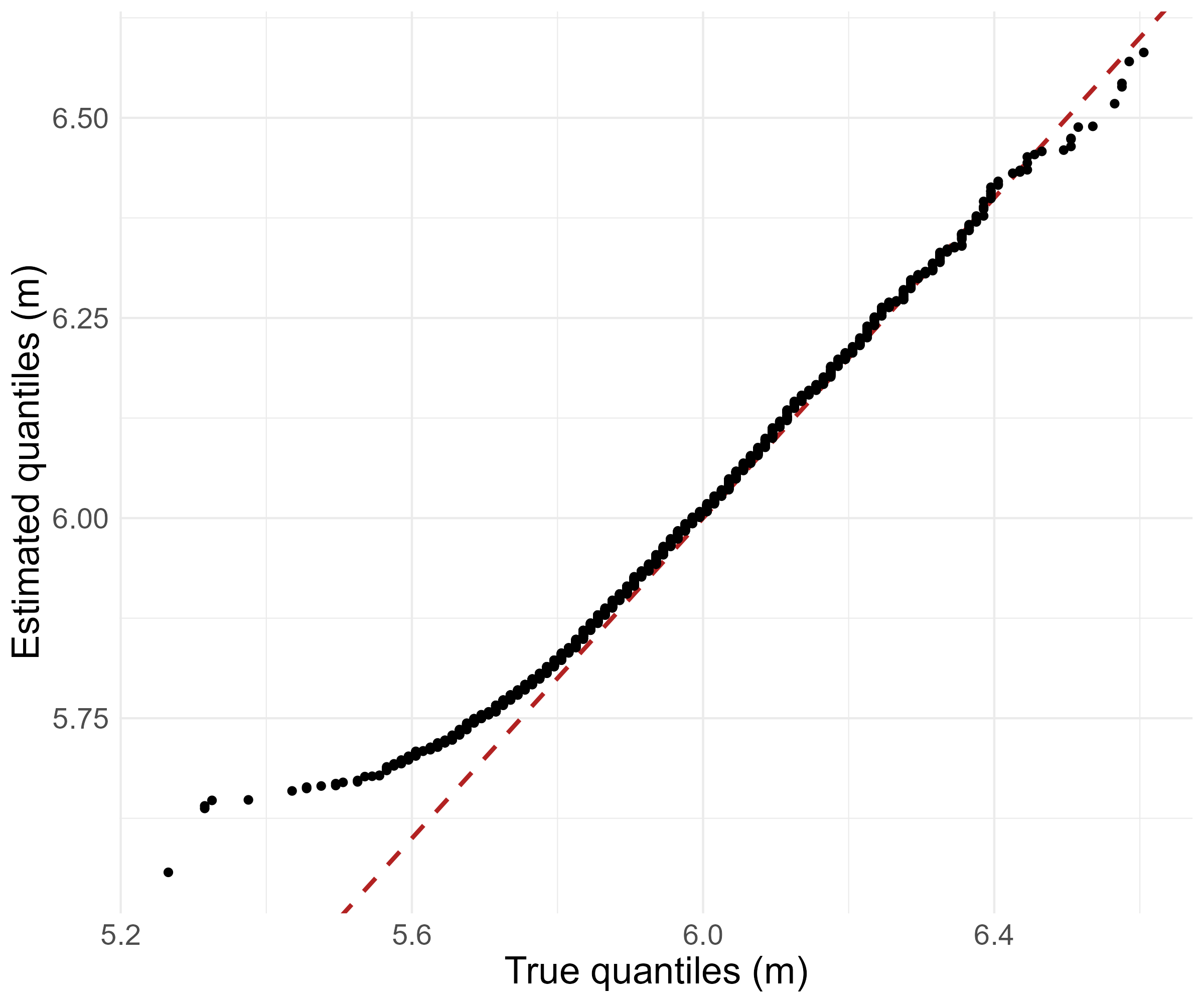}
  \hspace{0.2cm}
  \includegraphics[width=.315\textwidth]{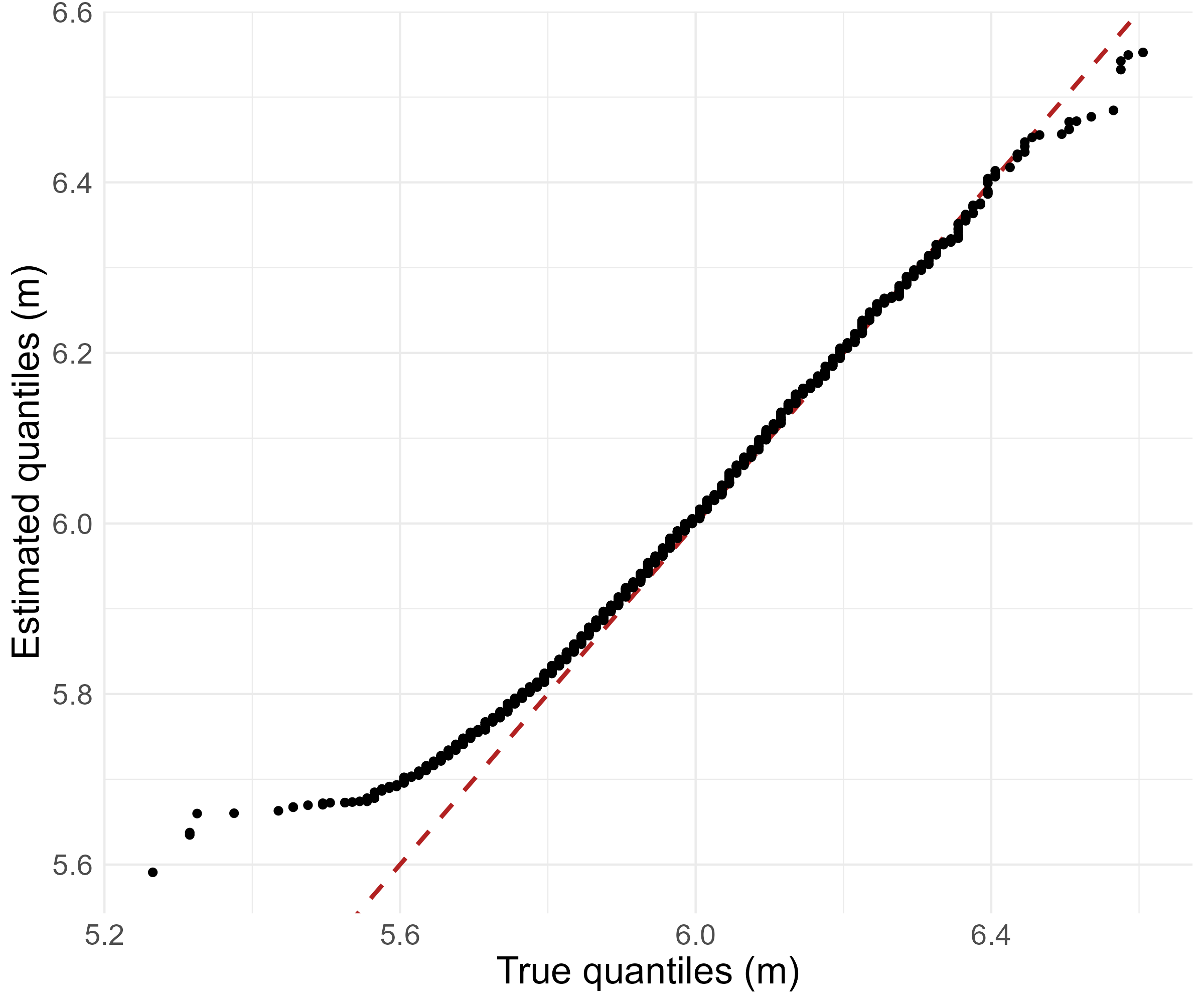}
  \hspace{0.2cm}
  \includegraphics[width=.315\textwidth]{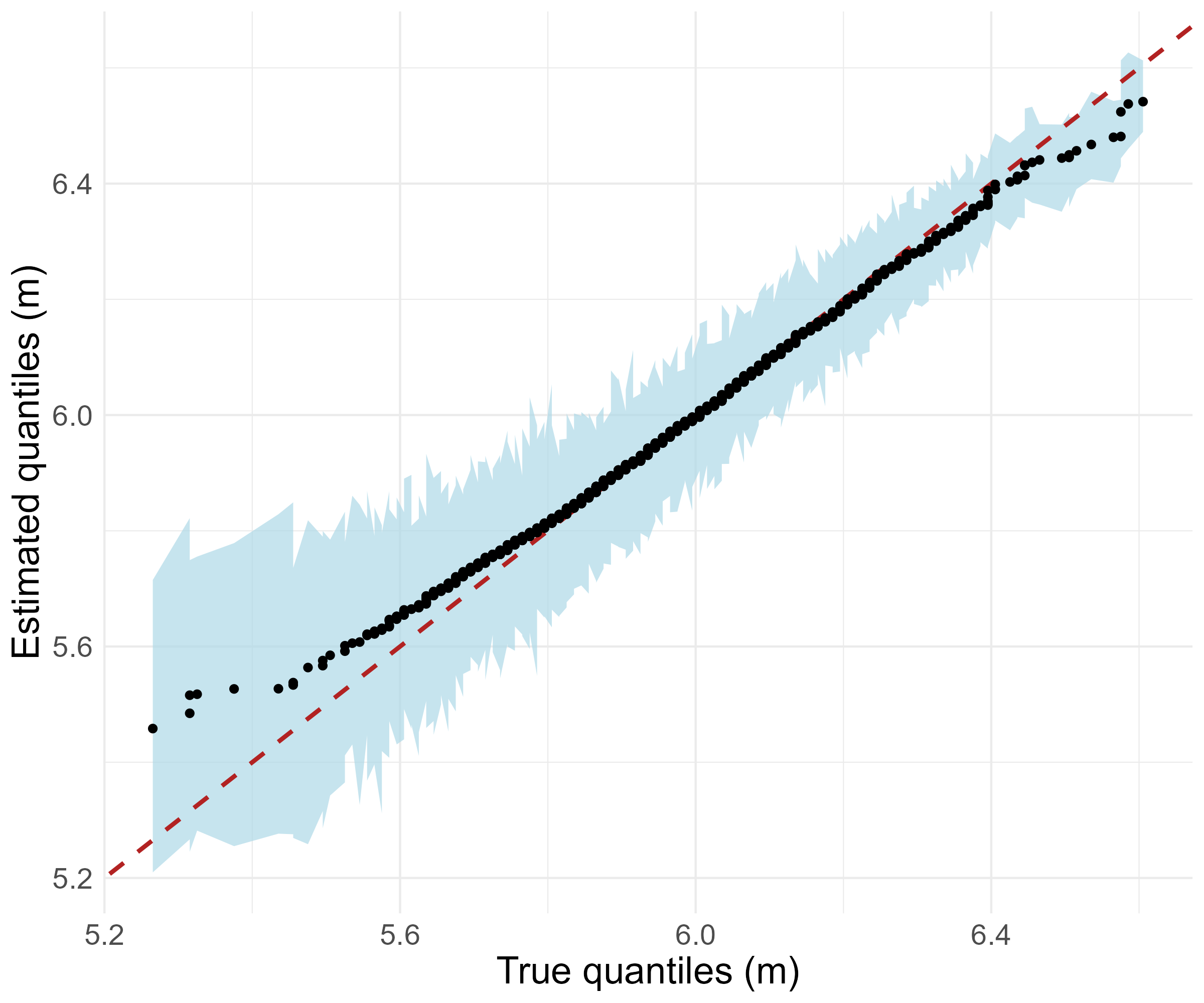}
\caption{QQ-plots comparing observed sea level exceedances of the Port Tudy test set (x-axis), ranging from 10/08/1966 to 31/12/1999, to predicted data (y-axis) from the algorithms of Sections~\ref{sec:reg_proc} and \ref{sec:mgp_proc}. The plots show results from the ROXANE procedure with RF regression (left), ROXANE procedure with OLS regression (middle), and MGPRED (right) with 0.95-confidence bands (lightblue). The dotted red line represents the identity line $x=y$.\label{fig:qqplot_tudy_sl}}
\end{figure}

\begin{figure}[ht!] 
  \centering
  \includegraphics[width=.29\textwidth]{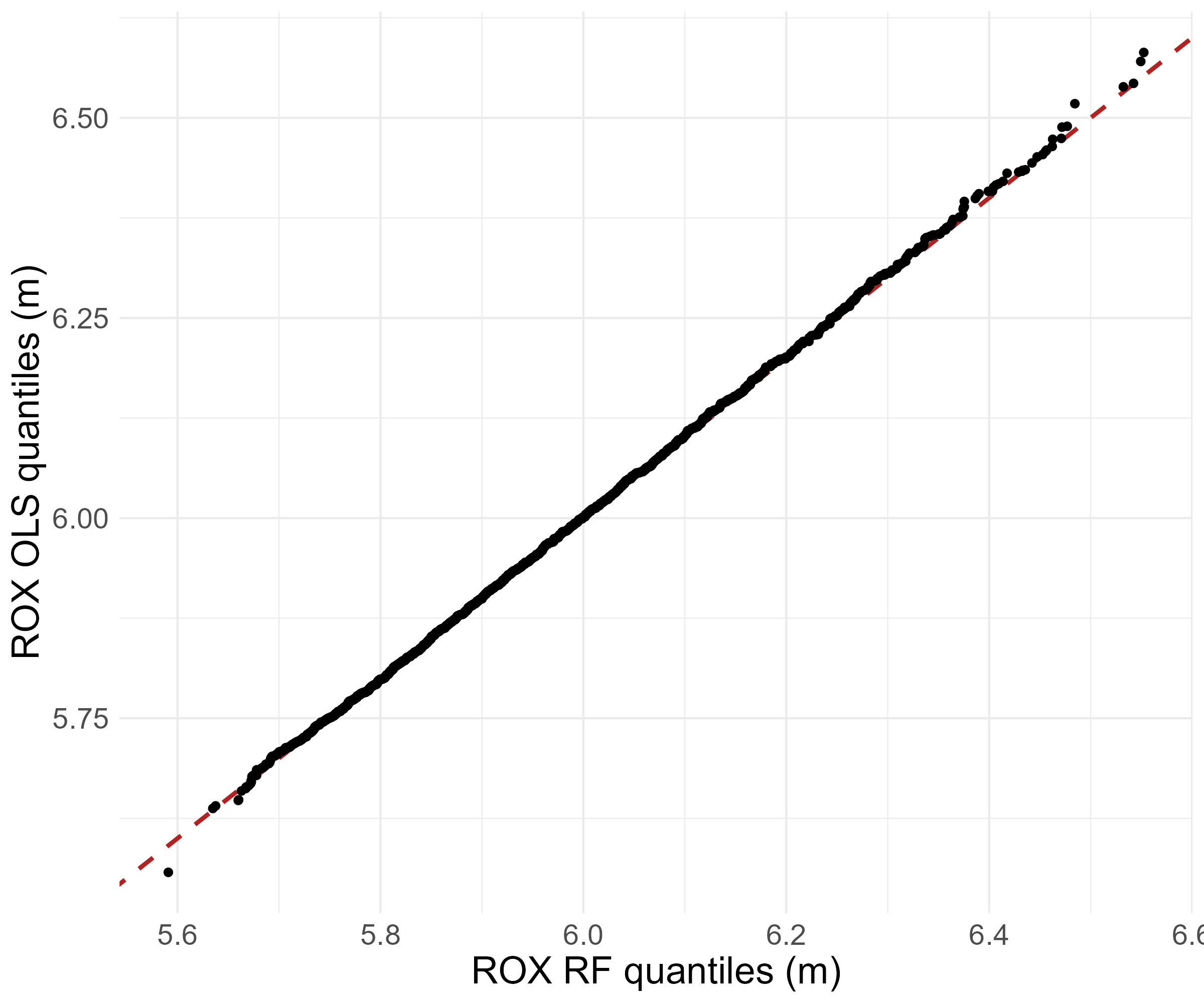}
  \hspace{0.7cm}
  \includegraphics[width=.29\textwidth]{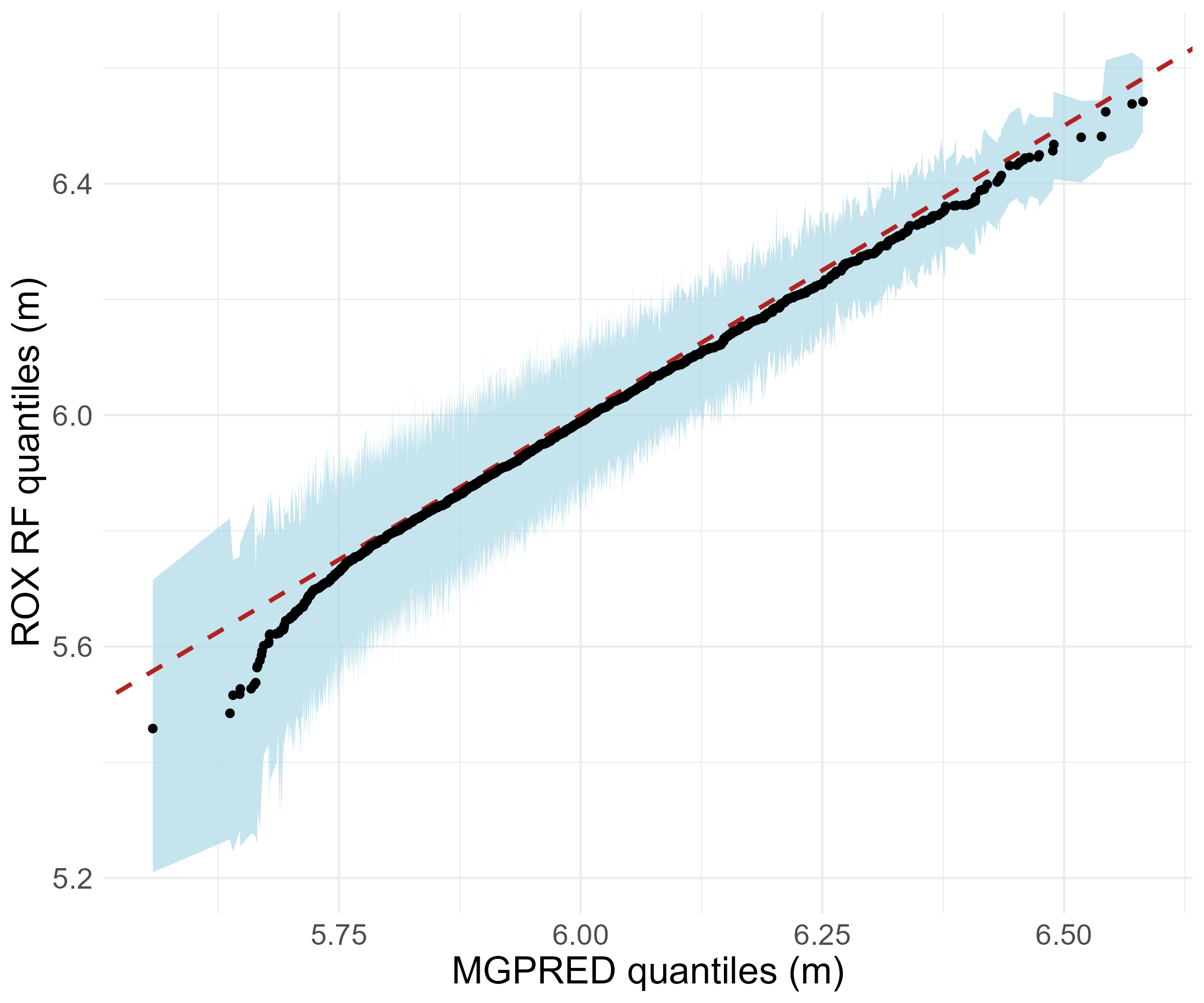}
  \hspace{0.7cm}
  \includegraphics[width=.29\textwidth]{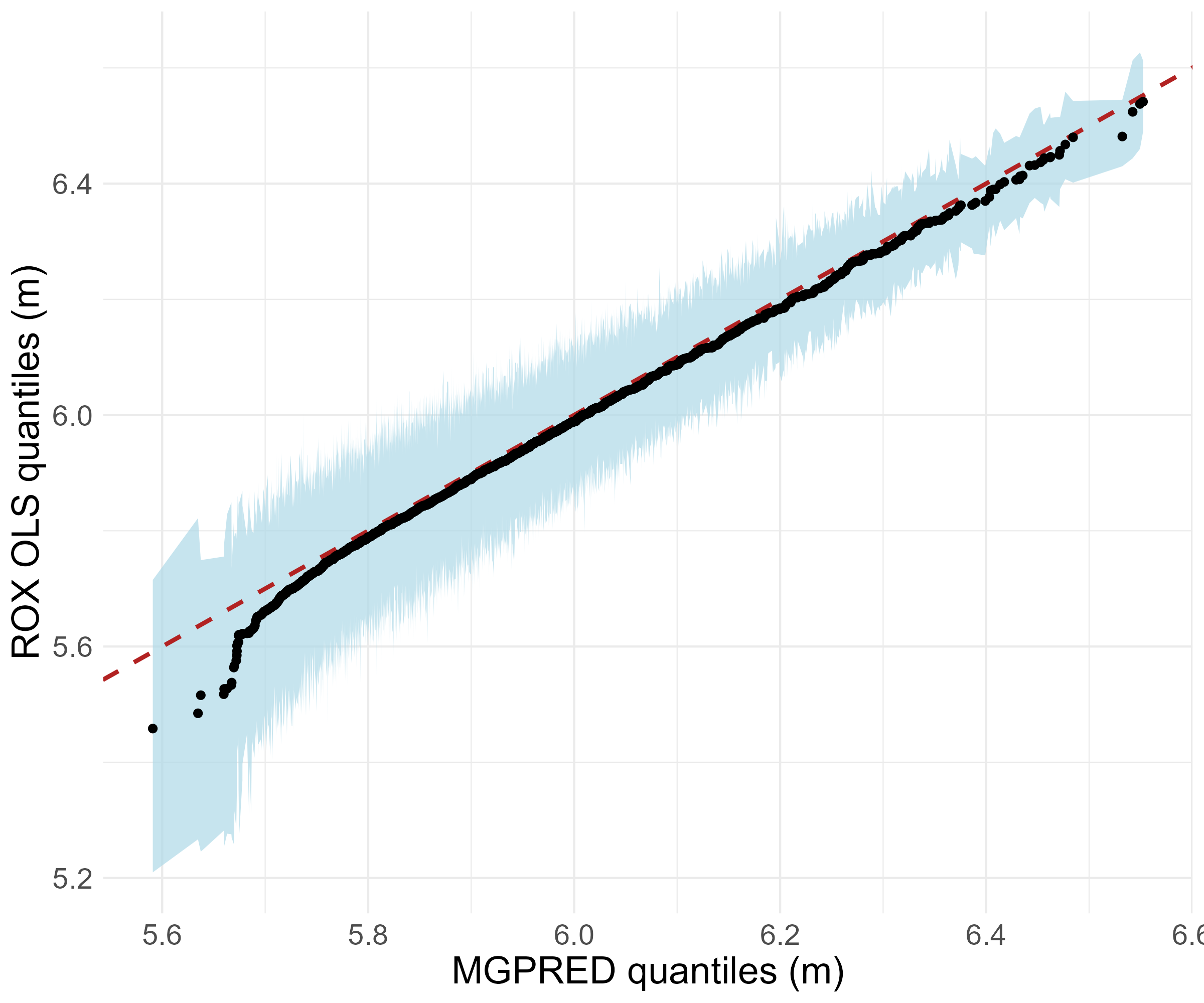}
\caption{QQ-plots comparing predicted sea level exceedances of the Port Tudy test set, ranging from 10/08/1966 to 31/12/1999, from the algorithms of Sections~\ref{sec:reg_proc} and \ref{sec:mgp_proc}: ROX RF \emph{vs} ROX RF (left), MGPRED \emph{vs} ROX RF (middle), and MGPRED \emph{vs} ROX OLS (middle), with 0.95-confidence bands (lightblue) from MGPRED. The dotted red line represents the identity line $x=y$.\label{fig:qqplot_tudy}}
\end{figure}

\begin{figure}[ht]
  \centering
  \includegraphics[width=.315\textwidth]{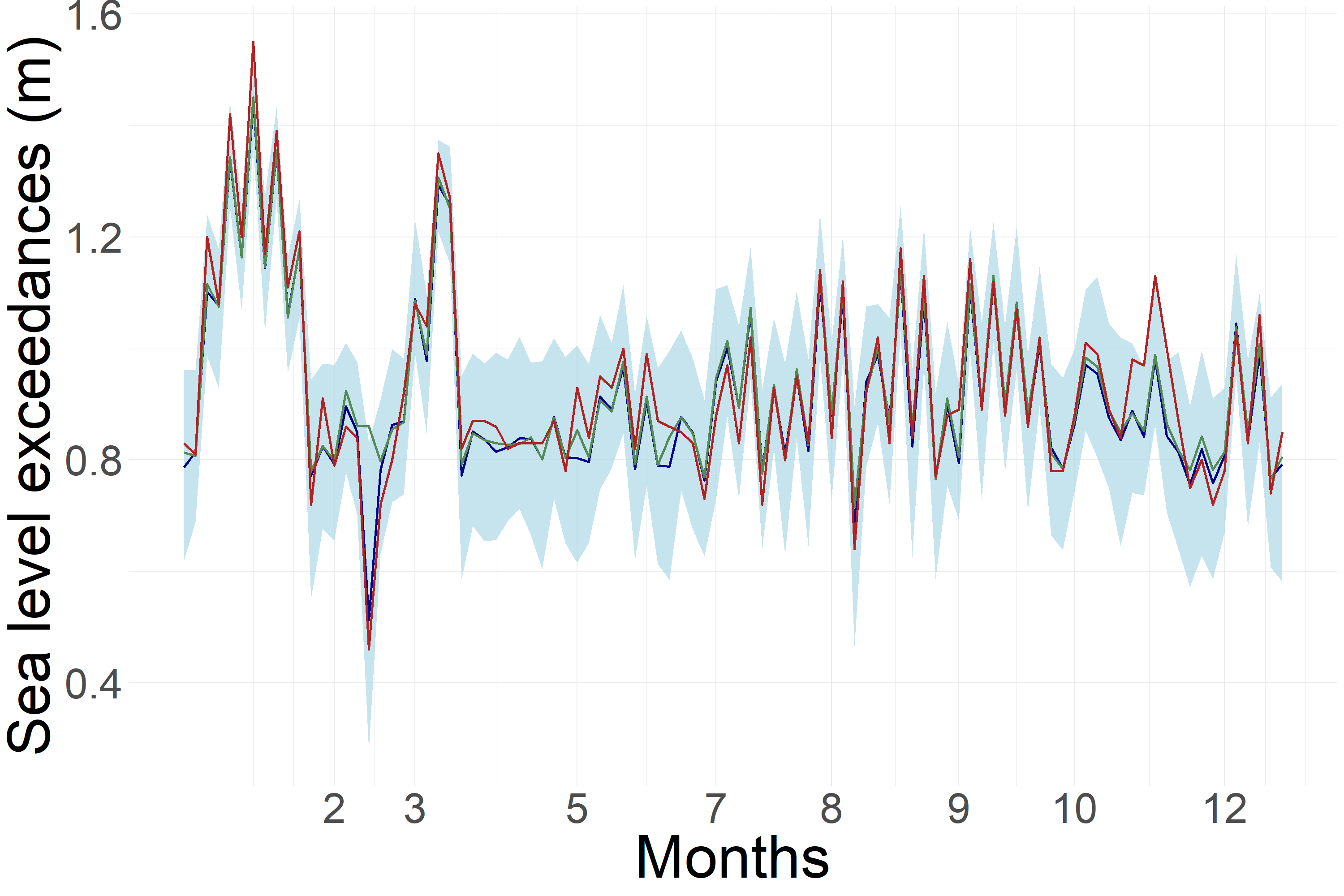}
  \hspace{0.2cm}
  \includegraphics[width=.315\textwidth]{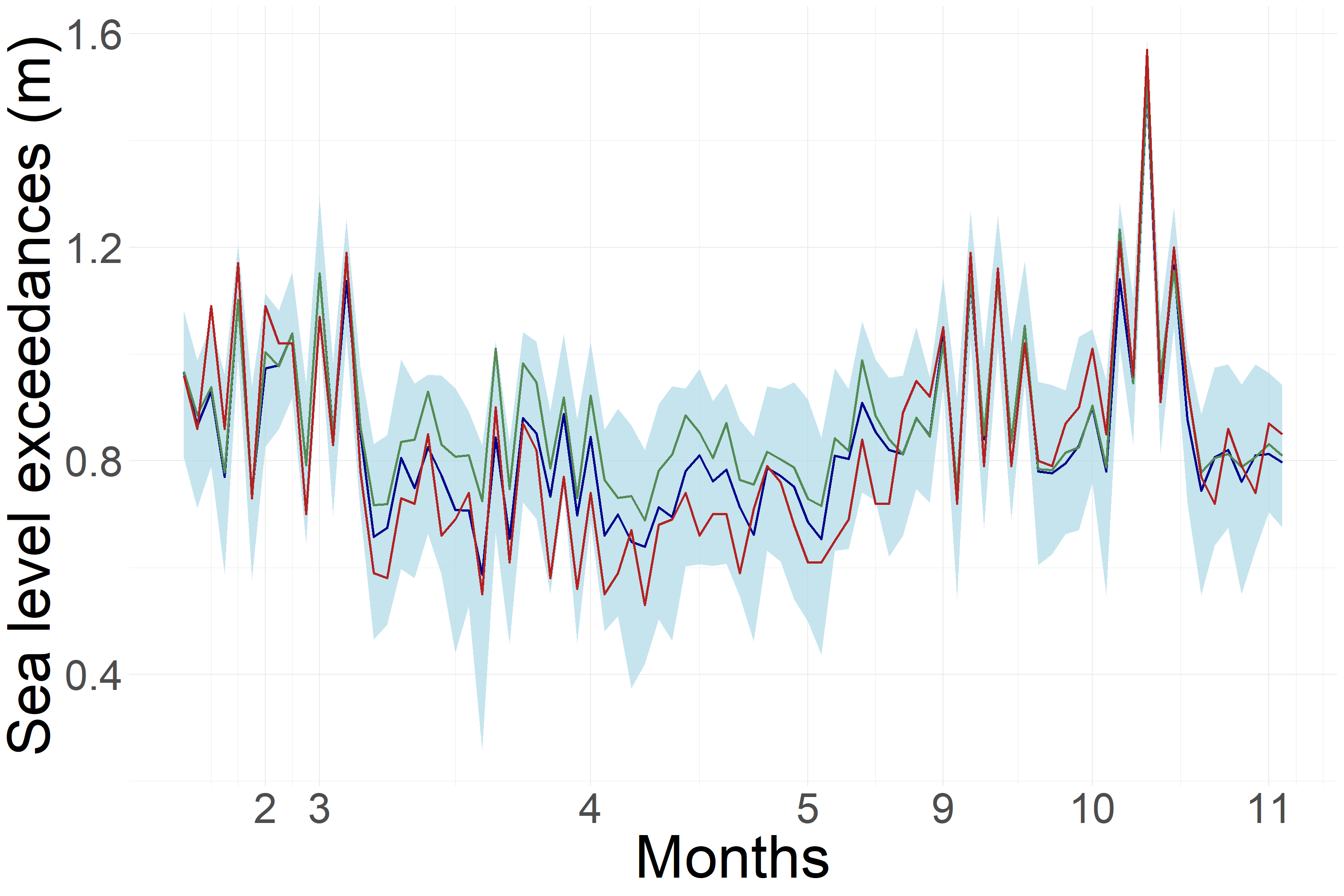}
  \hspace{0.2cm}
  \includegraphics[width=.315\textwidth]{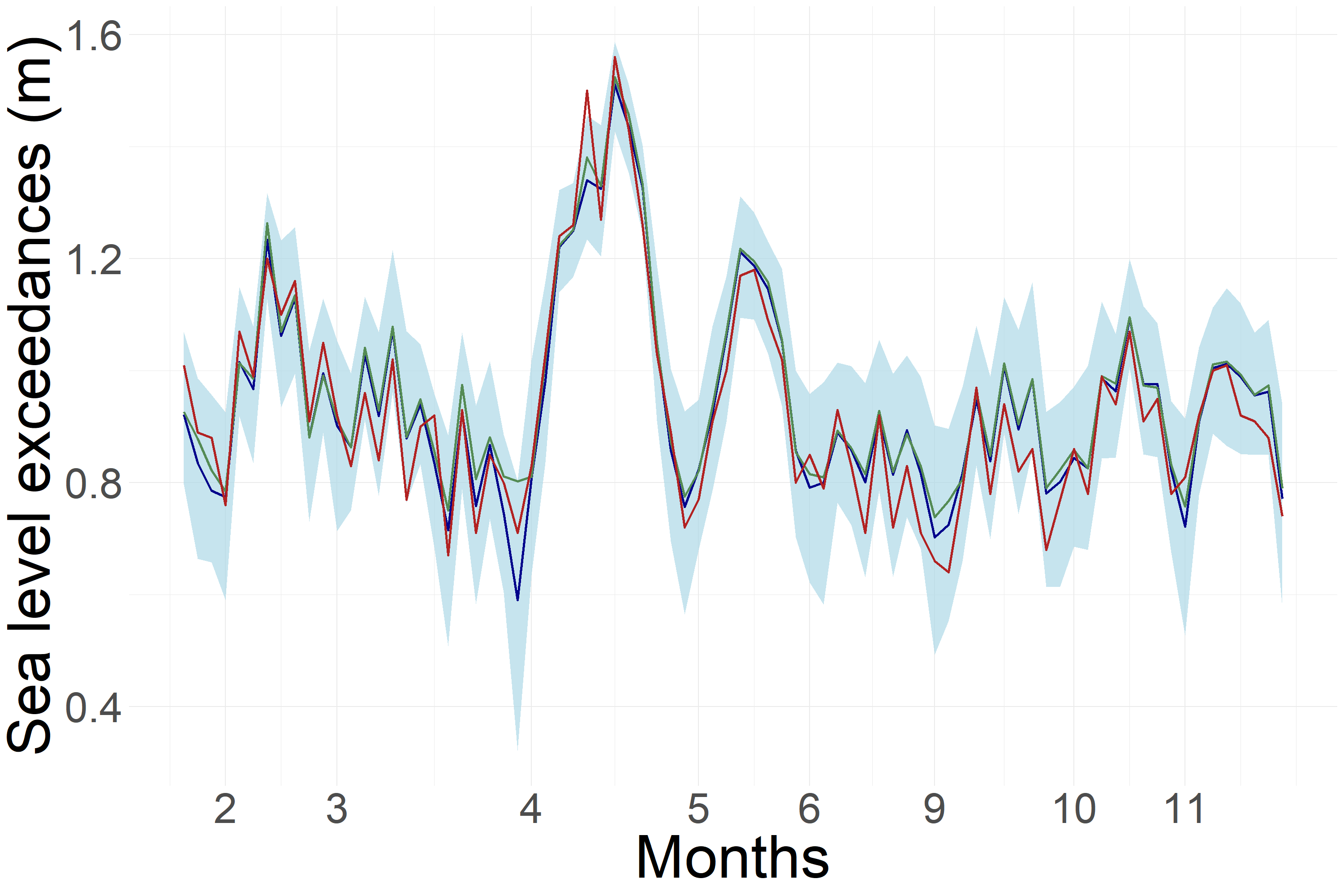}

  \caption{Predicted sea levels at Port Tudy station for the years 1996 (left), 1987 (middle), 1985 (right). These years are characterized by significant sea levels at the Port Tudy tide gauge. Red curves represent the true values on the test set; green curves represent the predicted values by the ROXANE procedure with OLS algorithm; blue curves represent the predicted values by MGPRED with 0.95-confidence intervals (lightblue). The global 0.95-coverage probability is 0.93.\label{fig:pred_tudy_sl}}
\end{figure}

\begin{figure}[ht!] 
  \centering
  \includegraphics[width=.315\textwidth]{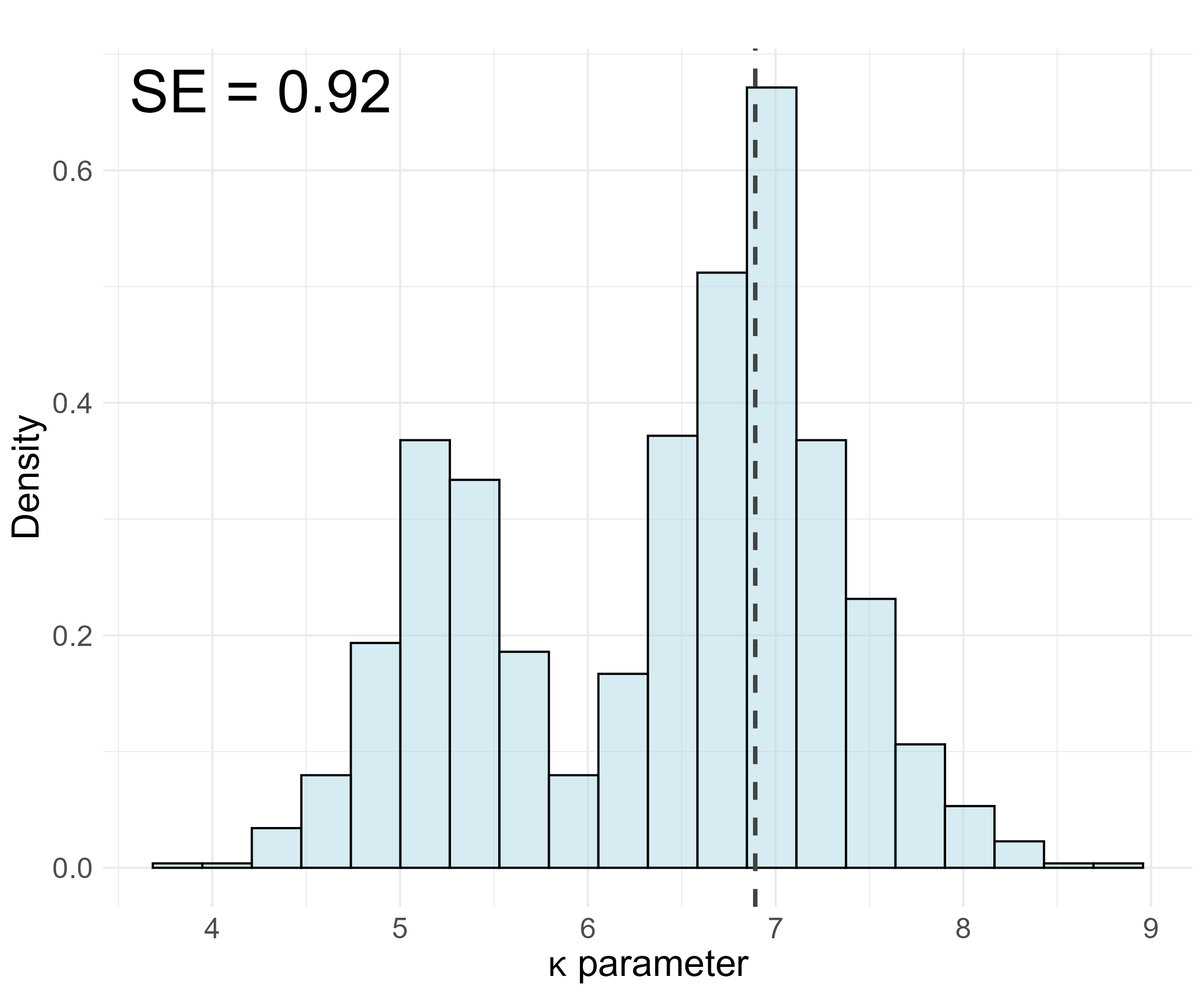}
  \hspace{0.2cm}
  \includegraphics[width=.315\textwidth]{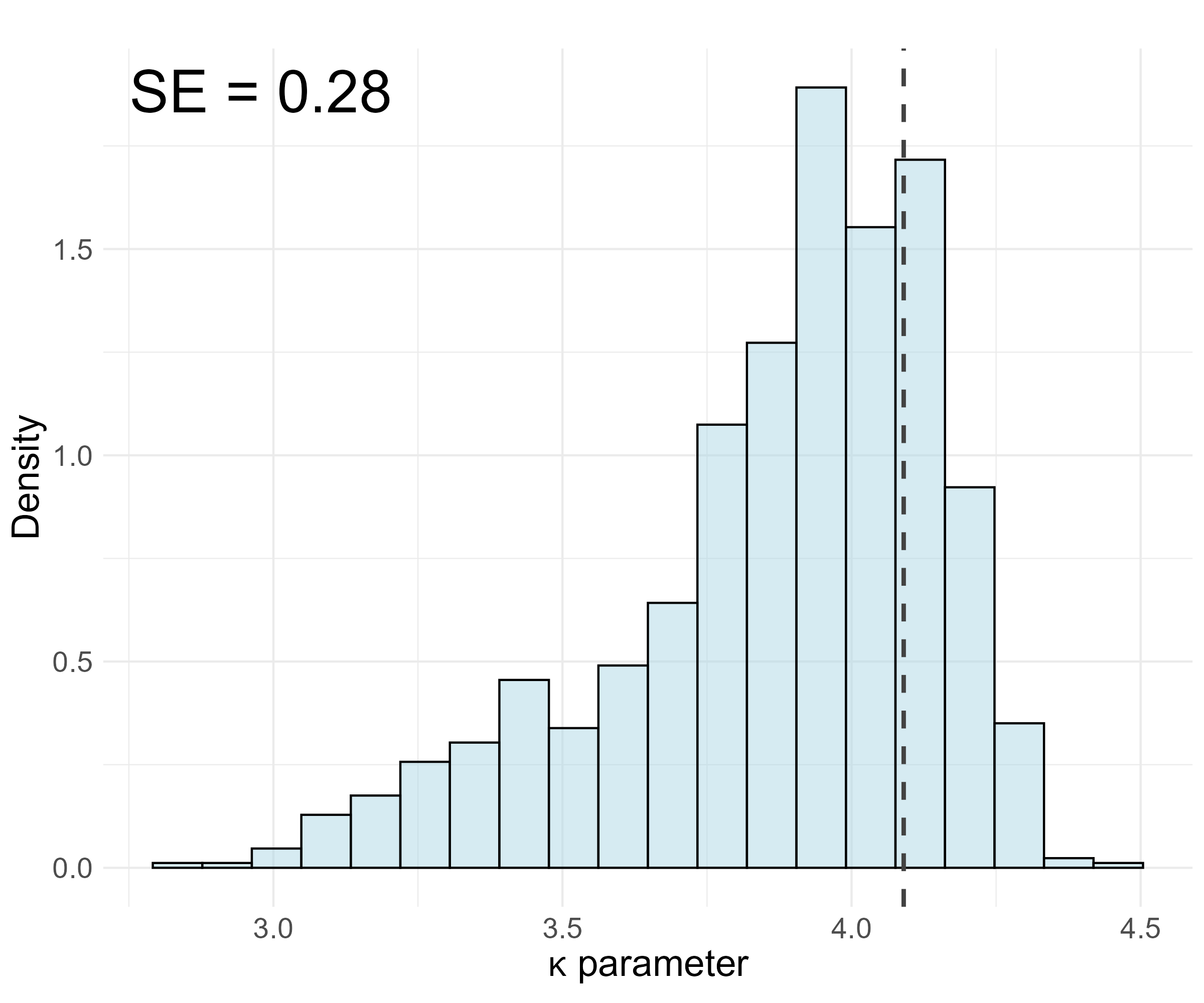}
  \hspace{0.2cm}
  \includegraphics[width=.315\textwidth]{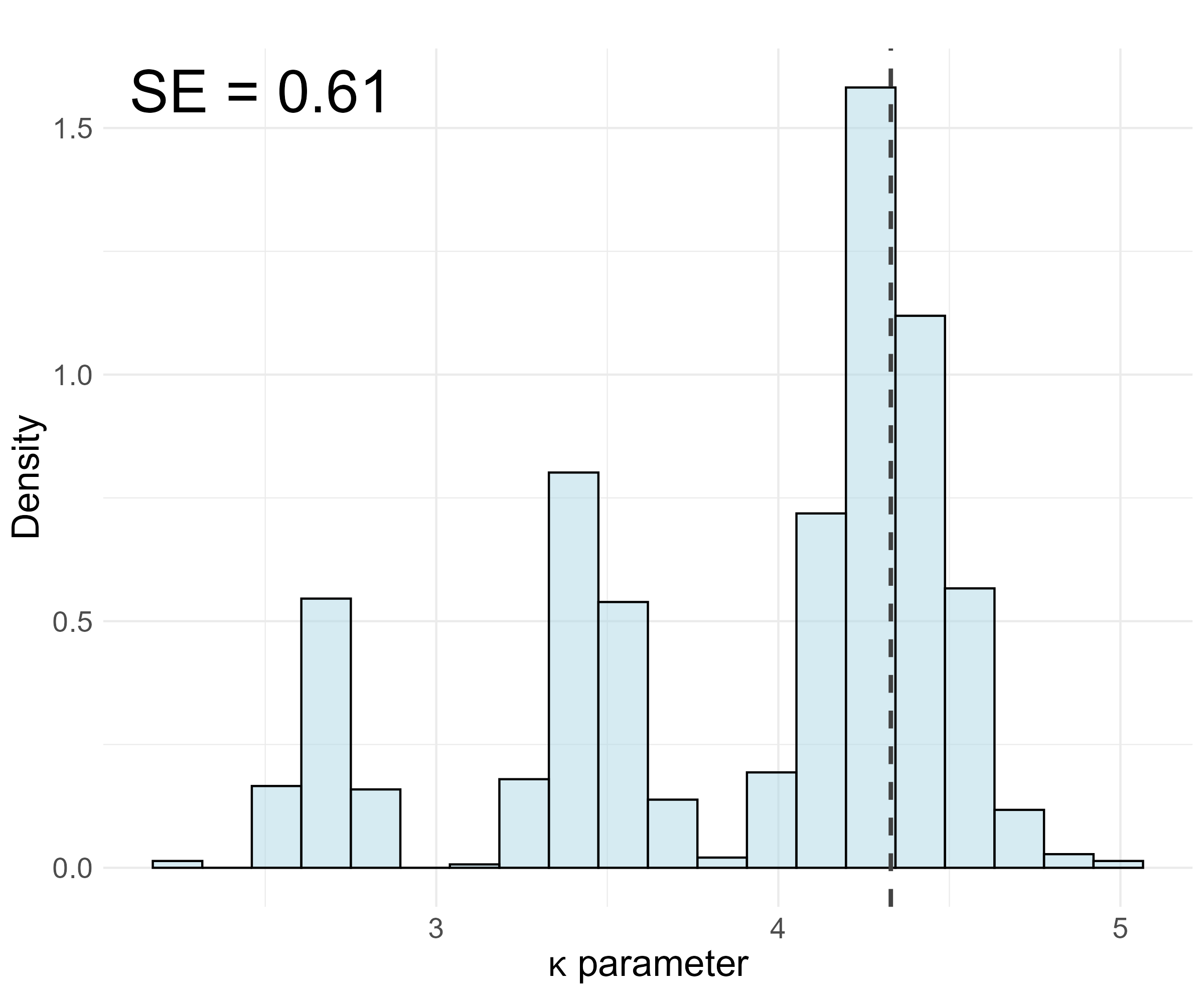}
  \vspace{0.2cm}
  \includegraphics[width=.315\textwidth]{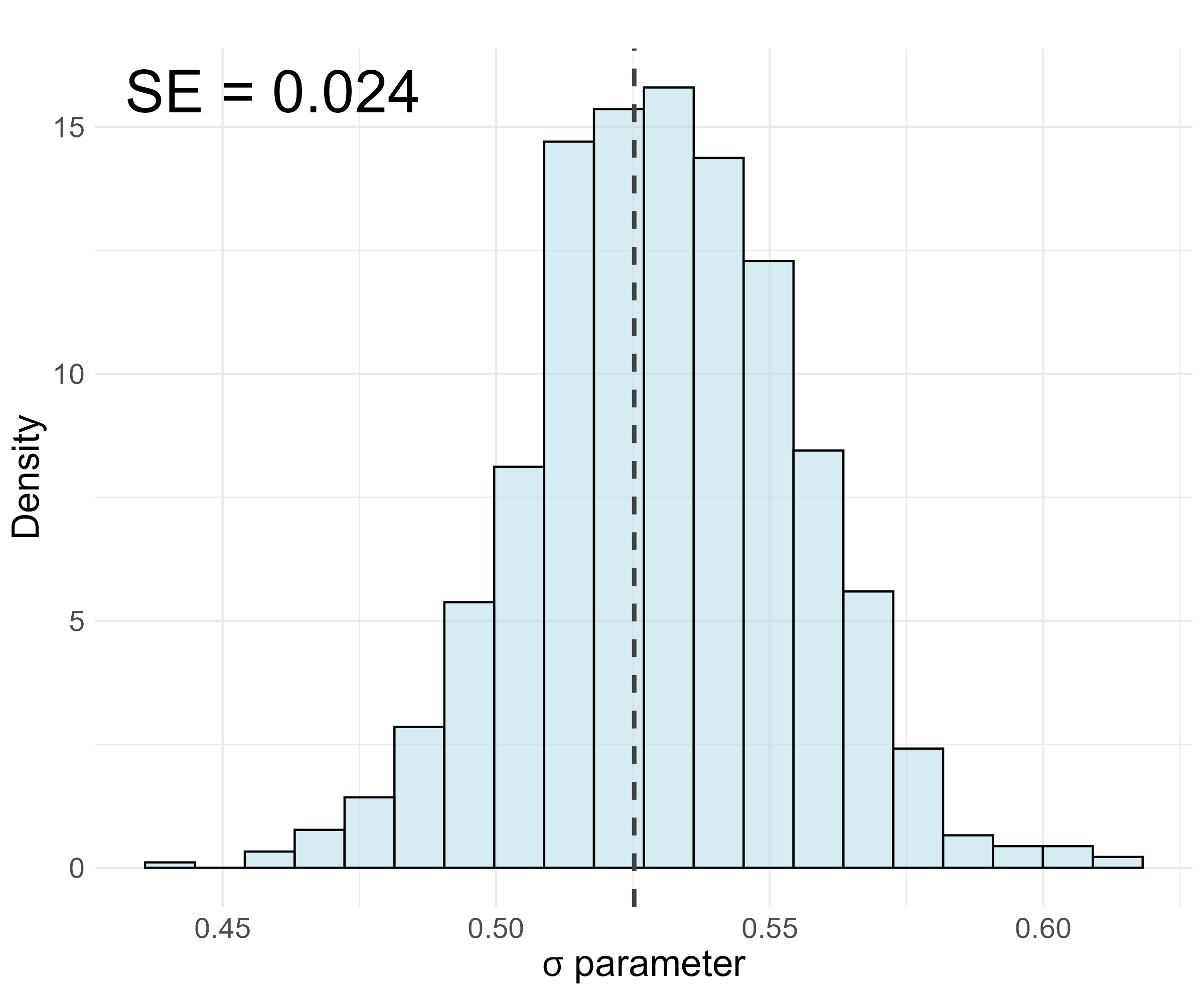}
  \hspace{0.2cm}
  \includegraphics[width=.315\textwidth]{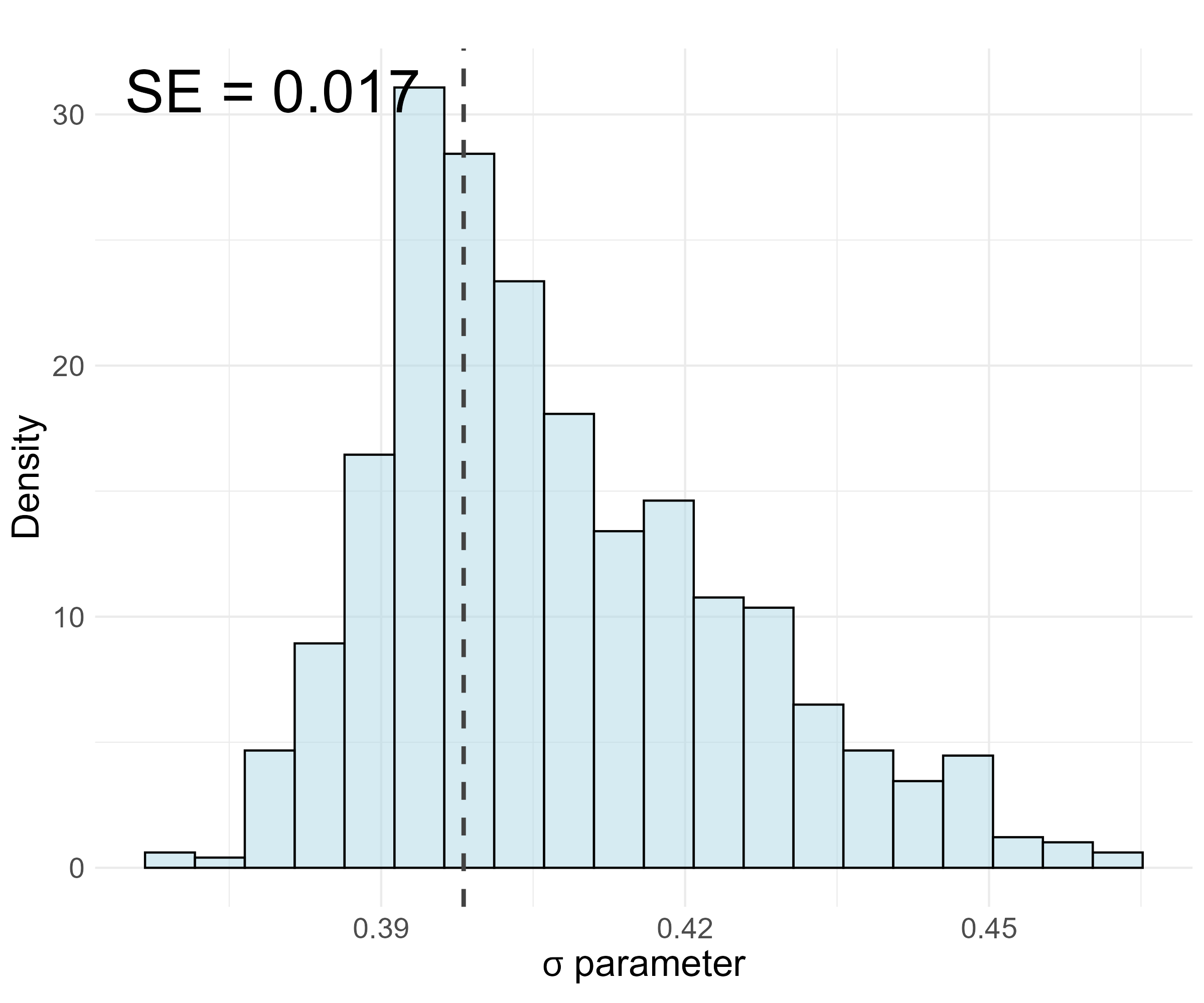}
  \hspace{0.2cm}
  \includegraphics[width=.315\textwidth]{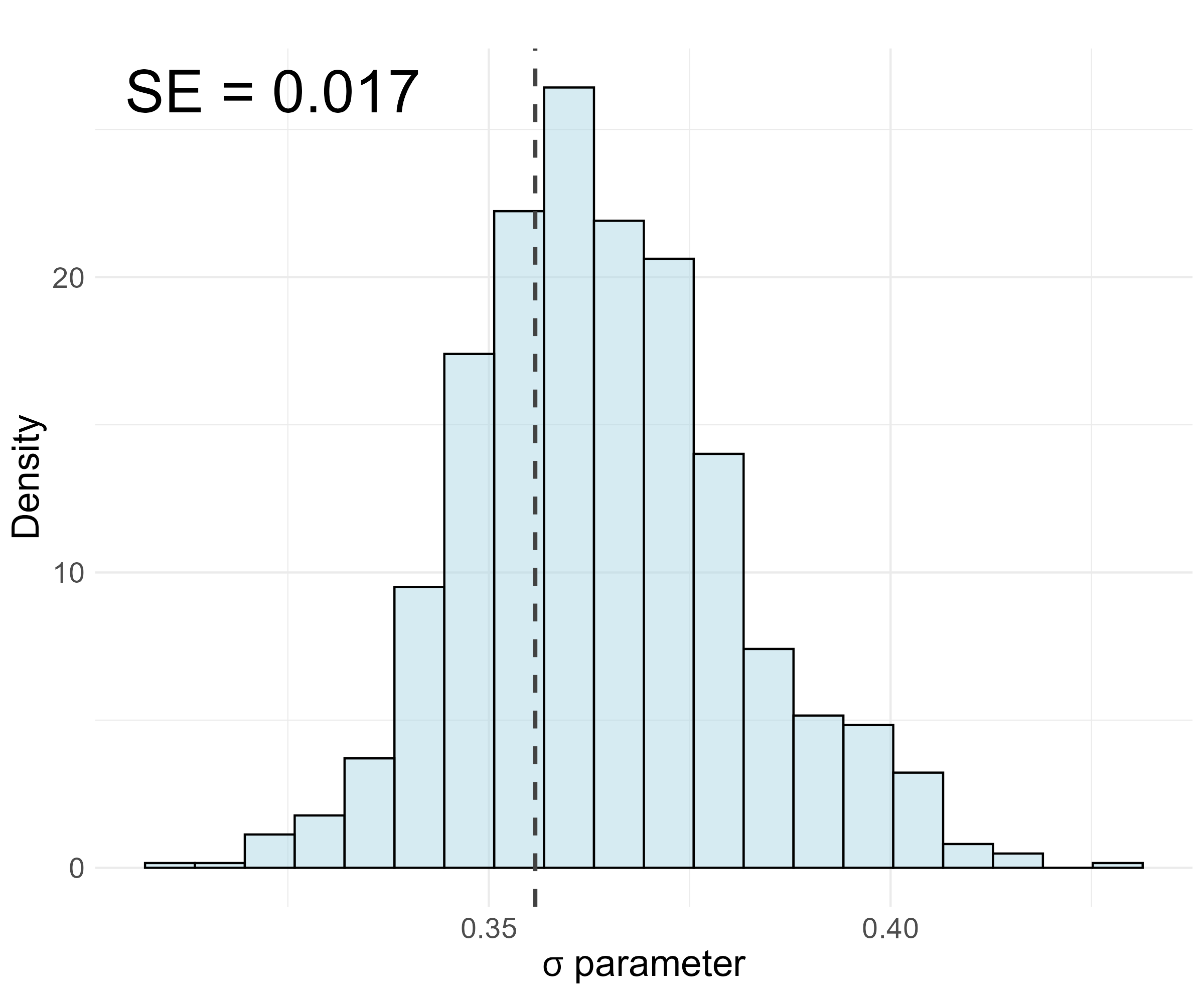}
  \vspace{0.2cm}
  \includegraphics[width=.315\textwidth]{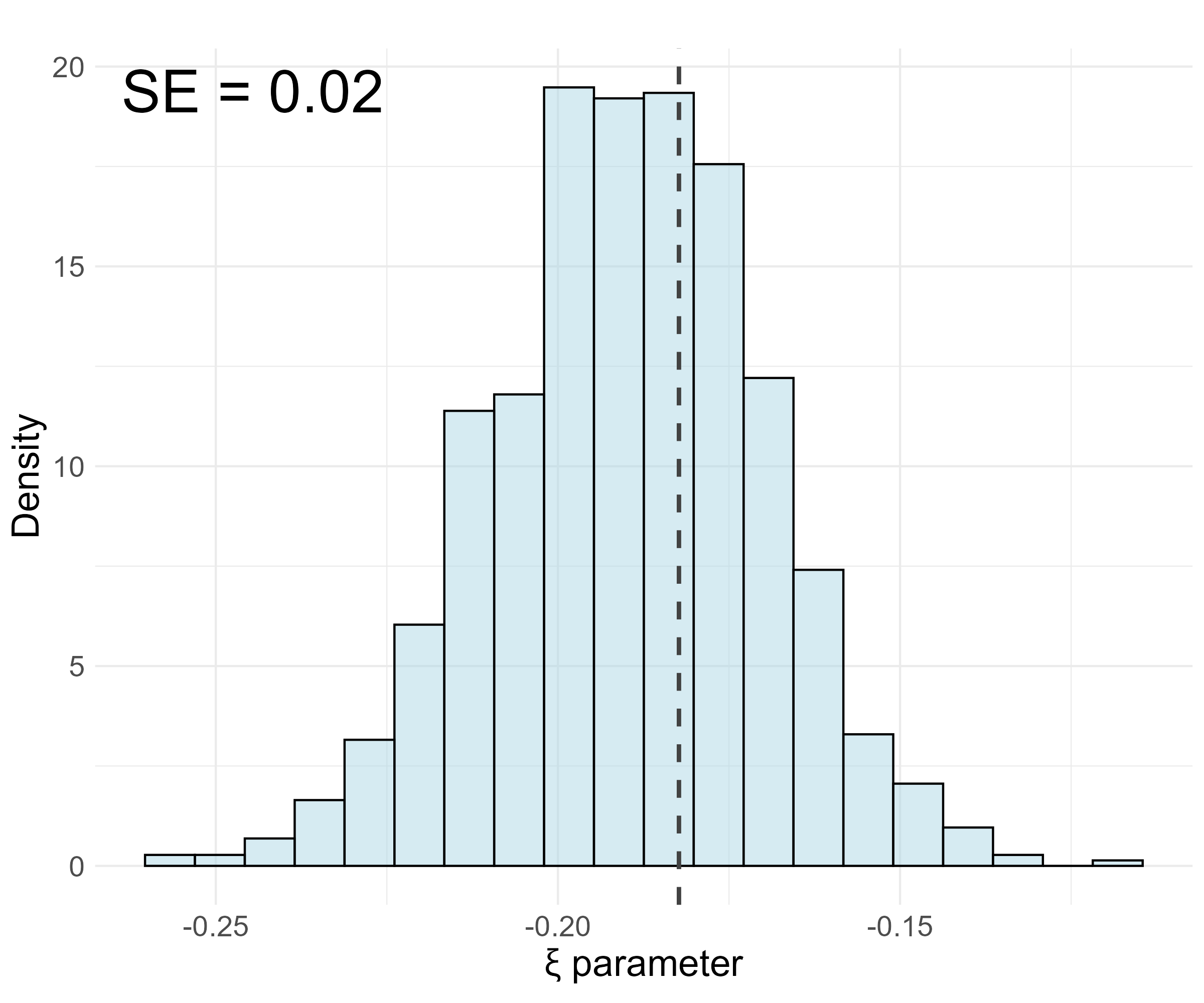}
  \hspace{0.2cm}
  \includegraphics[width=.315\textwidth]{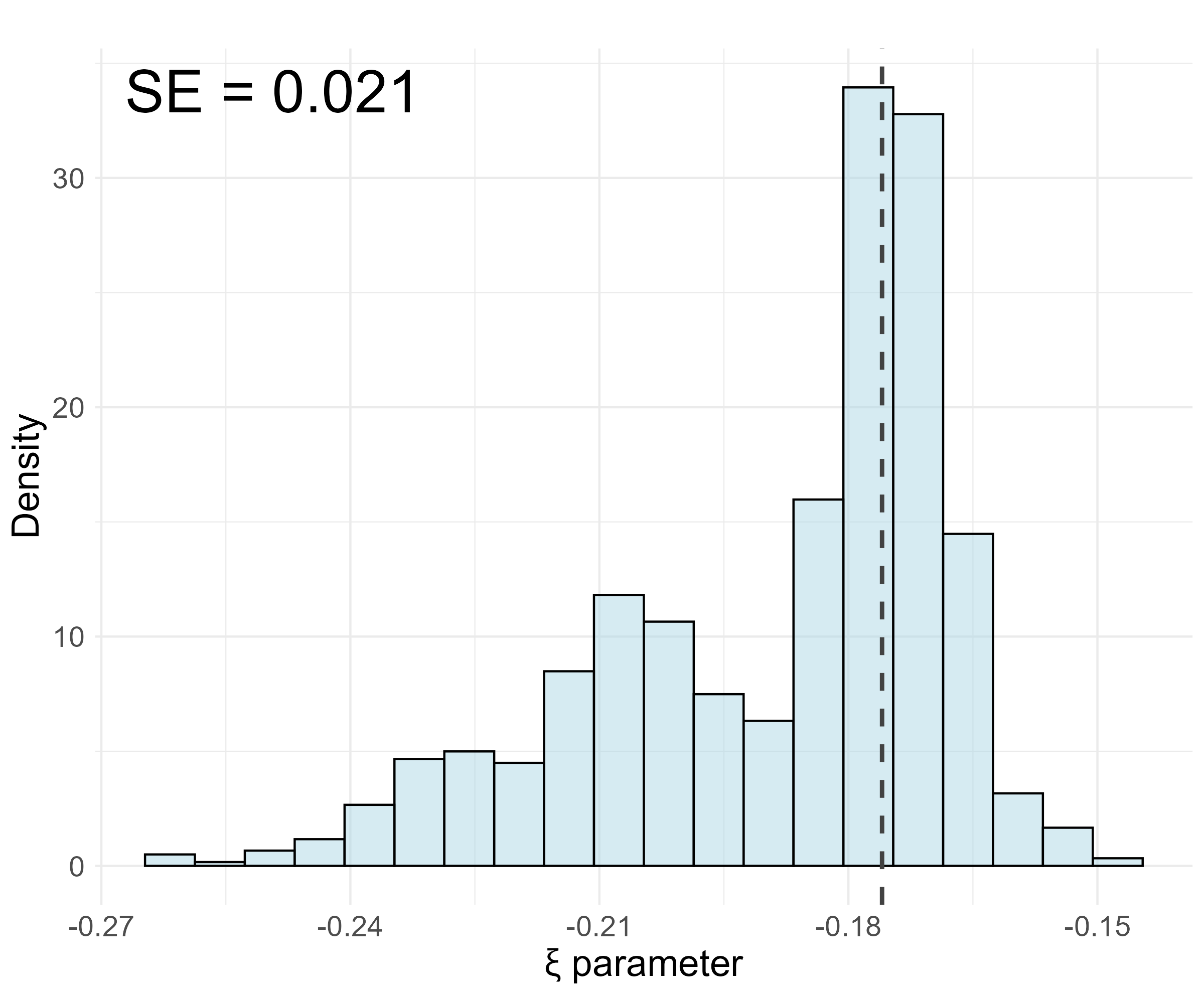}
  \hspace{0.2cm}
  \includegraphics[width=.315\textwidth]{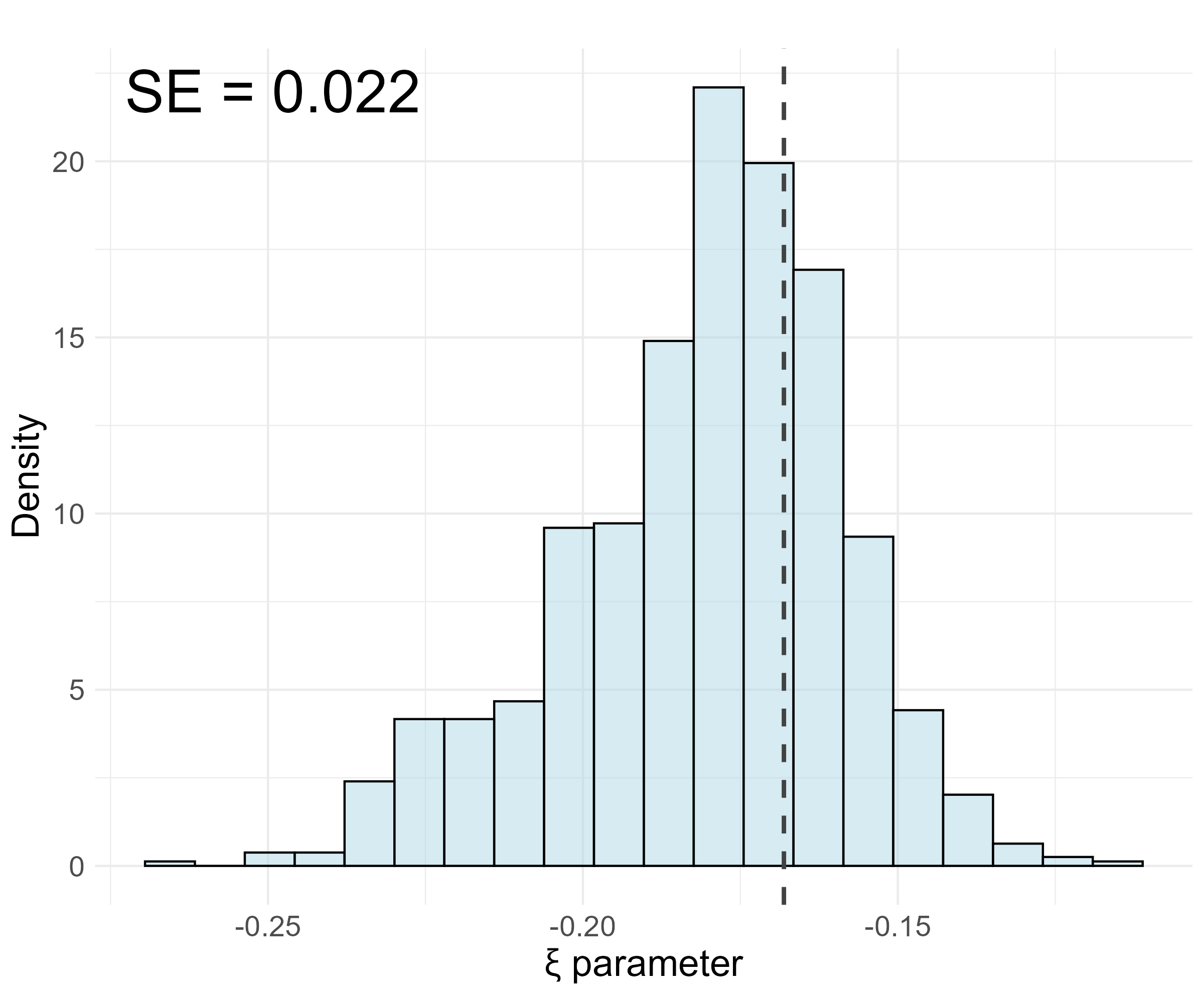}
\caption{Histograms of the EGP parameter estimators computed from $R = 1000$ bootstrap samples, using block
bootstrap with block length $l = 50$, for the sea level data at each station. The grey vertical dashed line represents the estimate obtained from the full dataset (without resampling). The standard error is indicated in top-left corner of each figure.\label{fig:stability_margins_sl}}
\end{figure}

\begin{figure}[ht!] 
  \centering
  \includegraphics[width=.23\textwidth]{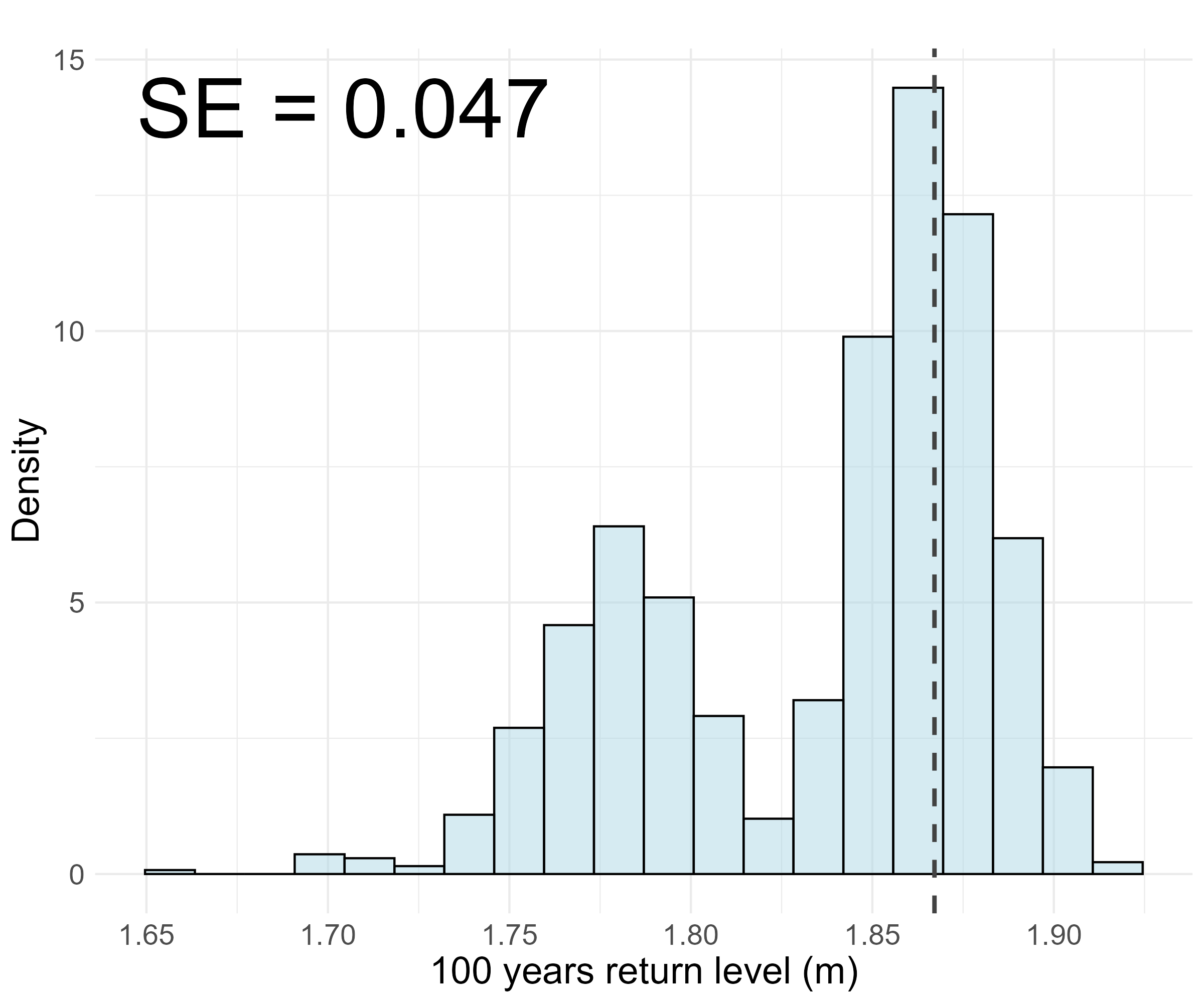}
  \hspace{0.2cm}
  \includegraphics[width=.23\textwidth]{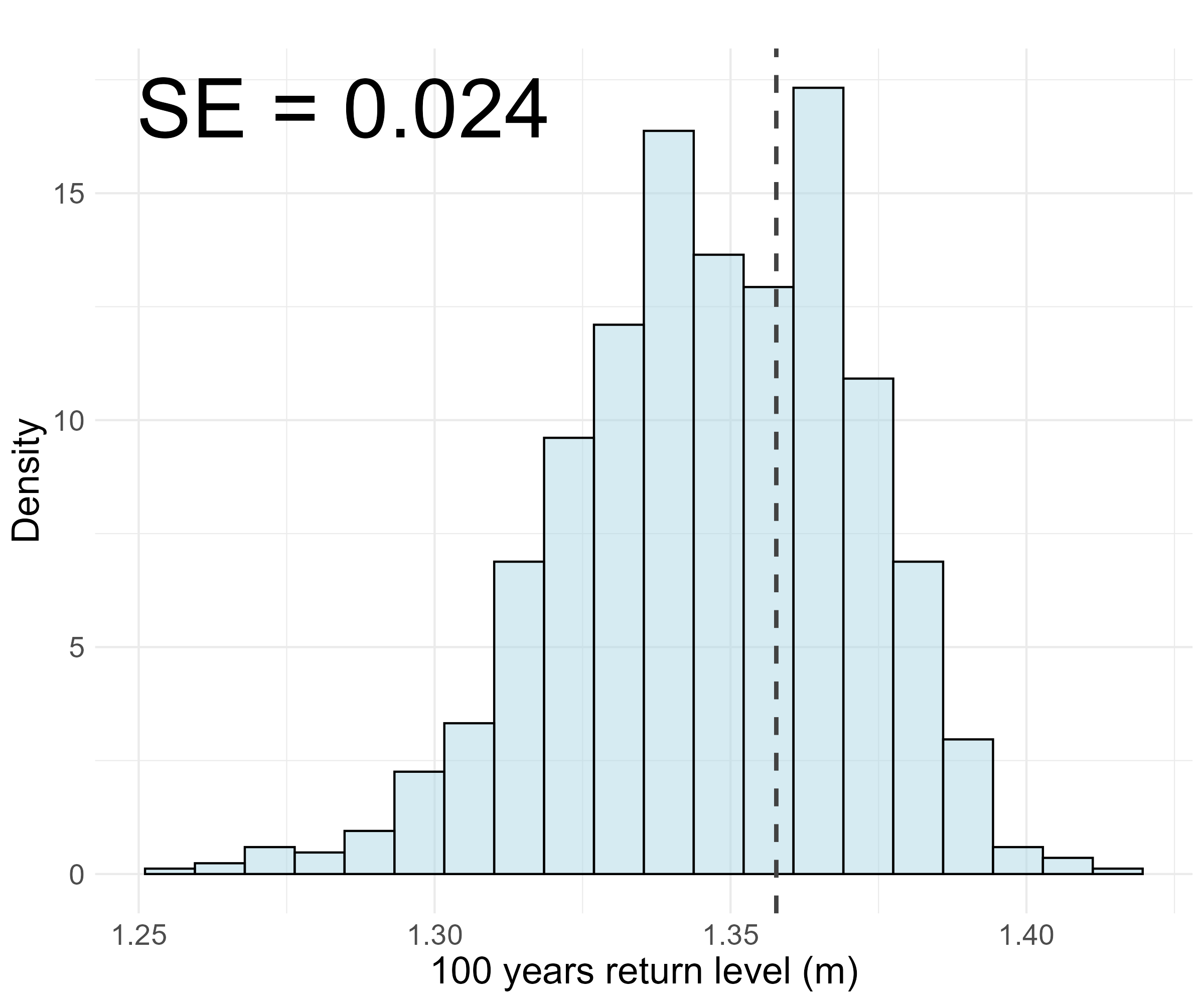}
  \hspace{0.2cm}
  \includegraphics[width=.23\textwidth]{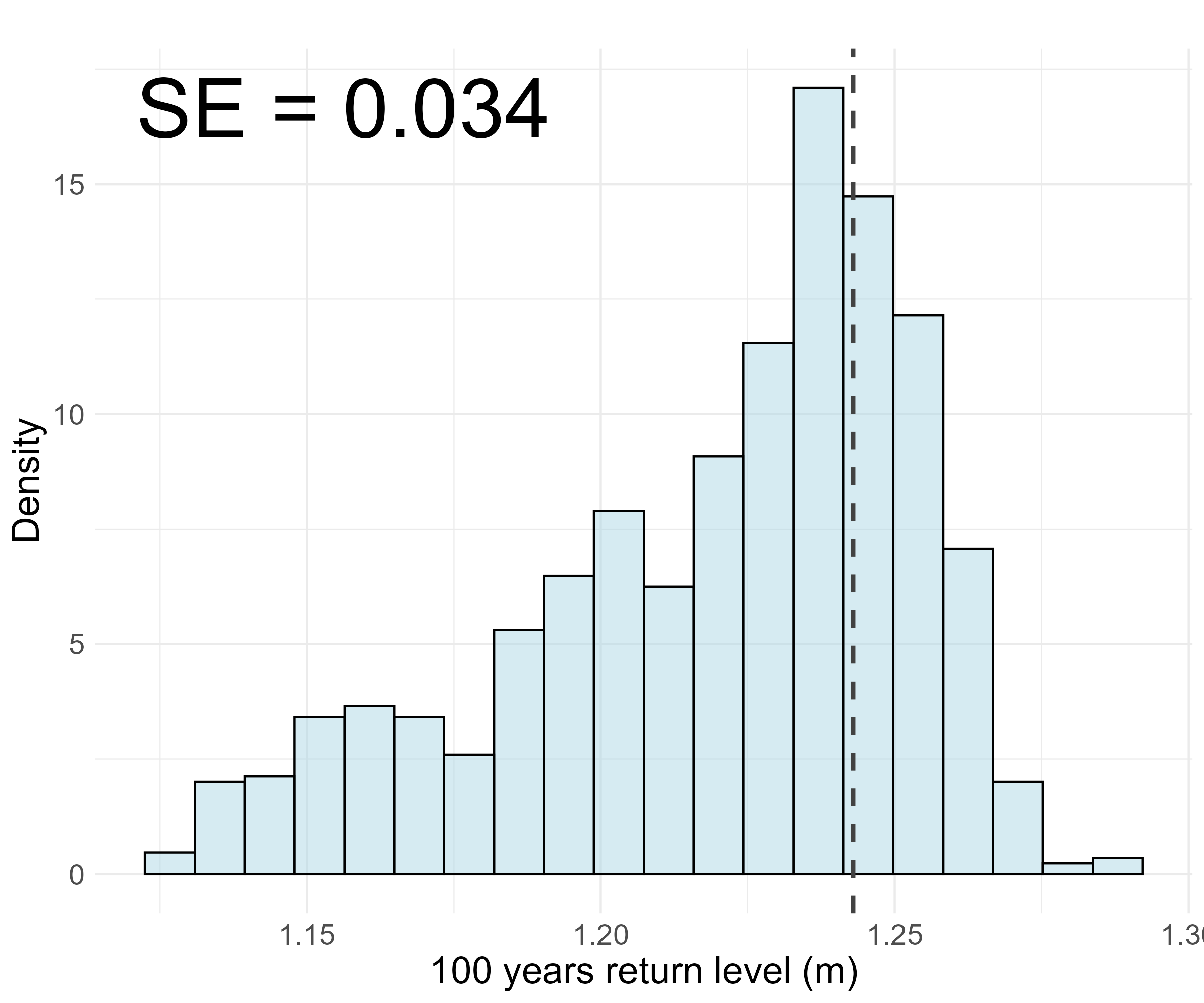}
\hspace{0.2cm}
  \includegraphics[width=.23\textwidth]{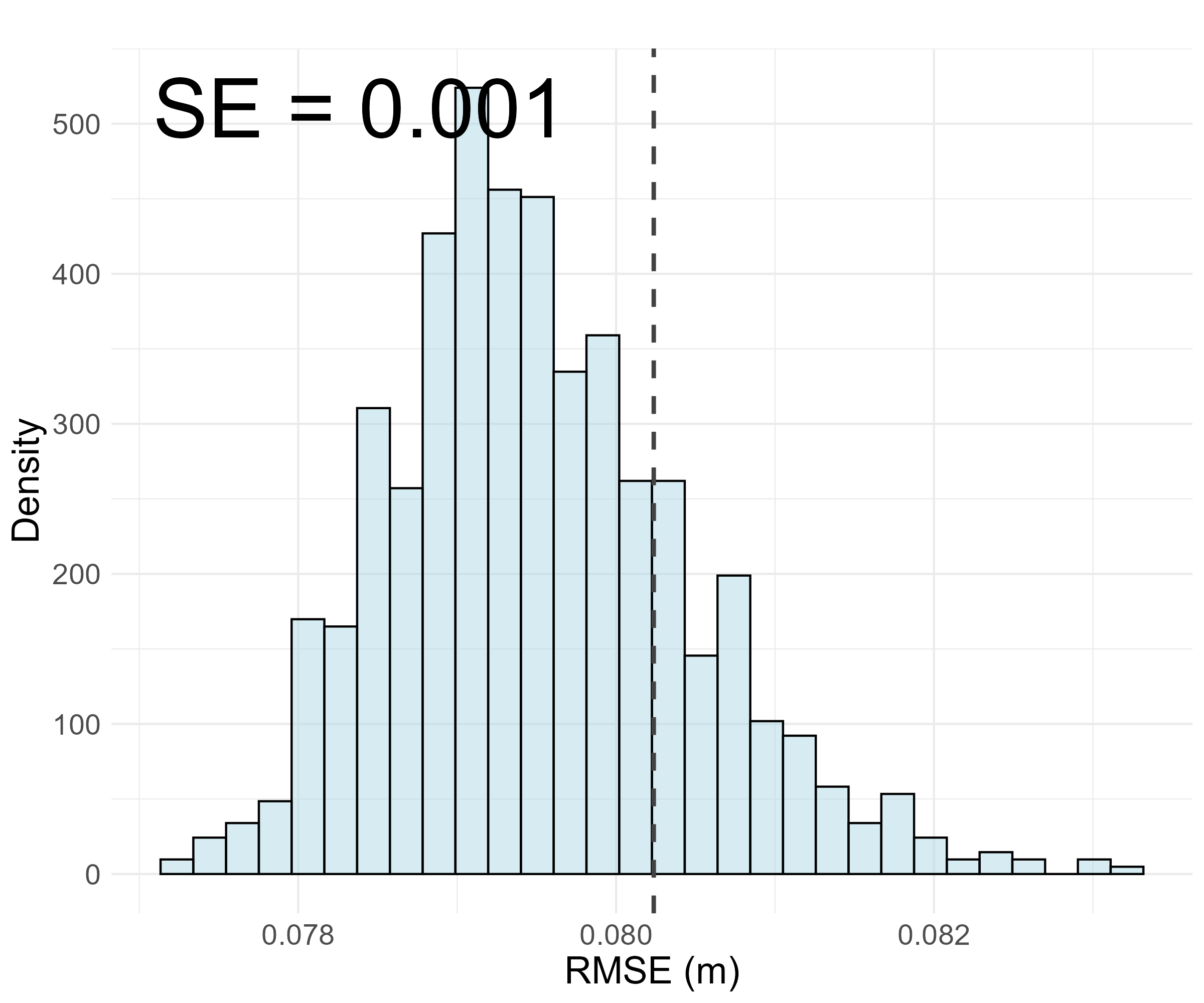}

\caption{Histograms of the 100-year return level (left and middle), computed using the EGP parameter estimates from $R = 1000$ bootstrap samples, using block
bootstrap with block length $l = 50$, for the sea level data at each station (Brest: left; Saint-Nazaire: middle-left; Port Tudy: middle-right), and histogram of the RMSE (right) computed on the fixed test set using the ROXANE routine with the OLS algorithm and the different bootstrap training samples. The grey vertical dashed lines represent the estimates obtained from the full dataset (without resampling). The standard error is indicated in top-left corner of each figure.\label{fig:stability_rl_sl}}
\end{figure}

\clearpage
\subsection{Concarneau study}

\begin{table}[ht!]
\caption{RMSE$\times10^2$(SSE$\times10^2$) and MAE$\times10^2$(ASE$\times10^2$) of predicted sea level and skew surge exceedances at Concarneau station from the ROXANE procedure with RF regression (ROX RF), ROXANE procedure with OLS regression (ROX OLS) and MGPRED. Errors are measured on the test set covering the period from 28/06/1999 to 31/12/2010. Errors are computed on the entire test set (columns: RMSE and MSE) and on its most extreme half, i.e., for observations with $Y_i$ exceeding their empirical median computed on the extreme test set (columns: $\mbox{RMSE}_{\mbox{EXT}}$ and $\mbox{MAE}_{\mbox{EXT}}$). \label{tab:concarneau}}
\vspace{0.2cm}
\begin{center}
\begin{small}
\begin{sc}
\begin{tabular}{lcccccr}
\toprule
Training models/Errors & RMSE & MAE & $\mbox{RMSE}_{\mbox{ext}}$ & $\mbox{MAE}_{\mbox{ext}}$ \\
\midrule
Sea levels: ROX RF & 7.3(0.4) & 5.2(0.2) & 7.9(0.5) & 6.0(0.2) \\
\phantom{Sea levels: }ROX OLS & \textbf{6.8(0.4)} & \textbf{5.0(0.2)} & \textbf{7.5(0.5)} & \textbf{5.8(0.2)} \\
\phantom{Sea levels: }MGPRED & 7.5(0.4) & 5.7(0.2) & 8.7(0.5) & 6.7(0.3) \\
\midrule
Skew surges: ROX RF & \textbf{6.6(0.4)} & \textbf{4.9(0.1)} & 6.6(0.3) & 5.3(0.1)\\
\phantom{Skew surges: }ROX OLS & \textbf{6.6(0.4)} & 5.0(0.1) & \textbf{6.4(0.3)} & \textbf{5.2(0.1)} \\
\phantom{Skew surges: }MGPRED & 6.7(0.2) & 5.3(0.1) & 8.0(0.3) & 6.6(0.2) \\
\bottomrule
\end{tabular}
\end{sc}
\end{small}
\end{center}
\end{table}

\begin{figure}[t!]
  \centering
  \includegraphics[width=.4\textwidth]{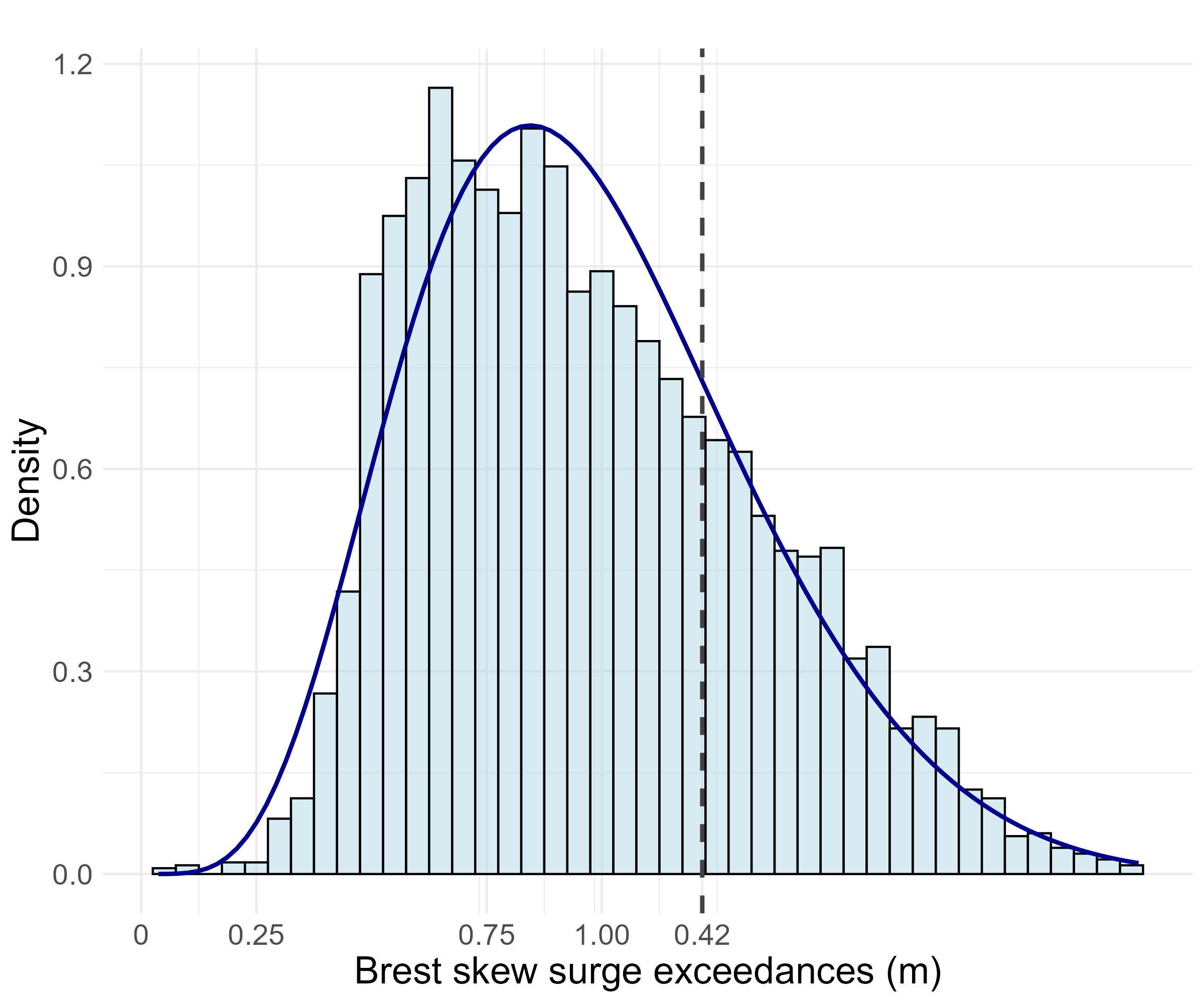}
  \hspace{1cm}
  \includegraphics[width=.4\textwidth]{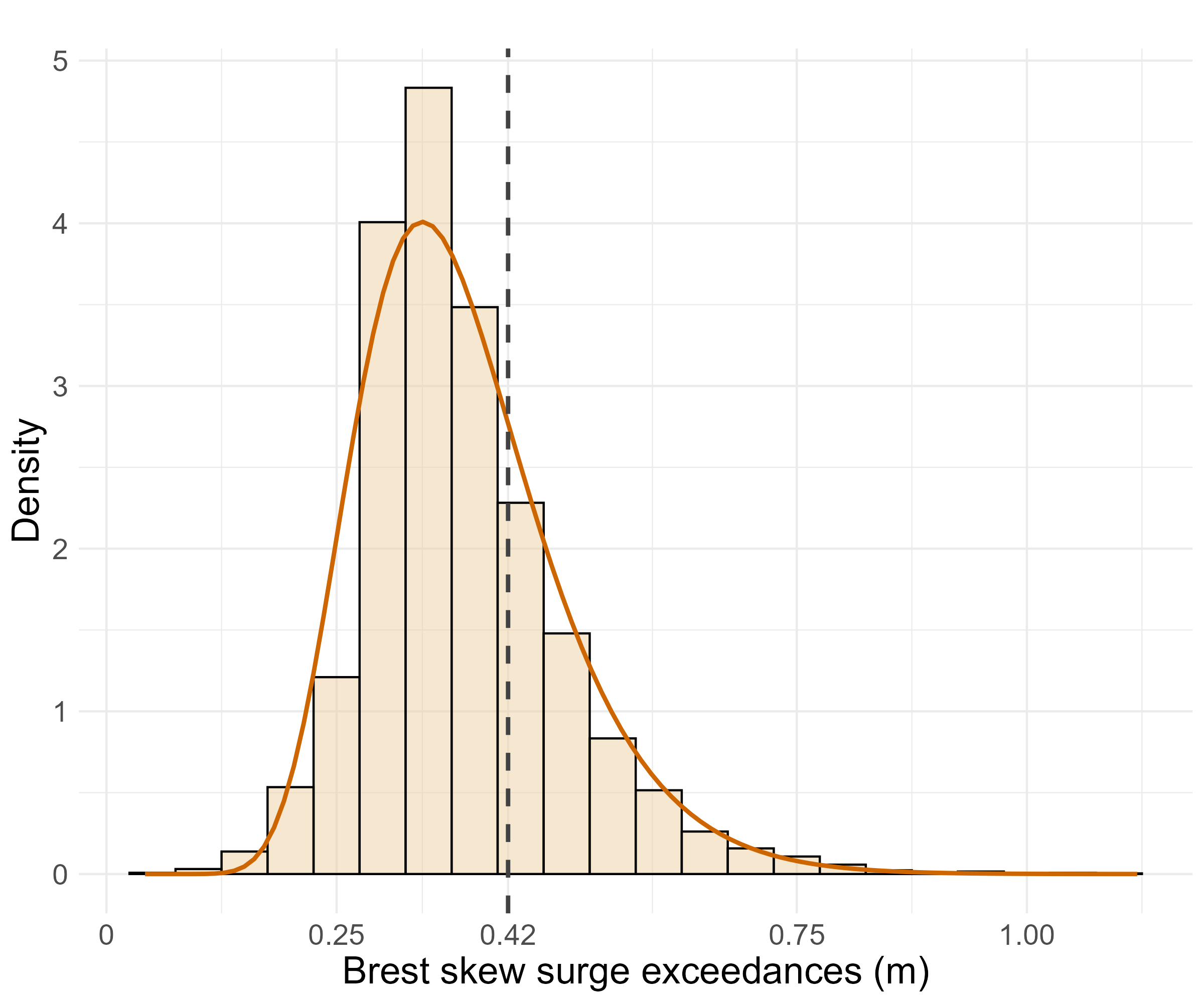}

  \vspace{1cm}

  \includegraphics[width=.4\textwidth]{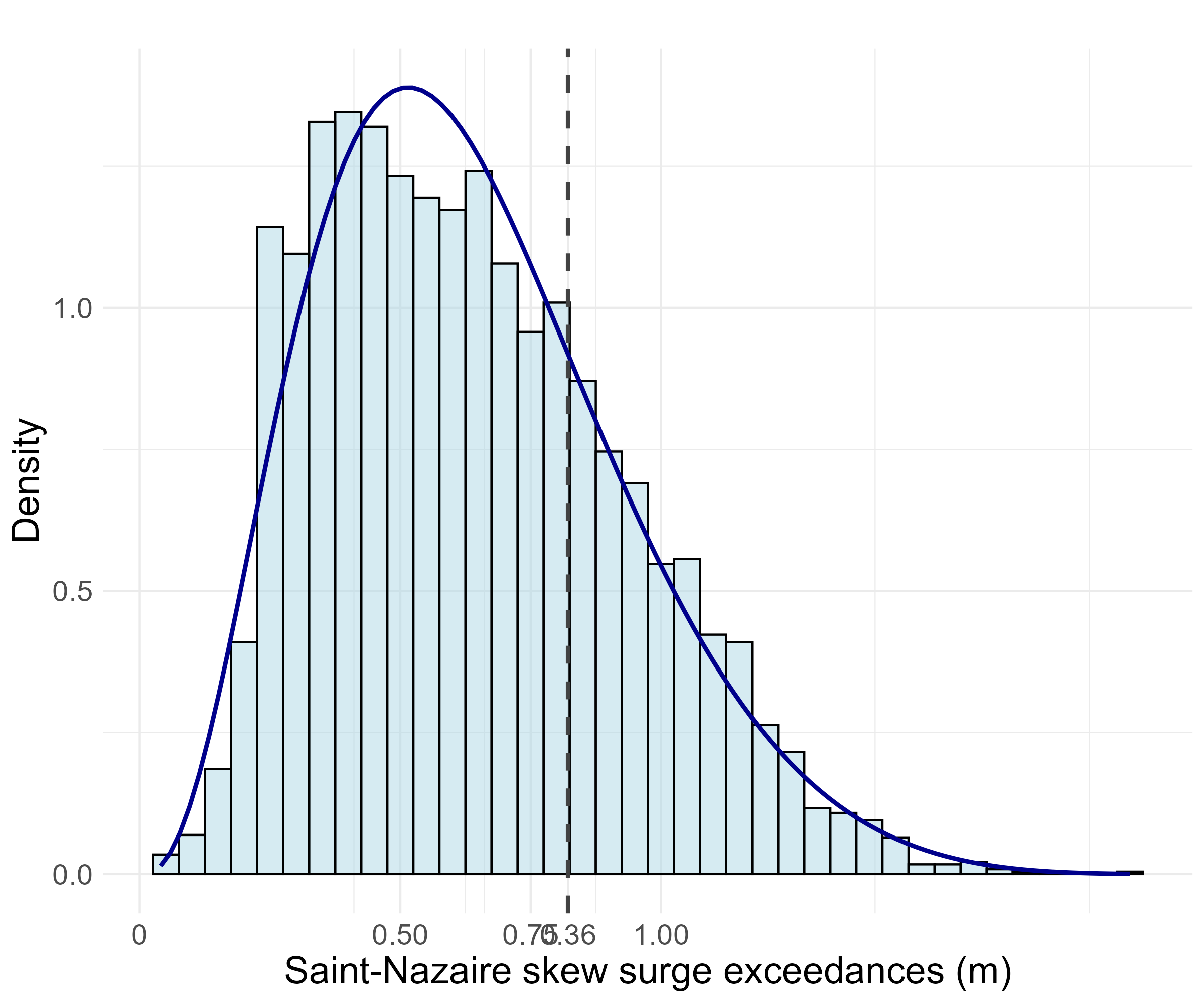}
  \hspace{1cm}
  \includegraphics[width=.4\textwidth]{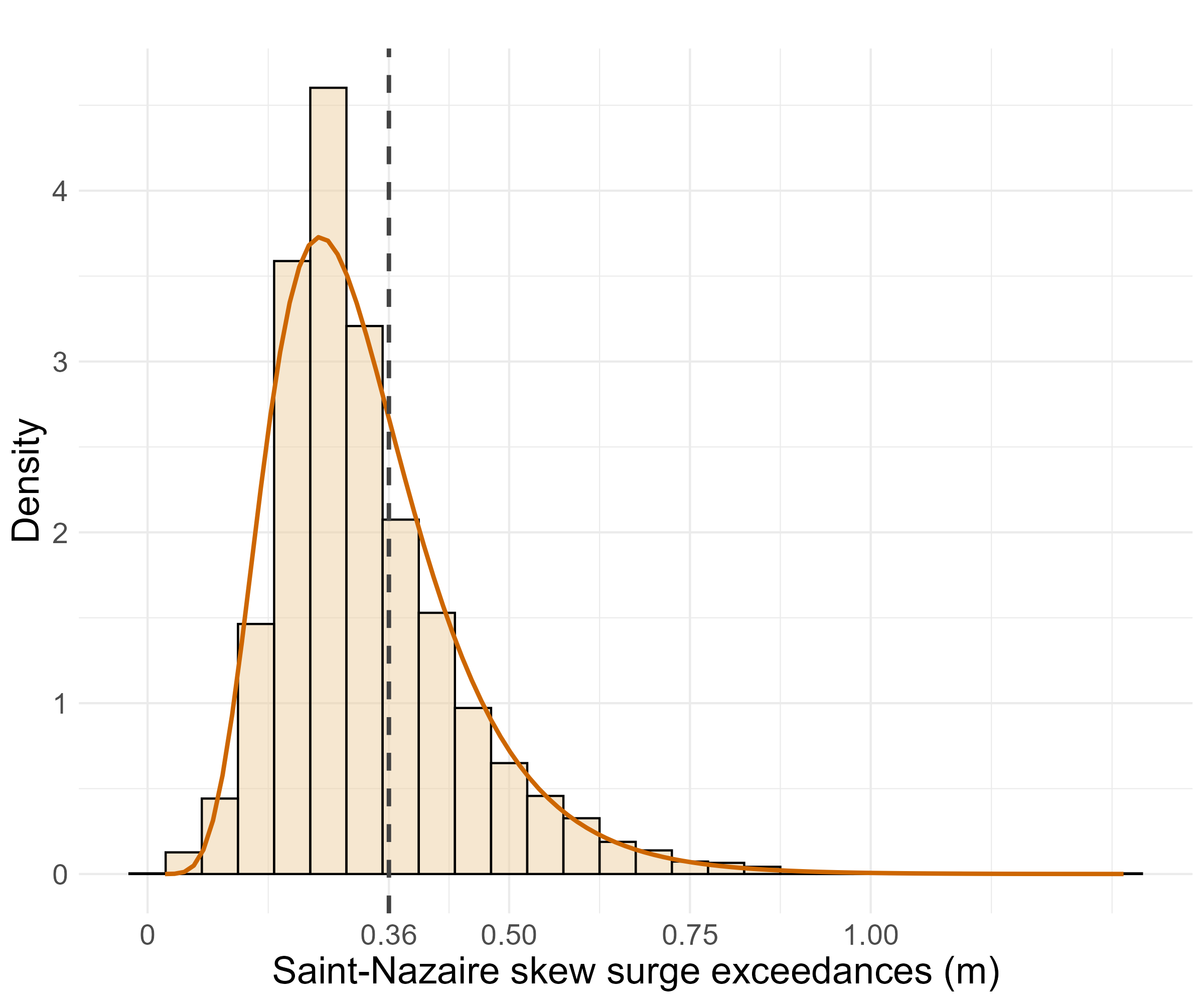}

  \vspace{1cm}

  \includegraphics[width=.4\textwidth]{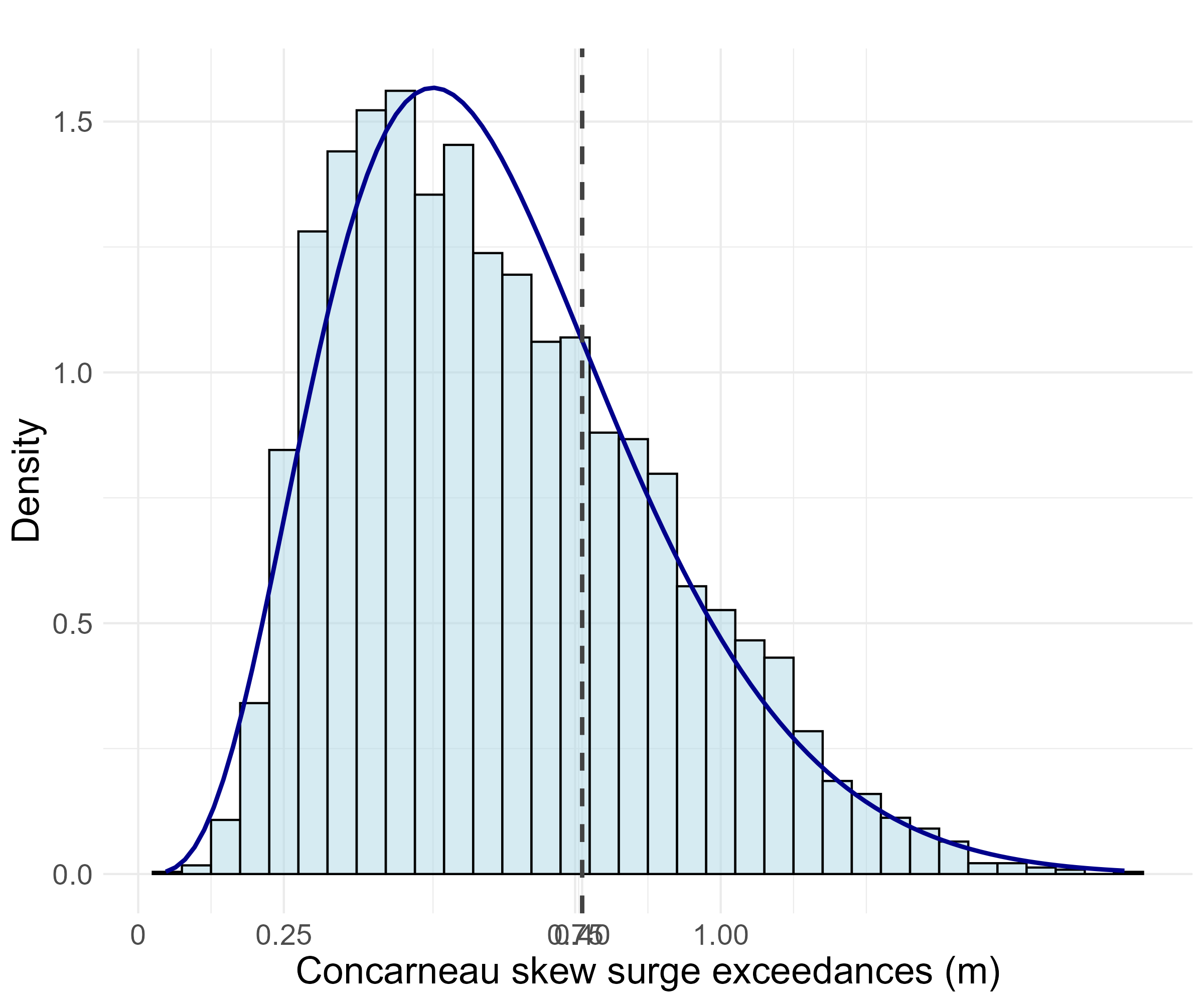}
  \hspace{1cm}
  \includegraphics[width=.4\textwidth]{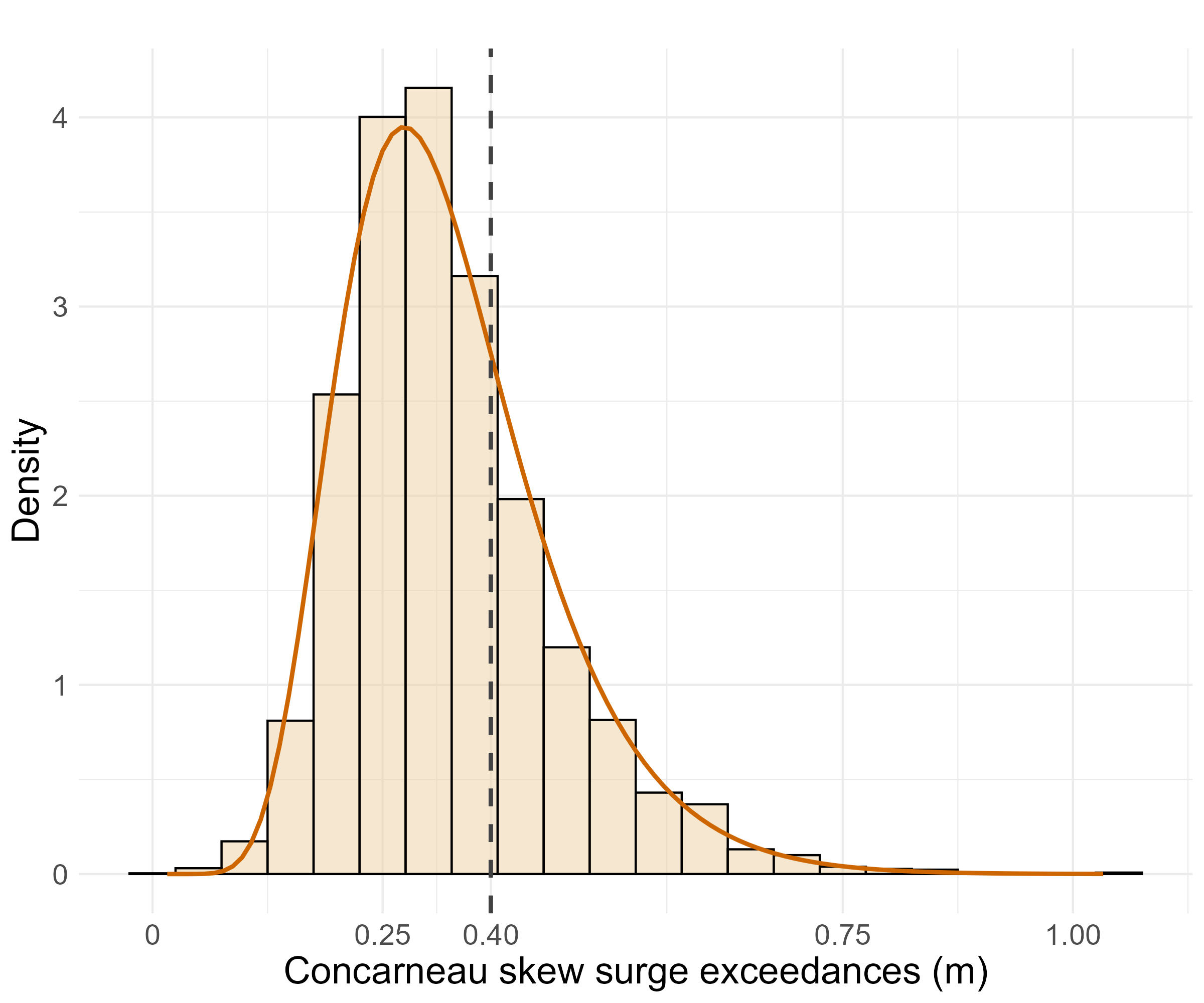}

  \caption{Histograms of sea level (right) and skew surge (right) exceedances at the three stations Brest (top row), Saint-Nazaire (middle row) and Concarneau (bottom row), from 28/06/1999 to 31/12/2010. The darkblue and darkorange curves represent the fitted EGP densities. The dotted vertical grey lines represent the smallest point above which each fitted density is convex, which  represent the chosen marginal thresholds via Algorithm~\ref{algo:th_select}. \label{fig:density_concarneau}}
\end{figure}

\begin{figure}[ht!]

  \centering
  \includegraphics[width=.4\textwidth]{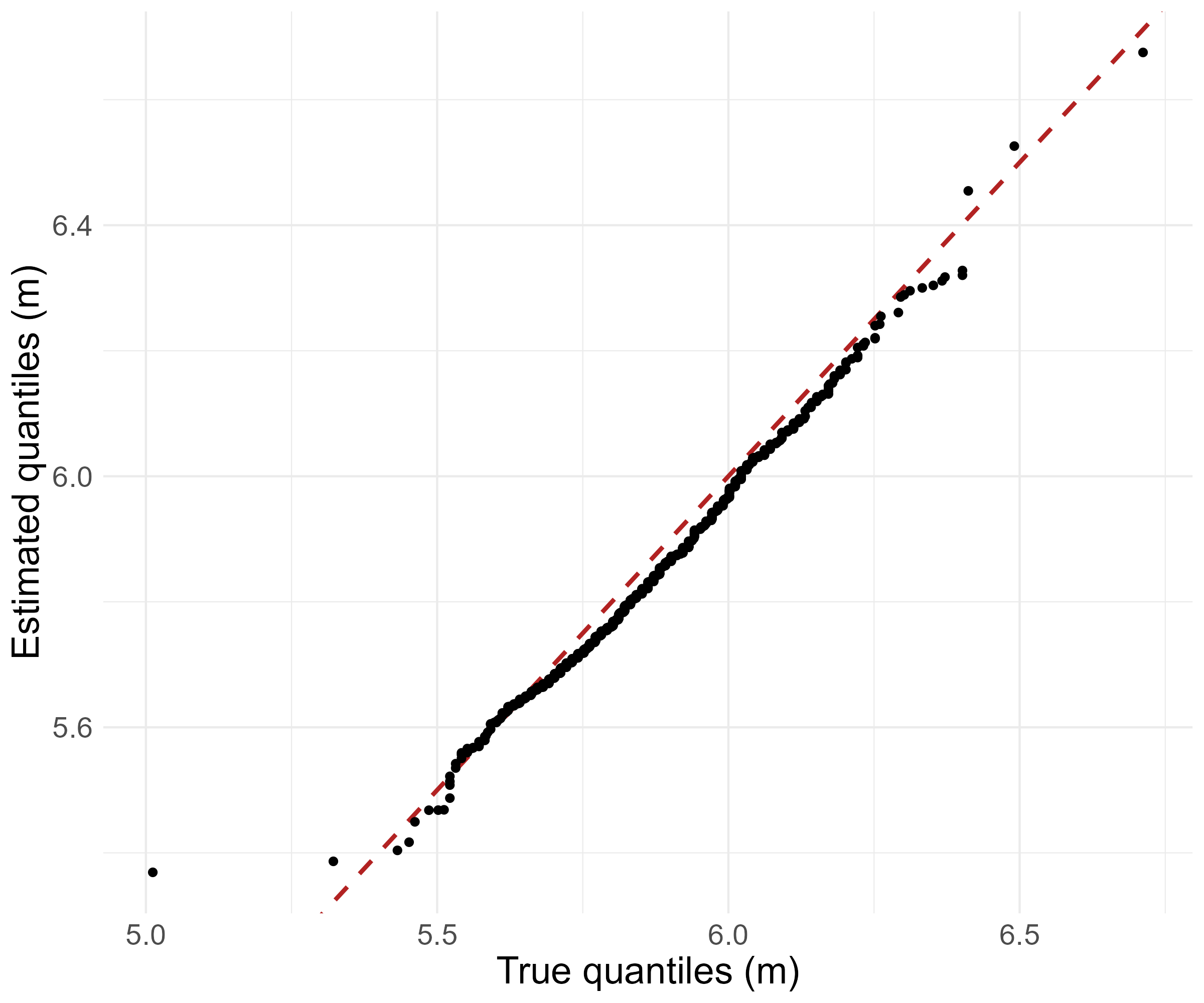}
  \hspace{1cm}
  \includegraphics[width=.4\textwidth]{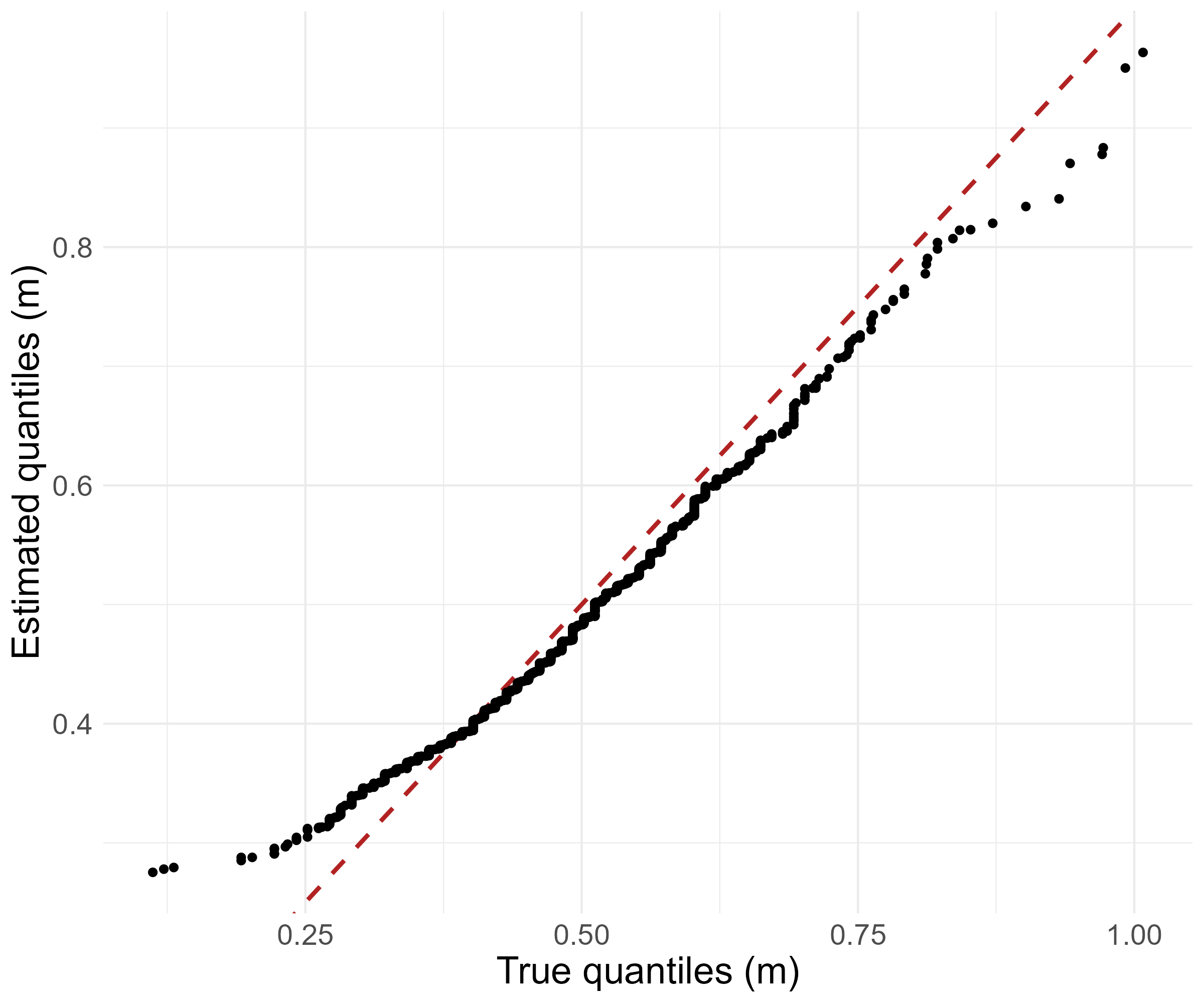}

  \vspace{1cm}

  \includegraphics[width=.4\textwidth]{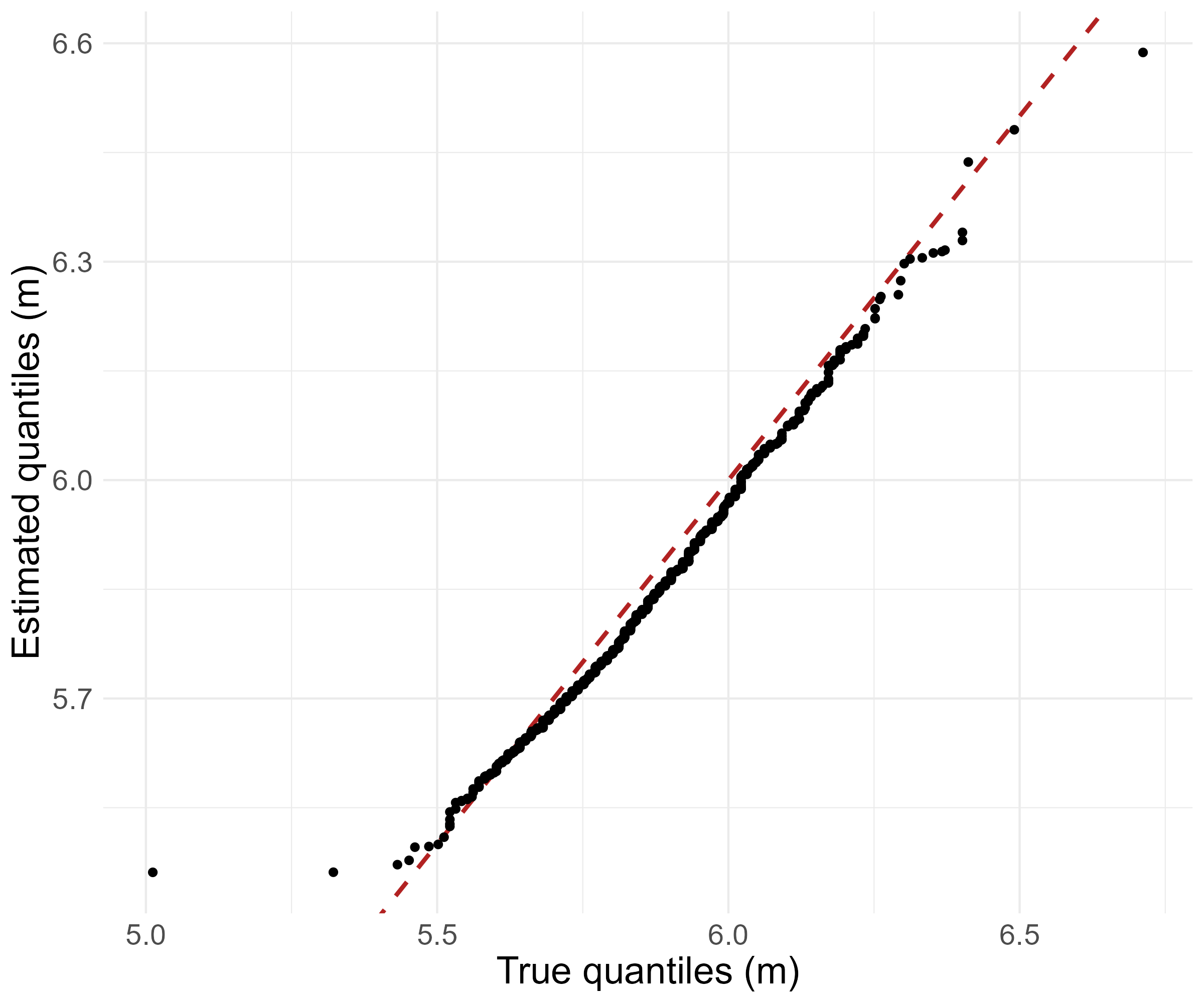}
  \hspace{1cm}
  \includegraphics[width=.4\textwidth]{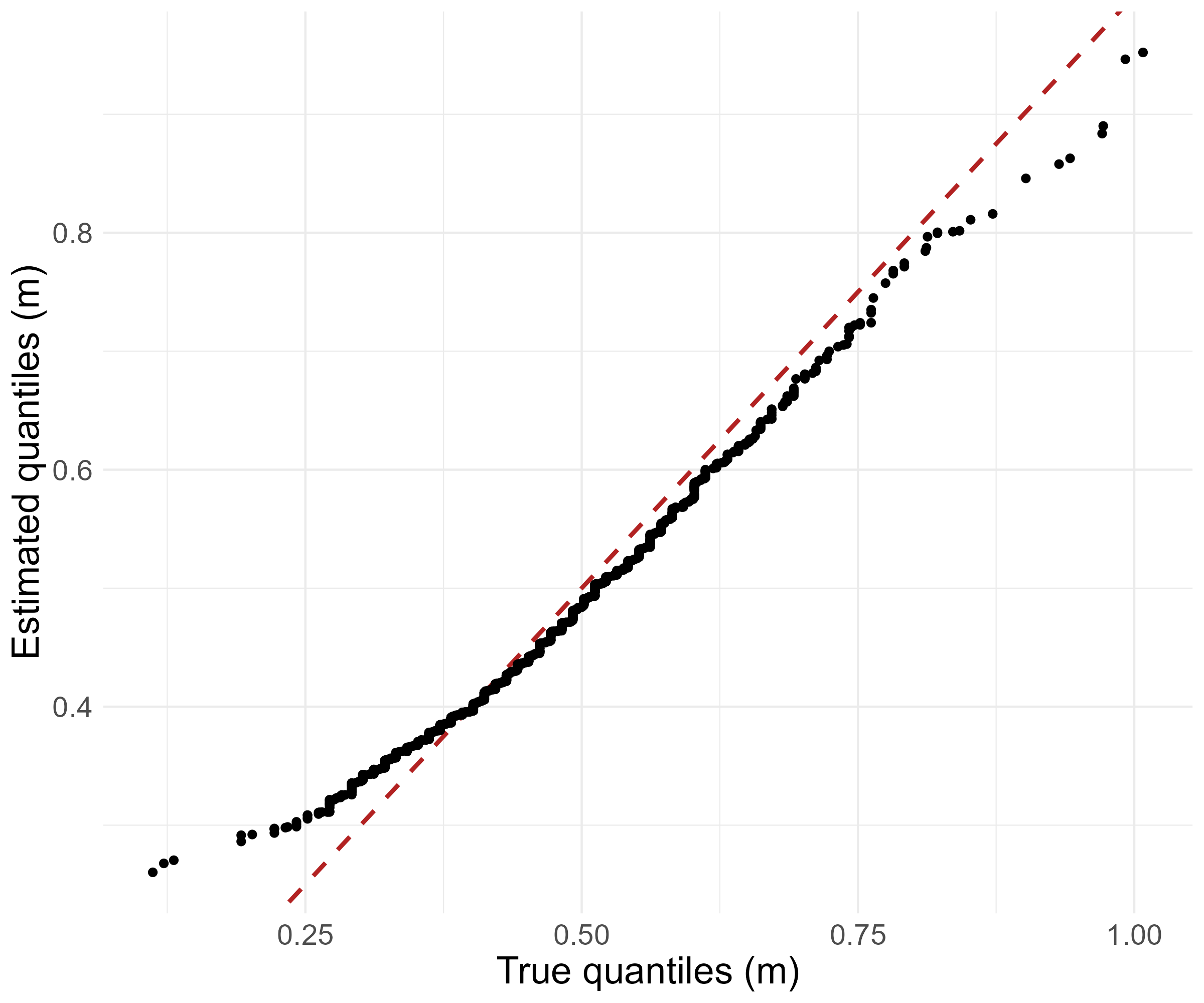}

  \vspace{1cm}

  \includegraphics[width=.4\textwidth]{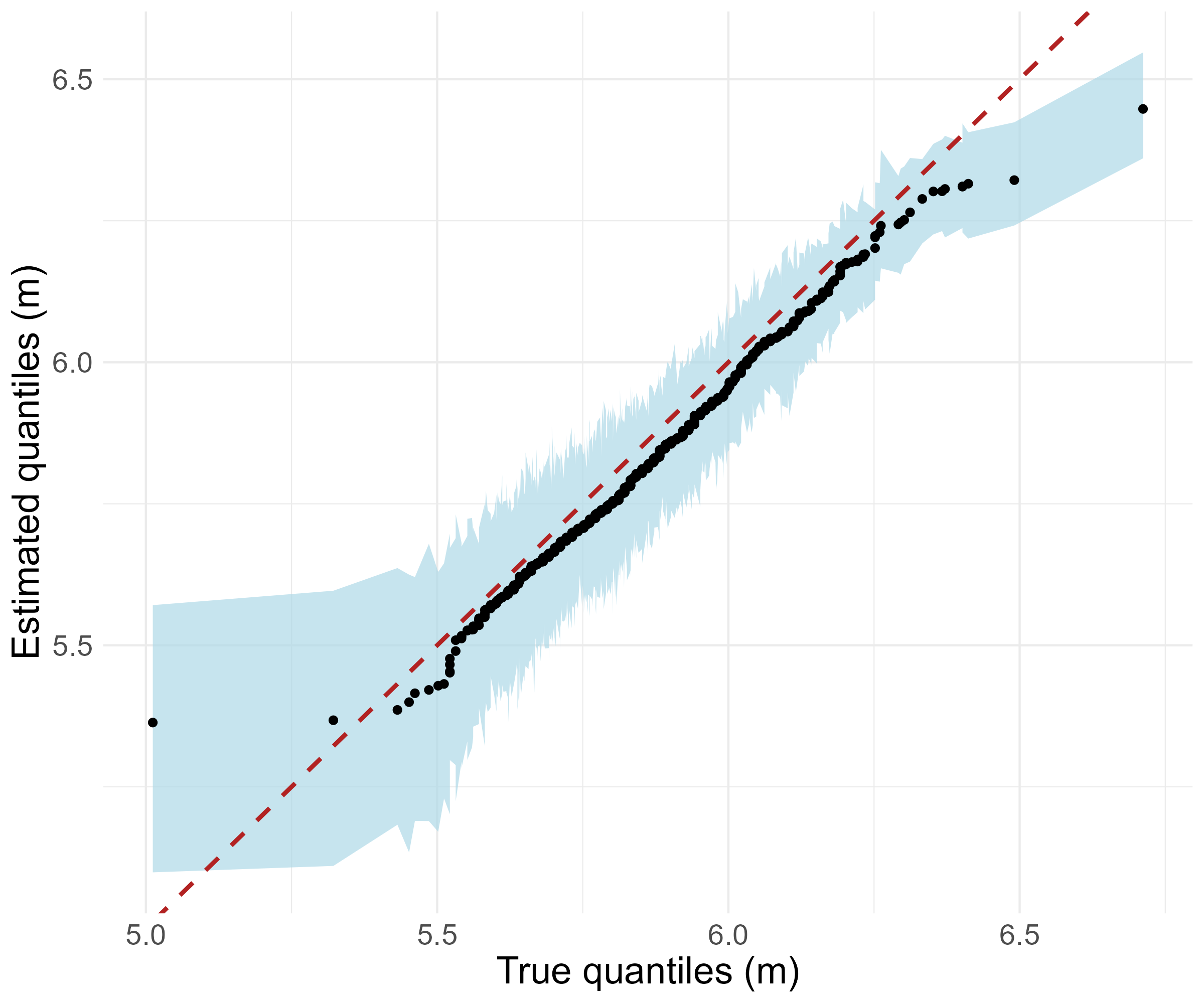}
  \hspace{1cm}
  \includegraphics[width=.4\textwidth]{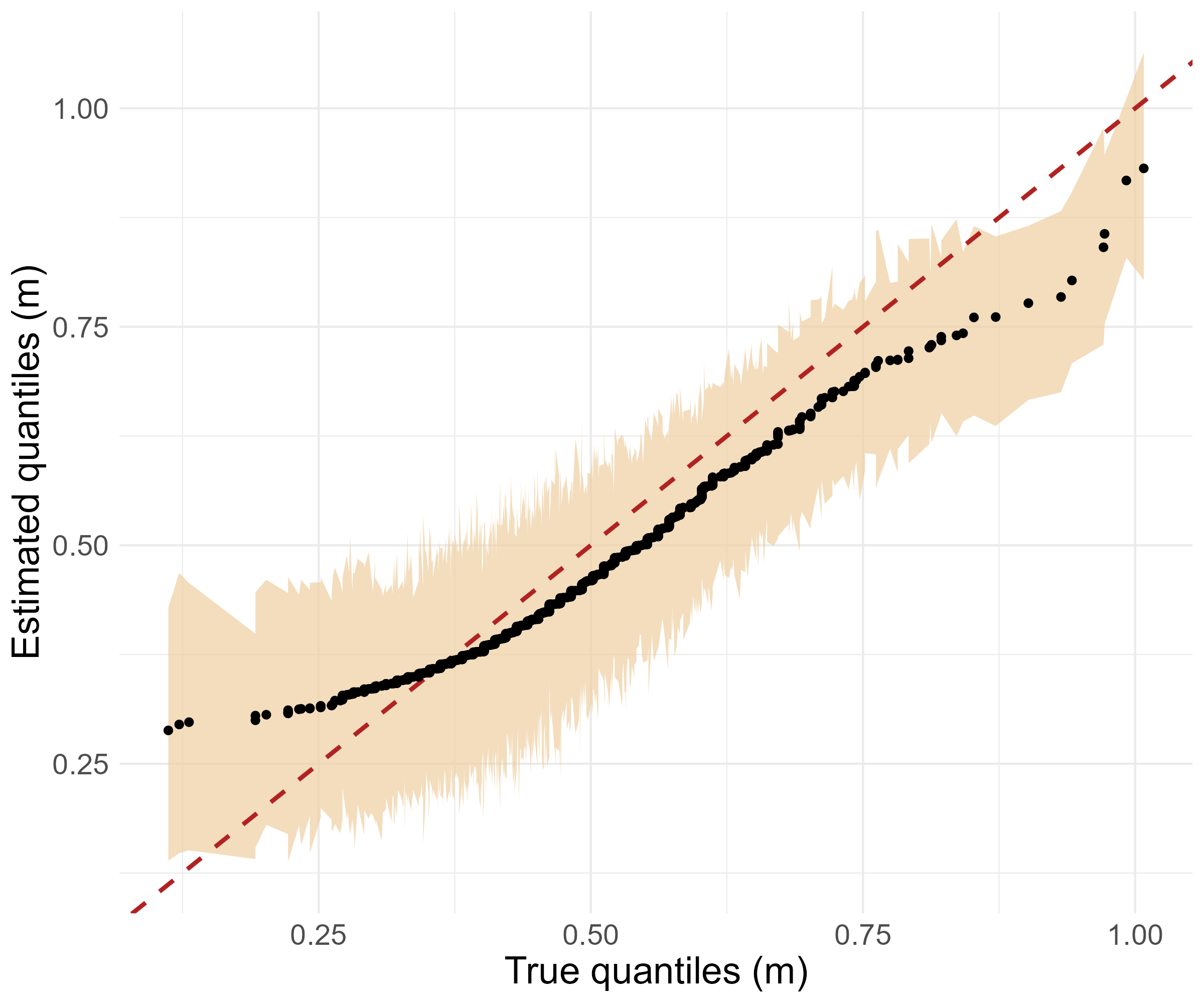}
\caption{QQ-plots comparing observed sea level (left) and skew surge (right) exceedances of the Concarneau test set (x-axis), ranging from  28/06/1999 to 31/12/2010, to predicted data (y-axis) from the algorithms of Sections~\ref{sec:reg_proc} and \ref{sec:mgp_proc}. The plots show results from the ROXANE procedure with RF regression (top row), ROXANE procedure with OLS regression (middle row), and MGPRED (bottom row) with 0.95-confidence bands (lightblue and lightorange). The dotted red line represents the identity line $x=y$. The global 0.95-coverage probability associated with MGPRED is 0.93 for sea levels and 0.96 for skew
surges. \label{fig:qqplot_concarneau}}
\end{figure}

\clearpage
\subsection{Le Crouesty study}

\begin{table}[ht!]
\vskip-0.4cm
\caption{RMSE$\times10^2$(SSE$\times10^2$) and MAE$\times10^2$(ASE$\times10^2$) of predicted sea level and skew surge exceedances at Le Crouesty station from the ROXANE procedure with RF regression (ROX RF), ROXANE procedure with OLS regression (ROX OLS) and MGPRED. Errors are measured on the test set covering the period from 14/03/1996 to 31/12/2014. Errors are computed on the entire test set (columns: RMSE and MSE) and on its most extreme half, i.e., for observations with $Y_i$ exceeding their empirical median computed on the extreme test set (columns: $\mbox{RMSE}_{\mbox{EXT}}$ and $\mbox{MAE}_{\mbox{EXT}}$). \label{tab:lecrouesty}}
\vspace{0.2cm}
\begin{center}
\begin{small}
\begin{sc}
\begin{tabular}{lcccccr}
\toprule
Training models/Errors & RMSE & MAE & $\mbox{RMSE}_{\mbox{ext}}$ & $\mbox{MAE}_{\mbox{ext}}$ \\
\midrule
Sea levels: ROX RF & \textbf{5.9(0.3)} & \textbf{4.5(0.1)} & \textbf{6.1(0.4)} & \textbf{4.9(0.2)} \\
\phantom{Sea levels: }ROX OLS & \textbf{5.9(0.3)} & \textbf{4.5(0.1)} & 6.2(0.4) & \textbf{4.9(0.2)} \\
\phantom{Sea levels: }MGPRED & 7.5(0.4) & 5.6(0.2) & 8.2(0.7) & 6.1(0.3) \\
\midrule
Skew surges: ROX RF & 5.7(0.2) & 4.3(0.1) & 5.5(0.4) & 4.0(0.1)\\
\phantom{Skew surges: }ROX OLS & 5.6(0.2) & 4.2(0.1) & 5.5(0.3) & \textbf{3.9(0.1)} \\
\phantom{Skew surges: }MGPRED & \textbf{5.2(0.2)} & \textbf{4.0(0.1)} & \textbf{5.4(0.3)} & 4.1(0.1)\\
\bottomrule
\end{tabular}
\end{sc}
\end{small}
\end{center}
\end{table}

\begin{figure}[t!]
  \centering
  \includegraphics[width=.4\textwidth]{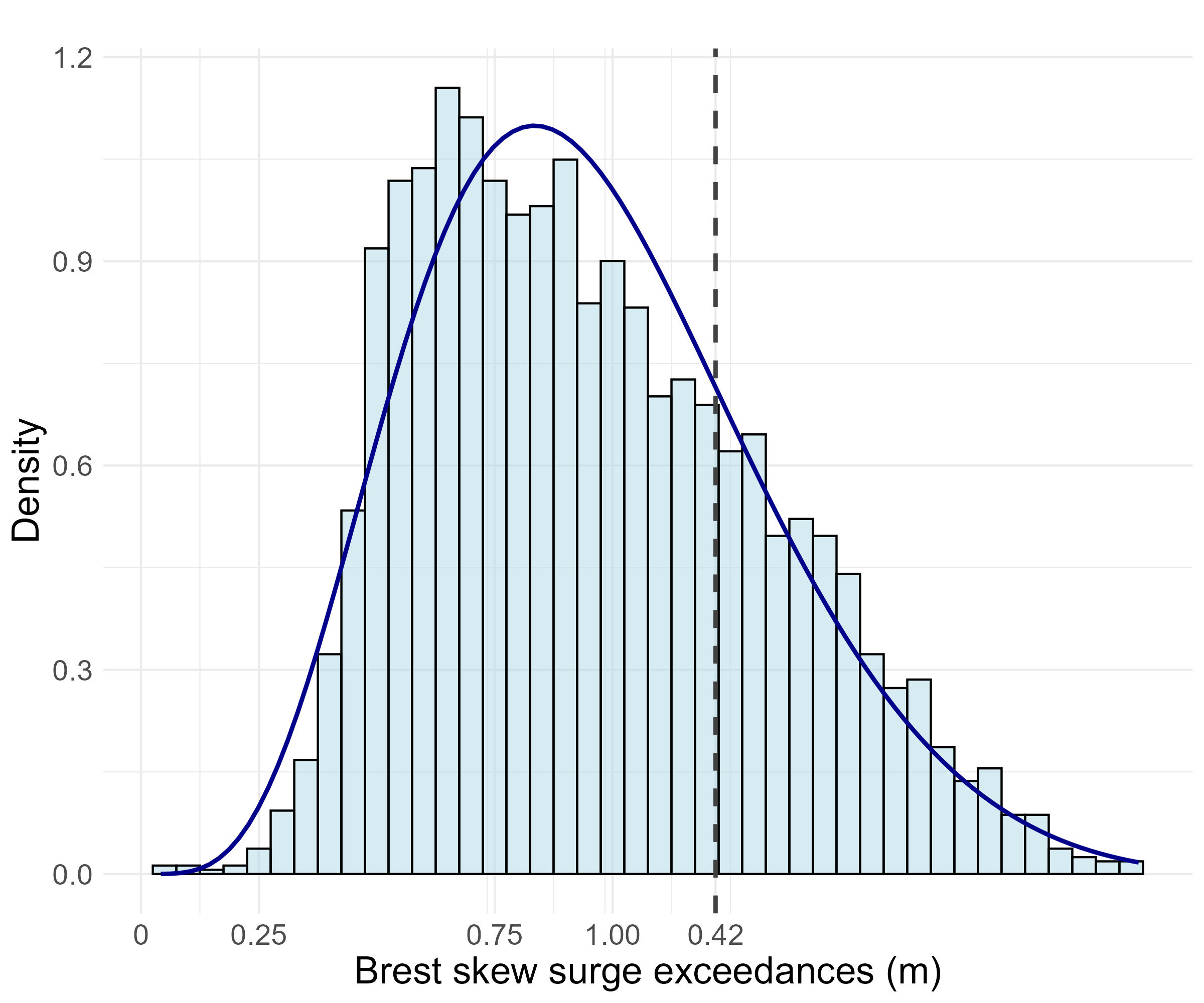}
  \hspace{1cm}
  \includegraphics[width=.4\textwidth]{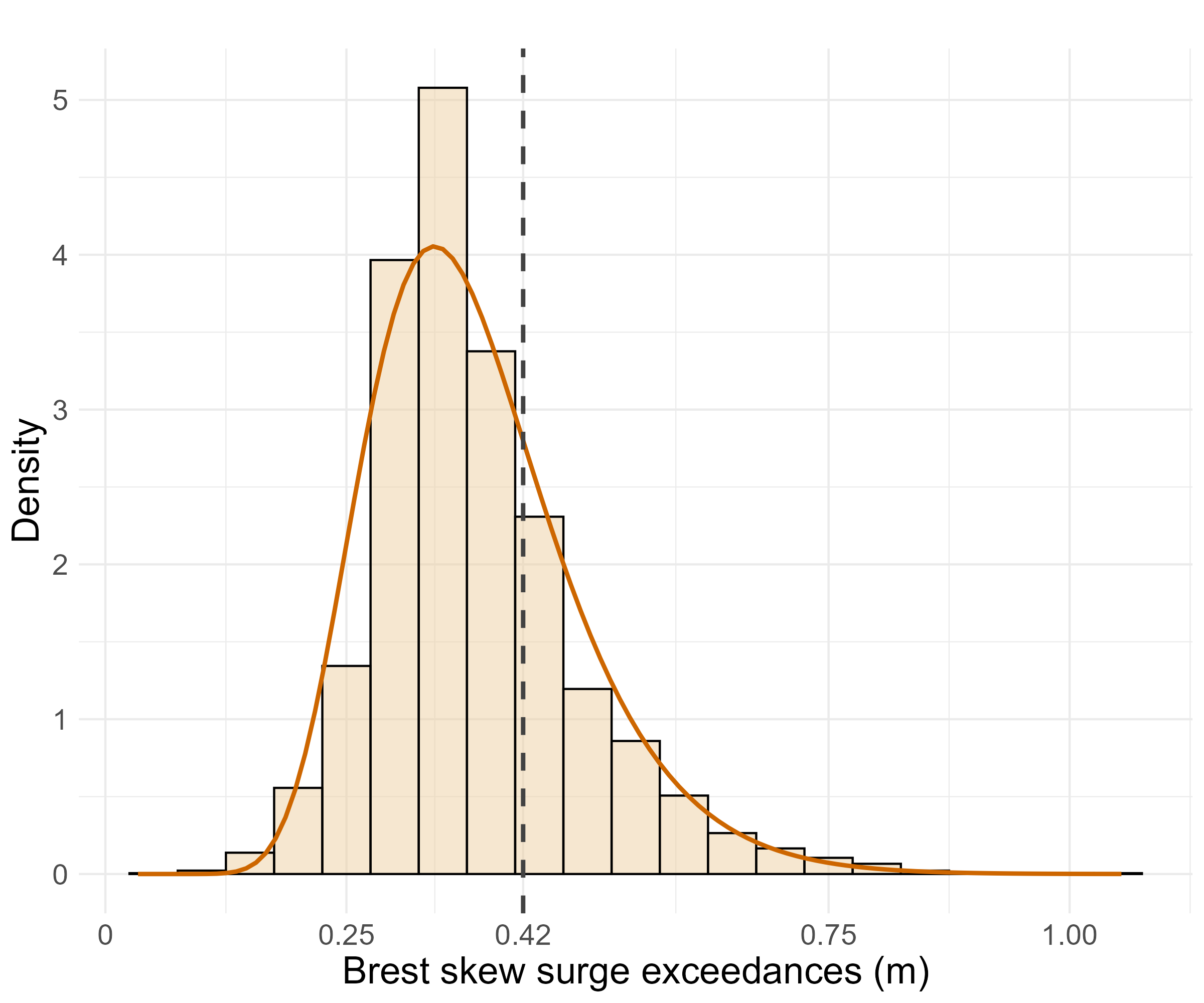}

  \vspace{1cm}

  \includegraphics[width=.4\textwidth]{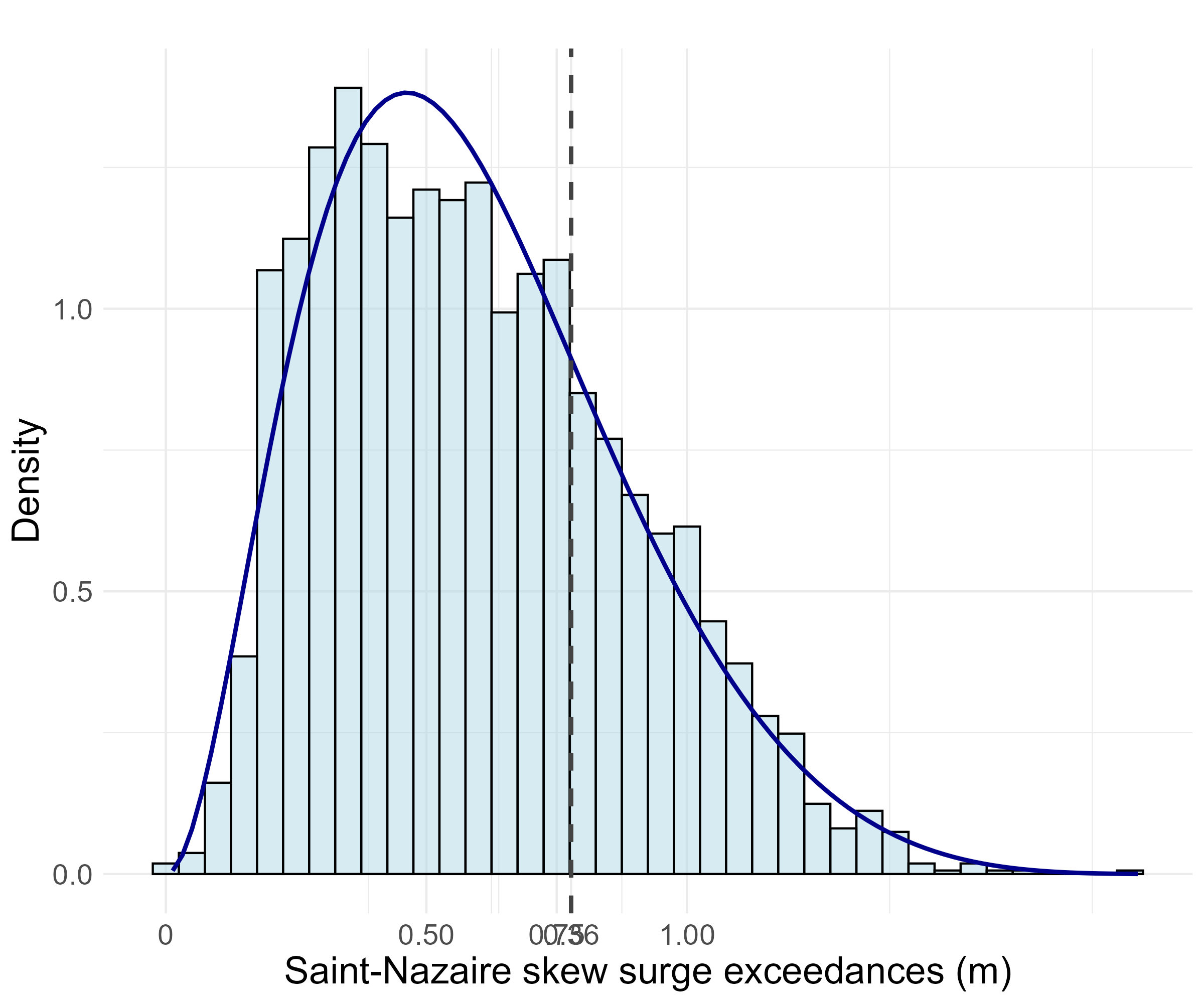}
  \hspace{1cm}
  \includegraphics[width=.4\textwidth]{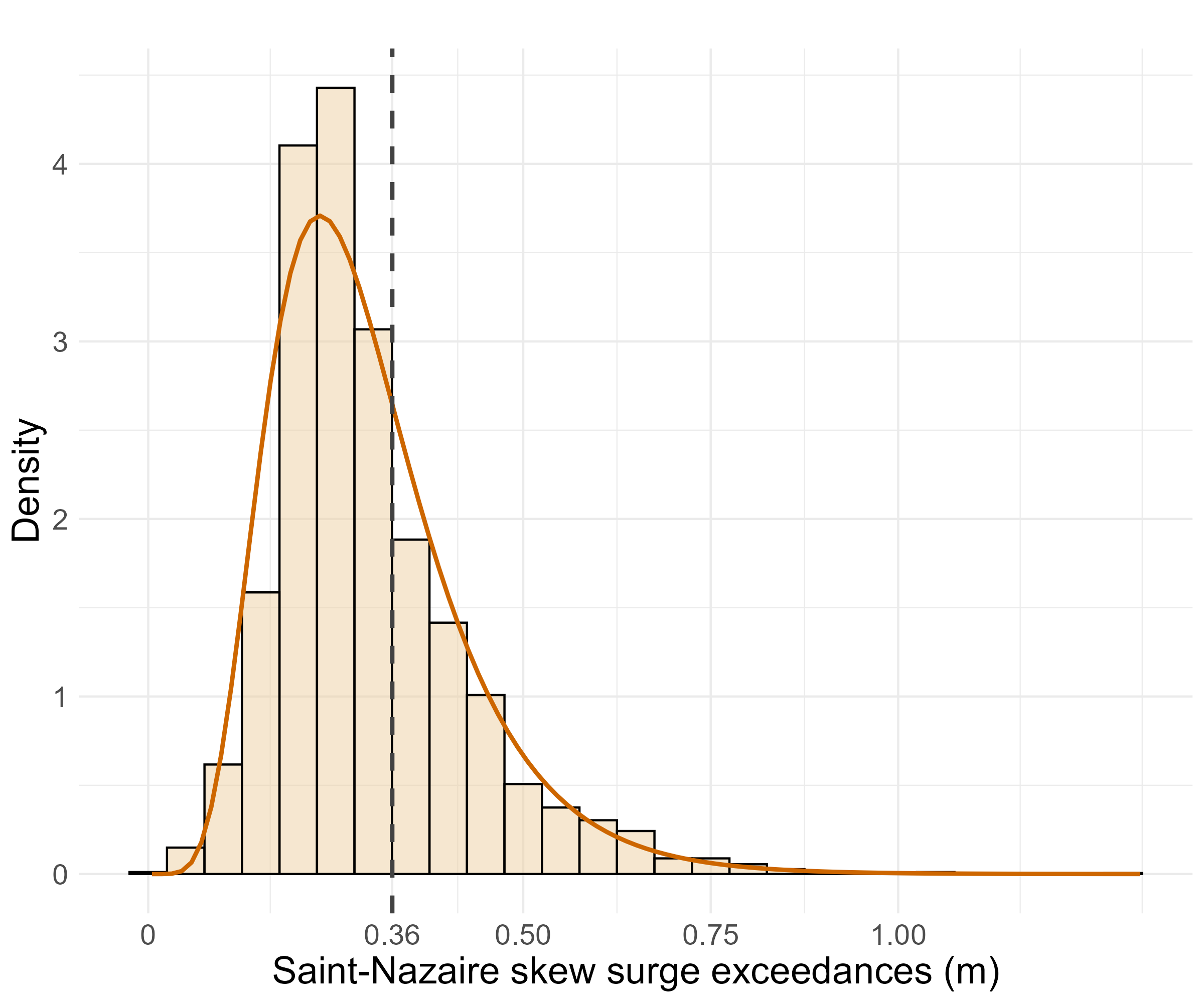}

  \vspace{1cm}

  \includegraphics[width=.4\textwidth]{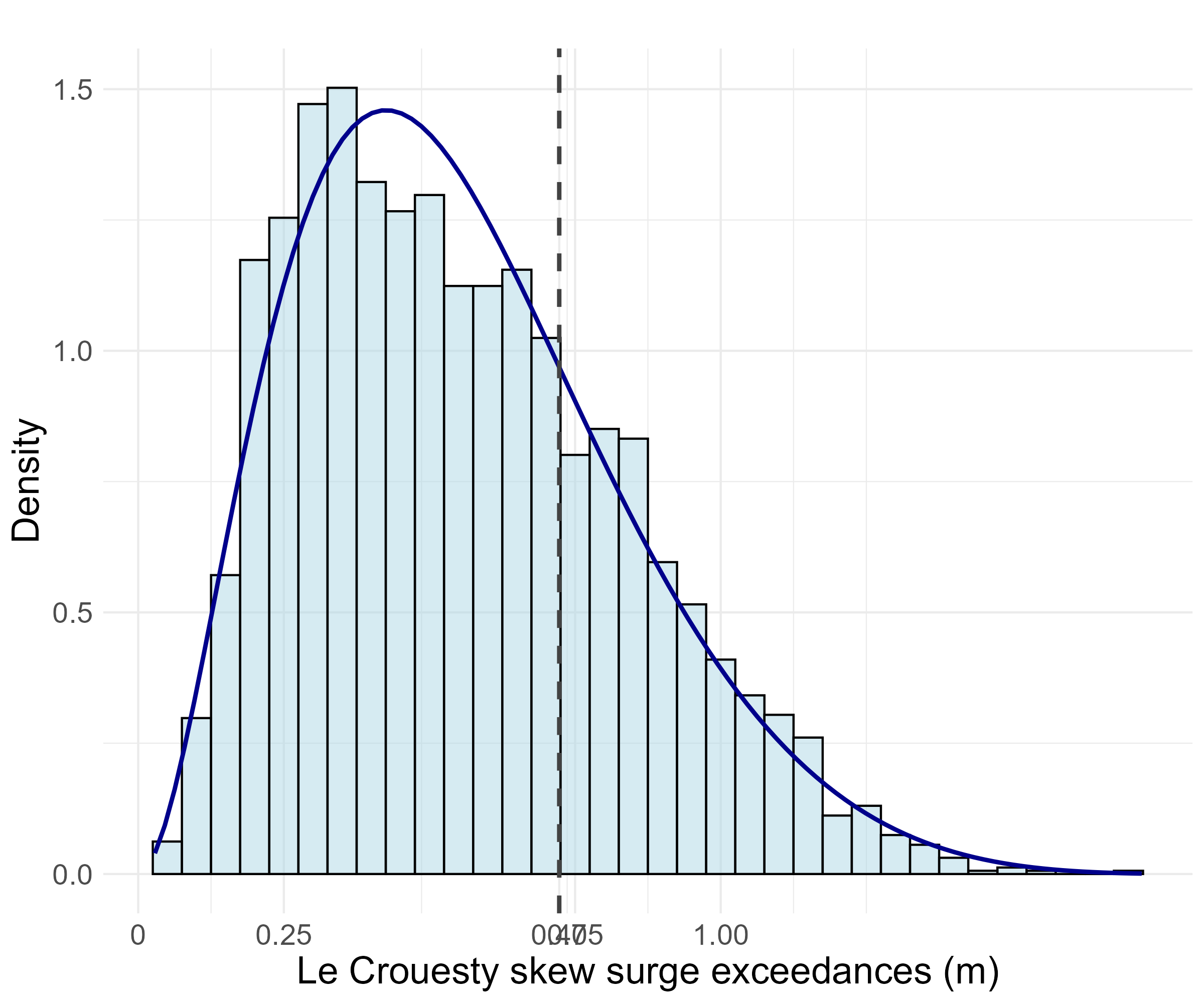}
  \hspace{1cm}
  \includegraphics[width=.4\textwidth]{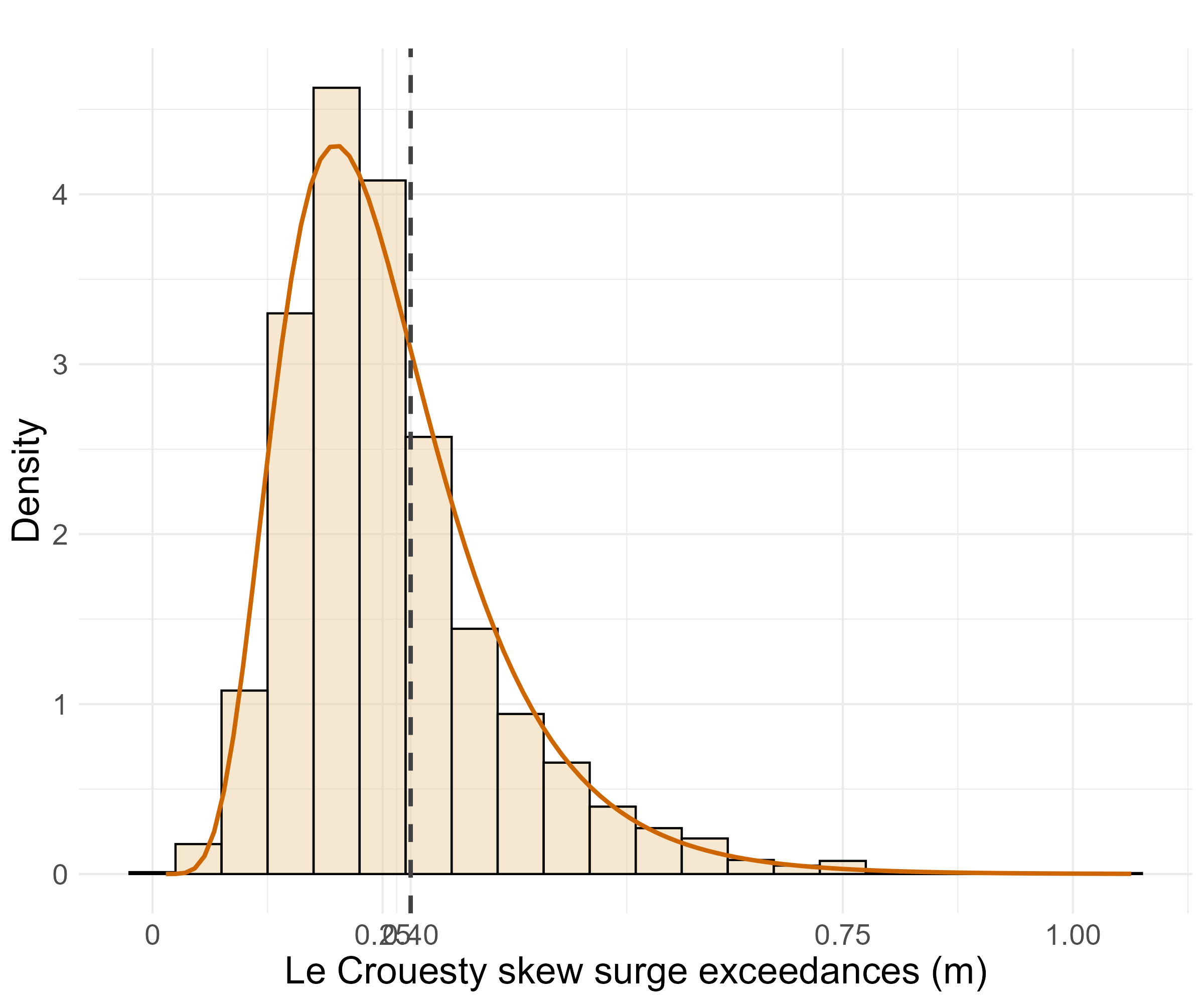}

  \caption{Histograms of sea level (right) and skew surge (right) exceedances at the three stations Brest (top row), Saint-Nazaire (middle row) and Le Crouesty (bottom row), from 14/03/1996 to 31/12/2014. The darkblue and darkorange curves represent the fitted EGP densities. The dotted vertical grey lines represent the smallest point above which each fitted density is convex, which  represent the chosen marginal thresholds via Algorithm~\ref{algo:th_select}. \label{fig:density_lecrouesty}}
\end{figure}

\begin{figure}[ht!]

  \centering
  \includegraphics[width=.4\textwidth]{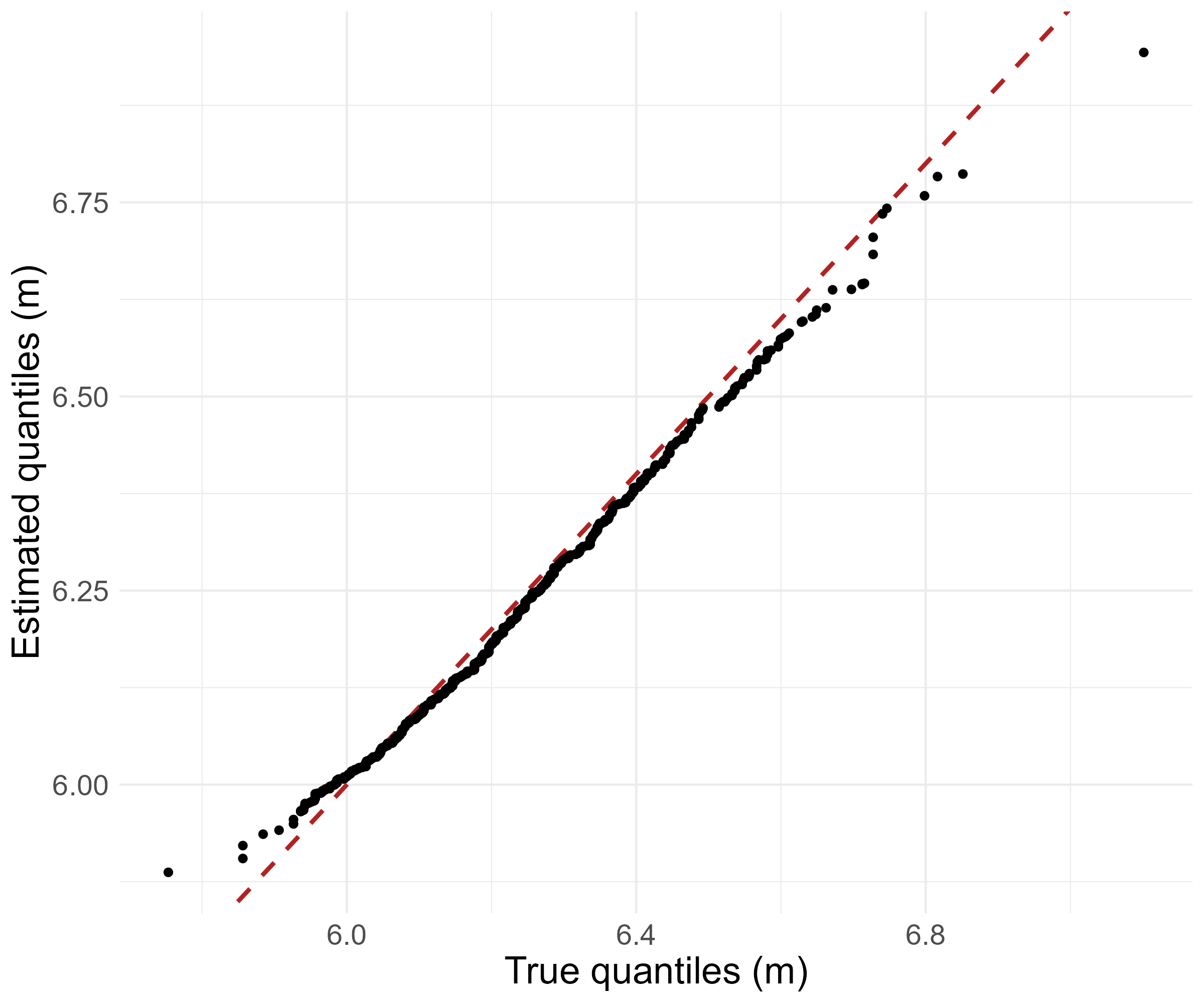}
  \hspace{1cm}
  \includegraphics[width=.4\textwidth]{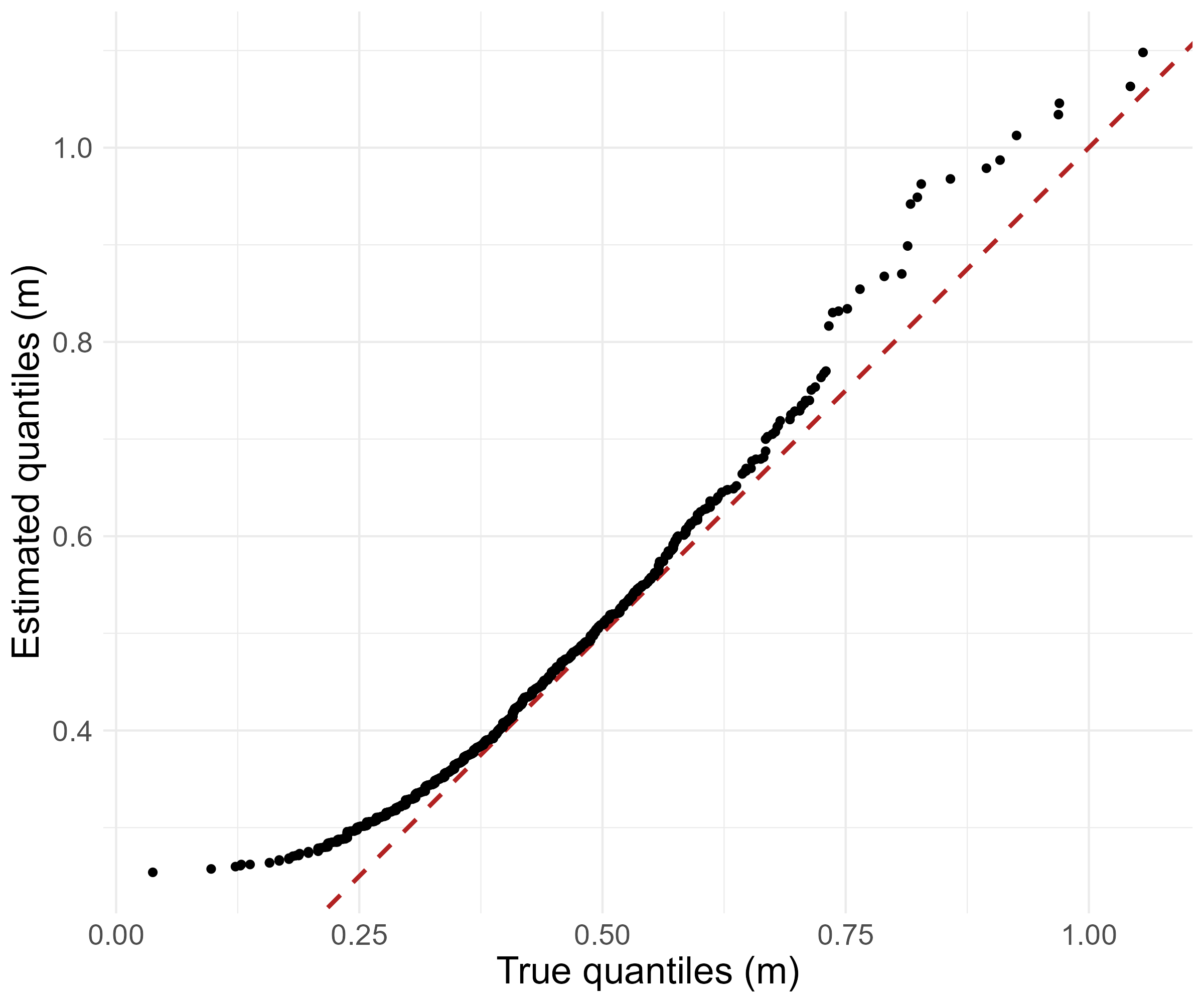}

  \vspace{1cm}

  \includegraphics[width=.4\textwidth]{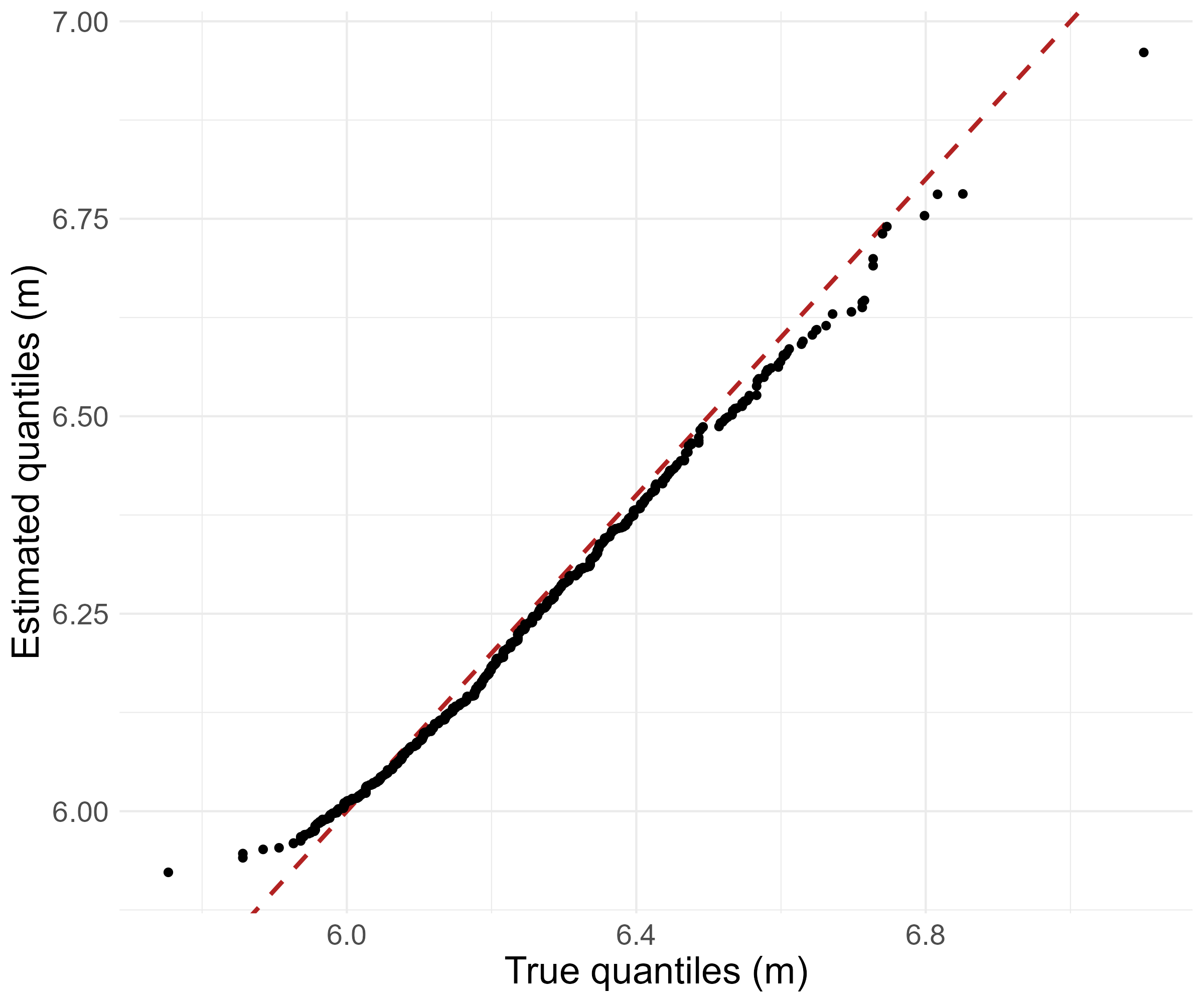}
  \hspace{1cm}
  \includegraphics[width=.4\textwidth]{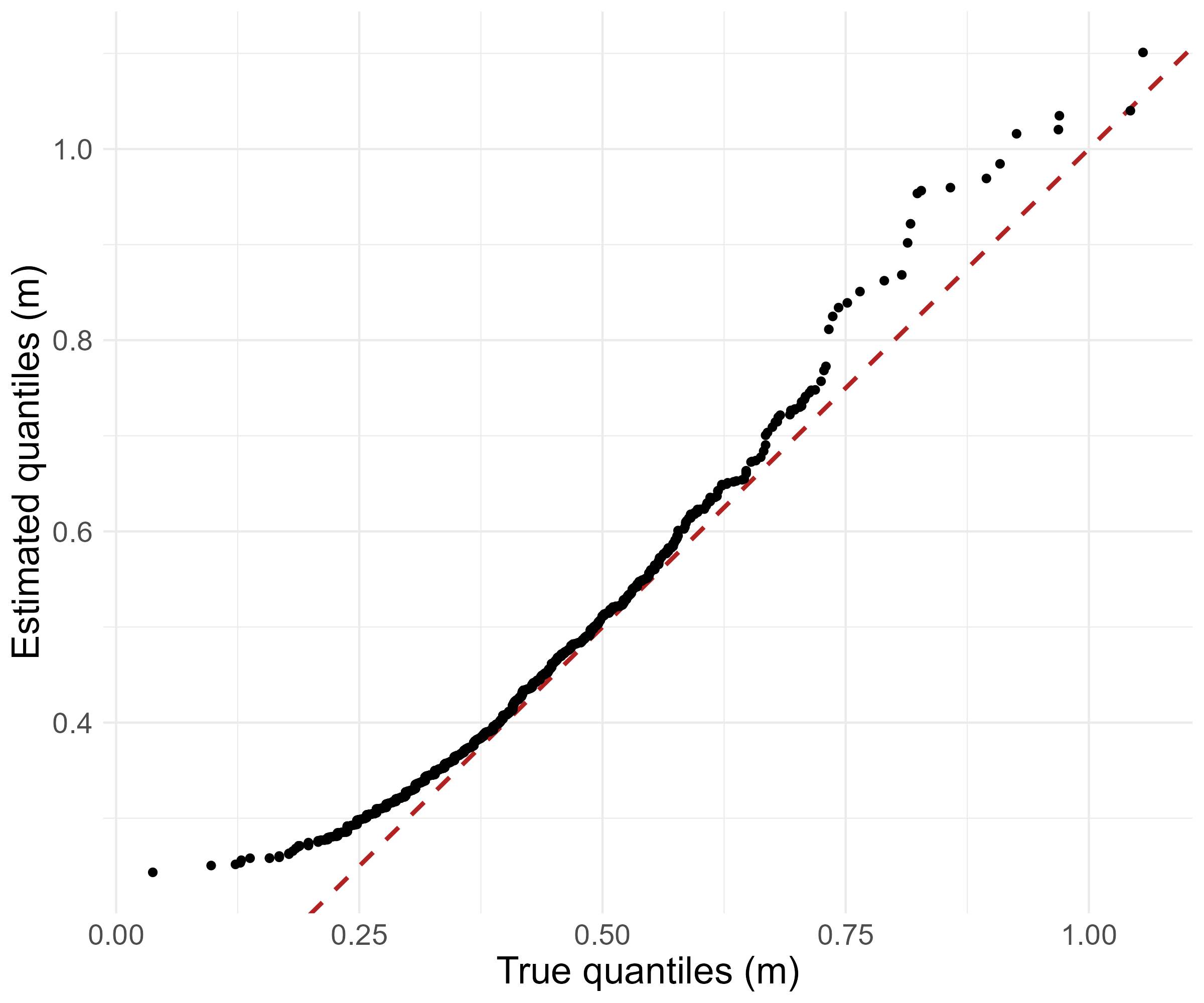}

  \vspace{1cm}

  \includegraphics[width=.4\textwidth]{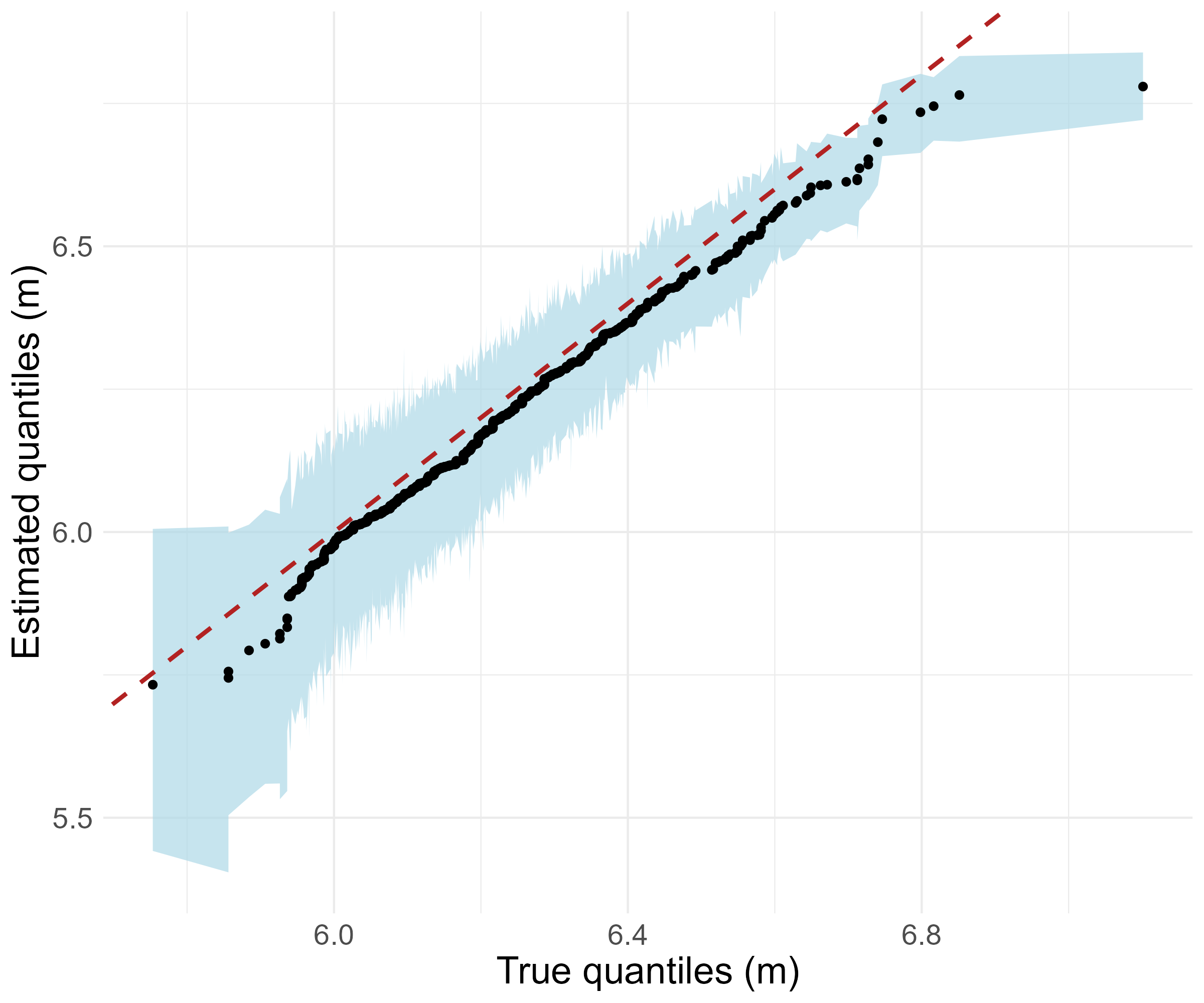}
  \hspace{1cm}
  \includegraphics[width=.4\textwidth]{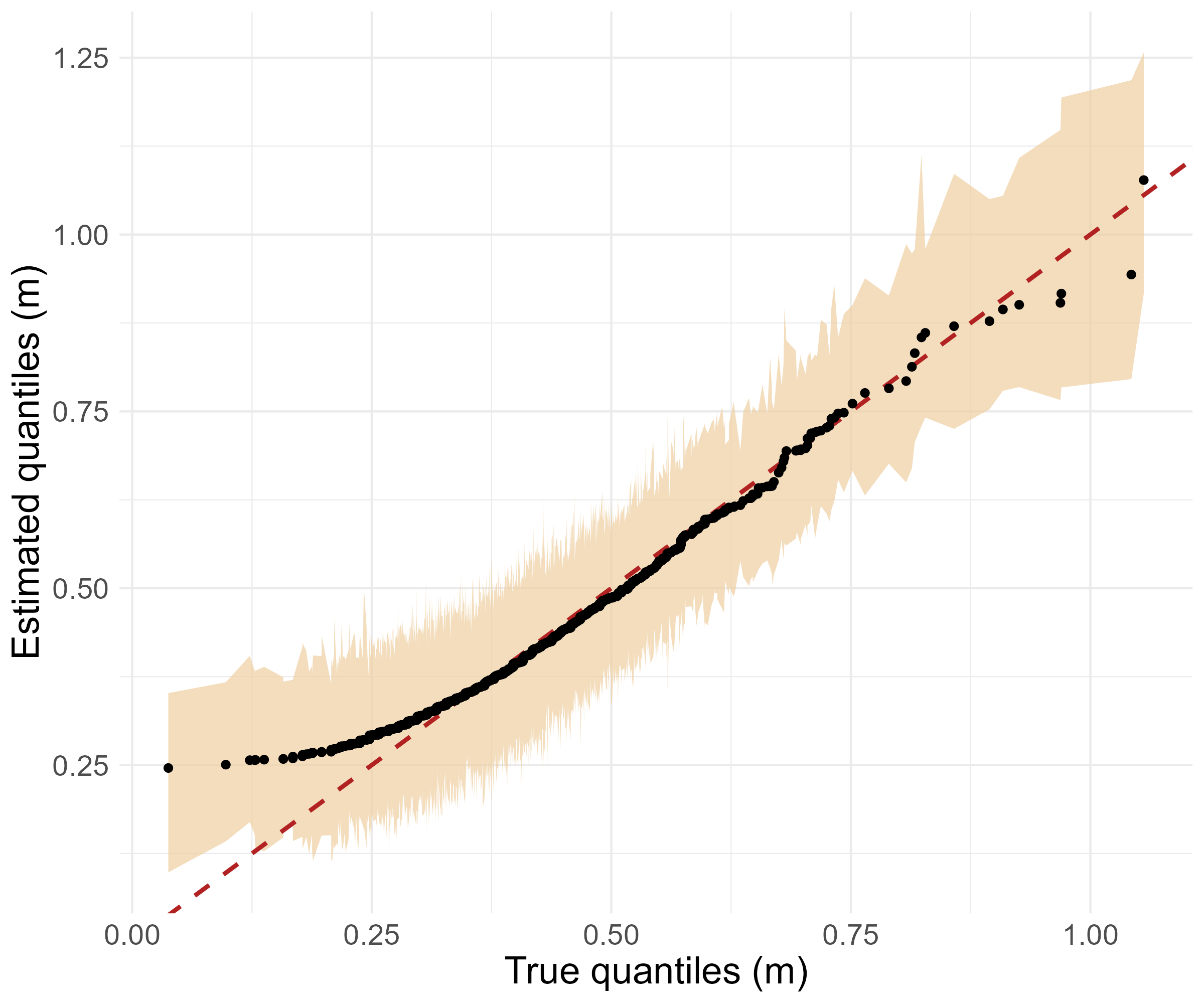}
\caption{QQ-plots comparing observed sea level (left) and skew surge (right) exceedances of the Le Crouesty test set (x-axis), ranging from  14/03/1996 to 31/12/2014, to predicted data (y-axis) from the algorithms of Sections~\ref{sec:reg_proc} and \ref{sec:mgp_proc}. The plots show results from the ROXANE procedure with RF regression (top row), ROXANE procedure with OLS regression (middle row), and MGPRED (bottom row) with 0.95-confidence bands (lightblue and lightorange). The dotted red line represents the identity line $x=y$. The global 0.95-coverage probability associated with MGPRED is 0.94 for sea levels and 0.98 for skew
surges.\label{fig:qqplot_lecrouesty}}
\end{figure}

\end{document}